\documentclass[manuscript,screen]{acmart}

\acmJournal{TOSEM}

\AtBeginDocument{%
  }

\usepackage{color}
\usepackage[table]{xcolor}
\usepackage{tikz}       
\usepackage{pgfplots}   
\pgfplotsset{compat=1.18}
\usepackage{hyperref}
\usepackage{dirtytalk} 

\newif\ifdraftmode  
\draftmodefalse

\definecolor{OUCHcol}{rgb}{1.0,0.0,1.0} 
\definecolor{shouldcitecol}{rgb}{0.0,0.0,0.8} 
\definecolor{gentagcol}{rgb}{0.0,1.0,0.0} 
\ifdraftmode
  \DeclareRobustCommand*{\obacht}[2][nobody]{\textcolor{OUCHcol}{OBACHT: #2 }}
\else
  \DeclareRobustCommand*{\obacht}[2][nobody]{}
\fi

\DeclareRobustCommand*{\gentag}[1]{%
  \ifdraftmode\textcolor{gentagcol}{\textsf{\scriptsize[#1]}}\space\fi}
\newcommand*{\Quote}[1]{\emph{``#1''}}  
\newcommand*{\Code}[1]{{\small\ttfamily\hyphenchar\font=`\-\relax{#1}}}  

\expandafter\gdef\csname svy@sample/records\endcsname{5380}
\expandafter\gdef\csname svy@sample/firstdate\endcsname{2026-08-04}
\expandafter\gdef\csname svy@sample/lastdate\endcsname{2026-09-15}
\expandafter\gdef\csname svy@sample/nonempty\endcsname{413}
\expandafter\gdef\csname svy@sample/eligible\endcsname{242}
\expandafter\gdef\csname svy@sample/eligiblepct\endcsname{59}
\expandafter\gdef\csname svy@funnel/empty\endcsname{4967}
\expandafter\gdef\csname svy@funnel/emptypct\endcsname{92}
\expandafter\gdef\csname svy@funnel/fewtasks\endcsname{132}
\expandafter\gdef\csname svy@funnel/fewtaskspct\endcsname{32}
\expandafter\gdef\csname svy@funnel/straightliners\endcsname{33}
\expandafter\gdef\csname svy@funnel/straightlinerspct\endcsname{12}
\expandafter\gdef\csname svy@funnel/attcheck\endcsname{6}
\expandafter\gdef\csname svy@funnel/attcheckpct\endcsname{2}
\expandafter\gdef\csname svy@sample/all\endcsname{242}
\expandafter\gdef\csname svy@sample/allpct\endcsname{100}
\expandafter\gdef\csname svy@sample/users\endcsname{171}
\expandafter\gdef\csname svy@sample/userspct\endcsname{71}
\expandafter\gdef\csname svy@sample/nonusers\endcsname{69}
\expandafter\gdef\csname svy@sample/nonuserspct\endcsname{29}
\expandafter\gdef\csname svy@sample/unclassified\endcsname{2}
\expandafter\gdef\csname svy@sample/unclassifiedpct\endcsname{1}
\expandafter\gdef\csname svy@sample/undefinedusers\endcsname{2}
\expandafter\gdef\csname svy@sample/subgroup/s1/n\endcsname{45}
\expandafter\gdef\csname svy@sample/subgroup/s1/pct\endcsname{19}
\expandafter\gdef\csname svy@sample/subgroup/s1/ofnonuserspct\endcsname{65}
\expandafter\gdef\csname svy@sample/subgroup/s1/label\endcsname{S1 consistent non-users}
\expandafter\gdef\csname svy@sample/subgroup/s2/n\endcsname{13}
\expandafter\gdef\csname svy@sample/subgroup/s2/pct\endcsname{5}
\expandafter\gdef\csname svy@sample/subgroup/s2/ofnonuserspct\endcsname{19}
\expandafter\gdef\csname svy@sample/subgroup/s2/label\endcsname{S2 frequent AI use, no agent tool}
\expandafter\gdef\csname svy@sample/subgroup/s3/n\endcsname{11}
\expandafter\gdef\csname svy@sample/subgroup/s3/pct\endcsname{5}
\expandafter\gdef\csname svy@sample/subgroup/s3/ofnonuserspct\endcsname{16}
\expandafter\gdef\csname svy@sample/subgroup/s3/label\endcsname{S3 agent tool, no current practice}
\expandafter\gdef\csname svy@item/t01/code\endcsname{T01}
\expandafter\gdef\csname svy@item/t01/pred\endcsname{H}
\expandafter\gdef\csname svy@item/t01/label\endcsname{what a task must achieve to be acceptable}
\expandafter\gdef\csname svy@item/t01/usage/diff\endcsname{0.44}
\expandafter\gdef\csname svy@item/t01/usage/difflow\endcsname{0.20}
\expandafter\gdef\csname svy@item/t01/usage/diffhigh\endcsname{0.67}
\expandafter\gdef\csname svy@item/t01/usage/g\endcsname{0.50}
\expandafter\gdef\csname svy@item/t01/all/n\endcsname{241}
\expandafter\gdef\csname svy@item/t01/all/miss\endcsname{1}
\expandafter\gdef\csname svy@item/t01/all/median\endcsname{2.0}
\expandafter\gdef\csname svy@item/t01/all/cilow\endcsname{1.0}
\expandafter\gdef\csname svy@item/t01/all/cihigh\endcsname{2.0}
\expandafter\gdef\csname svy@item/t01/all/mean\endcsname{1.75}
\expandafter\gdef\csname svy@item/t01/all/meanlow\endcsname{1.64}
\expandafter\gdef\csname svy@item/t01/all/meanhigh\endcsname{1.86}
\expandafter\gdef\csname svy@item/t01/all/se\endcsname{0.057}
\expandafter\gdef\csname svy@item/t01/all/sd\endcsname{0.89}
\expandafter\gdef\csname svy@item/t01/all/sdhigh\endcsname{0}
\expandafter\gdef\csname svy@item/t01/all/verdict\endcsname{CONFIRMED}
\expandafter\gdef\csname svy@item/t01/all/dip\endcsname{0}
\expandafter\gdef\csname svy@item/t01/all/shapelow\endcsname{88}
\expandafter\gdef\csname svy@item/t01/all/shapemid\endcsname{6}
\expandafter\gdef\csname svy@item/t01/all/shapehigh\endcsname{6}
\expandafter\gdef\csname svy@item/t01/users/n\endcsname{171}
\expandafter\gdef\csname svy@item/t01/users/miss\endcsname{0}
\expandafter\gdef\csname svy@item/t01/users/median\endcsname{2.0}
\expandafter\gdef\csname svy@item/t01/users/cilow\endcsname{2.0}
\expandafter\gdef\csname svy@item/t01/users/cihigh\endcsname{2.0}
\expandafter\gdef\csname svy@item/t01/users/mean\endcsname{1.88}
\expandafter\gdef\csname svy@item/t01/users/meanlow\endcsname{1.74}
\expandafter\gdef\csname svy@item/t01/users/meanhigh\endcsname{2.01}
\expandafter\gdef\csname svy@item/t01/users/se\endcsname{0.069}
\expandafter\gdef\csname svy@item/t01/users/sd\endcsname{0.90}
\expandafter\gdef\csname svy@item/t01/users/sdhigh\endcsname{0}
\expandafter\gdef\csname svy@item/t01/users/verdict\endcsname{CONFIRMED}
\expandafter\gdef\csname svy@item/t01/users/dip\endcsname{0}
\expandafter\gdef\csname svy@item/t01/users/shapelow\endcsname{83}
\expandafter\gdef\csname svy@item/t01/users/shapemid\endcsname{9}
\expandafter\gdef\csname svy@item/t01/users/shapehigh\endcsname{8}
\expandafter\gdef\csname svy@item/t01/nonusers/n\endcsname{68}
\expandafter\gdef\csname svy@item/t01/nonusers/miss\endcsname{1}
\expandafter\gdef\csname svy@item/t01/nonusers/median\endcsname{1.0}
\expandafter\gdef\csname svy@item/t01/nonusers/cilow\endcsname{1.0}
\expandafter\gdef\csname svy@item/t01/nonusers/cihigh\endcsname{1.0}
\expandafter\gdef\csname svy@item/t01/nonusers/mean\endcsname{1.44}
\expandafter\gdef\csname svy@item/t01/nonusers/meanlow\endcsname{1.25}
\expandafter\gdef\csname svy@item/t01/nonusers/meanhigh\endcsname{1.63}
\expandafter\gdef\csname svy@item/t01/nonusers/se\endcsname{0.095}
\expandafter\gdef\csname svy@item/t01/nonusers/sd\endcsname{0.78}
\expandafter\gdef\csname svy@item/t01/nonusers/sdhigh\endcsname{0}
\expandafter\gdef\csname svy@item/t01/nonusers/verdict\endcsname{CONFIRMED}
\expandafter\gdef\csname svy@item/t01/nonusers/dip\endcsname{0}
\expandafter\gdef\csname svy@item/t01/nonusers/shapelow\endcsname{95}
\expandafter\gdef\csname svy@item/t01/nonusers/shapemid\endcsname{0}
\expandafter\gdef\csname svy@item/t01/nonusers/shapehigh\endcsname{5}
\expandafter\gdef\csname svy@item/t03/code\endcsname{T03}
\expandafter\gdef\csname svy@item/t03/pred\endcsname{E}
\expandafter\gdef\csname svy@item/t03/label\endcsname{developing/critiquing ideas in a spec or plan}
\expandafter\gdef\csname svy@item/t03/usage/diff\endcsname{0.56}
\expandafter\gdef\csname svy@item/t03/usage/difflow\endcsname{0.30}
\expandafter\gdef\csname svy@item/t03/usage/diffhigh\endcsname{0.82}
\expandafter\gdef\csname svy@item/t03/usage/g\endcsname{0.61}
\expandafter\gdef\csname svy@item/t03/all/n\endcsname{241}
\expandafter\gdef\csname svy@item/t03/all/miss\endcsname{1}
\expandafter\gdef\csname svy@item/t03/all/median\endcsname{2.0}
\expandafter\gdef\csname svy@item/t03/all/cilow\endcsname{2.0}
\expandafter\gdef\csname svy@item/t03/all/cihigh\endcsname{2.0}
\expandafter\gdef\csname svy@item/t03/all/mean\endcsname{2.40}
\expandafter\gdef\csname svy@item/t03/all/meanlow\endcsname{2.28}
\expandafter\gdef\csname svy@item/t03/all/meanhigh\endcsname{2.52}
\expandafter\gdef\csname svy@item/t03/all/se\endcsname{0.061}
\expandafter\gdef\csname svy@item/t03/all/sd\endcsname{0.94}
\expandafter\gdef\csname svy@item/t03/all/sdhigh\endcsname{0}
\expandafter\gdef\csname svy@item/t03/all/verdict\endcsname{INCONCLUSIVE}
\expandafter\gdef\csname svy@item/t03/all/dip\endcsname{0}
\expandafter\gdef\csname svy@item/t03/all/shapelow\endcsname{59}
\expandafter\gdef\csname svy@item/t03/all/shapemid\endcsname{28}
\expandafter\gdef\csname svy@item/t03/all/shapehigh\endcsname{13}
\expandafter\gdef\csname svy@item/t03/users/n\endcsname{170}
\expandafter\gdef\csname svy@item/t03/users/miss\endcsname{1}
\expandafter\gdef\csname svy@item/t03/users/median\endcsname{2.0}
\expandafter\gdef\csname svy@item/t03/users/cilow\endcsname{2.0}
\expandafter\gdef\csname svy@item/t03/users/cihigh\endcsname{3.0}
\expandafter\gdef\csname svy@item/t03/users/mean\endcsname{2.57}
\expandafter\gdef\csname svy@item/t03/users/meanlow\endcsname{2.44}
\expandafter\gdef\csname svy@item/t03/users/meanhigh\endcsname{2.71}
\expandafter\gdef\csname svy@item/t03/users/se\endcsname{0.068}
\expandafter\gdef\csname svy@item/t03/users/sd\endcsname{0.89}
\expandafter\gdef\csname svy@item/t03/users/sdhigh\endcsname{0}
\expandafter\gdef\csname svy@item/t03/users/verdict\endcsname{INCONCLUSIVE}
\expandafter\gdef\csname svy@item/t03/users/dip\endcsname{0}
\expandafter\gdef\csname svy@item/t03/users/shapelow\endcsname{54}
\expandafter\gdef\csname svy@item/t03/users/shapemid\endcsname{31}
\expandafter\gdef\csname svy@item/t03/users/shapehigh\endcsname{15}
\expandafter\gdef\csname svy@item/t03/nonusers/n\endcsname{69}
\expandafter\gdef\csname svy@item/t03/nonusers/miss\endcsname{0}
\expandafter\gdef\csname svy@item/t03/nonusers/median\endcsname{2.0}
\expandafter\gdef\csname svy@item/t03/nonusers/cilow\endcsname{2.0}
\expandafter\gdef\csname svy@item/t03/nonusers/cihigh\endcsname{2.0}
\expandafter\gdef\csname svy@item/t03/nonusers/mean\endcsname{2.01}
\expandafter\gdef\csname svy@item/t03/nonusers/meanlow\endcsname{1.79}
\expandafter\gdef\csname svy@item/t03/nonusers/meanhigh\endcsname{2.24}
\expandafter\gdef\csname svy@item/t03/nonusers/se\endcsname{0.112}
\expandafter\gdef\csname svy@item/t03/nonusers/sd\endcsname{0.93}
\expandafter\gdef\csname svy@item/t03/nonusers/sdhigh\endcsname{0}
\expandafter\gdef\csname svy@item/t03/nonusers/verdict\endcsname{REJECTED}
\expandafter\gdef\csname svy@item/t03/nonusers/dip\endcsname{0}
\expandafter\gdef\csname svy@item/t03/nonusers/shapelow\endcsname{72}
\expandafter\gdef\csname svy@item/t03/nonusers/shapemid\endcsname{22}
\expandafter\gdef\csname svy@item/t03/nonusers/shapehigh\endcsname{6}
\expandafter\gdef\csname svy@item/t04/code\endcsname{T04}
\expandafter\gdef\csname svy@item/t04/pred\endcsname{X}
\expandafter\gdef\csname svy@item/t04/label\endcsname{understanding domain / unfamiliar codebase}
\expandafter\gdef\csname svy@item/t04/usage/diff\endcsname{0.65}
\expandafter\gdef\csname svy@item/t04/usage/difflow\endcsname{0.39}
\expandafter\gdef\csname svy@item/t04/usage/diffhigh\endcsname{0.90}
\expandafter\gdef\csname svy@item/t04/usage/g\endcsname{0.70}
\expandafter\gdef\csname svy@item/t04/all/n\endcsname{237}
\expandafter\gdef\csname svy@item/t04/all/miss\endcsname{5}
\expandafter\gdef\csname svy@item/t04/all/median\endcsname{2.0}
\expandafter\gdef\csname svy@item/t04/all/cilow\endcsname{2.0}
\expandafter\gdef\csname svy@item/t04/all/cihigh\endcsname{2.0}
\expandafter\gdef\csname svy@item/t04/all/mean\endcsname{2.48}
\expandafter\gdef\csname svy@item/t04/all/meanlow\endcsname{2.36}
\expandafter\gdef\csname svy@item/t04/all/meanhigh\endcsname{2.60}
\expandafter\gdef\csname svy@item/t04/all/se\endcsname{0.063}
\expandafter\gdef\csname svy@item/t04/all/sd\endcsname{0.97}
\expandafter\gdef\csname svy@item/t04/all/sdhigh\endcsname{0}
\expandafter\gdef\csname svy@item/t04/all/verdict\endcsname{NO_PREDICTION}
\expandafter\gdef\csname svy@item/t04/all/dip\endcsname{0}
\expandafter\gdef\csname svy@item/t04/all/shapelow\endcsname{57}
\expandafter\gdef\csname svy@item/t04/all/shapemid\endcsname{28}
\expandafter\gdef\csname svy@item/t04/all/shapehigh\endcsname{15}
\expandafter\gdef\csname svy@item/t04/users/n\endcsname{166}
\expandafter\gdef\csname svy@item/t04/users/miss\endcsname{5}
\expandafter\gdef\csname svy@item/t04/users/median\endcsname{3.0}
\expandafter\gdef\csname svy@item/t04/users/cilow\endcsname{2.0}
\expandafter\gdef\csname svy@item/t04/users/cihigh\endcsname{3.0}
\expandafter\gdef\csname svy@item/t04/users/mean\endcsname{2.67}
\expandafter\gdef\csname svy@item/t04/users/meanlow\endcsname{2.53}
\expandafter\gdef\csname svy@item/t04/users/meanhigh\endcsname{2.82}
\expandafter\gdef\csname svy@item/t04/users/se\endcsname{0.073}
\expandafter\gdef\csname svy@item/t04/users/sd\endcsname{0.94}
\expandafter\gdef\csname svy@item/t04/users/sdhigh\endcsname{0}
\expandafter\gdef\csname svy@item/t04/users/verdict\endcsname{NO_PREDICTION}
\expandafter\gdef\csname svy@item/t04/users/dip\endcsname{0}
\expandafter\gdef\csname svy@item/t04/users/shapelow\endcsname{49}
\expandafter\gdef\csname svy@item/t04/users/shapemid\endcsname{32}
\expandafter\gdef\csname svy@item/t04/users/shapehigh\endcsname{19}
\expandafter\gdef\csname svy@item/t04/nonusers/n\endcsname{69}
\expandafter\gdef\csname svy@item/t04/nonusers/miss\endcsname{0}
\expandafter\gdef\csname svy@item/t04/nonusers/median\endcsname{2.0}
\expandafter\gdef\csname svy@item/t04/nonusers/cilow\endcsname{2.0}
\expandafter\gdef\csname svy@item/t04/nonusers/cihigh\endcsname{2.0}
\expandafter\gdef\csname svy@item/t04/nonusers/mean\endcsname{2.03}
\expandafter\gdef\csname svy@item/t04/nonusers/meanlow\endcsname{1.82}
\expandafter\gdef\csname svy@item/t04/nonusers/meanhigh\endcsname{2.24}
\expandafter\gdef\csname svy@item/t04/nonusers/se\endcsname{0.105}
\expandafter\gdef\csname svy@item/t04/nonusers/sd\endcsname{0.87}
\expandafter\gdef\csname svy@item/t04/nonusers/sdhigh\endcsname{0}
\expandafter\gdef\csname svy@item/t04/nonusers/verdict\endcsname{NO_PREDICTION}
\expandafter\gdef\csname svy@item/t04/nonusers/dip\endcsname{0}
\expandafter\gdef\csname svy@item/t04/nonusers/shapelow\endcsname{75}
\expandafter\gdef\csname svy@item/t04/nonusers/shapemid\endcsname{18}
\expandafter\gdef\csname svy@item/t04/nonusers/shapehigh\endcsname{7}
\expandafter\gdef\csname svy@item/t05/code\endcsname{T05}
\expandafter\gdef\csname svy@item/t05/pred\endcsname{A}
\expandafter\gdef\csname svy@item/t05/label\endcsname{tracing code paths, summarizing how parts work}
\expandafter\gdef\csname svy@item/t05/usage/diff\endcsname{1.11}
\expandafter\gdef\csname svy@item/t05/usage/difflow\endcsname{0.81}
\expandafter\gdef\csname svy@item/t05/usage/diffhigh\endcsname{1.41}
\expandafter\gdef\csname svy@item/t05/usage/g\endcsname{1.18}
\expandafter\gdef\csname svy@item/t05/all/n\endcsname{237}
\expandafter\gdef\csname svy@item/t05/all/miss\endcsname{5}
\expandafter\gdef\csname svy@item/t05/all/median\endcsname{4.0}
\expandafter\gdef\csname svy@item/t05/all/cilow\endcsname{4.0}
\expandafter\gdef\csname svy@item/t05/all/cihigh\endcsname{4.0}
\expandafter\gdef\csname svy@item/t05/all/mean\endcsname{3.55}
\expandafter\gdef\csname svy@item/t05/all/meanlow\endcsname{3.41}
\expandafter\gdef\csname svy@item/t05/all/meanhigh\endcsname{3.68}
\expandafter\gdef\csname svy@item/t05/all/se\endcsname{0.069}
\expandafter\gdef\csname svy@item/t05/all/sd\endcsname{1.06}
\expandafter\gdef\csname svy@item/t05/all/sdhigh\endcsname{0}
\expandafter\gdef\csname svy@item/t05/all/verdict\endcsname{INCONCLUSIVE}
\expandafter\gdef\csname svy@item/t05/all/dip\endcsname{0}
\expandafter\gdef\csname svy@item/t05/all/shapelow\endcsname{17}
\expandafter\gdef\csname svy@item/t05/all/shapemid\endcsname{26}
\expandafter\gdef\csname svy@item/t05/all/shapehigh\endcsname{57}
\expandafter\gdef\csname svy@item/t05/users/n\endcsname{169}
\expandafter\gdef\csname svy@item/t05/users/miss\endcsname{2}
\expandafter\gdef\csname svy@item/t05/users/median\endcsname{4.0}
\expandafter\gdef\csname svy@item/t05/users/cilow\endcsname{4.0}
\expandafter\gdef\csname svy@item/t05/users/cihigh\endcsname{4.0}
\expandafter\gdef\csname svy@item/t05/users/mean\endcsname{3.87}
\expandafter\gdef\csname svy@item/t05/users/meanlow\endcsname{3.74}
\expandafter\gdef\csname svy@item/t05/users/meanhigh\endcsname{4.00}
\expandafter\gdef\csname svy@item/t05/users/se\endcsname{0.066}
\expandafter\gdef\csname svy@item/t05/users/sd\endcsname{0.86}
\expandafter\gdef\csname svy@item/t05/users/sdhigh\endcsname{0}
\expandafter\gdef\csname svy@item/t05/users/verdict\endcsname{CONFIRMED}
\expandafter\gdef\csname svy@item/t05/users/dip\endcsname{0}
\expandafter\gdef\csname svy@item/t05/users/shapelow\endcsname{7}
\expandafter\gdef\csname svy@item/t05/users/shapemid\endcsname{23}
\expandafter\gdef\csname svy@item/t05/users/shapehigh\endcsname{70}
\expandafter\gdef\csname svy@item/t05/nonusers/n\endcsname{66}
\expandafter\gdef\csname svy@item/t05/nonusers/miss\endcsname{3}
\expandafter\gdef\csname svy@item/t05/nonusers/median\endcsname{3.0}
\expandafter\gdef\csname svy@item/t05/nonusers/cilow\endcsname{2.0}
\expandafter\gdef\csname svy@item/t05/nonusers/cihigh\endcsname{3.0}
\expandafter\gdef\csname svy@item/t05/nonusers/mean\endcsname{2.76}
\expandafter\gdef\csname svy@item/t05/nonusers/meanlow\endcsname{2.48}
\expandafter\gdef\csname svy@item/t05/nonusers/meanhigh\endcsname{3.03}
\expandafter\gdef\csname svy@item/t05/nonusers/se\endcsname{0.137}
\expandafter\gdef\csname svy@item/t05/nonusers/sd\endcsname{1.11}
\expandafter\gdef\csname svy@item/t05/nonusers/sdhigh\endcsname{1}
\expandafter\gdef\csname svy@item/t05/nonusers/verdict\endcsname{INCONCLUSIVE}
\expandafter\gdef\csname svy@item/t05/nonusers/dip\endcsname{0}
\expandafter\gdef\csname svy@item/t05/nonusers/shapelow\endcsname{42}
\expandafter\gdef\csname svy@item/t05/nonusers/shapemid\endcsname{32}
\expandafter\gdef\csname svy@item/t05/nonusers/shapehigh\endcsname{26}
\expandafter\gdef\csname svy@item/t06/code\endcsname{T06}
\expandafter\gdef\csname svy@item/t06/pred\endcsname{A}
\expandafter\gdef\csname svy@item/t06/label\endcsname{throwaway prototypes for alternative approaches}
\expandafter\gdef\csname svy@item/t06/usage/diff\endcsname{1.25}
\expandafter\gdef\csname svy@item/t06/usage/difflow\endcsname{0.93}
\expandafter\gdef\csname svy@item/t06/usage/diffhigh\endcsname{1.56}
\expandafter\gdef\csname svy@item/t06/usage/g\endcsname{1.30}
\expandafter\gdef\csname svy@item/t06/all/n\endcsname{237}
\expandafter\gdef\csname svy@item/t06/all/miss\endcsname{5}
\expandafter\gdef\csname svy@item/t06/all/median\endcsname{4.0}
\expandafter\gdef\csname svy@item/t06/all/cilow\endcsname{4.0}
\expandafter\gdef\csname svy@item/t06/all/cihigh\endcsname{4.0}
\expandafter\gdef\csname svy@item/t06/all/mean\endcsname{3.83}
\expandafter\gdef\csname svy@item/t06/all/meanlow\endcsname{3.69}
\expandafter\gdef\csname svy@item/t06/all/meanhigh\endcsname{3.97}
\expandafter\gdef\csname svy@item/t06/all/se\endcsname{0.072}
\expandafter\gdef\csname svy@item/t06/all/sd\endcsname{1.11}
\expandafter\gdef\csname svy@item/t06/all/sdhigh\endcsname{0}
\expandafter\gdef\csname svy@item/t06/all/verdict\endcsname{CONFIRMED}
\expandafter\gdef\csname svy@item/t06/all/dip\endcsname{0}
\expandafter\gdef\csname svy@item/t06/all/shapelow\endcsname{15}
\expandafter\gdef\csname svy@item/t06/all/shapemid\endcsname{14}
\expandafter\gdef\csname svy@item/t06/all/shapehigh\endcsname{71}
\expandafter\gdef\csname svy@item/t06/users/n\endcsname{169}
\expandafter\gdef\csname svy@item/t06/users/miss\endcsname{2}
\expandafter\gdef\csname svy@item/t06/users/median\endcsname{4.0}
\expandafter\gdef\csname svy@item/t06/users/cilow\endcsname{4.0}
\expandafter\gdef\csname svy@item/t06/users/cihigh\endcsname{4.0}
\expandafter\gdef\csname svy@item/t06/users/mean\endcsname{4.17}
\expandafter\gdef\csname svy@item/t06/users/meanlow\endcsname{4.04}
\expandafter\gdef\csname svy@item/t06/users/meanhigh\endcsname{4.30}
\expandafter\gdef\csname svy@item/t06/users/se\endcsname{0.066}
\expandafter\gdef\csname svy@item/t06/users/sd\endcsname{0.86}
\expandafter\gdef\csname svy@item/t06/users/sdhigh\endcsname{0}
\expandafter\gdef\csname svy@item/t06/users/verdict\endcsname{CONFIRMED}
\expandafter\gdef\csname svy@item/t06/users/dip\endcsname{0}
\expandafter\gdef\csname svy@item/t06/users/shapelow\endcsname{4}
\expandafter\gdef\csname svy@item/t06/users/shapemid\endcsname{12}
\expandafter\gdef\csname svy@item/t06/users/shapehigh\endcsname{84}
\expandafter\gdef\csname svy@item/t06/nonusers/n\endcsname{66}
\expandafter\gdef\csname svy@item/t06/nonusers/miss\endcsname{3}
\expandafter\gdef\csname svy@item/t06/nonusers/median\endcsname{3.0}
\expandafter\gdef\csname svy@item/t06/nonusers/cilow\endcsname{2.0}
\expandafter\gdef\csname svy@item/t06/nonusers/cihigh\endcsname{4.0}
\expandafter\gdef\csname svy@item/t06/nonusers/mean\endcsname{2.92}
\expandafter\gdef\csname svy@item/t06/nonusers/meanlow\endcsname{2.63}
\expandafter\gdef\csname svy@item/t06/nonusers/meanhigh\endcsname{3.21}
\expandafter\gdef\csname svy@item/t06/nonusers/se\endcsname{0.145}
\expandafter\gdef\csname svy@item/t06/nonusers/sd\endcsname{1.18}
\expandafter\gdef\csname svy@item/t06/nonusers/sdhigh\endcsname{1}
\expandafter\gdef\csname svy@item/t06/nonusers/verdict\endcsname{INCONCLUSIVE}
\expandafter\gdef\csname svy@item/t06/nonusers/dip\endcsname{1}
\expandafter\gdef\csname svy@item/t06/nonusers/shapelow\endcsname{41}
\expandafter\gdef\csname svy@item/t06/nonusers/shapemid\endcsname{21}
\expandafter\gdef\csname svy@item/t06/nonusers/shapehigh\endcsname{38}
\expandafter\gdef\csname svy@item/t07/code\endcsname{T07}
\expandafter\gdef\csname svy@item/t07/pred\endcsname{H}
\expandafter\gdef\csname svy@item/t07/label\endcsname{judging which information matters for a task}
\expandafter\gdef\csname svy@item/t07/usage/diff\endcsname{0.90}
\expandafter\gdef\csname svy@item/t07/usage/difflow\endcsname{0.62}
\expandafter\gdef\csname svy@item/t07/usage/diffhigh\endcsname{1.18}
\expandafter\gdef\csname svy@item/t07/usage/g\endcsname{0.84}
\expandafter\gdef\csname svy@item/t07/all/n\endcsname{237}
\expandafter\gdef\csname svy@item/t07/all/miss\endcsname{5}
\expandafter\gdef\csname svy@item/t07/all/median\endcsname{2.0}
\expandafter\gdef\csname svy@item/t07/all/cilow\endcsname{2.0}
\expandafter\gdef\csname svy@item/t07/all/cihigh\endcsname{3.0}
\expandafter\gdef\csname svy@item/t07/all/mean\endcsname{2.50}
\expandafter\gdef\csname svy@item/t07/all/meanlow\endcsname{2.35}
\expandafter\gdef\csname svy@item/t07/all/meanhigh\endcsname{2.64}
\expandafter\gdef\csname svy@item/t07/all/se\endcsname{0.074}
\expandafter\gdef\csname svy@item/t07/all/sd\endcsname{1.14}
\expandafter\gdef\csname svy@item/t07/all/sdhigh\endcsname{0}
\expandafter\gdef\csname svy@item/t07/all/verdict\endcsname{INCONCLUSIVE}
\expandafter\gdef\csname svy@item/t07/all/dip\endcsname{0}
\expandafter\gdef\csname svy@item/t07/all/shapelow\endcsname{55}
\expandafter\gdef\csname svy@item/t07/all/shapemid\endcsname{22}
\expandafter\gdef\csname svy@item/t07/all/shapehigh\endcsname{23}
\expandafter\gdef\csname svy@item/t07/users/n\endcsname{168}
\expandafter\gdef\csname svy@item/t07/users/miss\endcsname{3}
\expandafter\gdef\csname svy@item/t07/users/median\endcsname{3.0}
\expandafter\gdef\csname svy@item/t07/users/cilow\endcsname{2.0}
\expandafter\gdef\csname svy@item/t07/users/cihigh\endcsname{3.0}
\expandafter\gdef\csname svy@item/t07/users/mean\endcsname{2.75}
\expandafter\gdef\csname svy@item/t07/users/meanlow\endcsname{2.58}
\expandafter\gdef\csname svy@item/t07/users/meanhigh\endcsname{2.92}
\expandafter\gdef\csname svy@item/t07/users/se\endcsname{0.086}
\expandafter\gdef\csname svy@item/t07/users/sd\endcsname{1.11}
\expandafter\gdef\csname svy@item/t07/users/sdhigh\endcsname{1}
\expandafter\gdef\csname svy@item/t07/users/verdict\endcsname{REJECTED}
\expandafter\gdef\csname svy@item/t07/users/dip\endcsname{0}
\expandafter\gdef\csname svy@item/t07/users/shapelow\endcsname{46}
\expandafter\gdef\csname svy@item/t07/users/shapemid\endcsname{26}
\expandafter\gdef\csname svy@item/t07/users/shapehigh\endcsname{28}
\expandafter\gdef\csname svy@item/t07/nonusers/n\endcsname{67}
\expandafter\gdef\csname svy@item/t07/nonusers/miss\endcsname{2}
\expandafter\gdef\csname svy@item/t07/nonusers/median\endcsname{2.0}
\expandafter\gdef\csname svy@item/t07/nonusers/cilow\endcsname{1.0}
\expandafter\gdef\csname svy@item/t07/nonusers/cihigh\endcsname{2.0}
\expandafter\gdef\csname svy@item/t07/nonusers/mean\endcsname{1.85}
\expandafter\gdef\csname svy@item/t07/nonusers/meanlow\endcsname{1.62}
\expandafter\gdef\csname svy@item/t07/nonusers/meanhigh\endcsname{2.08}
\expandafter\gdef\csname svy@item/t07/nonusers/se\endcsname{0.115}
\expandafter\gdef\csname svy@item/t07/nonusers/sd\endcsname{0.94}
\expandafter\gdef\csname svy@item/t07/nonusers/sdhigh\endcsname{0}
\expandafter\gdef\csname svy@item/t07/nonusers/verdict\endcsname{CONFIRMED}
\expandafter\gdef\csname svy@item/t07/nonusers/dip\endcsname{0}
\expandafter\gdef\csname svy@item/t07/nonusers/shapelow\endcsname{78}
\expandafter\gdef\csname svy@item/t07/nonusers/shapemid\endcsname{15}
\expandafter\gdef\csname svy@item/t07/nonusers/shapehigh\endcsname{7}
\expandafter\gdef\csname svy@item/t09/code\endcsname{T09}
\expandafter\gdef\csname svy@item/t09/pred\endcsname{H}
\expandafter\gdef\csname svy@item/t09/label\endcsname{writing and improving prompts / task descriptions}
\expandafter\gdef\csname svy@item/t09/usage/diff\endcsname{0.44}
\expandafter\gdef\csname svy@item/t09/usage/difflow\endcsname{0.12}
\expandafter\gdef\csname svy@item/t09/usage/diffhigh\endcsname{0.76}
\expandafter\gdef\csname svy@item/t09/usage/g\endcsname{0.41}
\expandafter\gdef\csname svy@item/t09/all/n\endcsname{228}
\expandafter\gdef\csname svy@item/t09/all/miss\endcsname{14}
\expandafter\gdef\csname svy@item/t09/all/median\endcsname{2.0}
\expandafter\gdef\csname svy@item/t09/all/cilow\endcsname{2.0}
\expandafter\gdef\csname svy@item/t09/all/cihigh\endcsname{2.0}
\expandafter\gdef\csname svy@item/t09/all/mean\endcsname{2.23}
\expandafter\gdef\csname svy@item/t09/all/meanlow\endcsname{2.09}
\expandafter\gdef\csname svy@item/t09/all/meanhigh\endcsname{2.37}
\expandafter\gdef\csname svy@item/t09/all/se\endcsname{0.072}
\expandafter\gdef\csname svy@item/t09/all/sd\endcsname{1.08}
\expandafter\gdef\csname svy@item/t09/all/sdhigh\endcsname{0}
\expandafter\gdef\csname svy@item/t09/all/verdict\endcsname{CONFIRMED}
\expandafter\gdef\csname svy@item/t09/all/dip\endcsname{0}
\expandafter\gdef\csname svy@item/t09/all/shapelow\endcsname{67}
\expandafter\gdef\csname svy@item/t09/all/shapemid\endcsname{19}
\expandafter\gdef\csname svy@item/t09/all/shapehigh\endcsname{14}
\expandafter\gdef\csname svy@item/t09/users/n\endcsname{167}
\expandafter\gdef\csname svy@item/t09/users/miss\endcsname{4}
\expandafter\gdef\csname svy@item/t09/users/median\endcsname{2.0}
\expandafter\gdef\csname svy@item/t09/users/cilow\endcsname{2.0}
\expandafter\gdef\csname svy@item/t09/users/cihigh\endcsname{2.0}
\expandafter\gdef\csname svy@item/t09/users/mean\endcsname{2.35}
\expandafter\gdef\csname svy@item/t09/users/meanlow\endcsname{2.19}
\expandafter\gdef\csname svy@item/t09/users/meanhigh\endcsname{2.52}
\expandafter\gdef\csname svy@item/t09/users/se\endcsname{0.083}
\expandafter\gdef\csname svy@item/t09/users/sd\endcsname{1.07}
\expandafter\gdef\csname svy@item/t09/users/sdhigh\endcsname{1}
\expandafter\gdef\csname svy@item/t09/users/verdict\endcsname{INCONCLUSIVE}
\expandafter\gdef\csname svy@item/t09/users/dip\endcsname{0}
\expandafter\gdef\csname svy@item/t09/users/shapelow\endcsname{64}
\expandafter\gdef\csname svy@item/t09/users/shapemid\endcsname{20}
\expandafter\gdef\csname svy@item/t09/users/shapehigh\endcsname{16}
\expandafter\gdef\csname svy@item/t09/nonusers/n\endcsname{59}
\expandafter\gdef\csname svy@item/t09/nonusers/miss\endcsname{10}
\expandafter\gdef\csname svy@item/t09/nonusers/median\endcsname{2.0}
\expandafter\gdef\csname svy@item/t09/nonusers/cilow\endcsname{1.0}
\expandafter\gdef\csname svy@item/t09/nonusers/cihigh\endcsname{2.0}
\expandafter\gdef\csname svy@item/t09/nonusers/mean\endcsname{1.92}
\expandafter\gdef\csname svy@item/t09/nonusers/meanlow\endcsname{1.64}
\expandafter\gdef\csname svy@item/t09/nonusers/meanhigh\endcsname{2.19}
\expandafter\gdef\csname svy@item/t09/nonusers/se\endcsname{0.139}
\expandafter\gdef\csname svy@item/t09/nonusers/sd\endcsname{1.07}
\expandafter\gdef\csname svy@item/t09/nonusers/sdhigh\endcsname{1}
\expandafter\gdef\csname svy@item/t09/nonusers/verdict\endcsname{CONFIRMED}
\expandafter\gdef\csname svy@item/t09/nonusers/dip\endcsname{0}
\expandafter\gdef\csname svy@item/t09/nonusers/shapelow\endcsname{72}
\expandafter\gdef\csname svy@item/t09/nonusers/shapemid\endcsname{17}
\expandafter\gdef\csname svy@item/t09/nonusers/shapehigh\endcsname{11}
\expandafter\gdef\csname svy@item/t10/code\endcsname{T10}
\expandafter\gdef\csname svy@item/t10/pred\endcsname{H}
\expandafter\gdef\csname svy@item/t10/label\endcsname{choosing which AI tools and agents to use}
\expandafter\gdef\csname svy@item/t10/usage/diff\endcsname{0.29}
\expandafter\gdef\csname svy@item/t10/usage/difflow\endcsname{0.02}
\expandafter\gdef\csname svy@item/t10/usage/diffhigh\endcsname{0.55}
\expandafter\gdef\csname svy@item/t10/usage/g\endcsname{0.31}
\expandafter\gdef\csname svy@item/t10/all/n\endcsname{232}
\expandafter\gdef\csname svy@item/t10/all/miss\endcsname{10}
\expandafter\gdef\csname svy@item/t10/all/median\endcsname{2.0}
\expandafter\gdef\csname svy@item/t10/all/cilow\endcsname{1.0}
\expandafter\gdef\csname svy@item/t10/all/cihigh\endcsname{2.0}
\expandafter\gdef\csname svy@item/t10/all/mean\endcsname{1.82}
\expandafter\gdef\csname svy@item/t10/all/meanlow\endcsname{1.70}
\expandafter\gdef\csname svy@item/t10/all/meanhigh\endcsname{1.94}
\expandafter\gdef\csname svy@item/t10/all/se\endcsname{0.061}
\expandafter\gdef\csname svy@item/t10/all/sd\endcsname{0.93}
\expandafter\gdef\csname svy@item/t10/all/sdhigh\endcsname{0}
\expandafter\gdef\csname svy@item/t10/all/verdict\endcsname{CONFIRMED}
\expandafter\gdef\csname svy@item/t10/all/dip\endcsname{0}
\expandafter\gdef\csname svy@item/t10/all/shapelow\endcsname{82}
\expandafter\gdef\csname svy@item/t10/all/shapemid\endcsname{11}
\expandafter\gdef\csname svy@item/t10/all/shapehigh\endcsname{7}
\expandafter\gdef\csname svy@item/t10/users/n\endcsname{169}
\expandafter\gdef\csname svy@item/t10/users/miss\endcsname{2}
\expandafter\gdef\csname svy@item/t10/users/median\endcsname{2.0}
\expandafter\gdef\csname svy@item/t10/users/cilow\endcsname{2.0}
\expandafter\gdef\csname svy@item/t10/users/cihigh\endcsname{2.0}
\expandafter\gdef\csname svy@item/t10/users/mean\endcsname{1.89}
\expandafter\gdef\csname svy@item/t10/users/meanlow\endcsname{1.75}
\expandafter\gdef\csname svy@item/t10/users/meanhigh\endcsname{2.04}
\expandafter\gdef\csname svy@item/t10/users/se\endcsname{0.072}
\expandafter\gdef\csname svy@item/t10/users/sd\endcsname{0.93}
\expandafter\gdef\csname svy@item/t10/users/sdhigh\endcsname{0}
\expandafter\gdef\csname svy@item/t10/users/verdict\endcsname{CONFIRMED}
\expandafter\gdef\csname svy@item/t10/users/dip\endcsname{0}
\expandafter\gdef\csname svy@item/t10/users/shapelow\endcsname{82}
\expandafter\gdef\csname svy@item/t10/users/shapemid\endcsname{10}
\expandafter\gdef\csname svy@item/t10/users/shapehigh\endcsname{8}
\expandafter\gdef\csname svy@item/t10/nonusers/n\endcsname{61}
\expandafter\gdef\csname svy@item/t10/nonusers/miss\endcsname{8}
\expandafter\gdef\csname svy@item/t10/nonusers/median\endcsname{1.0}
\expandafter\gdef\csname svy@item/t10/nonusers/cilow\endcsname{1.0}
\expandafter\gdef\csname svy@item/t10/nonusers/cihigh\endcsname{2.0}
\expandafter\gdef\csname svy@item/t10/nonusers/mean\endcsname{1.61}
\expandafter\gdef\csname svy@item/t10/nonusers/meanlow\endcsname{1.38}
\expandafter\gdef\csname svy@item/t10/nonusers/meanhigh\endcsname{1.83}
\expandafter\gdef\csname svy@item/t10/nonusers/se\endcsname{0.113}
\expandafter\gdef\csname svy@item/t10/nonusers/sd\endcsname{0.88}
\expandafter\gdef\csname svy@item/t10/nonusers/sdhigh\endcsname{0}
\expandafter\gdef\csname svy@item/t10/nonusers/verdict\endcsname{CONFIRMED}
\expandafter\gdef\csname svy@item/t10/nonusers/dip\endcsname{0}
\expandafter\gdef\csname svy@item/t10/nonusers/shapelow\endcsname{84}
\expandafter\gdef\csname svy@item/t10/nonusers/shapemid\endcsname{11}
\expandafter\gdef\csname svy@item/t10/nonusers/shapehigh\endcsname{5}
\expandafter\gdef\csname svy@item/t11/code\endcsname{T11}
\expandafter\gdef\csname svy@item/t11/pred\endcsname{X}
\expandafter\gdef\csname svy@item/t11/label\endcsname{designing environments, sandboxing, safety boundaries}
\expandafter\gdef\csname svy@item/t11/usage/diff\endcsname{0.70}
\expandafter\gdef\csname svy@item/t11/usage/difflow\endcsname{0.45}
\expandafter\gdef\csname svy@item/t11/usage/diffhigh\endcsname{0.96}
\expandafter\gdef\csname svy@item/t11/usage/g\endcsname{0.72}
\expandafter\gdef\csname svy@item/t11/all/n\endcsname{232}
\expandafter\gdef\csname svy@item/t11/all/miss\endcsname{10}
\expandafter\gdef\csname svy@item/t11/all/median\endcsname{2.0}
\expandafter\gdef\csname svy@item/t11/all/cilow\endcsname{2.0}
\expandafter\gdef\csname svy@item/t11/all/cihigh\endcsname{2.0}
\expandafter\gdef\csname svy@item/t11/all/mean\endcsname{2.05}
\expandafter\gdef\csname svy@item/t11/all/meanlow\endcsname{1.92}
\expandafter\gdef\csname svy@item/t11/all/meanhigh\endcsname{2.18}
\expandafter\gdef\csname svy@item/t11/all/se\endcsname{0.067}
\expandafter\gdef\csname svy@item/t11/all/sd\endcsname{1.02}
\expandafter\gdef\csname svy@item/t11/all/sdhigh\endcsname{0}
\expandafter\gdef\csname svy@item/t11/all/verdict\endcsname{NO_PREDICTION}
\expandafter\gdef\csname svy@item/t11/all/dip\endcsname{0}
\expandafter\gdef\csname svy@item/t11/all/shapelow\endcsname{75}
\expandafter\gdef\csname svy@item/t11/all/shapemid\endcsname{14}
\expandafter\gdef\csname svy@item/t11/all/shapehigh\endcsname{11}
\expandafter\gdef\csname svy@item/t11/users/n\endcsname{164}
\expandafter\gdef\csname svy@item/t11/users/miss\endcsname{7}
\expandafter\gdef\csname svy@item/t11/users/median\endcsname{2.0}
\expandafter\gdef\csname svy@item/t11/users/cilow\endcsname{2.0}
\expandafter\gdef\csname svy@item/t11/users/cihigh\endcsname{2.0}
\expandafter\gdef\csname svy@item/t11/users/mean\endcsname{2.26}
\expandafter\gdef\csname svy@item/t11/users/meanlow\endcsname{2.10}
\expandafter\gdef\csname svy@item/t11/users/meanhigh\endcsname{2.42}
\expandafter\gdef\csname svy@item/t11/users/se\endcsname{0.080}
\expandafter\gdef\csname svy@item/t11/users/sd\endcsname{1.02}
\expandafter\gdef\csname svy@item/t11/users/sdhigh\endcsname{1}
\expandafter\gdef\csname svy@item/t11/users/verdict\endcsname{NO_PREDICTION}
\expandafter\gdef\csname svy@item/t11/users/dip\endcsname{0}
\expandafter\gdef\csname svy@item/t11/users/shapelow\endcsname{69}
\expandafter\gdef\csname svy@item/t11/users/shapemid\endcsname{17}
\expandafter\gdef\csname svy@item/t11/users/shapehigh\endcsname{14}
\expandafter\gdef\csname svy@item/t11/nonusers/n\endcsname{66}
\expandafter\gdef\csname svy@item/t11/nonusers/miss\endcsname{3}
\expandafter\gdef\csname svy@item/t11/nonusers/median\endcsname{1.0}
\expandafter\gdef\csname svy@item/t11/nonusers/cilow\endcsname{1.0}
\expandafter\gdef\csname svy@item/t11/nonusers/cihigh\endcsname{2.0}
\expandafter\gdef\csname svy@item/t11/nonusers/mean\endcsname{1.56}
\expandafter\gdef\csname svy@item/t11/nonusers/meanlow\endcsname{1.36}
\expandafter\gdef\csname svy@item/t11/nonusers/meanhigh\endcsname{1.76}
\expandafter\gdef\csname svy@item/t11/nonusers/se\endcsname{0.102}
\expandafter\gdef\csname svy@item/t11/nonusers/sd\endcsname{0.83}
\expandafter\gdef\csname svy@item/t11/nonusers/sdhigh\endcsname{0}
\expandafter\gdef\csname svy@item/t11/nonusers/verdict\endcsname{NO_PREDICTION}
\expandafter\gdef\csname svy@item/t11/nonusers/dip\endcsname{0}
\expandafter\gdef\csname svy@item/t11/nonusers/shapelow\endcsname{88}
\expandafter\gdef\csname svy@item/t11/nonusers/shapemid\endcsname{8}
\expandafter\gdef\csname svy@item/t11/nonusers/shapehigh\endcsname{4}
\expandafter\gdef\csname svy@item/t12/code\endcsname{T12}
\expandafter\gdef\csname svy@item/t12/pred\endcsname{X}
\expandafter\gdef\csname svy@item/t12/label\endcsname{carrying out setup and configuration once chosen}
\expandafter\gdef\csname svy@item/t12/usage/diff\endcsname{1.22}
\expandafter\gdef\csname svy@item/t12/usage/difflow\endcsname{0.90}
\expandafter\gdef\csname svy@item/t12/usage/diffhigh\endcsname{1.54}
\expandafter\gdef\csname svy@item/t12/usage/g\endcsname{1.08}
\expandafter\gdef\csname svy@item/t12/all/n\endcsname{236}
\expandafter\gdef\csname svy@item/t12/all/miss\endcsname{6}
\expandafter\gdef\csname svy@item/t12/all/median\endcsname{3.0}
\expandafter\gdef\csname svy@item/t12/all/cilow\endcsname{3.0}
\expandafter\gdef\csname svy@item/t12/all/cihigh\endcsname{4.0}
\expandafter\gdef\csname svy@item/t12/all/mean\endcsname{3.14}
\expandafter\gdef\csname svy@item/t12/all/meanlow\endcsname{2.98}
\expandafter\gdef\csname svy@item/t12/all/meanhigh\endcsname{3.30}
\expandafter\gdef\csname svy@item/t12/all/se\endcsname{0.081}
\expandafter\gdef\csname svy@item/t12/all/sd\endcsname{1.25}
\expandafter\gdef\csname svy@item/t12/all/sdhigh\endcsname{0}
\expandafter\gdef\csname svy@item/t12/all/verdict\endcsname{NO_PREDICTION}
\expandafter\gdef\csname svy@item/t12/all/dip\endcsname{1}
\expandafter\gdef\csname svy@item/t12/all/shapelow\endcsname{34}
\expandafter\gdef\csname svy@item/t12/all/shapemid\endcsname{18}
\expandafter\gdef\csname svy@item/t12/all/shapehigh\endcsname{48}
\expandafter\gdef\csname svy@item/t12/users/n\endcsname{167}
\expandafter\gdef\csname svy@item/t12/users/miss\endcsname{4}
\expandafter\gdef\csname svy@item/t12/users/median\endcsname{4.0}
\expandafter\gdef\csname svy@item/t12/users/cilow\endcsname{4.0}
\expandafter\gdef\csname svy@item/t12/users/cihigh\endcsname{4.0}
\expandafter\gdef\csname svy@item/t12/users/mean\endcsname{3.49}
\expandafter\gdef\csname svy@item/t12/users/meanlow\endcsname{3.32}
\expandafter\gdef\csname svy@item/t12/users/meanhigh\endcsname{3.66}
\expandafter\gdef\csname svy@item/t12/users/se\endcsname{0.087}
\expandafter\gdef\csname svy@item/t12/users/sd\endcsname{1.12}
\expandafter\gdef\csname svy@item/t12/users/sdhigh\endcsname{1}
\expandafter\gdef\csname svy@item/t12/users/verdict\endcsname{NO_PREDICTION}
\expandafter\gdef\csname svy@item/t12/users/dip\endcsname{1}
\expandafter\gdef\csname svy@item/t12/users/shapelow\endcsname{22}
\expandafter\gdef\csname svy@item/t12/users/shapemid\endcsname{17}
\expandafter\gdef\csname svy@item/t12/users/shapehigh\endcsname{61}
\expandafter\gdef\csname svy@item/t12/nonusers/n\endcsname{67}
\expandafter\gdef\csname svy@item/t12/nonusers/miss\endcsname{2}
\expandafter\gdef\csname svy@item/t12/nonusers/median\endcsname{2.0}
\expandafter\gdef\csname svy@item/t12/nonusers/cilow\endcsname{2.0}
\expandafter\gdef\csname svy@item/t12/nonusers/cihigh\endcsname{2.0}
\expandafter\gdef\csname svy@item/t12/nonusers/mean\endcsname{2.27}
\expandafter\gdef\csname svy@item/t12/nonusers/meanlow\endcsname{1.99}
\expandafter\gdef\csname svy@item/t12/nonusers/meanhigh\endcsname{2.54}
\expandafter\gdef\csname svy@item/t12/nonusers/se\endcsname{0.137}
\expandafter\gdef\csname svy@item/t12/nonusers/sd\endcsname{1.12}
\expandafter\gdef\csname svy@item/t12/nonusers/sdhigh\endcsname{1}
\expandafter\gdef\csname svy@item/t12/nonusers/verdict\endcsname{NO_PREDICTION}
\expandafter\gdef\csname svy@item/t12/nonusers/dip\endcsname{0}
\expandafter\gdef\csname svy@item/t12/nonusers/shapelow\endcsname{63}
\expandafter\gdef\csname svy@item/t12/nonusers/shapemid\endcsname{21}
\expandafter\gdef\csname svy@item/t12/nonusers/shapehigh\endcsname{16}
\expandafter\gdef\csname svy@item/t13/code\endcsname{T13}
\expandafter\gdef\csname svy@item/t13/pred\endcsname{H}
\expandafter\gdef\csname svy@item/t13/label\endcsname{splitting work into human-reviewable chunks}
\expandafter\gdef\csname svy@item/t13/usage/diff\endcsname{0.85}
\expandafter\gdef\csname svy@item/t13/usage/difflow\endcsname{0.52}
\expandafter\gdef\csname svy@item/t13/usage/diffhigh\endcsname{1.19}
\expandafter\gdef\csname svy@item/t13/usage/g\endcsname{0.72}
\expandafter\gdef\csname svy@item/t13/all/n\endcsname{227}
\expandafter\gdef\csname svy@item/t13/all/miss\endcsname{15}
\expandafter\gdef\csname svy@item/t13/all/median\endcsname{3.0}
\expandafter\gdef\csname svy@item/t13/all/cilow\endcsname{3.0}
\expandafter\gdef\csname svy@item/t13/all/cihigh\endcsname{3.0}
\expandafter\gdef\csname svy@item/t13/all/mean\endcsname{2.97}
\expandafter\gdef\csname svy@item/t13/all/meanlow\endcsname{2.81}
\expandafter\gdef\csname svy@item/t13/all/meanhigh\endcsname{3.13}
\expandafter\gdef\csname svy@item/t13/all/se\endcsname{0.082}
\expandafter\gdef\csname svy@item/t13/all/sd\endcsname{1.24}
\expandafter\gdef\csname svy@item/t13/all/sdhigh\endcsname{0}
\expandafter\gdef\csname svy@item/t13/all/verdict\endcsname{REJECTED}
\expandafter\gdef\csname svy@item/t13/all/dip\endcsname{0}
\expandafter\gdef\csname svy@item/t13/all/shapelow\endcsname{35}
\expandafter\gdef\csname svy@item/t13/all/shapemid\endcsname{30}
\expandafter\gdef\csname svy@item/t13/all/shapehigh\endcsname{35}
\expandafter\gdef\csname svy@item/t13/users/n\endcsname{163}
\expandafter\gdef\csname svy@item/t13/users/miss\endcsname{8}
\expandafter\gdef\csname svy@item/t13/users/median\endcsname{3.0}
\expandafter\gdef\csname svy@item/t13/users/cilow\endcsname{3.0}
\expandafter\gdef\csname svy@item/t13/users/cihigh\endcsname{4.0}
\expandafter\gdef\csname svy@item/t13/users/mean\endcsname{3.21}
\expandafter\gdef\csname svy@item/t13/users/meanlow\endcsname{3.02}
\expandafter\gdef\csname svy@item/t13/users/meanhigh\endcsname{3.40}
\expandafter\gdef\csname svy@item/t13/users/se\endcsname{0.095}
\expandafter\gdef\csname svy@item/t13/users/sd\endcsname{1.21}
\expandafter\gdef\csname svy@item/t13/users/sdhigh\endcsname{1}
\expandafter\gdef\csname svy@item/t13/users/verdict\endcsname{REJECTED}
\expandafter\gdef\csname svy@item/t13/users/dip\endcsname{0}
\expandafter\gdef\csname svy@item/t13/users/shapelow\endcsname{27}
\expandafter\gdef\csname svy@item/t13/users/shapemid\endcsname{30}
\expandafter\gdef\csname svy@item/t13/users/shapehigh\endcsname{43}
\expandafter\gdef\csname svy@item/t13/nonusers/n\endcsname{62}
\expandafter\gdef\csname svy@item/t13/nonusers/miss\endcsname{7}
\expandafter\gdef\csname svy@item/t13/nonusers/median\endcsname{2.0}
\expandafter\gdef\csname svy@item/t13/nonusers/cilow\endcsname{2.0}
\expandafter\gdef\csname svy@item/t13/nonusers/cihigh\endcsname{3.0}
\expandafter\gdef\csname svy@item/t13/nonusers/mean\endcsname{2.35}
\expandafter\gdef\csname svy@item/t13/nonusers/meanlow\endcsname{2.07}
\expandafter\gdef\csname svy@item/t13/nonusers/meanhigh\endcsname{2.63}
\expandafter\gdef\csname svy@item/t13/nonusers/se\endcsname{0.140}
\expandafter\gdef\csname svy@item/t13/nonusers/sd\endcsname{1.10}
\expandafter\gdef\csname svy@item/t13/nonusers/sdhigh\endcsname{1}
\expandafter\gdef\csname svy@item/t13/nonusers/verdict\endcsname{INCONCLUSIVE}
\expandafter\gdef\csname svy@item/t13/nonusers/dip\endcsname{0}
\expandafter\gdef\csname svy@item/t13/nonusers/shapelow\endcsname{55}
\expandafter\gdef\csname svy@item/t13/nonusers/shapemid\endcsname{31}
\expandafter\gdef\csname svy@item/t13/nonusers/shapehigh\endcsname{14}
\expandafter\gdef\csname svy@item/t16/code\endcsname{T16}
\expandafter\gdef\csname svy@item/t16/pred\endcsname{H}
\expandafter\gdef\csname svy@item/t16/label\endcsname{defining what counts as evidence it works}
\expandafter\gdef\csname svy@item/t16/usage/diff\endcsname{0.58}
\expandafter\gdef\csname svy@item/t16/usage/difflow\endcsname{0.33}
\expandafter\gdef\csname svy@item/t16/usage/diffhigh\endcsname{0.84}
\expandafter\gdef\csname svy@item/t16/usage/g\endcsname{0.62}
\expandafter\gdef\csname svy@item/t16/all/n\endcsname{231}
\expandafter\gdef\csname svy@item/t16/all/miss\endcsname{11}
\expandafter\gdef\csname svy@item/t16/all/median\endcsname{2.0}
\expandafter\gdef\csname svy@item/t16/all/cilow\endcsname{2.0}
\expandafter\gdef\csname svy@item/t16/all/cihigh\endcsname{2.0}
\expandafter\gdef\csname svy@item/t16/all/mean\endcsname{1.96}
\expandafter\gdef\csname svy@item/t16/all/meanlow\endcsname{1.83}
\expandafter\gdef\csname svy@item/t16/all/meanhigh\endcsname{2.08}
\expandafter\gdef\csname svy@item/t16/all/se\endcsname{0.064}
\expandafter\gdef\csname svy@item/t16/all/sd\endcsname{0.97}
\expandafter\gdef\csname svy@item/t16/all/sdhigh\endcsname{0}
\expandafter\gdef\csname svy@item/t16/all/verdict\endcsname{CONFIRMED}
\expandafter\gdef\csname svy@item/t16/all/dip\endcsname{0}
\expandafter\gdef\csname svy@item/t16/all/shapelow\endcsname{78}
\expandafter\gdef\csname svy@item/t16/all/shapemid\endcsname{12}
\expandafter\gdef\csname svy@item/t16/all/shapehigh\endcsname{10}
\expandafter\gdef\csname svy@item/t16/users/n\endcsname{164}
\expandafter\gdef\csname svy@item/t16/users/miss\endcsname{7}
\expandafter\gdef\csname svy@item/t16/users/median\endcsname{2.0}
\expandafter\gdef\csname svy@item/t16/users/cilow\endcsname{2.0}
\expandafter\gdef\csname svy@item/t16/users/cihigh\endcsname{2.0}
\expandafter\gdef\csname svy@item/t16/users/mean\endcsname{2.12}
\expandafter\gdef\csname svy@item/t16/users/meanlow\endcsname{1.97}
\expandafter\gdef\csname svy@item/t16/users/meanhigh\endcsname{2.27}
\expandafter\gdef\csname svy@item/t16/users/se\endcsname{0.076}
\expandafter\gdef\csname svy@item/t16/users/sd\endcsname{0.98}
\expandafter\gdef\csname svy@item/t16/users/sdhigh\endcsname{0}
\expandafter\gdef\csname svy@item/t16/users/verdict\endcsname{CONFIRMED}
\expandafter\gdef\csname svy@item/t16/users/dip\endcsname{0}
\expandafter\gdef\csname svy@item/t16/users/shapelow\endcsname{73}
\expandafter\gdef\csname svy@item/t16/users/shapemid\endcsname{16}
\expandafter\gdef\csname svy@item/t16/users/shapehigh\endcsname{11}
\expandafter\gdef\csname svy@item/t16/nonusers/n\endcsname{65}
\expandafter\gdef\csname svy@item/t16/nonusers/miss\endcsname{4}
\expandafter\gdef\csname svy@item/t16/nonusers/median\endcsname{1.0}
\expandafter\gdef\csname svy@item/t16/nonusers/cilow\endcsname{1.0}
\expandafter\gdef\csname svy@item/t16/nonusers/cihigh\endcsname{1.0}
\expandafter\gdef\csname svy@item/t16/nonusers/mean\endcsname{1.54}
\expandafter\gdef\csname svy@item/t16/nonusers/meanlow\endcsname{1.33}
\expandafter\gdef\csname svy@item/t16/nonusers/meanhigh\endcsname{1.75}
\expandafter\gdef\csname svy@item/t16/nonusers/se\endcsname{0.105}
\expandafter\gdef\csname svy@item/t16/nonusers/sd\endcsname{0.85}
\expandafter\gdef\csname svy@item/t16/nonusers/sdhigh\endcsname{0}
\expandafter\gdef\csname svy@item/t16/nonusers/verdict\endcsname{CONFIRMED}
\expandafter\gdef\csname svy@item/t16/nonusers/dip\endcsname{0}
\expandafter\gdef\csname svy@item/t16/nonusers/shapelow\endcsname{89}
\expandafter\gdef\csname svy@item/t16/nonusers/shapemid\endcsname{5}
\expandafter\gdef\csname svy@item/t16/nonusers/shapehigh\endcsname{6}
\expandafter\gdef\csname svy@item/t17/code\endcsname{T17}
\expandafter\gdef\csname svy@item/t17/pred\endcsname{A}
\expandafter\gdef\csname svy@item/t17/label\endcsname{implementing and checking with tests/linters}
\expandafter\gdef\csname svy@item/t17/usage/diff\endcsname{1.31}
\expandafter\gdef\csname svy@item/t17/usage/difflow\endcsname{1.00}
\expandafter\gdef\csname svy@item/t17/usage/diffhigh\endcsname{1.63}
\expandafter\gdef\csname svy@item/t17/usage/g\endcsname{1.24}
\expandafter\gdef\csname svy@item/t17/all/n\endcsname{237}
\expandafter\gdef\csname svy@item/t17/all/miss\endcsname{5}
\expandafter\gdef\csname svy@item/t17/all/median\endcsname{4.0}
\expandafter\gdef\csname svy@item/t17/all/cilow\endcsname{3.0}
\expandafter\gdef\csname svy@item/t17/all/cihigh\endcsname{4.0}
\expandafter\gdef\csname svy@item/t17/all/mean\endcsname{3.35}
\expandafter\gdef\csname svy@item/t17/all/meanlow\endcsname{3.20}
\expandafter\gdef\csname svy@item/t17/all/meanhigh\endcsname{3.51}
\expandafter\gdef\csname svy@item/t17/all/se\endcsname{0.078}
\expandafter\gdef\csname svy@item/t17/all/sd\endcsname{1.20}
\expandafter\gdef\csname svy@item/t17/all/sdhigh\endcsname{0}
\expandafter\gdef\csname svy@item/t17/all/verdict\endcsname{INCONCLUSIVE}
\expandafter\gdef\csname svy@item/t17/all/dip\endcsname{0}
\expandafter\gdef\csname svy@item/t17/all/shapelow\endcsname{27}
\expandafter\gdef\csname svy@item/t17/all/shapemid\endcsname{20}
\expandafter\gdef\csname svy@item/t17/all/shapehigh\endcsname{53}
\expandafter\gdef\csname svy@item/t17/users/n\endcsname{169}
\expandafter\gdef\csname svy@item/t17/users/miss\endcsname{2}
\expandafter\gdef\csname svy@item/t17/users/median\endcsname{4.0}
\expandafter\gdef\csname svy@item/t17/users/cilow\endcsname{4.0}
\expandafter\gdef\csname svy@item/t17/users/cihigh\endcsname{4.0}
\expandafter\gdef\csname svy@item/t17/users/mean\endcsname{3.72}
\expandafter\gdef\csname svy@item/t17/users/meanlow\endcsname{3.57}
\expandafter\gdef\csname svy@item/t17/users/meanhigh\endcsname{3.88}
\expandafter\gdef\csname svy@item/t17/users/se\endcsname{0.078}
\expandafter\gdef\csname svy@item/t17/users/sd\endcsname{1.02}
\expandafter\gdef\csname svy@item/t17/users/sdhigh\endcsname{1}
\expandafter\gdef\csname svy@item/t17/users/verdict\endcsname{CONFIRMED}
\expandafter\gdef\csname svy@item/t17/users/dip\endcsname{0}
\expandafter\gdef\csname svy@item/t17/users/shapelow\endcsname{15}
\expandafter\gdef\csname svy@item/t17/users/shapemid\endcsname{19}
\expandafter\gdef\csname svy@item/t17/users/shapehigh\endcsname{66}
\expandafter\gdef\csname svy@item/t17/nonusers/n\endcsname{66}
\expandafter\gdef\csname svy@item/t17/nonusers/miss\endcsname{3}
\expandafter\gdef\csname svy@item/t17/nonusers/median\endcsname{2.0}
\expandafter\gdef\csname svy@item/t17/nonusers/cilow\endcsname{2.0}
\expandafter\gdef\csname svy@item/t17/nonusers/cihigh\endcsname{3.0}
\expandafter\gdef\csname svy@item/t17/nonusers/mean\endcsname{2.41}
\expandafter\gdef\csname svy@item/t17/nonusers/meanlow\endcsname{2.13}
\expandafter\gdef\csname svy@item/t17/nonusers/meanhigh\endcsname{2.69}
\expandafter\gdef\csname svy@item/t17/nonusers/se\endcsname{0.140}
\expandafter\gdef\csname svy@item/t17/nonusers/sd\endcsname{1.14}
\expandafter\gdef\csname svy@item/t17/nonusers/sdhigh\endcsname{1}
\expandafter\gdef\csname svy@item/t17/nonusers/verdict\endcsname{INCONCLUSIVE}
\expandafter\gdef\csname svy@item/t17/nonusers/dip\endcsname{0}
\expandafter\gdef\csname svy@item/t17/nonusers/shapelow\endcsname{59}
\expandafter\gdef\csname svy@item/t17/nonusers/shapemid\endcsname{20}
\expandafter\gdef\csname svy@item/t17/nonusers/shapehigh\endcsname{21}
\expandafter\gdef\csname svy@item/t18/code\endcsname{T18}
\expandafter\gdef\csname svy@item/t18/pred\endcsname{E}
\expandafter\gdef\csname svy@item/t18/label\endcsname{reviewing a change for gaps, risks, problems}
\expandafter\gdef\csname svy@item/t18/usage/diff\endcsname{0.47}
\expandafter\gdef\csname svy@item/t18/usage/difflow\endcsname{0.17}
\expandafter\gdef\csname svy@item/t18/usage/diffhigh\endcsname{0.77}
\expandafter\gdef\csname svy@item/t18/usage/g\endcsname{0.44}
\expandafter\gdef\csname svy@item/t18/all/n\endcsname{240}
\expandafter\gdef\csname svy@item/t18/all/miss\endcsname{2}
\expandafter\gdef\csname svy@item/t18/all/median\endcsname{3.0}
\expandafter\gdef\csname svy@item/t18/all/cilow\endcsname{3.0}
\expandafter\gdef\csname svy@item/t18/all/cihigh\endcsname{3.0}
\expandafter\gdef\csname svy@item/t18/all/mean\endcsname{2.88}
\expandafter\gdef\csname svy@item/t18/all/meanlow\endcsname{2.74}
\expandafter\gdef\csname svy@item/t18/all/meanhigh\endcsname{3.01}
\expandafter\gdef\csname svy@item/t18/all/se\endcsname{0.070}
\expandafter\gdef\csname svy@item/t18/all/sd\endcsname{1.08}
\expandafter\gdef\csname svy@item/t18/all/sdhigh\endcsname{0}
\expandafter\gdef\csname svy@item/t18/all/verdict\endcsname{CONFIRMED}
\expandafter\gdef\csname svy@item/t18/all/dip\endcsname{0}
\expandafter\gdef\csname svy@item/t18/all/shapelow\endcsname{40}
\expandafter\gdef\csname svy@item/t18/all/shapemid\endcsname{29}
\expandafter\gdef\csname svy@item/t18/all/shapehigh\endcsname{31}
\expandafter\gdef\csname svy@item/t18/users/n\endcsname{171}
\expandafter\gdef\csname svy@item/t18/users/miss\endcsname{0}
\expandafter\gdef\csname svy@item/t18/users/median\endcsname{3.0}
\expandafter\gdef\csname svy@item/t18/users/cilow\endcsname{3.0}
\expandafter\gdef\csname svy@item/t18/users/cihigh\endcsname{3.0}
\expandafter\gdef\csname svy@item/t18/users/mean\endcsname{3.01}
\expandafter\gdef\csname svy@item/t18/users/meanlow\endcsname{2.84}
\expandafter\gdef\csname svy@item/t18/users/meanhigh\endcsname{3.17}
\expandafter\gdef\csname svy@item/t18/users/se\endcsname{0.083}
\expandafter\gdef\csname svy@item/t18/users/sd\endcsname{1.08}
\expandafter\gdef\csname svy@item/t18/users/sdhigh\endcsname{1}
\expandafter\gdef\csname svy@item/t18/users/verdict\endcsname{CONFIRMED}
\expandafter\gdef\csname svy@item/t18/users/dip\endcsname{0}
\expandafter\gdef\csname svy@item/t18/users/shapelow\endcsname{36}
\expandafter\gdef\csname svy@item/t18/users/shapemid\endcsname{28}
\expandafter\gdef\csname svy@item/t18/users/shapehigh\endcsname{36}
\expandafter\gdef\csname svy@item/t18/nonusers/n\endcsname{67}
\expandafter\gdef\csname svy@item/t18/nonusers/miss\endcsname{2}
\expandafter\gdef\csname svy@item/t18/nonusers/median\endcsname{2.0}
\expandafter\gdef\csname svy@item/t18/nonusers/cilow\endcsname{2.0}
\expandafter\gdef\csname svy@item/t18/nonusers/cihigh\endcsname{3.0}
\expandafter\gdef\csname svy@item/t18/nonusers/mean\endcsname{2.54}
\expandafter\gdef\csname svy@item/t18/nonusers/meanlow\endcsname{2.28}
\expandafter\gdef\csname svy@item/t18/nonusers/meanhigh\endcsname{2.79}
\expandafter\gdef\csname svy@item/t18/nonusers/se\endcsname{0.126}
\expandafter\gdef\csname svy@item/t18/nonusers/sd\endcsname{1.03}
\expandafter\gdef\csname svy@item/t18/nonusers/sdhigh\endcsname{1}
\expandafter\gdef\csname svy@item/t18/nonusers/verdict\endcsname{INCONCLUSIVE}
\expandafter\gdef\csname svy@item/t18/nonusers/dip\endcsname{0}
\expandafter\gdef\csname svy@item/t18/nonusers/shapelow\endcsname{51}
\expandafter\gdef\csname svy@item/t18/nonusers/shapemid\endcsname{31}
\expandafter\gdef\csname svy@item/t18/nonusers/shapehigh\endcsname{18}
\expandafter\gdef\csname svy@item/t20/code\endcsname{T20}
\expandafter\gdef\csname svy@item/t20/pred\endcsname{H}
\expandafter\gdef\csname svy@item/t20/label\endcsname{deciding when and why to refactor or clean up}
\expandafter\gdef\csname svy@item/t20/usage/diff\endcsname{0.56}
\expandafter\gdef\csname svy@item/t20/usage/difflow\endcsname{0.34}
\expandafter\gdef\csname svy@item/t20/usage/diffhigh\endcsname{0.79}
\expandafter\gdef\csname svy@item/t20/usage/g\endcsname{0.63}
\expandafter\gdef\csname svy@item/t20/all/n\endcsname{242}
\expandafter\gdef\csname svy@item/t20/all/miss\endcsname{0}
\expandafter\gdef\csname svy@item/t20/all/median\endcsname{2.0}
\expandafter\gdef\csname svy@item/t20/all/cilow\endcsname{2.0}
\expandafter\gdef\csname svy@item/t20/all/cihigh\endcsname{2.0}
\expandafter\gdef\csname svy@item/t20/all/mean\endcsname{1.96}
\expandafter\gdef\csname svy@item/t20/all/meanlow\endcsname{1.85}
\expandafter\gdef\csname svy@item/t20/all/meanhigh\endcsname{2.08}
\expandafter\gdef\csname svy@item/t20/all/se\endcsname{0.060}
\expandafter\gdef\csname svy@item/t20/all/sd\endcsname{0.93}
\expandafter\gdef\csname svy@item/t20/all/sdhigh\endcsname{0}
\expandafter\gdef\csname svy@item/t20/all/verdict\endcsname{CONFIRMED}
\expandafter\gdef\csname svy@item/t20/all/dip\endcsname{0}
\expandafter\gdef\csname svy@item/t20/all/shapelow\endcsname{78}
\expandafter\gdef\csname svy@item/t20/all/shapemid\endcsname{15}
\expandafter\gdef\csname svy@item/t20/all/shapehigh\endcsname{7}
\expandafter\gdef\csname svy@item/t20/users/n\endcsname{171}
\expandafter\gdef\csname svy@item/t20/users/miss\endcsname{0}
\expandafter\gdef\csname svy@item/t20/users/median\endcsname{2.0}
\expandafter\gdef\csname svy@item/t20/users/cilow\endcsname{2.0}
\expandafter\gdef\csname svy@item/t20/users/cihigh\endcsname{2.0}
\expandafter\gdef\csname svy@item/t20/users/mean\endcsname{2.13}
\expandafter\gdef\csname svy@item/t20/users/meanlow\endcsname{1.99}
\expandafter\gdef\csname svy@item/t20/users/meanhigh\endcsname{2.27}
\expandafter\gdef\csname svy@item/t20/users/se\endcsname{0.073}
\expandafter\gdef\csname svy@item/t20/users/sd\endcsname{0.95}
\expandafter\gdef\csname svy@item/t20/users/sdhigh\endcsname{0}
\expandafter\gdef\csname svy@item/t20/users/verdict\endcsname{CONFIRMED}
\expandafter\gdef\csname svy@item/t20/users/dip\endcsname{0}
\expandafter\gdef\csname svy@item/t20/users/shapelow\endcsname{74}
\expandafter\gdef\csname svy@item/t20/users/shapemid\endcsname{17}
\expandafter\gdef\csname svy@item/t20/users/shapehigh\endcsname{9}
\expandafter\gdef\csname svy@item/t20/nonusers/n\endcsname{69}
\expandafter\gdef\csname svy@item/t20/nonusers/miss\endcsname{0}
\expandafter\gdef\csname svy@item/t20/nonusers/median\endcsname{1.0}
\expandafter\gdef\csname svy@item/t20/nonusers/cilow\endcsname{1.0}
\expandafter\gdef\csname svy@item/t20/nonusers/cihigh\endcsname{2.0}
\expandafter\gdef\csname svy@item/t20/nonusers/mean\endcsname{1.57}
\expandafter\gdef\csname svy@item/t20/nonusers/meanlow\endcsname{1.39}
\expandafter\gdef\csname svy@item/t20/nonusers/meanhigh\endcsname{1.74}
\expandafter\gdef\csname svy@item/t20/nonusers/se\endcsname{0.089}
\expandafter\gdef\csname svy@item/t20/nonusers/sd\endcsname{0.74}
\expandafter\gdef\csname svy@item/t20/nonusers/sdhigh\endcsname{0}
\expandafter\gdef\csname svy@item/t20/nonusers/verdict\endcsname{CONFIRMED}
\expandafter\gdef\csname svy@item/t20/nonusers/dip\endcsname{0}
\expandafter\gdef\csname svy@item/t20/nonusers/shapelow\endcsname{89}
\expandafter\gdef\csname svy@item/t20/nonusers/shapemid\endcsname{10}
\expandafter\gdef\csname svy@item/t20/nonusers/shapehigh\endcsname{1}
\expandafter\gdef\csname svy@item/t21/code\endcsname{T21}
\expandafter\gdef\csname svy@item/t21/pred\endcsname{A}
\expandafter\gdef\csname svy@item/t21/label\endcsname{carrying out mechanical refactors/cleanup/fixes}
\expandafter\gdef\csname svy@item/t21/usage/diff\endcsname{1.24}
\expandafter\gdef\csname svy@item/t21/usage/difflow\endcsname{0.91}
\expandafter\gdef\csname svy@item/t21/usage/diffhigh\endcsname{1.58}
\expandafter\gdef\csname svy@item/t21/usage/g\endcsname{1.24}
\expandafter\gdef\csname svy@item/t21/all/n\endcsname{235}
\expandafter\gdef\csname svy@item/t21/all/miss\endcsname{7}
\expandafter\gdef\csname svy@item/t21/all/median\endcsname{4.0}
\expandafter\gdef\csname svy@item/t21/all/cilow\endcsname{4.0}
\expandafter\gdef\csname svy@item/t21/all/cihigh\endcsname{4.0}
\expandafter\gdef\csname svy@item/t21/all/mean\endcsname{3.65}
\expandafter\gdef\csname svy@item/t21/all/meanlow\endcsname{3.50}
\expandafter\gdef\csname svy@item/t21/all/meanhigh\endcsname{3.80}
\expandafter\gdef\csname svy@item/t21/all/se\endcsname{0.075}
\expandafter\gdef\csname svy@item/t21/all/sd\endcsname{1.15}
\expandafter\gdef\csname svy@item/t21/all/sdhigh\endcsname{0}
\expandafter\gdef\csname svy@item/t21/all/verdict\endcsname{CONFIRMED}
\expandafter\gdef\csname svy@item/t21/all/dip\endcsname{0}
\expandafter\gdef\csname svy@item/t21/all/shapelow\endcsname{20}
\expandafter\gdef\csname svy@item/t21/all/shapemid\endcsname{14}
\expandafter\gdef\csname svy@item/t21/all/shapehigh\endcsname{66}
\expandafter\gdef\csname svy@item/t21/users/n\endcsname{168}
\expandafter\gdef\csname svy@item/t21/users/miss\endcsname{3}
\expandafter\gdef\csname svy@item/t21/users/median\endcsname{4.0}
\expandafter\gdef\csname svy@item/t21/users/cilow\endcsname{4.0}
\expandafter\gdef\csname svy@item/t21/users/cihigh\endcsname{4.0}
\expandafter\gdef\csname svy@item/t21/users/mean\endcsname{4.01}
\expandafter\gdef\csname svy@item/t21/users/meanlow\endcsname{3.88}
\expandafter\gdef\csname svy@item/t21/users/meanhigh\endcsname{4.15}
\expandafter\gdef\csname svy@item/t21/users/se\endcsname{0.069}
\expandafter\gdef\csname svy@item/t21/users/sd\endcsname{0.89}
\expandafter\gdef\csname svy@item/t21/users/sdhigh\endcsname{0}
\expandafter\gdef\csname svy@item/t21/users/verdict\endcsname{CONFIRMED}
\expandafter\gdef\csname svy@item/t21/users/dip\endcsname{0}
\expandafter\gdef\csname svy@item/t21/users/shapelow\endcsname{8}
\expandafter\gdef\csname svy@item/t21/users/shapemid\endcsname{14}
\expandafter\gdef\csname svy@item/t21/users/shapehigh\endcsname{78}
\expandafter\gdef\csname svy@item/t21/nonusers/n\endcsname{65}
\expandafter\gdef\csname svy@item/t21/nonusers/miss\endcsname{4}
\expandafter\gdef\csname svy@item/t21/nonusers/median\endcsname{3.0}
\expandafter\gdef\csname svy@item/t21/nonusers/cilow\endcsname{2.0}
\expandafter\gdef\csname svy@item/t21/nonusers/cihigh\endcsname{4.0}
\expandafter\gdef\csname svy@item/t21/nonusers/mean\endcsname{2.77}
\expandafter\gdef\csname svy@item/t21/nonusers/meanlow\endcsname{2.46}
\expandafter\gdef\csname svy@item/t21/nonusers/meanhigh\endcsname{3.08}
\expandafter\gdef\csname svy@item/t21/nonusers/se\endcsname{0.153}
\expandafter\gdef\csname svy@item/t21/nonusers/sd\endcsname{1.23}
\expandafter\gdef\csname svy@item/t21/nonusers/sdhigh\endcsname{1}
\expandafter\gdef\csname svy@item/t21/nonusers/verdict\endcsname{INCONCLUSIVE}
\expandafter\gdef\csname svy@item/t21/nonusers/dip\endcsname{1}
\expandafter\gdef\csname svy@item/t21/nonusers/shapelow\endcsname{46}
\expandafter\gdef\csname svy@item/t21/nonusers/shapemid\endcsname{14}
\expandafter\gdef\csname svy@item/t21/nonusers/shapehigh\endcsname{40}
\expandafter\gdef\csname svy@item/t22/code\endcsname{T22}
\expandafter\gdef\csname svy@item/t22/pred\endcsname{X}
\expandafter\gdef\csname svy@item/t22/label\endcsname{removing small imperfections near the end}
\expandafter\gdef\csname svy@item/t22/usage/diff\endcsname{1.03}
\expandafter\gdef\csname svy@item/t22/usage/difflow\endcsname{0.67}
\expandafter\gdef\csname svy@item/t22/usage/diffhigh\endcsname{1.40}
\expandafter\gdef\csname svy@item/t22/usage/g\endcsname{0.83}
\expandafter\gdef\csname svy@item/t22/all/n\endcsname{237}
\expandafter\gdef\csname svy@item/t22/all/miss\endcsname{5}
\expandafter\gdef\csname svy@item/t22/all/median\endcsname{3.0}
\expandafter\gdef\csname svy@item/t22/all/cilow\endcsname{3.0}
\expandafter\gdef\csname svy@item/t22/all/cihigh\endcsname{3.0}
\expandafter\gdef\csname svy@item/t22/all/mean\endcsname{3.00}
\expandafter\gdef\csname svy@item/t22/all/meanlow\endcsname{2.83}
\expandafter\gdef\csname svy@item/t22/all/meanhigh\endcsname{3.17}
\expandafter\gdef\csname svy@item/t22/all/se\endcsname{0.085}
\expandafter\gdef\csname svy@item/t22/all/sd\endcsname{1.31}
\expandafter\gdef\csname svy@item/t22/all/sdhigh\endcsname{0}
\expandafter\gdef\csname svy@item/t22/all/verdict\endcsname{NO_PREDICTION}
\expandafter\gdef\csname svy@item/t22/all/dip\endcsname{1}
\expandafter\gdef\csname svy@item/t22/all/shapelow\endcsname{40}
\expandafter\gdef\csname svy@item/t22/all/shapemid\endcsname{19}
\expandafter\gdef\csname svy@item/t22/all/shapehigh\endcsname{41}
\expandafter\gdef\csname svy@item/t22/users/n\endcsname{170}
\expandafter\gdef\csname svy@item/t22/users/miss\endcsname{1}
\expandafter\gdef\csname svy@item/t22/users/median\endcsname{4.0}
\expandafter\gdef\csname svy@item/t22/users/cilow\endcsname{3.0}
\expandafter\gdef\csname svy@item/t22/users/cihigh\endcsname{4.0}
\expandafter\gdef\csname svy@item/t22/users/mean\endcsname{3.29}
\expandafter\gdef\csname svy@item/t22/users/meanlow\endcsname{3.11}
\expandafter\gdef\csname svy@item/t22/users/meanhigh\endcsname{3.48}
\expandafter\gdef\csname svy@item/t22/users/se\endcsname{0.094}
\expandafter\gdef\csname svy@item/t22/users/sd\endcsname{1.22}
\expandafter\gdef\csname svy@item/t22/users/sdhigh\endcsname{1}
\expandafter\gdef\csname svy@item/t22/users/verdict\endcsname{NO_PREDICTION}
\expandafter\gdef\csname svy@item/t22/users/dip\endcsname{1}
\expandafter\gdef\csname svy@item/t22/users/shapelow\endcsname{30}
\expandafter\gdef\csname svy@item/t22/users/shapemid\endcsname{20}
\expandafter\gdef\csname svy@item/t22/users/shapehigh\endcsname{50}
\expandafter\gdef\csname svy@item/t22/nonusers/n\endcsname{65}
\expandafter\gdef\csname svy@item/t22/nonusers/miss\endcsname{4}
\expandafter\gdef\csname svy@item/t22/nonusers/median\endcsname{2.0}
\expandafter\gdef\csname svy@item/t22/nonusers/cilow\endcsname{2.0}
\expandafter\gdef\csname svy@item/t22/nonusers/cihigh\endcsname{2.0}
\expandafter\gdef\csname svy@item/t22/nonusers/mean\endcsname{2.26}
\expandafter\gdef\csname svy@item/t22/nonusers/meanlow\endcsname{1.95}
\expandafter\gdef\csname svy@item/t22/nonusers/meanhigh\endcsname{2.58}
\expandafter\gdef\csname svy@item/t22/nonusers/se\endcsname{0.157}
\expandafter\gdef\csname svy@item/t22/nonusers/sd\endcsname{1.27}
\expandafter\gdef\csname svy@item/t22/nonusers/sdhigh\endcsname{1}
\expandafter\gdef\csname svy@item/t22/nonusers/verdict\endcsname{NO_PREDICTION}
\expandafter\gdef\csname svy@item/t22/nonusers/dip\endcsname{0}
\expandafter\gdef\csname svy@item/t22/nonusers/shapelow\endcsname{64}
\expandafter\gdef\csname svy@item/t22/nonusers/shapemid\endcsname{17}
\expandafter\gdef\csname svy@item/t22/nonusers/shapehigh\endcsname{19}
\expandafter\gdef\csname svy@item/t23/code\endcsname{T23}
\expandafter\gdef\csname svy@item/t23/pred\endcsname{H}
\expandafter\gdef\csname svy@item/t23/label\endcsname{final judgment that a change is ready to ship}
\expandafter\gdef\csname svy@item/t23/usage/diff\endcsname{0.41}
\expandafter\gdef\csname svy@item/t23/usage/difflow\endcsname{0.21}
\expandafter\gdef\csname svy@item/t23/usage/diffhigh\endcsname{0.62}
\expandafter\gdef\csname svy@item/t23/usage/g\endcsname{0.51}
\expandafter\gdef\csname svy@item/t23/all/n\endcsname{239}
\expandafter\gdef\csname svy@item/t23/all/miss\endcsname{3}
\expandafter\gdef\csname svy@item/t23/all/median\endcsname{1.0}
\expandafter\gdef\csname svy@item/t23/all/cilow\endcsname{1.0}
\expandafter\gdef\csname svy@item/t23/all/cihigh\endcsname{2.0}
\expandafter\gdef\csname svy@item/t23/all/mean\endcsname{1.59}
\expandafter\gdef\csname svy@item/t23/all/meanlow\endcsname{1.49}
\expandafter\gdef\csname svy@item/t23/all/meanhigh\endcsname{1.70}
\expandafter\gdef\csname svy@item/t23/all/se\endcsname{0.053}
\expandafter\gdef\csname svy@item/t23/all/sd\endcsname{0.82}
\expandafter\gdef\csname svy@item/t23/all/sdhigh\endcsname{0}
\expandafter\gdef\csname svy@item/t23/all/verdict\endcsname{CONFIRMED}
\expandafter\gdef\csname svy@item/t23/all/dip\endcsname{0}
\expandafter\gdef\csname svy@item/t23/all/shapelow\endcsname{90}
\expandafter\gdef\csname svy@item/t23/all/shapemid\endcsname{6}
\expandafter\gdef\csname svy@item/t23/all/shapehigh\endcsname{4}
\expandafter\gdef\csname svy@item/t23/users/n\endcsname{170}
\expandafter\gdef\csname svy@item/t23/users/miss\endcsname{1}
\expandafter\gdef\csname svy@item/t23/users/median\endcsname{2.0}
\expandafter\gdef\csname svy@item/t23/users/cilow\endcsname{1.0}
\expandafter\gdef\csname svy@item/t23/users/cihigh\endcsname{2.0}
\expandafter\gdef\csname svy@item/t23/users/mean\endcsname{1.71}
\expandafter\gdef\csname svy@item/t23/users/meanlow\endcsname{1.58}
\expandafter\gdef\csname svy@item/t23/users/meanhigh\endcsname{1.84}
\expandafter\gdef\csname svy@item/t23/users/se\endcsname{0.065}
\expandafter\gdef\csname svy@item/t23/users/sd\endcsname{0.85}
\expandafter\gdef\csname svy@item/t23/users/sdhigh\endcsname{0}
\expandafter\gdef\csname svy@item/t23/users/verdict\endcsname{CONFIRMED}
\expandafter\gdef\csname svy@item/t23/users/dip\endcsname{0}
\expandafter\gdef\csname svy@item/t23/users/shapelow\endcsname{86}
\expandafter\gdef\csname svy@item/t23/users/shapemid\endcsname{8}
\expandafter\gdef\csname svy@item/t23/users/shapehigh\endcsname{6}
\expandafter\gdef\csname svy@item/t23/nonusers/n\endcsname{67}
\expandafter\gdef\csname svy@item/t23/nonusers/miss\endcsname{2}
\expandafter\gdef\csname svy@item/t23/nonusers/median\endcsname{1.0}
\expandafter\gdef\csname svy@item/t23/nonusers/cilow\endcsname{1.0}
\expandafter\gdef\csname svy@item/t23/nonusers/cihigh\endcsname{1.0}
\expandafter\gdef\csname svy@item/t23/nonusers/mean\endcsname{1.30}
\expandafter\gdef\csname svy@item/t23/nonusers/meanlow\endcsname{1.14}
\expandafter\gdef\csname svy@item/t23/nonusers/meanhigh\endcsname{1.46}
\expandafter\gdef\csname svy@item/t23/nonusers/se\endcsname{0.080}
\expandafter\gdef\csname svy@item/t23/nonusers/sd\endcsname{0.65}
\expandafter\gdef\csname svy@item/t23/nonusers/sdhigh\endcsname{0}
\expandafter\gdef\csname svy@item/t23/nonusers/verdict\endcsname{CONFIRMED}
\expandafter\gdef\csname svy@item/t23/nonusers/dip\endcsname{0}
\expandafter\gdef\csname svy@item/t23/nonusers/shapelow\endcsname{96}
\expandafter\gdef\csname svy@item/t23/nonusers/shapemid\endcsname{1}
\expandafter\gdef\csname svy@item/t23/nonusers/shapehigh\endcsname{3}
\expandafter\gdef\csname svy@item/t24/code\endcsname{T24}
\expandafter\gdef\csname svy@item/t24/pred\endcsname{H}
\expandafter\gdef\csname svy@item/t24/label\endcsname{what counts as acceptable use of AI}
\expandafter\gdef\csname svy@item/t24/usage/diff\endcsname{0.24}
\expandafter\gdef\csname svy@item/t24/usage/difflow\endcsname{0.05}
\expandafter\gdef\csname svy@item/t24/usage/diffhigh\endcsname{0.43}
\expandafter\gdef\csname svy@item/t24/usage/g\endcsname{0.34}
\expandafter\gdef\csname svy@item/t24/all/n\endcsname{228}
\expandafter\gdef\csname svy@item/t24/all/miss\endcsname{14}
\expandafter\gdef\csname svy@item/t24/all/median\endcsname{1.0}
\expandafter\gdef\csname svy@item/t24/all/cilow\endcsname{1.0}
\expandafter\gdef\csname svy@item/t24/all/cihigh\endcsname{1.0}
\expandafter\gdef\csname svy@item/t24/all/mean\endcsname{1.33}
\expandafter\gdef\csname svy@item/t24/all/meanlow\endcsname{1.24}
\expandafter\gdef\csname svy@item/t24/all/meanhigh\endcsname{1.42}
\expandafter\gdef\csname svy@item/t24/all/se\endcsname{0.047}
\expandafter\gdef\csname svy@item/t24/all/sd\endcsname{0.70}
\expandafter\gdef\csname svy@item/t24/all/sdhigh\endcsname{0}
\expandafter\gdef\csname svy@item/t24/all/verdict\endcsname{CONFIRMED}
\expandafter\gdef\csname svy@item/t24/all/dip\endcsname{0}
\expandafter\gdef\csname svy@item/t24/all/shapelow\endcsname{92}
\expandafter\gdef\csname svy@item/t24/all/shapemid\endcsname{6}
\expandafter\gdef\csname svy@item/t24/all/shapehigh\endcsname{2}
\expandafter\gdef\csname svy@item/t24/users/n\endcsname{159}
\expandafter\gdef\csname svy@item/t24/users/miss\endcsname{12}
\expandafter\gdef\csname svy@item/t24/users/median\endcsname{1.0}
\expandafter\gdef\csname svy@item/t24/users/cilow\endcsname{1.0}
\expandafter\gdef\csname svy@item/t24/users/cihigh\endcsname{1.0}
\expandafter\gdef\csname svy@item/t24/users/mean\endcsname{1.40}
\expandafter\gdef\csname svy@item/t24/users/meanlow\endcsname{1.29}
\expandafter\gdef\csname svy@item/t24/users/meanhigh\endcsname{1.52}
\expandafter\gdef\csname svy@item/t24/users/se\endcsname{0.057}
\expandafter\gdef\csname svy@item/t24/users/sd\endcsname{0.72}
\expandafter\gdef\csname svy@item/t24/users/sdhigh\endcsname{0}
\expandafter\gdef\csname svy@item/t24/users/verdict\endcsname{CONFIRMED}
\expandafter\gdef\csname svy@item/t24/users/dip\endcsname{0}
\expandafter\gdef\csname svy@item/t24/users/shapelow\endcsname{90}
\expandafter\gdef\csname svy@item/t24/users/shapemid\endcsname{8}
\expandafter\gdef\csname svy@item/t24/users/shapehigh\endcsname{2}
\expandafter\gdef\csname svy@item/t24/nonusers/n\endcsname{67}
\expandafter\gdef\csname svy@item/t24/nonusers/miss\endcsname{2}
\expandafter\gdef\csname svy@item/t24/nonusers/median\endcsname{1.0}
\expandafter\gdef\csname svy@item/t24/nonusers/cilow\endcsname{1.0}
\expandafter\gdef\csname svy@item/t24/nonusers/cihigh\endcsname{1.0}
\expandafter\gdef\csname svy@item/t24/nonusers/mean\endcsname{1.16}
\expandafter\gdef\csname svy@item/t24/nonusers/meanlow\endcsname{1.01}
\expandafter\gdef\csname svy@item/t24/nonusers/meanhigh\endcsname{1.32}
\expandafter\gdef\csname svy@item/t24/nonusers/se\endcsname{0.078}
\expandafter\gdef\csname svy@item/t24/nonusers/sd\endcsname{0.64}
\expandafter\gdef\csname svy@item/t24/nonusers/sdhigh\endcsname{0}
\expandafter\gdef\csname svy@item/t24/nonusers/verdict\endcsname{CONFIRMED}
\expandafter\gdef\csname svy@item/t24/nonusers/dip\endcsname{0}
\expandafter\gdef\csname svy@item/t24/nonusers/shapelow\endcsname{97}
\expandafter\gdef\csname svy@item/t24/nonusers/shapemid\endcsname{0}
\expandafter\gdef\csname svy@item/t24/nonusers/shapehigh\endcsname{3}
\expandafter\gdef\csname svy@item/accdev/code\endcsname{R01}
\expandafter\gdef\csname svy@item/accdev/pred\endcsname{Agree}
\expandafter\gdef\csname svy@item/accdev/label\endcsname{developer as accountable as if they wrote it}
\expandafter\gdef\csname svy@item/accdev/usage/diff\endcsname{-0.19}
\expandafter\gdef\csname svy@item/accdev/usage/difflow\endcsname{-0.41}
\expandafter\gdef\csname svy@item/accdev/usage/diffhigh\endcsname{0.03}
\expandafter\gdef\csname svy@item/accdev/usage/g\endcsname{-0.24}
\expandafter\gdef\csname svy@item/accdev/all/n\endcsname{231}
\expandafter\gdef\csname svy@item/accdev/all/miss\endcsname{11}
\expandafter\gdef\csname svy@item/accdev/all/median\endcsname{5.0}
\expandafter\gdef\csname svy@item/accdev/all/cilow\endcsname{5.0}
\expandafter\gdef\csname svy@item/accdev/all/cihigh\endcsname{5.0}
\expandafter\gdef\csname svy@item/accdev/all/mean\endcsname{4.52}
\expandafter\gdef\csname svy@item/accdev/all/meanlow\endcsname{4.42}
\expandafter\gdef\csname svy@item/accdev/all/meanhigh\endcsname{4.63}
\expandafter\gdef\csname svy@item/accdev/all/se\endcsname{0.052}
\expandafter\gdef\csname svy@item/accdev/all/sd\endcsname{0.78}
\expandafter\gdef\csname svy@item/accdev/all/sdhigh\endcsname{0}
\expandafter\gdef\csname svy@item/accdev/all/verdict\endcsname{CONFIRMED}
\expandafter\gdef\csname svy@item/accdev/all/dip\endcsname{0}
\expandafter\gdef\csname svy@item/accdev/all/shapelow\endcsname{2}
\expandafter\gdef\csname svy@item/accdev/all/shapemid\endcsname{7}
\expandafter\gdef\csname svy@item/accdev/all/shapehigh\endcsname{91}
\expandafter\gdef\csname svy@item/accdev/users/n\endcsname{162}
\expandafter\gdef\csname svy@item/accdev/users/miss\endcsname{9}
\expandafter\gdef\csname svy@item/accdev/users/median\endcsname{5.0}
\expandafter\gdef\csname svy@item/accdev/users/cilow\endcsname{5.0}
\expandafter\gdef\csname svy@item/accdev/users/cihigh\endcsname{5.0}
\expandafter\gdef\csname svy@item/accdev/users/mean\endcsname{4.47}
\expandafter\gdef\csname svy@item/accdev/users/meanlow\endcsname{4.35}
\expandafter\gdef\csname svy@item/accdev/users/meanhigh\endcsname{4.59}
\expandafter\gdef\csname svy@item/accdev/users/se\endcsname{0.063}
\expandafter\gdef\csname svy@item/accdev/users/sd\endcsname{0.80}
\expandafter\gdef\csname svy@item/accdev/users/sdhigh\endcsname{0}
\expandafter\gdef\csname svy@item/accdev/users/verdict\endcsname{CONFIRMED}
\expandafter\gdef\csname svy@item/accdev/users/dip\endcsname{0}
\expandafter\gdef\csname svy@item/accdev/users/shapelow\endcsname{2}
\expandafter\gdef\csname svy@item/accdev/users/shapemid\endcsname{8}
\expandafter\gdef\csname svy@item/accdev/users/shapehigh\endcsname{90}
\expandafter\gdef\csname svy@item/accdev/nonusers/n\endcsname{67}
\expandafter\gdef\csname svy@item/accdev/nonusers/miss\endcsname{2}
\expandafter\gdef\csname svy@item/accdev/nonusers/median\endcsname{5.0}
\expandafter\gdef\csname svy@item/accdev/nonusers/cilow\endcsname{5.0}
\expandafter\gdef\csname svy@item/accdev/nonusers/cihigh\endcsname{5.0}
\expandafter\gdef\csname svy@item/accdev/nonusers/mean\endcsname{4.66}
\expandafter\gdef\csname svy@item/accdev/nonusers/meanlow\endcsname{4.47}
\expandafter\gdef\csname svy@item/accdev/nonusers/meanhigh\endcsname{4.84}
\expandafter\gdef\csname svy@item/accdev/nonusers/se\endcsname{0.092}
\expandafter\gdef\csname svy@item/accdev/nonusers/sd\endcsname{0.75}
\expandafter\gdef\csname svy@item/accdev/nonusers/sdhigh\endcsname{0}
\expandafter\gdef\csname svy@item/accdev/nonusers/verdict\endcsname{CONFIRMED}
\expandafter\gdef\csname svy@item/accdev/nonusers/dip\endcsname{0}
\expandafter\gdef\csname svy@item/accdev/nonusers/shapelow\endcsname{3}
\expandafter\gdef\csname svy@item/accdev/nonusers/shapemid\endcsname{3}
\expandafter\gdef\csname svy@item/accdev/nonusers/shapehigh\endcsname{94}
\expandafter\gdef\csname svy@item/accreason/code\endcsname{R02}
\expandafter\gdef\csname svy@item/accreason/pred\endcsname{Agree}
\expandafter\gdef\csname svy@item/accreason/label\endcsname{failure cause can be AI error, human error, or both}
\expandafter\gdef\csname svy@item/accreason/usage/diff\endcsname{0.31}
\expandafter\gdef\csname svy@item/accreason/usage/difflow\endcsname{-0.03}
\expandafter\gdef\csname svy@item/accreason/usage/diffhigh\endcsname{0.65}
\expandafter\gdef\csname svy@item/accreason/usage/g\endcsname{0.30}
\expandafter\gdef\csname svy@item/accreason/all/n\endcsname{227}
\expandafter\gdef\csname svy@item/accreason/all/miss\endcsname{15}
\expandafter\gdef\csname svy@item/accreason/all/median\endcsname{4.0}
\expandafter\gdef\csname svy@item/accreason/all/cilow\endcsname{4.0}
\expandafter\gdef\csname svy@item/accreason/all/cihigh\endcsname{5.0}
\expandafter\gdef\csname svy@item/accreason/all/mean\endcsname{4.19}
\expandafter\gdef\csname svy@item/accreason/all/meanlow\endcsname{4.05}
\expandafter\gdef\csname svy@item/accreason/all/meanhigh\endcsname{4.32}
\expandafter\gdef\csname svy@item/accreason/all/se\endcsname{0.069}
\expandafter\gdef\csname svy@item/accreason/all/sd\endcsname{1.04}
\expandafter\gdef\csname svy@item/accreason/all/sdhigh\endcsname{0}
\expandafter\gdef\csname svy@item/accreason/all/verdict\endcsname{CONFIRMED}
\expandafter\gdef\csname svy@item/accreason/all/dip\endcsname{0}
\expandafter\gdef\csname svy@item/accreason/all/shapelow\endcsname{8}
\expandafter\gdef\csname svy@item/accreason/all/shapemid\endcsname{9}
\expandafter\gdef\csname svy@item/accreason/all/shapehigh\endcsname{83}
\expandafter\gdef\csname svy@item/accreason/users/n\endcsname{162}
\expandafter\gdef\csname svy@item/accreason/users/miss\endcsname{9}
\expandafter\gdef\csname svy@item/accreason/users/median\endcsname{5.0}
\expandafter\gdef\csname svy@item/accreason/users/cilow\endcsname{4.0}
\expandafter\gdef\csname svy@item/accreason/users/cihigh\endcsname{5.0}
\expandafter\gdef\csname svy@item/accreason/users/mean\endcsname{4.28}
\expandafter\gdef\csname svy@item/accreason/users/meanlow\endcsname{4.13}
\expandafter\gdef\csname svy@item/accreason/users/meanhigh\endcsname{4.42}
\expandafter\gdef\csname svy@item/accreason/users/se\endcsname{0.073}
\expandafter\gdef\csname svy@item/accreason/users/sd\endcsname{0.93}
\expandafter\gdef\csname svy@item/accreason/users/sdhigh\endcsname{0}
\expandafter\gdef\csname svy@item/accreason/users/verdict\endcsname{CONFIRMED}
\expandafter\gdef\csname svy@item/accreason/users/dip\endcsname{0}
\expandafter\gdef\csname svy@item/accreason/users/shapelow\endcsname{5}
\expandafter\gdef\csname svy@item/accreason/users/shapemid\endcsname{9}
\expandafter\gdef\csname svy@item/accreason/users/shapehigh\endcsname{86}
\expandafter\gdef\csname svy@item/accreason/nonusers/n\endcsname{63}
\expandafter\gdef\csname svy@item/accreason/nonusers/miss\endcsname{6}
\expandafter\gdef\csname svy@item/accreason/nonusers/median\endcsname{4.0}
\expandafter\gdef\csname svy@item/accreason/nonusers/cilow\endcsname{4.0}
\expandafter\gdef\csname svy@item/accreason/nonusers/cihigh\endcsname{5.0}
\expandafter\gdef\csname svy@item/accreason/nonusers/mean\endcsname{3.97}
\expandafter\gdef\csname svy@item/accreason/nonusers/meanlow\endcsname{3.66}
\expandafter\gdef\csname svy@item/accreason/nonusers/meanhigh\endcsname{4.28}
\expandafter\gdef\csname svy@item/accreason/nonusers/se\endcsname{0.155}
\expandafter\gdef\csname svy@item/accreason/nonusers/sd\endcsname{1.23}
\expandafter\gdef\csname svy@item/accreason/nonusers/sdhigh\endcsname{0}
\expandafter\gdef\csname svy@item/accreason/nonusers/verdict\endcsname{CONFIRMED}
\expandafter\gdef\csname svy@item/accreason/nonusers/dip\endcsname{0}
\expandafter\gdef\csname svy@item/accreason/nonusers/shapelow\endcsname{13}
\expandafter\gdef\csname svy@item/accreason/nonusers/shapemid\endcsname{10}
\expandafter\gdef\csname svy@item/accreason/nonusers/shapehigh\endcsname{77}
\expandafter\gdef\csname svy@item/t01/country/diff\endcsname{-0.04}
\expandafter\gdef\csname svy@item/t01/country/difflow\endcsname{-0.31}
\expandafter\gdef\csname svy@item/t01/country/diffhigh\endcsname{0.23}
\expandafter\gdef\csname svy@item/t01/country/g\endcsname{-0.04}
\expandafter\gdef\csname svy@item/t01/germany/n\endcsname{124}
\expandafter\gdef\csname svy@item/t01/germany/miss\endcsname{0}
\expandafter\gdef\csname svy@item/t01/germany/median\endcsname{2.0}
\expandafter\gdef\csname svy@item/t01/germany/cilow\endcsname{1.0}
\expandafter\gdef\csname svy@item/t01/germany/cihigh\endcsname{2.0}
\expandafter\gdef\csname svy@item/t01/germany/mean\endcsname{1.69}
\expandafter\gdef\csname svy@item/t01/germany/meanlow\endcsname{1.55}
\expandafter\gdef\csname svy@item/t01/germany/meanhigh\endcsname{1.84}
\expandafter\gdef\csname svy@item/t01/germany/se\endcsname{0.072}
\expandafter\gdef\csname svy@item/t01/germany/sd\endcsname{0.80}
\expandafter\gdef\csname svy@item/t01/germany/sdhigh\endcsname{0}
\expandafter\gdef\csname svy@item/t01/germany/verdict\endcsname{CONFIRMED}
\expandafter\gdef\csname svy@item/t01/germany/dip\endcsname{0}
\expandafter\gdef\csname svy@item/t01/germany/shapelow\endcsname{91}
\expandafter\gdef\csname svy@item/t01/germany/shapemid\endcsname{4}
\expandafter\gdef\csname svy@item/t01/germany/shapehigh\endcsname{5}
\expandafter\gdef\csname svy@item/t01/elsewhere/n\endcsname{71}
\expandafter\gdef\csname svy@item/t01/elsewhere/miss\endcsname{1}
\expandafter\gdef\csname svy@item/t01/elsewhere/median\endcsname{1.0}
\expandafter\gdef\csname svy@item/t01/elsewhere/cilow\endcsname{1.0}
\expandafter\gdef\csname svy@item/t01/elsewhere/cihigh\endcsname{2.0}
\expandafter\gdef\csname svy@item/t01/elsewhere/mean\endcsname{1.73}
\expandafter\gdef\csname svy@item/t01/elsewhere/meanlow\endcsname{1.50}
\expandafter\gdef\csname svy@item/t01/elsewhere/meanhigh\endcsname{1.97}
\expandafter\gdef\csname svy@item/t01/elsewhere/se\endcsname{0.117}
\expandafter\gdef\csname svy@item/t01/elsewhere/sd\endcsname{0.98}
\expandafter\gdef\csname svy@item/t01/elsewhere/sdhigh\endcsname{0}
\expandafter\gdef\csname svy@item/t01/elsewhere/verdict\endcsname{CONFIRMED}
\expandafter\gdef\csname svy@item/t01/elsewhere/dip\endcsname{0}
\expandafter\gdef\csname svy@item/t01/elsewhere/shapelow\endcsname{84}
\expandafter\gdef\csname svy@item/t01/elsewhere/shapemid\endcsname{8}
\expandafter\gdef\csname svy@item/t01/elsewhere/shapehigh\endcsname{8}
\expandafter\gdef\csname svy@item/t03/country/diff\endcsname{0.07}
\expandafter\gdef\csname svy@item/t03/country/difflow\endcsname{-0.24}
\expandafter\gdef\csname svy@item/t03/country/diffhigh\endcsname{0.38}
\expandafter\gdef\csname svy@item/t03/country/g\endcsname{0.07}
\expandafter\gdef\csname svy@item/t03/germany/n\endcsname{124}
\expandafter\gdef\csname svy@item/t03/germany/miss\endcsname{0}
\expandafter\gdef\csname svy@item/t03/germany/median\endcsname{2.0}
\expandafter\gdef\csname svy@item/t03/germany/cilow\endcsname{2.0}
\expandafter\gdef\csname svy@item/t03/germany/cihigh\endcsname{3.0}
\expandafter\gdef\csname svy@item/t03/germany/mean\endcsname{2.45}
\expandafter\gdef\csname svy@item/t03/germany/meanlow\endcsname{2.29}
\expandafter\gdef\csname svy@item/t03/germany/meanhigh\endcsname{2.61}
\expandafter\gdef\csname svy@item/t03/germany/se\endcsname{0.081}
\expandafter\gdef\csname svy@item/t03/germany/sd\endcsname{0.90}
\expandafter\gdef\csname svy@item/t03/germany/sdhigh\endcsname{0}
\expandafter\gdef\csname svy@item/t03/germany/verdict\endcsname{INCONCLUSIVE}
\expandafter\gdef\csname svy@item/t03/germany/dip\endcsname{0}
\expandafter\gdef\csname svy@item/t03/germany/shapelow\endcsname{55}
\expandafter\gdef\csname svy@item/t03/germany/shapemid\endcsname{35}
\expandafter\gdef\csname svy@item/t03/germany/shapehigh\endcsname{10}
\expandafter\gdef\csname svy@item/t03/elsewhere/n\endcsname{71}
\expandafter\gdef\csname svy@item/t03/elsewhere/miss\endcsname{1}
\expandafter\gdef\csname svy@item/t03/elsewhere/median\endcsname{2.0}
\expandafter\gdef\csname svy@item/t03/elsewhere/cilow\endcsname{2.0}
\expandafter\gdef\csname svy@item/t03/elsewhere/cihigh\endcsname{2.0}
\expandafter\gdef\csname svy@item/t03/elsewhere/mean\endcsname{2.38}
\expandafter\gdef\csname svy@item/t03/elsewhere/meanlow\endcsname{2.12}
\expandafter\gdef\csname svy@item/t03/elsewhere/meanhigh\endcsname{2.64}
\expandafter\gdef\csname svy@item/t03/elsewhere/se\endcsname{0.132}
\expandafter\gdef\csname svy@item/t03/elsewhere/sd\endcsname{1.11}
\expandafter\gdef\csname svy@item/t03/elsewhere/sdhigh\endcsname{0}
\expandafter\gdef\csname svy@item/t03/elsewhere/verdict\endcsname{INCONCLUSIVE}
\expandafter\gdef\csname svy@item/t03/elsewhere/dip\endcsname{1}
\expandafter\gdef\csname svy@item/t03/elsewhere/shapelow\endcsname{64}
\expandafter\gdef\csname svy@item/t03/elsewhere/shapemid\endcsname{15}
\expandafter\gdef\csname svy@item/t03/elsewhere/shapehigh\endcsname{21}
\expandafter\gdef\csname svy@item/t04/country/diff\endcsname{0.00}
\expandafter\gdef\csname svy@item/t04/country/difflow\endcsname{-0.31}
\expandafter\gdef\csname svy@item/t04/country/diffhigh\endcsname{0.32}
\expandafter\gdef\csname svy@item/t04/country/g\endcsname{0.00}
\expandafter\gdef\csname svy@item/t04/germany/n\endcsname{122}
\expandafter\gdef\csname svy@item/t04/germany/miss\endcsname{2}
\expandafter\gdef\csname svy@item/t04/germany/median\endcsname{2.0}
\expandafter\gdef\csname svy@item/t04/germany/cilow\endcsname{2.0}
\expandafter\gdef\csname svy@item/t04/germany/cihigh\endcsname{3.0}
\expandafter\gdef\csname svy@item/t04/germany/mean\endcsname{2.46}
\expandafter\gdef\csname svy@item/t04/germany/meanlow\endcsname{2.30}
\expandafter\gdef\csname svy@item/t04/germany/meanhigh\endcsname{2.62}
\expandafter\gdef\csname svy@item/t04/germany/se\endcsname{0.079}
\expandafter\gdef\csname svy@item/t04/germany/sd\endcsname{0.87}
\expandafter\gdef\csname svy@item/t04/germany/sdhigh\endcsname{0}
\expandafter\gdef\csname svy@item/t04/germany/verdict\endcsname{NO_PREDICTION}
\expandafter\gdef\csname svy@item/t04/germany/dip\endcsname{0}
\expandafter\gdef\csname svy@item/t04/germany/shapelow\endcsname{58}
\expandafter\gdef\csname svy@item/t04/germany/shapemid\endcsname{29}
\expandafter\gdef\csname svy@item/t04/germany/shapehigh\endcsname{13}
\expandafter\gdef\csname svy@item/t04/elsewhere/n\endcsname{70}
\expandafter\gdef\csname svy@item/t04/elsewhere/miss\endcsname{2}
\expandafter\gdef\csname svy@item/t04/elsewhere/median\endcsname{2.0}
\expandafter\gdef\csname svy@item/t04/elsewhere/cilow\endcsname{2.0}
\expandafter\gdef\csname svy@item/t04/elsewhere/cihigh\endcsname{3.0}
\expandafter\gdef\csname svy@item/t04/elsewhere/mean\endcsname{2.46}
\expandafter\gdef\csname svy@item/t04/elsewhere/meanlow\endcsname{2.18}
\expandafter\gdef\csname svy@item/t04/elsewhere/meanhigh\endcsname{2.73}
\expandafter\gdef\csname svy@item/t04/elsewhere/se\endcsname{0.139}
\expandafter\gdef\csname svy@item/t04/elsewhere/sd\endcsname{1.16}
\expandafter\gdef\csname svy@item/t04/elsewhere/sdhigh\endcsname{0}
\expandafter\gdef\csname svy@item/t04/elsewhere/verdict\endcsname{NO_PREDICTION}
\expandafter\gdef\csname svy@item/t04/elsewhere/dip\endcsname{0}
\expandafter\gdef\csname svy@item/t04/elsewhere/shapelow\endcsname{61}
\expandafter\gdef\csname svy@item/t04/elsewhere/shapemid\endcsname{19}
\expandafter\gdef\csname svy@item/t04/elsewhere/shapehigh\endcsname{20}
\expandafter\gdef\csname svy@item/t05/country/diff\endcsname{0.04}
\expandafter\gdef\csname svy@item/t05/country/difflow\endcsname{-0.29}
\expandafter\gdef\csname svy@item/t05/country/diffhigh\endcsname{0.38}
\expandafter\gdef\csname svy@item/t05/country/g\endcsname{0.04}
\expandafter\gdef\csname svy@item/t05/germany/n\endcsname{120}
\expandafter\gdef\csname svy@item/t05/germany/miss\endcsname{4}
\expandafter\gdef\csname svy@item/t05/germany/median\endcsname{4.0}
\expandafter\gdef\csname svy@item/t05/germany/cilow\endcsname{4.0}
\expandafter\gdef\csname svy@item/t05/germany/cihigh\endcsname{4.0}
\expandafter\gdef\csname svy@item/t05/germany/mean\endcsname{3.61}
\expandafter\gdef\csname svy@item/t05/germany/meanlow\endcsname{3.43}
\expandafter\gdef\csname svy@item/t05/germany/meanhigh\endcsname{3.79}
\expandafter\gdef\csname svy@item/t05/germany/se\endcsname{0.090}
\expandafter\gdef\csname svy@item/t05/germany/sd\endcsname{0.99}
\expandafter\gdef\csname svy@item/t05/germany/sdhigh\endcsname{0}
\expandafter\gdef\csname svy@item/t05/germany/verdict\endcsname{INCONCLUSIVE}
\expandafter\gdef\csname svy@item/t05/germany/dip\endcsname{0}
\expandafter\gdef\csname svy@item/t05/germany/shapelow\endcsname{15}
\expandafter\gdef\csname svy@item/t05/germany/shapemid\endcsname{26}
\expandafter\gdef\csname svy@item/t05/germany/shapehigh\endcsname{59}
\expandafter\gdef\csname svy@item/t05/elsewhere/n\endcsname{71}
\expandafter\gdef\csname svy@item/t05/elsewhere/miss\endcsname{1}
\expandafter\gdef\csname svy@item/t05/elsewhere/median\endcsname{4.0}
\expandafter\gdef\csname svy@item/t05/elsewhere/cilow\endcsname{3.0}
\expandafter\gdef\csname svy@item/t05/elsewhere/cihigh\endcsname{4.0}
\expandafter\gdef\csname svy@item/t05/elsewhere/mean\endcsname{3.56}
\expandafter\gdef\csname svy@item/t05/elsewhere/meanlow\endcsname{3.28}
\expandafter\gdef\csname svy@item/t05/elsewhere/meanhigh\endcsname{3.85}
\expandafter\gdef\csname svy@item/t05/elsewhere/se\endcsname{0.141}
\expandafter\gdef\csname svy@item/t05/elsewhere/sd\endcsname{1.19}
\expandafter\gdef\csname svy@item/t05/elsewhere/sdhigh\endcsname{0}
\expandafter\gdef\csname svy@item/t05/elsewhere/verdict\endcsname{INCONCLUSIVE}
\expandafter\gdef\csname svy@item/t05/elsewhere/dip\endcsname{0}
\expandafter\gdef\csname svy@item/t05/elsewhere/shapelow\endcsname{20}
\expandafter\gdef\csname svy@item/t05/elsewhere/shapemid\endcsname{21}
\expandafter\gdef\csname svy@item/t05/elsewhere/shapehigh\endcsname{59}
\expandafter\gdef\csname svy@item/t06/country/diff\endcsname{0.32}
\expandafter\gdef\csname svy@item/t06/country/difflow\endcsname{-0.03}
\expandafter\gdef\csname svy@item/t06/country/diffhigh\endcsname{0.67}
\expandafter\gdef\csname svy@item/t06/country/g\endcsname{0.28}
\expandafter\gdef\csname svy@item/t06/germany/n\endcsname{121}
\expandafter\gdef\csname svy@item/t06/germany/miss\endcsname{3}
\expandafter\gdef\csname svy@item/t06/germany/median\endcsname{4.0}
\expandafter\gdef\csname svy@item/t06/germany/cilow\endcsname{4.0}
\expandafter\gdef\csname svy@item/t06/germany/cihigh\endcsname{4.0}
\expandafter\gdef\csname svy@item/t06/germany/mean\endcsname{3.97}
\expandafter\gdef\csname svy@item/t06/germany/meanlow\endcsname{3.78}
\expandafter\gdef\csname svy@item/t06/germany/meanhigh\endcsname{4.15}
\expandafter\gdef\csname svy@item/t06/germany/se\endcsname{0.095}
\expandafter\gdef\csname svy@item/t06/germany/sd\endcsname{1.04}
\expandafter\gdef\csname svy@item/t06/germany/sdhigh\endcsname{0}
\expandafter\gdef\csname svy@item/t06/germany/verdict\endcsname{CONFIRMED}
\expandafter\gdef\csname svy@item/t06/germany/dip\endcsname{0}
\expandafter\gdef\csname svy@item/t06/germany/shapelow\endcsname{11}
\expandafter\gdef\csname svy@item/t06/germany/shapemid\endcsname{13}
\expandafter\gdef\csname svy@item/t06/germany/shapehigh\endcsname{76}
\expandafter\gdef\csname svy@item/t06/elsewhere/n\endcsname{71}
\expandafter\gdef\csname svy@item/t06/elsewhere/miss\endcsname{1}
\expandafter\gdef\csname svy@item/t06/elsewhere/median\endcsname{4.0}
\expandafter\gdef\csname svy@item/t06/elsewhere/cilow\endcsname{4.0}
\expandafter\gdef\csname svy@item/t06/elsewhere/cihigh\endcsname{4.0}
\expandafter\gdef\csname svy@item/t06/elsewhere/mean\endcsname{3.65}
\expandafter\gdef\csname svy@item/t06/elsewhere/meanlow\endcsname{3.35}
\expandafter\gdef\csname svy@item/t06/elsewhere/meanhigh\endcsname{3.94}
\expandafter\gdef\csname svy@item/t06/elsewhere/se\endcsname{0.149}
\expandafter\gdef\csname svy@item/t06/elsewhere/sd\endcsname{1.25}
\expandafter\gdef\csname svy@item/t06/elsewhere/sdhigh\endcsname{0}
\expandafter\gdef\csname svy@item/t06/elsewhere/verdict\endcsname{INCONCLUSIVE}
\expandafter\gdef\csname svy@item/t06/elsewhere/dip\endcsname{0}
\expandafter\gdef\csname svy@item/t06/elsewhere/shapelow\endcsname{21}
\expandafter\gdef\csname svy@item/t06/elsewhere/shapemid\endcsname{13}
\expandafter\gdef\csname svy@item/t06/elsewhere/shapehigh\endcsname{66}
\expandafter\gdef\csname svy@item/t07/country/diff\endcsname{0.00}
\expandafter\gdef\csname svy@item/t07/country/difflow\endcsname{-0.37}
\expandafter\gdef\csname svy@item/t07/country/diffhigh\endcsname{0.37}
\expandafter\gdef\csname svy@item/t07/country/g\endcsname{0.00}
\expandafter\gdef\csname svy@item/t07/germany/n\endcsname{122}
\expandafter\gdef\csname svy@item/t07/germany/miss\endcsname{2}
\expandafter\gdef\csname svy@item/t07/germany/median\endcsname{2.0}
\expandafter\gdef\csname svy@item/t07/germany/cilow\endcsname{2.0}
\expandafter\gdef\csname svy@item/t07/germany/cihigh\endcsname{3.0}
\expandafter\gdef\csname svy@item/t07/germany/mean\endcsname{2.52}
\expandafter\gdef\csname svy@item/t07/germany/meanlow\endcsname{2.32}
\expandafter\gdef\csname svy@item/t07/germany/meanhigh\endcsname{2.71}
\expandafter\gdef\csname svy@item/t07/germany/se\endcsname{0.097}
\expandafter\gdef\csname svy@item/t07/germany/sd\endcsname{1.07}
\expandafter\gdef\csname svy@item/t07/germany/sdhigh\endcsname{0}
\expandafter\gdef\csname svy@item/t07/germany/verdict\endcsname{INCONCLUSIVE}
\expandafter\gdef\csname svy@item/t07/germany/dip\endcsname{0}
\expandafter\gdef\csname svy@item/t07/germany/shapelow\endcsname{53}
\expandafter\gdef\csname svy@item/t07/germany/shapemid\endcsname{27}
\expandafter\gdef\csname svy@item/t07/germany/shapehigh\endcsname{20}
\expandafter\gdef\csname svy@item/t07/elsewhere/n\endcsname{70}
\expandafter\gdef\csname svy@item/t07/elsewhere/miss\endcsname{2}
\expandafter\gdef\csname svy@item/t07/elsewhere/median\endcsname{2.0}
\expandafter\gdef\csname svy@item/t07/elsewhere/cilow\endcsname{2.0}
\expandafter\gdef\csname svy@item/t07/elsewhere/cihigh\endcsname{3.0}
\expandafter\gdef\csname svy@item/t07/elsewhere/mean\endcsname{2.51}
\expandafter\gdef\csname svy@item/t07/elsewhere/meanlow\endcsname{2.20}
\expandafter\gdef\csname svy@item/t07/elsewhere/meanhigh\endcsname{2.83}
\expandafter\gdef\csname svy@item/t07/elsewhere/se\endcsname{0.159}
\expandafter\gdef\csname svy@item/t07/elsewhere/sd\endcsname{1.33}
\expandafter\gdef\csname svy@item/t07/elsewhere/sdhigh\endcsname{0}
\expandafter\gdef\csname svy@item/t07/elsewhere/verdict\endcsname{INCONCLUSIVE}
\expandafter\gdef\csname svy@item/t07/elsewhere/dip\endcsname{1}
\expandafter\gdef\csname svy@item/t07/elsewhere/shapelow\endcsname{58}
\expandafter\gdef\csname svy@item/t07/elsewhere/shapemid\endcsname{14}
\expandafter\gdef\csname svy@item/t07/elsewhere/shapehigh\endcsname{28}
\expandafter\gdef\csname svy@item/t09/country/diff\endcsname{-0.12}
\expandafter\gdef\csname svy@item/t09/country/difflow\endcsname{-0.49}
\expandafter\gdef\csname svy@item/t09/country/diffhigh\endcsname{0.25}
\expandafter\gdef\csname svy@item/t09/country/g\endcsname{-0.11}
\expandafter\gdef\csname svy@item/t09/germany/n\endcsname{121}
\expandafter\gdef\csname svy@item/t09/germany/miss\endcsname{3}
\expandafter\gdef\csname svy@item/t09/germany/median\endcsname{2.0}
\expandafter\gdef\csname svy@item/t09/germany/cilow\endcsname{2.0}
\expandafter\gdef\csname svy@item/t09/germany/cihigh\endcsname{2.0}
\expandafter\gdef\csname svy@item/t09/germany/mean\endcsname{2.21}
\expandafter\gdef\csname svy@item/t09/germany/meanlow\endcsname{2.03}
\expandafter\gdef\csname svy@item/t09/germany/meanhigh\endcsname{2.39}
\expandafter\gdef\csname svy@item/t09/germany/se\endcsname{0.092}
\expandafter\gdef\csname svy@item/t09/germany/sd\endcsname{1.01}
\expandafter\gdef\csname svy@item/t09/germany/sdhigh\endcsname{0}
\expandafter\gdef\csname svy@item/t09/germany/verdict\endcsname{CONFIRMED}
\expandafter\gdef\csname svy@item/t09/germany/dip\endcsname{0}
\expandafter\gdef\csname svy@item/t09/germany/shapelow\endcsname{65}
\expandafter\gdef\csname svy@item/t09/germany/shapemid\endcsname{23}
\expandafter\gdef\csname svy@item/t09/germany/shapehigh\endcsname{12}
\expandafter\gdef\csname svy@item/t09/elsewhere/n\endcsname{64}
\expandafter\gdef\csname svy@item/t09/elsewhere/miss\endcsname{8}
\expandafter\gdef\csname svy@item/t09/elsewhere/median\endcsname{2.0}
\expandafter\gdef\csname svy@item/t09/elsewhere/cilow\endcsname{2.0}
\expandafter\gdef\csname svy@item/t09/elsewhere/cihigh\endcsname{2.0}
\expandafter\gdef\csname svy@item/t09/elsewhere/mean\endcsname{2.33}
\expandafter\gdef\csname svy@item/t09/elsewhere/meanlow\endcsname{2.00}
\expandafter\gdef\csname svy@item/t09/elsewhere/meanhigh\endcsname{2.65}
\expandafter\gdef\csname svy@item/t09/elsewhere/se\endcsname{0.162}
\expandafter\gdef\csname svy@item/t09/elsewhere/sd\endcsname{1.30}
\expandafter\gdef\csname svy@item/t09/elsewhere/sdhigh\endcsname{0}
\expandafter\gdef\csname svy@item/t09/elsewhere/verdict\endcsname{INCONCLUSIVE}
\expandafter\gdef\csname svy@item/t09/elsewhere/dip\endcsname{1}
\expandafter\gdef\csname svy@item/t09/elsewhere/shapelow\endcsname{67}
\expandafter\gdef\csname svy@item/t09/elsewhere/shapemid\endcsname{11}
\expandafter\gdef\csname svy@item/t09/elsewhere/shapehigh\endcsname{22}
\expandafter\gdef\csname svy@item/t10/country/diff\endcsname{0.27}
\expandafter\gdef\csname svy@item/t10/country/difflow\endcsname{-0.00}
\expandafter\gdef\csname svy@item/t10/country/diffhigh\endcsname{0.55}
\expandafter\gdef\csname svy@item/t10/country/g\endcsname{0.30}
\expandafter\gdef\csname svy@item/t10/germany/n\endcsname{122}
\expandafter\gdef\csname svy@item/t10/germany/miss\endcsname{2}
\expandafter\gdef\csname svy@item/t10/germany/median\endcsname{2.0}
\expandafter\gdef\csname svy@item/t10/germany/cilow\endcsname{2.0}
\expandafter\gdef\csname svy@item/t10/germany/cihigh\endcsname{2.0}
\expandafter\gdef\csname svy@item/t10/germany/mean\endcsname{1.95}
\expandafter\gdef\csname svy@item/t10/germany/meanlow\endcsname{1.78}
\expandafter\gdef\csname svy@item/t10/germany/meanhigh\endcsname{2.12}
\expandafter\gdef\csname svy@item/t10/germany/se\endcsname{0.085}
\expandafter\gdef\csname svy@item/t10/germany/sd\endcsname{0.93}
\expandafter\gdef\csname svy@item/t10/germany/sdhigh\endcsname{0}
\expandafter\gdef\csname svy@item/t10/germany/verdict\endcsname{CONFIRMED}
\expandafter\gdef\csname svy@item/t10/germany/dip\endcsname{0}
\expandafter\gdef\csname svy@item/t10/germany/shapelow\endcsname{78}
\expandafter\gdef\csname svy@item/t10/germany/shapemid\endcsname{14}
\expandafter\gdef\csname svy@item/t10/germany/shapehigh\endcsname{8}
\expandafter\gdef\csname svy@item/t10/elsewhere/n\endcsname{65}
\expandafter\gdef\csname svy@item/t10/elsewhere/miss\endcsname{7}
\expandafter\gdef\csname svy@item/t10/elsewhere/median\endcsname{1.0}
\expandafter\gdef\csname svy@item/t10/elsewhere/cilow\endcsname{1.0}
\expandafter\gdef\csname svy@item/t10/elsewhere/cihigh\endcsname{2.0}
\expandafter\gdef\csname svy@item/t10/elsewhere/mean\endcsname{1.68}
\expandafter\gdef\csname svy@item/t10/elsewhere/meanlow\endcsname{1.45}
\expandafter\gdef\csname svy@item/t10/elsewhere/meanhigh\endcsname{1.90}
\expandafter\gdef\csname svy@item/t10/elsewhere/se\endcsname{0.112}
\expandafter\gdef\csname svy@item/t10/elsewhere/sd\endcsname{0.90}
\expandafter\gdef\csname svy@item/t10/elsewhere/sdhigh\endcsname{0}
\expandafter\gdef\csname svy@item/t10/elsewhere/verdict\endcsname{CONFIRMED}
\expandafter\gdef\csname svy@item/t10/elsewhere/dip\endcsname{0}
\expandafter\gdef\csname svy@item/t10/elsewhere/shapelow\endcsname{87}
\expandafter\gdef\csname svy@item/t10/elsewhere/shapemid\endcsname{6}
\expandafter\gdef\csname svy@item/t10/elsewhere/shapehigh\endcsname{7}
\expandafter\gdef\csname svy@item/t11/country/diff\endcsname{-0.14}
\expandafter\gdef\csname svy@item/t11/country/difflow\endcsname{-0.46}
\expandafter\gdef\csname svy@item/t11/country/diffhigh\endcsname{0.19}
\expandafter\gdef\csname svy@item/t11/country/g\endcsname{-0.13}
\expandafter\gdef\csname svy@item/t11/germany/n\endcsname{117}
\expandafter\gdef\csname svy@item/t11/germany/miss\endcsname{7}
\expandafter\gdef\csname svy@item/t11/germany/median\endcsname{2.0}
\expandafter\gdef\csname svy@item/t11/germany/cilow\endcsname{2.0}
\expandafter\gdef\csname svy@item/t11/germany/cihigh\endcsname{2.0}
\expandafter\gdef\csname svy@item/t11/germany/mean\endcsname{2.03}
\expandafter\gdef\csname svy@item/t11/germany/meanlow\endcsname{1.85}
\expandafter\gdef\csname svy@item/t11/germany/meanhigh\endcsname{2.22}
\expandafter\gdef\csname svy@item/t11/germany/se\endcsname{0.093}
\expandafter\gdef\csname svy@item/t11/germany/sd\endcsname{1.01}
\expandafter\gdef\csname svy@item/t11/germany/sdhigh\endcsname{0}
\expandafter\gdef\csname svy@item/t11/germany/verdict\endcsname{NO_PREDICTION}
\expandafter\gdef\csname svy@item/t11/germany/dip\endcsname{0}
\expandafter\gdef\csname svy@item/t11/germany/shapelow\endcsname{74}
\expandafter\gdef\csname svy@item/t11/germany/shapemid\endcsname{16}
\expandafter\gdef\csname svy@item/t11/germany/shapehigh\endcsname{10}
\expandafter\gdef\csname svy@item/t11/elsewhere/n\endcsname{70}
\expandafter\gdef\csname svy@item/t11/elsewhere/miss\endcsname{2}
\expandafter\gdef\csname svy@item/t11/elsewhere/median\endcsname{2.0}
\expandafter\gdef\csname svy@item/t11/elsewhere/cilow\endcsname{2.0}
\expandafter\gdef\csname svy@item/t11/elsewhere/cihigh\endcsname{2.0}
\expandafter\gdef\csname svy@item/t11/elsewhere/mean\endcsname{2.17}
\expandafter\gdef\csname svy@item/t11/elsewhere/meanlow\endcsname{1.90}
\expandafter\gdef\csname svy@item/t11/elsewhere/meanhigh\endcsname{2.44}
\expandafter\gdef\csname svy@item/t11/elsewhere/se\endcsname{0.136}
\expandafter\gdef\csname svy@item/t11/elsewhere/sd\endcsname{1.14}
\expandafter\gdef\csname svy@item/t11/elsewhere/sdhigh\endcsname{0}
\expandafter\gdef\csname svy@item/t11/elsewhere/verdict\endcsname{NO_PREDICTION}
\expandafter\gdef\csname svy@item/t11/elsewhere/dip\endcsname{0}
\expandafter\gdef\csname svy@item/t11/elsewhere/shapelow\endcsname{73}
\expandafter\gdef\csname svy@item/t11/elsewhere/shapemid\endcsname{10}
\expandafter\gdef\csname svy@item/t11/elsewhere/shapehigh\endcsname{17}
\expandafter\gdef\csname svy@item/t12/country/diff\endcsname{-0.27}
\expandafter\gdef\csname svy@item/t12/country/difflow\endcsname{-0.66}
\expandafter\gdef\csname svy@item/t12/country/diffhigh\endcsname{0.12}
\expandafter\gdef\csname svy@item/t12/country/g\endcsname{-0.22}
\expandafter\gdef\csname svy@item/t12/germany/n\endcsname{120}
\expandafter\gdef\csname svy@item/t12/germany/miss\endcsname{4}
\expandafter\gdef\csname svy@item/t12/germany/median\endcsname{3.0}
\expandafter\gdef\csname svy@item/t12/germany/cilow\endcsname{3.0}
\expandafter\gdef\csname svy@item/t12/germany/cihigh\endcsname{4.0}
\expandafter\gdef\csname svy@item/t12/germany/mean\endcsname{3.08}
\expandafter\gdef\csname svy@item/t12/germany/meanlow\endcsname{2.87}
\expandafter\gdef\csname svy@item/t12/germany/meanhigh\endcsname{3.30}
\expandafter\gdef\csname svy@item/t12/germany/se\endcsname{0.108}
\expandafter\gdef\csname svy@item/t12/germany/sd\endcsname{1.19}
\expandafter\gdef\csname svy@item/t12/germany/sdhigh\endcsname{0}
\expandafter\gdef\csname svy@item/t12/germany/verdict\endcsname{NO_PREDICTION}
\expandafter\gdef\csname svy@item/t12/germany/dip\endcsname{0}
\expandafter\gdef\csname svy@item/t12/germany/shapelow\endcsname{34}
\expandafter\gdef\csname svy@item/t12/germany/shapemid\endcsname{22}
\expandafter\gdef\csname svy@item/t12/germany/shapehigh\endcsname{44}
\expandafter\gdef\csname svy@item/t12/elsewhere/n\endcsname{70}
\expandafter\gdef\csname svy@item/t12/elsewhere/miss\endcsname{2}
\expandafter\gdef\csname svy@item/t12/elsewhere/median\endcsname{4.0}
\expandafter\gdef\csname svy@item/t12/elsewhere/cilow\endcsname{3.0}
\expandafter\gdef\csname svy@item/t12/elsewhere/cihigh\endcsname{4.0}
\expandafter\gdef\csname svy@item/t12/elsewhere/mean\endcsname{3.36}
\expandafter\gdef\csname svy@item/t12/elsewhere/meanlow\endcsname{3.03}
\expandafter\gdef\csname svy@item/t12/elsewhere/meanhigh\endcsname{3.69}
\expandafter\gdef\csname svy@item/t12/elsewhere/se\endcsname{0.165}
\expandafter\gdef\csname svy@item/t12/elsewhere/sd\endcsname{1.38}
\expandafter\gdef\csname svy@item/t12/elsewhere/sdhigh\endcsname{0}
\expandafter\gdef\csname svy@item/t12/elsewhere/verdict\endcsname{NO_PREDICTION}
\expandafter\gdef\csname svy@item/t12/elsewhere/dip\endcsname{1}
\expandafter\gdef\csname svy@item/t12/elsewhere/shapelow\endcsname{31}
\expandafter\gdef\csname svy@item/t12/elsewhere/shapemid\endcsname{10}
\expandafter\gdef\csname svy@item/t12/elsewhere/shapehigh\endcsname{59}
\expandafter\gdef\csname svy@item/t13/country/diff\endcsname{-0.04}
\expandafter\gdef\csname svy@item/t13/country/difflow\endcsname{-0.45}
\expandafter\gdef\csname svy@item/t13/country/diffhigh\endcsname{0.38}
\expandafter\gdef\csname svy@item/t13/country/g\endcsname{-0.03}
\expandafter\gdef\csname svy@item/t13/germany/n\endcsname{114}
\expandafter\gdef\csname svy@item/t13/germany/miss\endcsname{10}
\expandafter\gdef\csname svy@item/t13/germany/median\endcsname{3.0}
\expandafter\gdef\csname svy@item/t13/germany/cilow\endcsname{3.0}
\expandafter\gdef\csname svy@item/t13/germany/cihigh\endcsname{3.0}
\expandafter\gdef\csname svy@item/t13/germany/mean\endcsname{2.96}
\expandafter\gdef\csname svy@item/t13/germany/meanlow\endcsname{2.74}
\expandafter\gdef\csname svy@item/t13/germany/meanhigh\endcsname{3.19}
\expandafter\gdef\csname svy@item/t13/germany/se\endcsname{0.113}
\expandafter\gdef\csname svy@item/t13/germany/sd\endcsname{1.20}
\expandafter\gdef\csname svy@item/t13/germany/sdhigh\endcsname{0}
\expandafter\gdef\csname svy@item/t13/germany/verdict\endcsname{REJECTED}
\expandafter\gdef\csname svy@item/t13/germany/dip\endcsname{0}
\expandafter\gdef\csname svy@item/t13/germany/shapelow\endcsname{34}
\expandafter\gdef\csname svy@item/t13/germany/shapemid\endcsname{35}
\expandafter\gdef\csname svy@item/t13/germany/shapehigh\endcsname{31}
\expandafter\gdef\csname svy@item/t13/elsewhere/n\endcsname{67}
\expandafter\gdef\csname svy@item/t13/elsewhere/miss\endcsname{5}
\expandafter\gdef\csname svy@item/t13/elsewhere/median\endcsname{3.0}
\expandafter\gdef\csname svy@item/t13/elsewhere/cilow\endcsname{2.0}
\expandafter\gdef\csname svy@item/t13/elsewhere/cihigh\endcsname{4.0}
\expandafter\gdef\csname svy@item/t13/elsewhere/mean\endcsname{3.00}
\expandafter\gdef\csname svy@item/t13/elsewhere/meanlow\endcsname{2.65}
\expandafter\gdef\csname svy@item/t13/elsewhere/meanhigh\endcsname{3.35}
\expandafter\gdef\csname svy@item/t13/elsewhere/se\endcsname{0.175}
\expandafter\gdef\csname svy@item/t13/elsewhere/sd\endcsname{1.44}
\expandafter\gdef\csname svy@item/t13/elsewhere/sdhigh\endcsname{0}
\expandafter\gdef\csname svy@item/t13/elsewhere/verdict\endcsname{INCONCLUSIVE}
\expandafter\gdef\csname svy@item/t13/elsewhere/dip\endcsname{1}
\expandafter\gdef\csname svy@item/t13/elsewhere/shapelow\endcsname{42}
\expandafter\gdef\csname svy@item/t13/elsewhere/shapemid\endcsname{13}
\expandafter\gdef\csname svy@item/t13/elsewhere/shapehigh\endcsname{45}
\expandafter\gdef\csname svy@item/t16/country/diff\endcsname{-0.06}
\expandafter\gdef\csname svy@item/t16/country/difflow\endcsname{-0.37}
\expandafter\gdef\csname svy@item/t16/country/diffhigh\endcsname{0.25}
\expandafter\gdef\csname svy@item/t16/country/g\endcsname{-0.06}
\expandafter\gdef\csname svy@item/t16/germany/n\endcsname{118}
\expandafter\gdef\csname svy@item/t16/germany/miss\endcsname{6}
\expandafter\gdef\csname svy@item/t16/germany/median\endcsname{2.0}
\expandafter\gdef\csname svy@item/t16/germany/cilow\endcsname{2.0}
\expandafter\gdef\csname svy@item/t16/germany/cihigh\endcsname{2.0}
\expandafter\gdef\csname svy@item/t16/germany/mean\endcsname{1.90}
\expandafter\gdef\csname svy@item/t16/germany/meanlow\endcsname{1.73}
\expandafter\gdef\csname svy@item/t16/germany/meanhigh\endcsname{2.07}
\expandafter\gdef\csname svy@item/t16/germany/se\endcsname{0.085}
\expandafter\gdef\csname svy@item/t16/germany/sd\endcsname{0.92}
\expandafter\gdef\csname svy@item/t16/germany/sdhigh\endcsname{0}
\expandafter\gdef\csname svy@item/t16/germany/verdict\endcsname{CONFIRMED}
\expandafter\gdef\csname svy@item/t16/germany/dip\endcsname{0}
\expandafter\gdef\csname svy@item/t16/germany/shapelow\endcsname{80}
\expandafter\gdef\csname svy@item/t16/germany/shapemid\endcsname{12}
\expandafter\gdef\csname svy@item/t16/germany/shapehigh\endcsname{8}
\expandafter\gdef\csname svy@item/t16/elsewhere/n\endcsname{70}
\expandafter\gdef\csname svy@item/t16/elsewhere/miss\endcsname{2}
\expandafter\gdef\csname svy@item/t16/elsewhere/median\endcsname{2.0}
\expandafter\gdef\csname svy@item/t16/elsewhere/cilow\endcsname{1.0}
\expandafter\gdef\csname svy@item/t16/elsewhere/cihigh\endcsname{2.0}
\expandafter\gdef\csname svy@item/t16/elsewhere/mean\endcsname{1.96}
\expandafter\gdef\csname svy@item/t16/elsewhere/meanlow\endcsname{1.70}
\expandafter\gdef\csname svy@item/t16/elsewhere/meanhigh\endcsname{2.22}
\expandafter\gdef\csname svy@item/t16/elsewhere/se\endcsname{0.131}
\expandafter\gdef\csname svy@item/t16/elsewhere/sd\endcsname{1.10}
\expandafter\gdef\csname svy@item/t16/elsewhere/sdhigh\endcsname{0}
\expandafter\gdef\csname svy@item/t16/elsewhere/verdict\endcsname{CONFIRMED}
\expandafter\gdef\csname svy@item/t16/elsewhere/dip\endcsname{0}
\expandafter\gdef\csname svy@item/t16/elsewhere/shapelow\endcsname{73}
\expandafter\gdef\csname svy@item/t16/elsewhere/shapemid\endcsname{14}
\expandafter\gdef\csname svy@item/t16/elsewhere/shapehigh\endcsname{13}
\expandafter\gdef\csname svy@item/t17/country/diff\endcsname{-0.09}
\expandafter\gdef\csname svy@item/t17/country/difflow\endcsname{-0.46}
\expandafter\gdef\csname svy@item/t17/country/diffhigh\endcsname{0.29}
\expandafter\gdef\csname svy@item/t17/country/g\endcsname{-0.07}
\expandafter\gdef\csname svy@item/t17/germany/n\endcsname{120}
\expandafter\gdef\csname svy@item/t17/germany/miss\endcsname{4}
\expandafter\gdef\csname svy@item/t17/germany/median\endcsname{3.0}
\expandafter\gdef\csname svy@item/t17/germany/cilow\endcsname{3.0}
\expandafter\gdef\csname svy@item/t17/germany/cihigh\endcsname{4.0}
\expandafter\gdef\csname svy@item/t17/germany/mean\endcsname{3.32}
\expandafter\gdef\csname svy@item/t17/germany/meanlow\endcsname{3.12}
\expandafter\gdef\csname svy@item/t17/germany/meanhigh\endcsname{3.51}
\expandafter\gdef\csname svy@item/t17/germany/se\endcsname{0.098}
\expandafter\gdef\csname svy@item/t17/germany/sd\endcsname{1.07}
\expandafter\gdef\csname svy@item/t17/germany/sdhigh\endcsname{0}
\expandafter\gdef\csname svy@item/t17/germany/verdict\endcsname{INCONCLUSIVE}
\expandafter\gdef\csname svy@item/t17/germany/dip\endcsname{0}
\expandafter\gdef\csname svy@item/t17/germany/shapelow\endcsname{26}
\expandafter\gdef\csname svy@item/t17/germany/shapemid\endcsname{27}
\expandafter\gdef\csname svy@item/t17/germany/shapehigh\endcsname{47}
\expandafter\gdef\csname svy@item/t17/elsewhere/n\endcsname{72}
\expandafter\gdef\csname svy@item/t17/elsewhere/miss\endcsname{0}
\expandafter\gdef\csname svy@item/t17/elsewhere/median\endcsname{4.0}
\expandafter\gdef\csname svy@item/t17/elsewhere/cilow\endcsname{3.0}
\expandafter\gdef\csname svy@item/t17/elsewhere/cihigh\endcsname{4.0}
\expandafter\gdef\csname svy@item/t17/elsewhere/mean\endcsname{3.40}
\expandafter\gdef\csname svy@item/t17/elsewhere/meanlow\endcsname{3.08}
\expandafter\gdef\csname svy@item/t17/elsewhere/meanhigh\endcsname{3.73}
\expandafter\gdef\csname svy@item/t17/elsewhere/se\endcsname{0.163}
\expandafter\gdef\csname svy@item/t17/elsewhere/sd\endcsname{1.38}
\expandafter\gdef\csname svy@item/t17/elsewhere/sdhigh\endcsname{0}
\expandafter\gdef\csname svy@item/t17/elsewhere/verdict\endcsname{INCONCLUSIVE}
\expandafter\gdef\csname svy@item/t17/elsewhere/dip\endcsname{0}
\expandafter\gdef\csname svy@item/t17/elsewhere/shapelow\endcsname{25}
\expandafter\gdef\csname svy@item/t17/elsewhere/shapemid\endcsname{17}
\expandafter\gdef\csname svy@item/t17/elsewhere/shapehigh\endcsname{58}
\expandafter\gdef\csname svy@item/t18/country/diff\endcsname{-0.20}
\expandafter\gdef\csname svy@item/t18/country/difflow\endcsname{-0.54}
\expandafter\gdef\csname svy@item/t18/country/diffhigh\endcsname{0.14}
\expandafter\gdef\csname svy@item/t18/country/g\endcsname{-0.18}
\expandafter\gdef\csname svy@item/t18/germany/n\endcsname{122}
\expandafter\gdef\csname svy@item/t18/germany/miss\endcsname{2}
\expandafter\gdef\csname svy@item/t18/germany/median\endcsname{3.0}
\expandafter\gdef\csname svy@item/t18/germany/cilow\endcsname{3.0}
\expandafter\gdef\csname svy@item/t18/germany/cihigh\endcsname{3.0}
\expandafter\gdef\csname svy@item/t18/germany/mean\endcsname{2.81}
\expandafter\gdef\csname svy@item/t18/germany/meanlow\endcsname{2.62}
\expandafter\gdef\csname svy@item/t18/germany/meanhigh\endcsname{3.00}
\expandafter\gdef\csname svy@item/t18/germany/se\endcsname{0.096}
\expandafter\gdef\csname svy@item/t18/germany/sd\endcsname{1.05}
\expandafter\gdef\csname svy@item/t18/germany/sdhigh\endcsname{0}
\expandafter\gdef\csname svy@item/t18/germany/verdict\endcsname{CONFIRMED}
\expandafter\gdef\csname svy@item/t18/germany/dip\endcsname{0}
\expandafter\gdef\csname svy@item/t18/germany/shapelow\endcsname{39}
\expandafter\gdef\csname svy@item/t18/germany/shapemid\endcsname{33}
\expandafter\gdef\csname svy@item/t18/germany/shapehigh\endcsname{28}
\expandafter\gdef\csname svy@item/t18/elsewhere/n\endcsname{72}
\expandafter\gdef\csname svy@item/t18/elsewhere/miss\endcsname{0}
\expandafter\gdef\csname svy@item/t18/elsewhere/median\endcsname{3.0}
\expandafter\gdef\csname svy@item/t18/elsewhere/cilow\endcsname{2.0}
\expandafter\gdef\csname svy@item/t18/elsewhere/cihigh\endcsname{3.0}
\expandafter\gdef\csname svy@item/t18/elsewhere/mean\endcsname{3.01}
\expandafter\gdef\csname svy@item/t18/elsewhere/meanlow\endcsname{2.73}
\expandafter\gdef\csname svy@item/t18/elsewhere/meanhigh\endcsname{3.30}
\expandafter\gdef\csname svy@item/t18/elsewhere/se\endcsname{0.143}
\expandafter\gdef\csname svy@item/t18/elsewhere/sd\endcsname{1.22}
\expandafter\gdef\csname svy@item/t18/elsewhere/sdhigh\endcsname{0}
\expandafter\gdef\csname svy@item/t18/elsewhere/verdict\endcsname{INCONCLUSIVE}
\expandafter\gdef\csname svy@item/t18/elsewhere/dip\endcsname{1}
\expandafter\gdef\csname svy@item/t18/elsewhere/shapelow\endcsname{41}
\expandafter\gdef\csname svy@item/t18/elsewhere/shapemid\endcsname{21}
\expandafter\gdef\csname svy@item/t18/elsewhere/shapehigh\endcsname{38}
\expandafter\gdef\csname svy@item/t20/country/diff\endcsname{0.02}
\expandafter\gdef\csname svy@item/t20/country/difflow\endcsname{-0.26}
\expandafter\gdef\csname svy@item/t20/country/diffhigh\endcsname{0.31}
\expandafter\gdef\csname svy@item/t20/country/g\endcsname{0.02}
\expandafter\gdef\csname svy@item/t20/germany/n\endcsname{124}
\expandafter\gdef\csname svy@item/t20/germany/miss\endcsname{0}
\expandafter\gdef\csname svy@item/t20/germany/median\endcsname{2.0}
\expandafter\gdef\csname svy@item/t20/germany/cilow\endcsname{2.0}
\expandafter\gdef\csname svy@item/t20/germany/cihigh\endcsname{2.0}
\expandafter\gdef\csname svy@item/t20/germany/mean\endcsname{1.97}
\expandafter\gdef\csname svy@item/t20/germany/meanlow\endcsname{1.81}
\expandafter\gdef\csname svy@item/t20/germany/meanhigh\endcsname{2.12}
\expandafter\gdef\csname svy@item/t20/germany/se\endcsname{0.079}
\expandafter\gdef\csname svy@item/t20/germany/sd\endcsname{0.88}
\expandafter\gdef\csname svy@item/t20/germany/sdhigh\endcsname{0}
\expandafter\gdef\csname svy@item/t20/germany/verdict\endcsname{CONFIRMED}
\expandafter\gdef\csname svy@item/t20/germany/dip\endcsname{0}
\expandafter\gdef\csname svy@item/t20/germany/shapelow\endcsname{80}
\expandafter\gdef\csname svy@item/t20/germany/shapemid\endcsname{15}
\expandafter\gdef\csname svy@item/t20/germany/shapehigh\endcsname{5}
\expandafter\gdef\csname svy@item/t20/elsewhere/n\endcsname{72}
\expandafter\gdef\csname svy@item/t20/elsewhere/miss\endcsname{0}
\expandafter\gdef\csname svy@item/t20/elsewhere/median\endcsname{2.0}
\expandafter\gdef\csname svy@item/t20/elsewhere/cilow\endcsname{1.0}
\expandafter\gdef\csname svy@item/t20/elsewhere/cihigh\endcsname{2.0}
\expandafter\gdef\csname svy@item/t20/elsewhere/mean\endcsname{1.94}
\expandafter\gdef\csname svy@item/t20/elsewhere/meanlow\endcsname{1.70}
\expandafter\gdef\csname svy@item/t20/elsewhere/meanhigh\endcsname{2.19}
\expandafter\gdef\csname svy@item/t20/elsewhere/se\endcsname{0.122}
\expandafter\gdef\csname svy@item/t20/elsewhere/sd\endcsname{1.03}
\expandafter\gdef\csname svy@item/t20/elsewhere/sdhigh\endcsname{0}
\expandafter\gdef\csname svy@item/t20/elsewhere/verdict\endcsname{CONFIRMED}
\expandafter\gdef\csname svy@item/t20/elsewhere/dip\endcsname{0}
\expandafter\gdef\csname svy@item/t20/elsewhere/shapelow\endcsname{77}
\expandafter\gdef\csname svy@item/t20/elsewhere/shapemid\endcsname{12}
\expandafter\gdef\csname svy@item/t20/elsewhere/shapehigh\endcsname{11}
\expandafter\gdef\csname svy@item/t21/country/diff\endcsname{0.05}
\expandafter\gdef\csname svy@item/t21/country/difflow\endcsname{-0.31}
\expandafter\gdef\csname svy@item/t21/country/diffhigh\endcsname{0.41}
\expandafter\gdef\csname svy@item/t21/country/g\endcsname{0.04}
\expandafter\gdef\csname svy@item/t21/germany/n\endcsname{119}
\expandafter\gdef\csname svy@item/t21/germany/miss\endcsname{5}
\expandafter\gdef\csname svy@item/t21/germany/median\endcsname{4.0}
\expandafter\gdef\csname svy@item/t21/germany/cilow\endcsname{4.0}
\expandafter\gdef\csname svy@item/t21/germany/cihigh\endcsname{4.0}
\expandafter\gdef\csname svy@item/t21/germany/mean\endcsname{3.70}
\expandafter\gdef\csname svy@item/t21/germany/meanlow\endcsname{3.51}
\expandafter\gdef\csname svy@item/t21/germany/meanhigh\endcsname{3.88}
\expandafter\gdef\csname svy@item/t21/germany/se\endcsname{0.093}
\expandafter\gdef\csname svy@item/t21/germany/sd\endcsname{1.01}
\expandafter\gdef\csname svy@item/t21/germany/sdhigh\endcsname{0}
\expandafter\gdef\csname svy@item/t21/germany/verdict\endcsname{CONFIRMED}
\expandafter\gdef\csname svy@item/t21/germany/dip\endcsname{0}
\expandafter\gdef\csname svy@item/t21/germany/shapelow\endcsname{15}
\expandafter\gdef\csname svy@item/t21/germany/shapemid\endcsname{18}
\expandafter\gdef\csname svy@item/t21/germany/shapehigh\endcsname{67}
\expandafter\gdef\csname svy@item/t21/elsewhere/n\endcsname{71}
\expandafter\gdef\csname svy@item/t21/elsewhere/miss\endcsname{1}
\expandafter\gdef\csname svy@item/t21/elsewhere/median\endcsname{4.0}
\expandafter\gdef\csname svy@item/t21/elsewhere/cilow\endcsname{4.0}
\expandafter\gdef\csname svy@item/t21/elsewhere/cihigh\endcsname{4.0}
\expandafter\gdef\csname svy@item/t21/elsewhere/mean\endcsname{3.65}
\expandafter\gdef\csname svy@item/t21/elsewhere/meanlow\endcsname{3.34}
\expandafter\gdef\csname svy@item/t21/elsewhere/meanhigh\endcsname{3.96}
\expandafter\gdef\csname svy@item/t21/elsewhere/se\endcsname{0.157}
\expandafter\gdef\csname svy@item/t21/elsewhere/sd\endcsname{1.32}
\expandafter\gdef\csname svy@item/t21/elsewhere/sdhigh\endcsname{0}
\expandafter\gdef\csname svy@item/t21/elsewhere/verdict\endcsname{INCONCLUSIVE}
\expandafter\gdef\csname svy@item/t21/elsewhere/dip\endcsname{1}
\expandafter\gdef\csname svy@item/t21/elsewhere/shapelow\endcsname{24}
\expandafter\gdef\csname svy@item/t21/elsewhere/shapemid\endcsname{8}
\expandafter\gdef\csname svy@item/t21/elsewhere/shapehigh\endcsname{68}
\expandafter\gdef\csname svy@item/t22/country/diff\endcsname{0.13}
\expandafter\gdef\csname svy@item/t22/country/difflow\endcsname{-0.28}
\expandafter\gdef\csname svy@item/t22/country/diffhigh\endcsname{0.53}
\expandafter\gdef\csname svy@item/t22/country/g\endcsname{0.09}
\expandafter\gdef\csname svy@item/t22/germany/n\endcsname{120}
\expandafter\gdef\csname svy@item/t22/germany/miss\endcsname{4}
\expandafter\gdef\csname svy@item/t22/germany/median\endcsname{3.0}
\expandafter\gdef\csname svy@item/t22/germany/cilow\endcsname{3.0}
\expandafter\gdef\csname svy@item/t22/germany/cihigh\endcsname{3.0}
\expandafter\gdef\csname svy@item/t22/germany/mean\endcsname{3.00}
\expandafter\gdef\csname svy@item/t22/germany/meanlow\endcsname{2.77}
\expandafter\gdef\csname svy@item/t22/germany/meanhigh\endcsname{3.23}
\expandafter\gdef\csname svy@item/t22/germany/se\endcsname{0.117}
\expandafter\gdef\csname svy@item/t22/germany/sd\endcsname{1.28}
\expandafter\gdef\csname svy@item/t22/germany/sdhigh\endcsname{0}
\expandafter\gdef\csname svy@item/t22/germany/verdict\endcsname{NO_PREDICTION}
\expandafter\gdef\csname svy@item/t22/germany/dip\endcsname{1}
\expandafter\gdef\csname svy@item/t22/germany/shapelow\endcsname{38}
\expandafter\gdef\csname svy@item/t22/germany/shapemid\endcsname{24}
\expandafter\gdef\csname svy@item/t22/germany/shapehigh\endcsname{38}
\expandafter\gdef\csname svy@item/t22/elsewhere/n\endcsname{71}
\expandafter\gdef\csname svy@item/t22/elsewhere/miss\endcsname{1}
\expandafter\gdef\csname svy@item/t22/elsewhere/median\endcsname{3.0}
\expandafter\gdef\csname svy@item/t22/elsewhere/cilow\endcsname{2.0}
\expandafter\gdef\csname svy@item/t22/elsewhere/cihigh\endcsname{4.0}
\expandafter\gdef\csname svy@item/t22/elsewhere/mean\endcsname{2.87}
\expandafter\gdef\csname svy@item/t22/elsewhere/meanlow\endcsname{2.54}
\expandafter\gdef\csname svy@item/t22/elsewhere/meanhigh\endcsname{3.21}
\expandafter\gdef\csname svy@item/t22/elsewhere/se\endcsname{0.168}
\expandafter\gdef\csname svy@item/t22/elsewhere/sd\endcsname{1.41}
\expandafter\gdef\csname svy@item/t22/elsewhere/sdhigh\endcsname{0}
\expandafter\gdef\csname svy@item/t22/elsewhere/verdict\endcsname{NO_PREDICTION}
\expandafter\gdef\csname svy@item/t22/elsewhere/dip\endcsname{1}
\expandafter\gdef\csname svy@item/t22/elsewhere/shapelow\endcsname{47}
\expandafter\gdef\csname svy@item/t22/elsewhere/shapemid\endcsname{11}
\expandafter\gdef\csname svy@item/t22/elsewhere/shapehigh\endcsname{42}
\expandafter\gdef\csname svy@item/t23/country/diff\endcsname{-0.23}
\expandafter\gdef\csname svy@item/t23/country/difflow\endcsname{-0.49}
\expandafter\gdef\csname svy@item/t23/country/diffhigh\endcsname{0.03}
\expandafter\gdef\csname svy@item/t23/country/g\endcsname{-0.29}
\expandafter\gdef\csname svy@item/t23/germany/n\endcsname{122}
\expandafter\gdef\csname svy@item/t23/germany/miss\endcsname{2}
\expandafter\gdef\csname svy@item/t23/germany/median\endcsname{1.0}
\expandafter\gdef\csname svy@item/t23/germany/cilow\endcsname{1.0}
\expandafter\gdef\csname svy@item/t23/germany/cihigh\endcsname{2.0}
\expandafter\gdef\csname svy@item/t23/germany/mean\endcsname{1.52}
\expandafter\gdef\csname svy@item/t23/germany/meanlow\endcsname{1.39}
\expandafter\gdef\csname svy@item/t23/germany/meanhigh\endcsname{1.64}
\expandafter\gdef\csname svy@item/t23/germany/se\endcsname{0.064}
\expandafter\gdef\csname svy@item/t23/germany/sd\endcsname{0.71}
\expandafter\gdef\csname svy@item/t23/germany/sdhigh\endcsname{0}
\expandafter\gdef\csname svy@item/t23/germany/verdict\endcsname{CONFIRMED}
\expandafter\gdef\csname svy@item/t23/germany/dip\endcsname{0}
\expandafter\gdef\csname svy@item/t23/germany/shapelow\endcsname{94}
\expandafter\gdef\csname svy@item/t23/germany/shapemid\endcsname{3}
\expandafter\gdef\csname svy@item/t23/germany/shapehigh\endcsname{3}
\expandafter\gdef\csname svy@item/t23/elsewhere/n\endcsname{72}
\expandafter\gdef\csname svy@item/t23/elsewhere/miss\endcsname{0}
\expandafter\gdef\csname svy@item/t23/elsewhere/median\endcsname{1.0}
\expandafter\gdef\csname svy@item/t23/elsewhere/cilow\endcsname{1.0}
\expandafter\gdef\csname svy@item/t23/elsewhere/cihigh\endcsname{2.0}
\expandafter\gdef\csname svy@item/t23/elsewhere/mean\endcsname{1.75}
\expandafter\gdef\csname svy@item/t23/elsewhere/meanlow\endcsname{1.52}
\expandafter\gdef\csname svy@item/t23/elsewhere/meanhigh\endcsname{1.98}
\expandafter\gdef\csname svy@item/t23/elsewhere/se\endcsname{0.115}
\expandafter\gdef\csname svy@item/t23/elsewhere/sd\endcsname{0.98}
\expandafter\gdef\csname svy@item/t23/elsewhere/sdhigh\endcsname{0}
\expandafter\gdef\csname svy@item/t23/elsewhere/verdict\endcsname{CONFIRMED}
\expandafter\gdef\csname svy@item/t23/elsewhere/dip\endcsname{0}
\expandafter\gdef\csname svy@item/t23/elsewhere/shapelow\endcsname{81}
\expandafter\gdef\csname svy@item/t23/elsewhere/shapemid\endcsname{12}
\expandafter\gdef\csname svy@item/t23/elsewhere/shapehigh\endcsname{7}
\expandafter\gdef\csname svy@item/t24/country/diff\endcsname{-0.09}
\expandafter\gdef\csname svy@item/t24/country/difflow\endcsname{-0.31}
\expandafter\gdef\csname svy@item/t24/country/diffhigh\endcsname{0.12}
\expandafter\gdef\csname svy@item/t24/country/g\endcsname{-0.15}
\expandafter\gdef\csname svy@item/t24/germany/n\endcsname{119}
\expandafter\gdef\csname svy@item/t24/germany/miss\endcsname{5}
\expandafter\gdef\csname svy@item/t24/germany/median\endcsname{1.0}
\expandafter\gdef\csname svy@item/t24/germany/cilow\endcsname{1.0}
\expandafter\gdef\csname svy@item/t24/germany/cihigh\endcsname{1.0}
\expandafter\gdef\csname svy@item/t24/germany/mean\endcsname{1.27}
\expandafter\gdef\csname svy@item/t24/germany/meanlow\endcsname{1.18}
\expandafter\gdef\csname svy@item/t24/germany/meanhigh\endcsname{1.36}
\expandafter\gdef\csname svy@item/t24/germany/se\endcsname{0.046}
\expandafter\gdef\csname svy@item/t24/germany/sd\endcsname{0.50}
\expandafter\gdef\csname svy@item/t24/germany/sdhigh\endcsname{0}
\expandafter\gdef\csname svy@item/t24/germany/verdict\endcsname{CONFIRMED}
\expandafter\gdef\csname svy@item/t24/germany/dip\endcsname{0}
\expandafter\gdef\csname svy@item/t24/germany/shapelow\endcsname{98}
\expandafter\gdef\csname svy@item/t24/germany/shapemid\endcsname{2}
\expandafter\gdef\csname svy@item/t24/germany/shapehigh\endcsname{0}
\expandafter\gdef\csname svy@item/t24/elsewhere/n\endcsname{66}
\expandafter\gdef\csname svy@item/t24/elsewhere/miss\endcsname{6}
\expandafter\gdef\csname svy@item/t24/elsewhere/median\endcsname{1.0}
\expandafter\gdef\csname svy@item/t24/elsewhere/cilow\endcsname{1.0}
\expandafter\gdef\csname svy@item/t24/elsewhere/cihigh\endcsname{1.0}
\expandafter\gdef\csname svy@item/t24/elsewhere/mean\endcsname{1.36}
\expandafter\gdef\csname svy@item/t24/elsewhere/meanlow\endcsname{1.16}
\expandafter\gdef\csname svy@item/t24/elsewhere/meanhigh\endcsname{1.56}
\expandafter\gdef\csname svy@item/t24/elsewhere/se\endcsname{0.100}
\expandafter\gdef\csname svy@item/t24/elsewhere/sd\endcsname{0.82}
\expandafter\gdef\csname svy@item/t24/elsewhere/sdhigh\endcsname{0}
\expandafter\gdef\csname svy@item/t24/elsewhere/verdict\endcsname{CONFIRMED}
\expandafter\gdef\csname svy@item/t24/elsewhere/dip\endcsname{0}
\expandafter\gdef\csname svy@item/t24/elsewhere/shapelow\endcsname{90}
\expandafter\gdef\csname svy@item/t24/elsewhere/shapemid\endcsname{8}
\expandafter\gdef\csname svy@item/t24/elsewhere/shapehigh\endcsname{2}
\expandafter\gdef\csname svy@item/accdev/country/diff\endcsname{0.06}
\expandafter\gdef\csname svy@item/accdev/country/difflow\endcsname{-0.19}
\expandafter\gdef\csname svy@item/accdev/country/diffhigh\endcsname{0.31}
\expandafter\gdef\csname svy@item/accdev/country/g\endcsname{0.08}
\expandafter\gdef\csname svy@item/accdev/germany/n\endcsname{122}
\expandafter\gdef\csname svy@item/accdev/germany/miss\endcsname{2}
\expandafter\gdef\csname svy@item/accdev/germany/median\endcsname{5.0}
\expandafter\gdef\csname svy@item/accdev/germany/cilow\endcsname{5.0}
\expandafter\gdef\csname svy@item/accdev/germany/cihigh\endcsname{5.0}
\expandafter\gdef\csname svy@item/accdev/germany/mean\endcsname{4.53}
\expandafter\gdef\csname svy@item/accdev/germany/meanlow\endcsname{4.39}
\expandafter\gdef\csname svy@item/accdev/germany/meanhigh\endcsname{4.67}
\expandafter\gdef\csname svy@item/accdev/germany/se\endcsname{0.070}
\expandafter\gdef\csname svy@item/accdev/germany/sd\endcsname{0.77}
\expandafter\gdef\csname svy@item/accdev/germany/sdhigh\endcsname{0}
\expandafter\gdef\csname svy@item/accdev/germany/verdict\endcsname{CONFIRMED}
\expandafter\gdef\csname svy@item/accdev/germany/dip\endcsname{0}
\expandafter\gdef\csname svy@item/accdev/germany/shapelow\endcsname{2}
\expandafter\gdef\csname svy@item/accdev/germany/shapemid\endcsname{5}
\expandafter\gdef\csname svy@item/accdev/germany/shapehigh\endcsname{93}
\expandafter\gdef\csname svy@item/accdev/elsewhere/n\endcsname{70}
\expandafter\gdef\csname svy@item/accdev/elsewhere/miss\endcsname{2}
\expandafter\gdef\csname svy@item/accdev/elsewhere/median\endcsname{5.0}
\expandafter\gdef\csname svy@item/accdev/elsewhere/cilow\endcsname{5.0}
\expandafter\gdef\csname svy@item/accdev/elsewhere/cihigh\endcsname{5.0}
\expandafter\gdef\csname svy@item/accdev/elsewhere/mean\endcsname{4.47}
\expandafter\gdef\csname svy@item/accdev/elsewhere/meanlow\endcsname{4.27}
\expandafter\gdef\csname svy@item/accdev/elsewhere/meanhigh\endcsname{4.68}
\expandafter\gdef\csname svy@item/accdev/elsewhere/se\endcsname{0.103}
\expandafter\gdef\csname svy@item/accdev/elsewhere/sd\endcsname{0.86}
\expandafter\gdef\csname svy@item/accdev/elsewhere/sdhigh\endcsname{0}
\expandafter\gdef\csname svy@item/accdev/elsewhere/verdict\endcsname{CONFIRMED}
\expandafter\gdef\csname svy@item/accdev/elsewhere/dip\endcsname{0}
\expandafter\gdef\csname svy@item/accdev/elsewhere/shapelow\endcsname{2}
\expandafter\gdef\csname svy@item/accdev/elsewhere/shapemid\endcsname{12}
\expandafter\gdef\csname svy@item/accdev/elsewhere/shapehigh\endcsname{86}
\expandafter\gdef\csname svy@item/accreason/country/diff\endcsname{0.02}
\expandafter\gdef\csname svy@item/accreason/country/difflow\endcsname{-0.31}
\expandafter\gdef\csname svy@item/accreason/country/diffhigh\endcsname{0.34}
\expandafter\gdef\csname svy@item/accreason/country/g\endcsname{0.02}
\expandafter\gdef\csname svy@item/accreason/germany/n\endcsname{119}
\expandafter\gdef\csname svy@item/accreason/germany/miss\endcsname{5}
\expandafter\gdef\csname svy@item/accreason/germany/median\endcsname{4.0}
\expandafter\gdef\csname svy@item/accreason/germany/cilow\endcsname{4.0}
\expandafter\gdef\csname svy@item/accreason/germany/cihigh\endcsname{5.0}
\expandafter\gdef\csname svy@item/accreason/germany/mean\endcsname{4.18}
\expandafter\gdef\csname svy@item/accreason/germany/meanlow\endcsname{4.00}
\expandafter\gdef\csname svy@item/accreason/germany/meanhigh\endcsname{4.36}
\expandafter\gdef\csname svy@item/accreason/germany/se\endcsname{0.091}
\expandafter\gdef\csname svy@item/accreason/germany/sd\endcsname{1.00}
\expandafter\gdef\csname svy@item/accreason/germany/sdhigh\endcsname{0}
\expandafter\gdef\csname svy@item/accreason/germany/verdict\endcsname{CONFIRMED}
\expandafter\gdef\csname svy@item/accreason/germany/dip\endcsname{0}
\expandafter\gdef\csname svy@item/accreason/germany/shapelow\endcsname{7}
\expandafter\gdef\csname svy@item/accreason/germany/shapemid\endcsname{9}
\expandafter\gdef\csname svy@item/accreason/germany/shapehigh\endcsname{84}
\expandafter\gdef\csname svy@item/accreason/elsewhere/n\endcsname{69}
\expandafter\gdef\csname svy@item/accreason/elsewhere/miss\endcsname{3}
\expandafter\gdef\csname svy@item/accreason/elsewhere/median\endcsname{5.0}
\expandafter\gdef\csname svy@item/accreason/elsewhere/cilow\endcsname{4.0}
\expandafter\gdef\csname svy@item/accreason/elsewhere/cihigh\endcsname{5.0}
\expandafter\gdef\csname svy@item/accreason/elsewhere/mean\endcsname{4.16}
\expandafter\gdef\csname svy@item/accreason/elsewhere/meanlow\endcsname{3.89}
\expandafter\gdef\csname svy@item/accreason/elsewhere/meanhigh\endcsname{4.43}
\expandafter\gdef\csname svy@item/accreason/elsewhere/se\endcsname{0.136}
\expandafter\gdef\csname svy@item/accreason/elsewhere/sd\endcsname{1.13}
\expandafter\gdef\csname svy@item/accreason/elsewhere/sdhigh\endcsname{0}
\expandafter\gdef\csname svy@item/accreason/elsewhere/verdict\endcsname{CONFIRMED}
\expandafter\gdef\csname svy@item/accreason/elsewhere/dip\endcsname{0}
\expandafter\gdef\csname svy@item/accreason/elsewhere/shapelow\endcsname{9}
\expandafter\gdef\csname svy@item/accreason/elsewhere/shapemid\endcsname{13}
\expandafter\gdef\csname svy@item/accreason/elsewhere/shapehigh\endcsname{78}
\expandafter\gdef\csname svy@item/t01/exposure/trend\endcsname{0.32}
\expandafter\gdef\csname svy@item/t01/exposure/trendlow\endcsname{0.21}
\expandafter\gdef\csname svy@item/t01/exposure/trendhigh\endcsname{0.44}
\expandafter\gdef\csname svy@item/t01/e0/n\endcsname{12}
\expandafter\gdef\csname svy@item/t01/e0/miss\endcsname{1}
\expandafter\gdef\csname svy@item/t01/e0/median\endcsname{1.0}
\expandafter\gdef\csname svy@item/t01/e0/cilow\endcsname{1.0}
\expandafter\gdef\csname svy@item/t01/e0/cihigh\endcsname{1.0}
\expandafter\gdef\csname svy@item/t01/e0/mean\endcsname{1.08}
\expandafter\gdef\csname svy@item/t01/e0/meanlow\endcsname{0.90}
\expandafter\gdef\csname svy@item/t01/e0/meanhigh\endcsname{1.27}
\expandafter\gdef\csname svy@item/t01/e0/se\endcsname{0.083}
\expandafter\gdef\csname svy@item/t01/e0/sd\endcsname{0.29}
\expandafter\gdef\csname svy@item/t01/e0/sdhigh\endcsname{0}
\expandafter\gdef\csname svy@item/t01/e0/verdict\endcsname{CONFIRMED}
\expandafter\gdef\csname svy@item/t01/e0/dip\endcsname{0}
\expandafter\gdef\csname svy@item/t01/e0/shapelow\endcsname{100}
\expandafter\gdef\csname svy@item/t01/e0/shapemid\endcsname{0}
\expandafter\gdef\csname svy@item/t01/e0/shapehigh\endcsname{0}
\expandafter\gdef\csname svy@item/t01/e1/n\endcsname{49}
\expandafter\gdef\csname svy@item/t01/e1/miss\endcsname{0}
\expandafter\gdef\csname svy@item/t01/e1/median\endcsname{1.0}
\expandafter\gdef\csname svy@item/t01/e1/cilow\endcsname{1.0}
\expandafter\gdef\csname svy@item/t01/e1/cihigh\endcsname{2.0}
\expandafter\gdef\csname svy@item/t01/e1/mean\endcsname{1.51}
\expandafter\gdef\csname svy@item/t01/e1/meanlow\endcsname{1.31}
\expandafter\gdef\csname svy@item/t01/e1/meanhigh\endcsname{1.71}
\expandafter\gdef\csname svy@item/t01/e1/se\endcsname{0.102}
\expandafter\gdef\csname svy@item/t01/e1/sd\endcsname{0.71}
\expandafter\gdef\csname svy@item/t01/e1/sdhigh\endcsname{0}
\expandafter\gdef\csname svy@item/t01/e1/verdict\endcsname{CONFIRMED}
\expandafter\gdef\csname svy@item/t01/e1/dip\endcsname{0}
\expandafter\gdef\csname svy@item/t01/e1/shapelow\endcsname{96}
\expandafter\gdef\csname svy@item/t01/e1/shapemid\endcsname{0}
\expandafter\gdef\csname svy@item/t01/e1/shapehigh\endcsname{4}
\expandafter\gdef\csname svy@item/t01/e2/n\endcsname{77}
\expandafter\gdef\csname svy@item/t01/e2/miss\endcsname{0}
\expandafter\gdef\csname svy@item/t01/e2/median\endcsname{1.0}
\expandafter\gdef\csname svy@item/t01/e2/cilow\endcsname{1.0}
\expandafter\gdef\csname svy@item/t01/e2/cihigh\endcsname{2.0}
\expandafter\gdef\csname svy@item/t01/e2/mean\endcsname{1.60}
\expandafter\gdef\csname svy@item/t01/e2/meanlow\endcsname{1.43}
\expandafter\gdef\csname svy@item/t01/e2/meanhigh\endcsname{1.77}
\expandafter\gdef\csname svy@item/t01/e2/se\endcsname{0.085}
\expandafter\gdef\csname svy@item/t01/e2/sd\endcsname{0.75}
\expandafter\gdef\csname svy@item/t01/e2/sdhigh\endcsname{0}
\expandafter\gdef\csname svy@item/t01/e2/verdict\endcsname{CONFIRMED}
\expandafter\gdef\csname svy@item/t01/e2/dip\endcsname{0}
\expandafter\gdef\csname svy@item/t01/e2/shapelow\endcsname{94}
\expandafter\gdef\csname svy@item/t01/e2/shapemid\endcsname{4}
\expandafter\gdef\csname svy@item/t01/e2/shapehigh\endcsname{2}
\expandafter\gdef\csname svy@item/t01/e3/n\endcsname{94}
\expandafter\gdef\csname svy@item/t01/e3/miss\endcsname{0}
\expandafter\gdef\csname svy@item/t01/e3/median\endcsname{2.0}
\expandafter\gdef\csname svy@item/t01/e3/cilow\endcsname{2.0}
\expandafter\gdef\csname svy@item/t01/e3/cihigh\endcsname{2.0}
\expandafter\gdef\csname svy@item/t01/e3/mean\endcsname{2.11}
\expandafter\gdef\csname svy@item/t01/e3/meanlow\endcsname{1.91}
\expandafter\gdef\csname svy@item/t01/e3/meanhigh\endcsname{2.30}
\expandafter\gdef\csname svy@item/t01/e3/se\endcsname{0.099}
\expandafter\gdef\csname svy@item/t01/e3/sd\endcsname{0.96}
\expandafter\gdef\csname svy@item/t01/e3/sdhigh\endcsname{0}
\expandafter\gdef\csname svy@item/t01/e3/verdict\endcsname{CONFIRMED}
\expandafter\gdef\csname svy@item/t01/e3/dip\endcsname{0}
\expandafter\gdef\csname svy@item/t01/e3/shapelow\endcsname{75}
\expandafter\gdef\csname svy@item/t01/e3/shapemid\endcsname{13}
\expandafter\gdef\csname svy@item/t01/e3/shapehigh\endcsname{12}
\expandafter\gdef\csname svy@item/t03/exposure/trend\endcsname{0.31}
\expandafter\gdef\csname svy@item/t03/exposure/trendlow\endcsname{0.19}
\expandafter\gdef\csname svy@item/t03/exposure/trendhigh\endcsname{0.44}
\expandafter\gdef\csname svy@item/t03/e0/n\endcsname{13}
\expandafter\gdef\csname svy@item/t03/e0/miss\endcsname{0}
\expandafter\gdef\csname svy@item/t03/e0/median\endcsname{1.0}
\expandafter\gdef\csname svy@item/t03/e0/cilow\endcsname{1.0}
\expandafter\gdef\csname svy@item/t03/e0/cihigh\endcsname{2.0}
\expandafter\gdef\csname svy@item/t03/e0/mean\endcsname{1.31}
\expandafter\gdef\csname svy@item/t03/e0/meanlow\endcsname{1.02}
\expandafter\gdef\csname svy@item/t03/e0/meanhigh\endcsname{1.60}
\expandafter\gdef\csname svy@item/t03/e0/se\endcsname{0.133}
\expandafter\gdef\csname svy@item/t03/e0/sd\endcsname{0.48}
\expandafter\gdef\csname svy@item/t03/e0/sdhigh\endcsname{0}
\expandafter\gdef\csname svy@item/t03/e0/verdict\endcsname{REJECTED}
\expandafter\gdef\csname svy@item/t03/e0/dip\endcsname{0}
\expandafter\gdef\csname svy@item/t03/e0/shapelow\endcsname{100}
\expandafter\gdef\csname svy@item/t03/e0/shapemid\endcsname{0}
\expandafter\gdef\csname svy@item/t03/e0/shapehigh\endcsname{0}
\expandafter\gdef\csname svy@item/t03/e1/n\endcsname{49}
\expandafter\gdef\csname svy@item/t03/e1/miss\endcsname{0}
\expandafter\gdef\csname svy@item/t03/e1/median\endcsname{2.0}
\expandafter\gdef\csname svy@item/t03/e1/cilow\endcsname{2.0}
\expandafter\gdef\csname svy@item/t03/e1/cihigh\endcsname{3.0}
\expandafter\gdef\csname svy@item/t03/e1/mean\endcsname{2.33}
\expandafter\gdef\csname svy@item/t03/e1/meanlow\endcsname{2.07}
\expandafter\gdef\csname svy@item/t03/e1/meanhigh\endcsname{2.58}
\expandafter\gdef\csname svy@item/t03/e1/se\endcsname{0.128}
\expandafter\gdef\csname svy@item/t03/e1/sd\endcsname{0.90}
\expandafter\gdef\csname svy@item/t03/e1/sdhigh\endcsname{0}
\expandafter\gdef\csname svy@item/t03/e1/verdict\endcsname{INCONCLUSIVE}
\expandafter\gdef\csname svy@item/t03/e1/dip\endcsname{0}
\expandafter\gdef\csname svy@item/t03/e1/shapelow\endcsname{61}
\expandafter\gdef\csname svy@item/t03/e1/shapemid\endcsname{31}
\expandafter\gdef\csname svy@item/t03/e1/shapehigh\endcsname{8}
\expandafter\gdef\csname svy@item/t03/e2/n\endcsname{77}
\expandafter\gdef\csname svy@item/t03/e2/miss\endcsname{0}
\expandafter\gdef\csname svy@item/t03/e2/median\endcsname{2.0}
\expandafter\gdef\csname svy@item/t03/e2/cilow\endcsname{2.0}
\expandafter\gdef\csname svy@item/t03/e2/cihigh\endcsname{3.0}
\expandafter\gdef\csname svy@item/t03/e2/mean\endcsname{2.40}
\expandafter\gdef\csname svy@item/t03/e2/meanlow\endcsname{2.22}
\expandafter\gdef\csname svy@item/t03/e2/meanhigh\endcsname{2.58}
\expandafter\gdef\csname svy@item/t03/e2/se\endcsname{0.091}
\expandafter\gdef\csname svy@item/t03/e2/sd\endcsname{0.80}
\expandafter\gdef\csname svy@item/t03/e2/sdhigh\endcsname{0}
\expandafter\gdef\csname svy@item/t03/e2/verdict\endcsname{INCONCLUSIVE}
\expandafter\gdef\csname svy@item/t03/e2/dip\endcsname{0}
\expandafter\gdef\csname svy@item/t03/e2/shapelow\endcsname{56}
\expandafter\gdef\csname svy@item/t03/e2/shapemid\endcsname{36}
\expandafter\gdef\csname svy@item/t03/e2/shapehigh\endcsname{8}
\expandafter\gdef\csname svy@item/t03/e3/n\endcsname{93}
\expandafter\gdef\csname svy@item/t03/e3/miss\endcsname{1}
\expandafter\gdef\csname svy@item/t03/e3/median\endcsname{2.0}
\expandafter\gdef\csname svy@item/t03/e3/cilow\endcsname{2.0}
\expandafter\gdef\csname svy@item/t03/e3/cihigh\endcsname{3.0}
\expandafter\gdef\csname svy@item/t03/e3/mean\endcsname{2.71}
\expandafter\gdef\csname svy@item/t03/e3/meanlow\endcsname{2.52}
\expandafter\gdef\csname svy@item/t03/e3/meanhigh\endcsname{2.90}
\expandafter\gdef\csname svy@item/t03/e3/se\endcsname{0.097}
\expandafter\gdef\csname svy@item/t03/e3/sd\endcsname{0.94}
\expandafter\gdef\csname svy@item/t03/e3/sdhigh\endcsname{0}
\expandafter\gdef\csname svy@item/t03/e3/verdict\endcsname{CONFIRMED}
\expandafter\gdef\csname svy@item/t03/e3/dip\endcsname{0}
\expandafter\gdef\csname svy@item/t03/e3/shapelow\endcsname{52}
\expandafter\gdef\csname svy@item/t03/e3/shapemid\endcsname{27}
\expandafter\gdef\csname svy@item/t03/e3/shapehigh\endcsname{21}
\expandafter\gdef\csname svy@item/t04/exposure/trend\endcsname{0.37}
\expandafter\gdef\csname svy@item/t04/exposure/trendlow\endcsname{0.24}
\expandafter\gdef\csname svy@item/t04/exposure/trendhigh\endcsname{0.50}
\expandafter\gdef\csname svy@item/t04/e0/n\endcsname{13}
\expandafter\gdef\csname svy@item/t04/e0/miss\endcsname{0}
\expandafter\gdef\csname svy@item/t04/e0/median\endcsname{1.0}
\expandafter\gdef\csname svy@item/t04/e0/cilow\endcsname{1.0}
\expandafter\gdef\csname svy@item/t04/e0/cihigh\endcsname{2.0}
\expandafter\gdef\csname svy@item/t04/e0/mean\endcsname{1.46}
\expandafter\gdef\csname svy@item/t04/e0/meanlow\endcsname{1.06}
\expandafter\gdef\csname svy@item/t04/e0/meanhigh\endcsname{1.86}
\expandafter\gdef\csname svy@item/t04/e0/se\endcsname{0.183}
\expandafter\gdef\csname svy@item/t04/e0/sd\endcsname{0.66}
\expandafter\gdef\csname svy@item/t04/e0/sdhigh\endcsname{0}
\expandafter\gdef\csname svy@item/t04/e0/verdict\endcsname{NO_PREDICTION}
\expandafter\gdef\csname svy@item/t04/e0/dip\endcsname{0}
\expandafter\gdef\csname svy@item/t04/e0/shapelow\endcsname{92}
\expandafter\gdef\csname svy@item/t04/e0/shapemid\endcsname{8}
\expandafter\gdef\csname svy@item/t04/e0/shapehigh\endcsname{0}
\expandafter\gdef\csname svy@item/t04/e1/n\endcsname{49}
\expandafter\gdef\csname svy@item/t04/e1/miss\endcsname{0}
\expandafter\gdef\csname svy@item/t04/e1/median\endcsname{2.0}
\expandafter\gdef\csname svy@item/t04/e1/cilow\endcsname{2.0}
\expandafter\gdef\csname svy@item/t04/e1/cihigh\endcsname{2.0}
\expandafter\gdef\csname svy@item/t04/e1/mean\endcsname{2.22}
\expandafter\gdef\csname svy@item/t04/e1/meanlow\endcsname{1.97}
\expandafter\gdef\csname svy@item/t04/e1/meanhigh\endcsname{2.48}
\expandafter\gdef\csname svy@item/t04/e1/se\endcsname{0.128}
\expandafter\gdef\csname svy@item/t04/e1/sd\endcsname{0.90}
\expandafter\gdef\csname svy@item/t04/e1/sdhigh\endcsname{0}
\expandafter\gdef\csname svy@item/t04/e1/verdict\endcsname{NO_PREDICTION}
\expandafter\gdef\csname svy@item/t04/e1/dip\endcsname{0}
\expandafter\gdef\csname svy@item/t04/e1/shapelow\endcsname{67}
\expandafter\gdef\csname svy@item/t04/e1/shapemid\endcsname{23}
\expandafter\gdef\csname svy@item/t04/e1/shapehigh\endcsname{10}
\expandafter\gdef\csname svy@item/t04/e2/n\endcsname{74}
\expandafter\gdef\csname svy@item/t04/e2/miss\endcsname{3}
\expandafter\gdef\csname svy@item/t04/e2/median\endcsname{2.0}
\expandafter\gdef\csname svy@item/t04/e2/cilow\endcsname{2.0}
\expandafter\gdef\csname svy@item/t04/e2/cihigh\endcsname{3.0}
\expandafter\gdef\csname svy@item/t04/e2/mean\endcsname{2.47}
\expandafter\gdef\csname svy@item/t04/e2/meanlow\endcsname{2.28}
\expandafter\gdef\csname svy@item/t04/e2/meanhigh\endcsname{2.67}
\expandafter\gdef\csname svy@item/t04/e2/se\endcsname{0.099}
\expandafter\gdef\csname svy@item/t04/e2/sd\endcsname{0.85}
\expandafter\gdef\csname svy@item/t04/e2/sdhigh\endcsname{0}
\expandafter\gdef\csname svy@item/t04/e2/verdict\endcsname{NO_PREDICTION}
\expandafter\gdef\csname svy@item/t04/e2/dip\endcsname{0}
\expandafter\gdef\csname svy@item/t04/e2/shapelow\endcsname{56}
\expandafter\gdef\csname svy@item/t04/e2/shapemid\endcsname{30}
\expandafter\gdef\csname svy@item/t04/e2/shapehigh\endcsname{14}
\expandafter\gdef\csname svy@item/t04/e3/n\endcsname{92}
\expandafter\gdef\csname svy@item/t04/e3/miss\endcsname{2}
\expandafter\gdef\csname svy@item/t04/e3/median\endcsname{3.0}
\expandafter\gdef\csname svy@item/t04/e3/cilow\endcsname{2.0}
\expandafter\gdef\csname svy@item/t04/e3/cihigh\endcsname{3.0}
\expandafter\gdef\csname svy@item/t04/e3/mean\endcsname{2.84}
\expandafter\gdef\csname svy@item/t04/e3/meanlow\endcsname{2.63}
\expandafter\gdef\csname svy@item/t04/e3/meanhigh\endcsname{3.04}
\expandafter\gdef\csname svy@item/t04/e3/se\endcsname{0.103}
\expandafter\gdef\csname svy@item/t04/e3/sd\endcsname{0.99}
\expandafter\gdef\csname svy@item/t04/e3/sdhigh\endcsname{0}
\expandafter\gdef\csname svy@item/t04/e3/verdict\endcsname{NO_PREDICTION}
\expandafter\gdef\csname svy@item/t04/e3/dip\endcsname{0}
\expandafter\gdef\csname svy@item/t04/e3/shapelow\endcsname{43}
\expandafter\gdef\csname svy@item/t04/e3/shapemid\endcsname{34}
\expandafter\gdef\csname svy@item/t04/e3/shapehigh\endcsname{23}
\expandafter\gdef\csname svy@item/t05/exposure/trend\endcsname{0.59}
\expandafter\gdef\csname svy@item/t05/exposure/trendlow\endcsname{0.46}
\expandafter\gdef\csname svy@item/t05/exposure/trendhigh\endcsname{0.71}
\expandafter\gdef\csname svy@item/t05/e0/n\endcsname{12}
\expandafter\gdef\csname svy@item/t05/e0/miss\endcsname{1}
\expandafter\gdef\csname svy@item/t05/e0/median\endcsname{2.0}
\expandafter\gdef\csname svy@item/t05/e0/cilow\endcsname{1.0}
\expandafter\gdef\csname svy@item/t05/e0/cihigh\endcsname{3.0}
\expandafter\gdef\csname svy@item/t05/e0/mean\endcsname{2.00}
\expandafter\gdef\csname svy@item/t05/e0/meanlow\endcsname{1.53}
\expandafter\gdef\csname svy@item/t05/e0/meanhigh\endcsname{2.47}
\expandafter\gdef\csname svy@item/t05/e0/se\endcsname{0.213}
\expandafter\gdef\csname svy@item/t05/e0/sd\endcsname{0.74}
\expandafter\gdef\csname svy@item/t05/e0/sdhigh\endcsname{0}
\expandafter\gdef\csname svy@item/t05/e0/verdict\endcsname{REJECTED}
\expandafter\gdef\csname svy@item/t05/e0/dip\endcsname{0}
\expandafter\gdef\csname svy@item/t05/e0/shapelow\endcsname{75}
\expandafter\gdef\csname svy@item/t05/e0/shapemid\endcsname{25}
\expandafter\gdef\csname svy@item/t05/e0/shapehigh\endcsname{0}
\expandafter\gdef\csname svy@item/t05/e1/n\endcsname{47}
\expandafter\gdef\csname svy@item/t05/e1/miss\endcsname{2}
\expandafter\gdef\csname svy@item/t05/e1/median\endcsname{3.0}
\expandafter\gdef\csname svy@item/t05/e1/cilow\endcsname{3.0}
\expandafter\gdef\csname svy@item/t05/e1/cihigh\endcsname{3.0}
\expandafter\gdef\csname svy@item/t05/e1/mean\endcsname{3.02}
\expandafter\gdef\csname svy@item/t05/e1/meanlow\endcsname{2.71}
\expandafter\gdef\csname svy@item/t05/e1/meanhigh\endcsname{3.34}
\expandafter\gdef\csname svy@item/t05/e1/se\endcsname{0.157}
\expandafter\gdef\csname svy@item/t05/e1/sd\endcsname{1.07}
\expandafter\gdef\csname svy@item/t05/e1/sdhigh\endcsname{0}
\expandafter\gdef\csname svy@item/t05/e1/verdict\endcsname{REJECTED}
\expandafter\gdef\csname svy@item/t05/e1/dip\endcsname{0}
\expandafter\gdef\csname svy@item/t05/e1/shapelow\endcsname{34}
\expandafter\gdef\csname svy@item/t05/e1/shapemid\endcsname{32}
\expandafter\gdef\csname svy@item/t05/e1/shapehigh\endcsname{34}
\expandafter\gdef\csname svy@item/t05/e2/n\endcsname{76}
\expandafter\gdef\csname svy@item/t05/e2/miss\endcsname{1}
\expandafter\gdef\csname svy@item/t05/e2/median\endcsname{4.0}
\expandafter\gdef\csname svy@item/t05/e2/cilow\endcsname{3.0}
\expandafter\gdef\csname svy@item/t05/e2/cihigh\endcsname{4.0}
\expandafter\gdef\csname svy@item/t05/e2/mean\endcsname{3.63}
\expandafter\gdef\csname svy@item/t05/e2/meanlow\endcsname{3.44}
\expandafter\gdef\csname svy@item/t05/e2/meanhigh\endcsname{3.82}
\expandafter\gdef\csname svy@item/t05/e2/se\endcsname{0.095}
\expandafter\gdef\csname svy@item/t05/e2/sd\endcsname{0.83}
\expandafter\gdef\csname svy@item/t05/e2/sdhigh\endcsname{0}
\expandafter\gdef\csname svy@item/t05/e2/verdict\endcsname{INCONCLUSIVE}
\expandafter\gdef\csname svy@item/t05/e2/dip\endcsname{0}
\expandafter\gdef\csname svy@item/t05/e2/shapelow\endcsname{9}
\expandafter\gdef\csname svy@item/t05/e2/shapemid\endcsname{32}
\expandafter\gdef\csname svy@item/t05/e2/shapehigh\endcsname{59}
\expandafter\gdef\csname svy@item/t05/e3/n\endcsname{93}
\expandafter\gdef\csname svy@item/t05/e3/miss\endcsname{1}
\expandafter\gdef\csname svy@item/t05/e3/median\endcsname{4.0}
\expandafter\gdef\csname svy@item/t05/e3/cilow\endcsname{4.0}
\expandafter\gdef\csname svy@item/t05/e3/cihigh\endcsname{4.0}
\expandafter\gdef\csname svy@item/t05/e3/mean\endcsname{4.06}
\expandafter\gdef\csname svy@item/t05/e3/meanlow\endcsname{3.89}
\expandafter\gdef\csname svy@item/t05/e3/meanhigh\endcsname{4.24}
\expandafter\gdef\csname svy@item/t05/e3/se\endcsname{0.088}
\expandafter\gdef\csname svy@item/t05/e3/sd\endcsname{0.84}
\expandafter\gdef\csname svy@item/t05/e3/sdhigh\endcsname{0}
\expandafter\gdef\csname svy@item/t05/e3/verdict\endcsname{CONFIRMED}
\expandafter\gdef\csname svy@item/t05/e3/dip\endcsname{0}
\expandafter\gdef\csname svy@item/t05/e3/shapelow\endcsname{6}
\expandafter\gdef\csname svy@item/t05/e3/shapemid\endcsname{16}
\expandafter\gdef\csname svy@item/t05/e3/shapehigh\endcsname{78}
\expandafter\gdef\csname svy@item/t06/exposure/trend\endcsname{0.64}
\expandafter\gdef\csname svy@item/t06/exposure/trendlow\endcsname{0.51}
\expandafter\gdef\csname svy@item/t06/exposure/trendhigh\endcsname{0.77}
\expandafter\gdef\csname svy@item/t06/e0/n\endcsname{12}
\expandafter\gdef\csname svy@item/t06/e0/miss\endcsname{1}
\expandafter\gdef\csname svy@item/t06/e0/median\endcsname{2.0}
\expandafter\gdef\csname svy@item/t06/e0/cilow\endcsname{1.0}
\expandafter\gdef\csname svy@item/t06/e0/cihigh\endcsname{2.0}
\expandafter\gdef\csname svy@item/t06/e0/mean\endcsname{1.83}
\expandafter\gdef\csname svy@item/t06/e0/meanlow\endcsname{1.38}
\expandafter\gdef\csname svy@item/t06/e0/meanhigh\endcsname{2.29}
\expandafter\gdef\csname svy@item/t06/e0/se\endcsname{0.207}
\expandafter\gdef\csname svy@item/t06/e0/sd\endcsname{0.72}
\expandafter\gdef\csname svy@item/t06/e0/sdhigh\endcsname{0}
\expandafter\gdef\csname svy@item/t06/e0/verdict\endcsname{REJECTED}
\expandafter\gdef\csname svy@item/t06/e0/dip\endcsname{0}
\expandafter\gdef\csname svy@item/t06/e0/shapelow\endcsname{83}
\expandafter\gdef\csname svy@item/t06/e0/shapemid\endcsname{17}
\expandafter\gdef\csname svy@item/t06/e0/shapehigh\endcsname{0}
\expandafter\gdef\csname svy@item/t06/e1/n\endcsname{47}
\expandafter\gdef\csname svy@item/t06/e1/miss\endcsname{2}
\expandafter\gdef\csname svy@item/t06/e1/median\endcsname{4.0}
\expandafter\gdef\csname svy@item/t06/e1/cilow\endcsname{3.0}
\expandafter\gdef\csname svy@item/t06/e1/cihigh\endcsname{4.0}
\expandafter\gdef\csname svy@item/t06/e1/mean\endcsname{3.28}
\expandafter\gdef\csname svy@item/t06/e1/meanlow\endcsname{2.95}
\expandafter\gdef\csname svy@item/t06/e1/meanhigh\endcsname{3.60}
\expandafter\gdef\csname svy@item/t06/e1/se\endcsname{0.160}
\expandafter\gdef\csname svy@item/t06/e1/sd\endcsname{1.10}
\expandafter\gdef\csname svy@item/t06/e1/sdhigh\endcsname{0}
\expandafter\gdef\csname svy@item/t06/e1/verdict\endcsname{INCONCLUSIVE}
\expandafter\gdef\csname svy@item/t06/e1/dip\endcsname{1}
\expandafter\gdef\csname svy@item/t06/e1/shapelow\endcsname{30}
\expandafter\gdef\csname svy@item/t06/e1/shapemid\endcsname{19}
\expandafter\gdef\csname svy@item/t06/e1/shapehigh\endcsname{51}
\expandafter\gdef\csname svy@item/t06/e2/n\endcsname{76}
\expandafter\gdef\csname svy@item/t06/e2/miss\endcsname{1}
\expandafter\gdef\csname svy@item/t06/e2/median\endcsname{4.0}
\expandafter\gdef\csname svy@item/t06/e2/cilow\endcsname{4.0}
\expandafter\gdef\csname svy@item/t06/e2/cihigh\endcsname{4.0}
\expandafter\gdef\csname svy@item/t06/e2/mean\endcsname{3.99}
\expandafter\gdef\csname svy@item/t06/e2/meanlow\endcsname{3.77}
\expandafter\gdef\csname svy@item/t06/e2/meanhigh\endcsname{4.20}
\expandafter\gdef\csname svy@item/t06/e2/se\endcsname{0.108}
\expandafter\gdef\csname svy@item/t06/e2/sd\endcsname{0.95}
\expandafter\gdef\csname svy@item/t06/e2/sdhigh\endcsname{0}
\expandafter\gdef\csname svy@item/t06/e2/verdict\endcsname{CONFIRMED}
\expandafter\gdef\csname svy@item/t06/e2/dip\endcsname{0}
\expandafter\gdef\csname svy@item/t06/e2/shapelow\endcsname{8}
\expandafter\gdef\csname svy@item/t06/e2/shapemid\endcsname{13}
\expandafter\gdef\csname svy@item/t06/e2/shapehigh\endcsname{79}
\expandafter\gdef\csname svy@item/t06/e3/n\endcsname{93}
\expandafter\gdef\csname svy@item/t06/e3/miss\endcsname{1}
\expandafter\gdef\csname svy@item/t06/e3/median\endcsname{4.0}
\expandafter\gdef\csname svy@item/t06/e3/cilow\endcsname{4.0}
\expandafter\gdef\csname svy@item/t06/e3/cihigh\endcsname{5.0}
\expandafter\gdef\csname svy@item/t06/e3/mean\endcsname{4.32}
\expandafter\gdef\csname svy@item/t06/e3/meanlow\endcsname{4.17}
\expandafter\gdef\csname svy@item/t06/e3/meanhigh\endcsname{4.48}
\expandafter\gdef\csname svy@item/t06/e3/se\endcsname{0.078}
\expandafter\gdef\csname svy@item/t06/e3/sd\endcsname{0.75}
\expandafter\gdef\csname svy@item/t06/e3/sdhigh\endcsname{0}
\expandafter\gdef\csname svy@item/t06/e3/verdict\endcsname{CONFIRMED}
\expandafter\gdef\csname svy@item/t06/e3/dip\endcsname{0}
\expandafter\gdef\csname svy@item/t06/e3/shapelow\endcsname{2}
\expandafter\gdef\csname svy@item/t06/e3/shapemid\endcsname{11}
\expandafter\gdef\csname svy@item/t06/e3/shapehigh\endcsname{87}
\expandafter\gdef\csname svy@item/t07/exposure/trend\endcsname{0.51}
\expandafter\gdef\csname svy@item/t07/exposure/trendlow\endcsname{0.36}
\expandafter\gdef\csname svy@item/t07/exposure/trendhigh\endcsname{0.66}
\expandafter\gdef\csname svy@item/t07/e0/n\endcsname{13}
\expandafter\gdef\csname svy@item/t07/e0/miss\endcsname{0}
\expandafter\gdef\csname svy@item/t07/e0/median\endcsname{1.0}
\expandafter\gdef\csname svy@item/t07/e0/cilow\endcsname{1.0}
\expandafter\gdef\csname svy@item/t07/e0/cihigh\endcsname{2.0}
\expandafter\gdef\csname svy@item/t07/e0/mean\endcsname{1.31}
\expandafter\gdef\csname svy@item/t07/e0/meanlow\endcsname{1.02}
\expandafter\gdef\csname svy@item/t07/e0/meanhigh\endcsname{1.60}
\expandafter\gdef\csname svy@item/t07/e0/se\endcsname{0.133}
\expandafter\gdef\csname svy@item/t07/e0/sd\endcsname{0.48}
\expandafter\gdef\csname svy@item/t07/e0/sdhigh\endcsname{0}
\expandafter\gdef\csname svy@item/t07/e0/verdict\endcsname{CONFIRMED}
\expandafter\gdef\csname svy@item/t07/e0/dip\endcsname{0}
\expandafter\gdef\csname svy@item/t07/e0/shapelow\endcsname{100}
\expandafter\gdef\csname svy@item/t07/e0/shapemid\endcsname{0}
\expandafter\gdef\csname svy@item/t07/e0/shapehigh\endcsname{0}
\expandafter\gdef\csname svy@item/t07/e1/n\endcsname{47}
\expandafter\gdef\csname svy@item/t07/e1/miss\endcsname{2}
\expandafter\gdef\csname svy@item/t07/e1/median\endcsname{2.0}
\expandafter\gdef\csname svy@item/t07/e1/cilow\endcsname{2.0}
\expandafter\gdef\csname svy@item/t07/e1/cihigh\endcsname{2.0}
\expandafter\gdef\csname svy@item/t07/e1/mean\endcsname{2.09}
\expandafter\gdef\csname svy@item/t07/e1/meanlow\endcsname{1.79}
\expandafter\gdef\csname svy@item/t07/e1/meanhigh\endcsname{2.38}
\expandafter\gdef\csname svy@item/t07/e1/se\endcsname{0.145}
\expandafter\gdef\csname svy@item/t07/e1/sd\endcsname{1.00}
\expandafter\gdef\csname svy@item/t07/e1/sdhigh\endcsname{0}
\expandafter\gdef\csname svy@item/t07/e1/verdict\endcsname{CONFIRMED}
\expandafter\gdef\csname svy@item/t07/e1/dip\endcsname{0}
\expandafter\gdef\csname svy@item/t07/e1/shapelow\endcsname{68}
\expandafter\gdef\csname svy@item/t07/e1/shapemid\endcsname{21}
\expandafter\gdef\csname svy@item/t07/e1/shapehigh\endcsname{11}
\expandafter\gdef\csname svy@item/t07/e2/n\endcsname{77}
\expandafter\gdef\csname svy@item/t07/e2/miss\endcsname{0}
\expandafter\gdef\csname svy@item/t07/e2/median\endcsname{2.0}
\expandafter\gdef\csname svy@item/t07/e2/cilow\endcsname{2.0}
\expandafter\gdef\csname svy@item/t07/e2/cihigh\endcsname{3.0}
\expandafter\gdef\csname svy@item/t07/e2/mean\endcsname{2.44}
\expandafter\gdef\csname svy@item/t07/e2/meanlow\endcsname{2.20}
\expandafter\gdef\csname svy@item/t07/e2/meanhigh\endcsname{2.68}
\expandafter\gdef\csname svy@item/t07/e2/se\endcsname{0.122}
\expandafter\gdef\csname svy@item/t07/e2/sd\endcsname{1.07}
\expandafter\gdef\csname svy@item/t07/e2/sdhigh\endcsname{0}
\expandafter\gdef\csname svy@item/t07/e2/verdict\endcsname{INCONCLUSIVE}
\expandafter\gdef\csname svy@item/t07/e2/dip\endcsname{0}
\expandafter\gdef\csname svy@item/t07/e2/shapelow\endcsname{53}
\expandafter\gdef\csname svy@item/t07/e2/shapemid\endcsname{30}
\expandafter\gdef\csname svy@item/t07/e2/shapehigh\endcsname{17}
\expandafter\gdef\csname svy@item/t07/e3/n\endcsname{91}
\expandafter\gdef\csname svy@item/t07/e3/miss\endcsname{3}
\expandafter\gdef\csname svy@item/t07/e3/median\endcsname{3.0}
\expandafter\gdef\csname svy@item/t07/e3/cilow\endcsname{2.0}
\expandafter\gdef\csname svy@item/t07/e3/cihigh\endcsname{3.0}
\expandafter\gdef\csname svy@item/t07/e3/mean\endcsname{3.01}
\expandafter\gdef\csname svy@item/t07/e3/meanlow\endcsname{2.78}
\expandafter\gdef\csname svy@item/t07/e3/meanhigh\endcsname{3.24}
\expandafter\gdef\csname svy@item/t07/e3/se\endcsname{0.114}
\expandafter\gdef\csname svy@item/t07/e3/sd\endcsname{1.09}
\expandafter\gdef\csname svy@item/t07/e3/sdhigh\endcsname{0}
\expandafter\gdef\csname svy@item/t07/e3/verdict\endcsname{INCONCLUSIVE}
\expandafter\gdef\csname svy@item/t07/e3/dip\endcsname{1}
\expandafter\gdef\csname svy@item/t07/e3/shapelow\endcsname{39}
\expandafter\gdef\csname svy@item/t07/e3/shapemid\endcsname{22}
\expandafter\gdef\csname svy@item/t07/e3/shapehigh\endcsname{39}
\expandafter\gdef\csname svy@item/t09/exposure/trend\endcsname{0.25}
\expandafter\gdef\csname svy@item/t09/exposure/trendlow\endcsname{0.08}
\expandafter\gdef\csname svy@item/t09/exposure/trendhigh\endcsname{0.41}
\expandafter\gdef\csname svy@item/t09/e0/n\endcsname{7}
\expandafter\gdef\csname svy@item/t09/e0/miss\endcsname{6}
\expandafter\gdef\csname svy@item/t09/e0/median\endcsname{1.0}
\expandafter\gdef\csname svy@item/t09/e0/cilow\endcsname{1.0}
\expandafter\gdef\csname svy@item/t09/e0/cihigh\endcsname{1.0}
\expandafter\gdef\csname svy@item/t09/e0/mean\endcsname{1.00}
\expandafter\gdef\csname svy@item/t09/e0/meanlow\endcsname{1.00}
\expandafter\gdef\csname svy@item/t09/e0/meanhigh\endcsname{1.00}
\expandafter\gdef\csname svy@item/t09/e0/se\endcsname{0.000}
\expandafter\gdef\csname svy@item/t09/e0/sd\endcsname{0.00}
\expandafter\gdef\csname svy@item/t09/e0/sdhigh\endcsname{0}
\expandafter\gdef\csname svy@item/t09/e0/verdict\endcsname{CONFIRMED}
\expandafter\gdef\csname svy@item/t09/e0/dip\endcsname{0}
\expandafter\gdef\csname svy@item/t09/e0/shapelow\endcsname{100}
\expandafter\gdef\csname svy@item/t09/e0/shapemid\endcsname{0}
\expandafter\gdef\csname svy@item/t09/e0/shapehigh\endcsname{0}
\expandafter\gdef\csname svy@item/t09/e1/n\endcsname{47}
\expandafter\gdef\csname svy@item/t09/e1/miss\endcsname{2}
\expandafter\gdef\csname svy@item/t09/e1/median\endcsname{2.0}
\expandafter\gdef\csname svy@item/t09/e1/cilow\endcsname{1.0}
\expandafter\gdef\csname svy@item/t09/e1/cihigh\endcsname{2.0}
\expandafter\gdef\csname svy@item/t09/e1/mean\endcsname{2.11}
\expandafter\gdef\csname svy@item/t09/e1/meanlow\endcsname{1.79}
\expandafter\gdef\csname svy@item/t09/e1/meanhigh\endcsname{2.43}
\expandafter\gdef\csname svy@item/t09/e1/se\endcsname{0.159}
\expandafter\gdef\csname svy@item/t09/e1/sd\endcsname{1.09}
\expandafter\gdef\csname svy@item/t09/e1/sdhigh\endcsname{0}
\expandafter\gdef\csname svy@item/t09/e1/verdict\endcsname{CONFIRMED}
\expandafter\gdef\csname svy@item/t09/e1/dip\endcsname{0}
\expandafter\gdef\csname svy@item/t09/e1/shapelow\endcsname{68}
\expandafter\gdef\csname svy@item/t09/e1/shapemid\endcsname{19}
\expandafter\gdef\csname svy@item/t09/e1/shapehigh\endcsname{13}
\expandafter\gdef\csname svy@item/t09/e2/n\endcsname{77}
\expandafter\gdef\csname svy@item/t09/e2/miss\endcsname{0}
\expandafter\gdef\csname svy@item/t09/e2/median\endcsname{2.0}
\expandafter\gdef\csname svy@item/t09/e2/cilow\endcsname{2.0}
\expandafter\gdef\csname svy@item/t09/e2/cihigh\endcsname{3.0}
\expandafter\gdef\csname svy@item/t09/e2/mean\endcsname{2.27}
\expandafter\gdef\csname svy@item/t09/e2/meanlow\endcsname{2.02}
\expandafter\gdef\csname svy@item/t09/e2/meanhigh\endcsname{2.52}
\expandafter\gdef\csname svy@item/t09/e2/se\endcsname{0.125}
\expandafter\gdef\csname svy@item/t09/e2/sd\endcsname{1.10}
\expandafter\gdef\csname svy@item/t09/e2/sdhigh\endcsname{0}
\expandafter\gdef\csname svy@item/t09/e2/verdict\endcsname{INCONCLUSIVE}
\expandafter\gdef\csname svy@item/t09/e2/dip\endcsname{0}
\expandafter\gdef\csname svy@item/t09/e2/shapelow\endcsname{61}
\expandafter\gdef\csname svy@item/t09/e2/shapemid\endcsname{26}
\expandafter\gdef\csname svy@item/t09/e2/shapehigh\endcsname{13}
\expandafter\gdef\csname svy@item/t09/e3/n\endcsname{90}
\expandafter\gdef\csname svy@item/t09/e3/miss\endcsname{4}
\expandafter\gdef\csname svy@item/t09/e3/median\endcsname{2.0}
\expandafter\gdef\csname svy@item/t09/e3/cilow\endcsname{2.0}
\expandafter\gdef\csname svy@item/t09/e3/cihigh\endcsname{2.0}
\expandafter\gdef\csname svy@item/t09/e3/mean\endcsname{2.42}
\expandafter\gdef\csname svy@item/t09/e3/meanlow\endcsname{2.20}
\expandafter\gdef\csname svy@item/t09/e3/meanhigh\endcsname{2.64}
\expandafter\gdef\csname svy@item/t09/e3/se\endcsname{0.111}
\expandafter\gdef\csname svy@item/t09/e3/sd\endcsname{1.05}
\expandafter\gdef\csname svy@item/t09/e3/sdhigh\endcsname{0}
\expandafter\gdef\csname svy@item/t09/e3/verdict\endcsname{INCONCLUSIVE}
\expandafter\gdef\csname svy@item/t09/e3/dip\endcsname{0}
\expandafter\gdef\csname svy@item/t09/e3/shapelow\endcsname{67}
\expandafter\gdef\csname svy@item/t09/e3/shapemid\endcsname{15}
\expandafter\gdef\csname svy@item/t09/e3/shapehigh\endcsname{18}
\expandafter\gdef\csname svy@item/t10/exposure/trend\endcsname{0.18}
\expandafter\gdef\csname svy@item/t10/exposure/trendlow\endcsname{0.05}
\expandafter\gdef\csname svy@item/t10/exposure/trendhigh\endcsname{0.32}
\expandafter\gdef\csname svy@item/t10/e0/n\endcsname{9}
\expandafter\gdef\csname svy@item/t10/e0/miss\endcsname{4}
\expandafter\gdef\csname svy@item/t10/e0/median\endcsname{1.0}
\expandafter\gdef\csname svy@item/t10/e0/cilow\endcsname{1.0}
\expandafter\gdef\csname svy@item/t10/e0/cihigh\endcsname{2.0}
\expandafter\gdef\csname svy@item/t10/e0/mean\endcsname{1.22}
\expandafter\gdef\csname svy@item/t10/e0/meanlow\endcsname{0.88}
\expandafter\gdef\csname svy@item/t10/e0/meanhigh\endcsname{1.56}
\expandafter\gdef\csname svy@item/t10/e0/se\endcsname{0.147}
\expandafter\gdef\csname svy@item/t10/e0/sd\endcsname{0.44}
\expandafter\gdef\csname svy@item/t10/e0/sdhigh\endcsname{0}
\expandafter\gdef\csname svy@item/t10/e0/verdict\endcsname{CONFIRMED}
\expandafter\gdef\csname svy@item/t10/e0/dip\endcsname{0}
\expandafter\gdef\csname svy@item/t10/e0/shapelow\endcsname{100}
\expandafter\gdef\csname svy@item/t10/e0/shapemid\endcsname{0}
\expandafter\gdef\csname svy@item/t10/e0/shapehigh\endcsname{0}
\expandafter\gdef\csname svy@item/t10/e1/n\endcsname{47}
\expandafter\gdef\csname svy@item/t10/e1/miss\endcsname{2}
\expandafter\gdef\csname svy@item/t10/e1/median\endcsname{1.0}
\expandafter\gdef\csname svy@item/t10/e1/cilow\endcsname{1.0}
\expandafter\gdef\csname svy@item/t10/e1/cihigh\endcsname{2.0}
\expandafter\gdef\csname svy@item/t10/e1/mean\endcsname{1.70}
\expandafter\gdef\csname svy@item/t10/e1/meanlow\endcsname{1.43}
\expandafter\gdef\csname svy@item/t10/e1/meanhigh\endcsname{1.98}
\expandafter\gdef\csname svy@item/t10/e1/se\endcsname{0.136}
\expandafter\gdef\csname svy@item/t10/e1/sd\endcsname{0.93}
\expandafter\gdef\csname svy@item/t10/e1/sdhigh\endcsname{0}
\expandafter\gdef\csname svy@item/t10/e1/verdict\endcsname{CONFIRMED}
\expandafter\gdef\csname svy@item/t10/e1/dip\endcsname{0}
\expandafter\gdef\csname svy@item/t10/e1/shapelow\endcsname{81}
\expandafter\gdef\csname svy@item/t10/e1/shapemid\endcsname{13}
\expandafter\gdef\csname svy@item/t10/e1/shapehigh\endcsname{6}
\expandafter\gdef\csname svy@item/t10/e2/n\endcsname{76}
\expandafter\gdef\csname svy@item/t10/e2/miss\endcsname{1}
\expandafter\gdef\csname svy@item/t10/e2/median\endcsname{2.0}
\expandafter\gdef\csname svy@item/t10/e2/cilow\endcsname{1.0}
\expandafter\gdef\csname svy@item/t10/e2/cihigh\endcsname{2.0}
\expandafter\gdef\csname svy@item/t10/e2/mean\endcsname{1.78}
\expandafter\gdef\csname svy@item/t10/e2/meanlow\endcsname{1.57}
\expandafter\gdef\csname svy@item/t10/e2/meanhigh\endcsname{1.99}
\expandafter\gdef\csname svy@item/t10/e2/se\endcsname{0.105}
\expandafter\gdef\csname svy@item/t10/e2/sd\endcsname{0.92}
\expandafter\gdef\csname svy@item/t10/e2/sdhigh\endcsname{0}
\expandafter\gdef\csname svy@item/t10/e2/verdict\endcsname{CONFIRMED}
\expandafter\gdef\csname svy@item/t10/e2/dip\endcsname{0}
\expandafter\gdef\csname svy@item/t10/e2/shapelow\endcsname{84}
\expandafter\gdef\csname svy@item/t10/e2/shapemid\endcsname{9}
\expandafter\gdef\csname svy@item/t10/e2/shapehigh\endcsname{7}
\expandafter\gdef\csname svy@item/t10/e3/n\endcsname{93}
\expandafter\gdef\csname svy@item/t10/e3/miss\endcsname{1}
\expandafter\gdef\csname svy@item/t10/e3/median\endcsname{2.0}
\expandafter\gdef\csname svy@item/t10/e3/cilow\endcsname{2.0}
\expandafter\gdef\csname svy@item/t10/e3/cihigh\endcsname{2.0}
\expandafter\gdef\csname svy@item/t10/e3/mean\endcsname{1.99}
\expandafter\gdef\csname svy@item/t10/e3/meanlow\endcsname{1.80}
\expandafter\gdef\csname svy@item/t10/e3/meanhigh\endcsname{2.18}
\expandafter\gdef\csname svy@item/t10/e3/se\endcsname{0.097}
\expandafter\gdef\csname svy@item/t10/e3/sd\endcsname{0.94}
\expandafter\gdef\csname svy@item/t10/e3/sdhigh\endcsname{0}
\expandafter\gdef\csname svy@item/t10/e3/verdict\endcsname{CONFIRMED}
\expandafter\gdef\csname svy@item/t10/e3/dip\endcsname{0}
\expandafter\gdef\csname svy@item/t10/e3/shapelow\endcsname{79}
\expandafter\gdef\csname svy@item/t10/e3/shapemid\endcsname{11}
\expandafter\gdef\csname svy@item/t10/e3/shapehigh\endcsname{10}
\expandafter\gdef\csname svy@item/t11/exposure/trend\endcsname{0.42}
\expandafter\gdef\csname svy@item/t11/exposure/trendlow\endcsname{0.29}
\expandafter\gdef\csname svy@item/t11/exposure/trendhigh\endcsname{0.56}
\expandafter\gdef\csname svy@item/t11/e0/n\endcsname{13}
\expandafter\gdef\csname svy@item/t11/e0/miss\endcsname{0}
\expandafter\gdef\csname svy@item/t11/e0/median\endcsname{1.0}
\expandafter\gdef\csname svy@item/t11/e0/cilow\endcsname{1.0}
\expandafter\gdef\csname svy@item/t11/e0/cihigh\endcsname{2.0}
\expandafter\gdef\csname svy@item/t11/e0/mean\endcsname{1.23}
\expandafter\gdef\csname svy@item/t11/e0/meanlow\endcsname{0.97}
\expandafter\gdef\csname svy@item/t11/e0/meanhigh\endcsname{1.50}
\expandafter\gdef\csname svy@item/t11/e0/se\endcsname{0.122}
\expandafter\gdef\csname svy@item/t11/e0/sd\endcsname{0.44}
\expandafter\gdef\csname svy@item/t11/e0/sdhigh\endcsname{0}
\expandafter\gdef\csname svy@item/t11/e0/verdict\endcsname{NO_PREDICTION}
\expandafter\gdef\csname svy@item/t11/e0/dip\endcsname{0}
\expandafter\gdef\csname svy@item/t11/e0/shapelow\endcsname{100}
\expandafter\gdef\csname svy@item/t11/e0/shapemid\endcsname{0}
\expandafter\gdef\csname svy@item/t11/e0/shapehigh\endcsname{0}
\expandafter\gdef\csname svy@item/t11/e1/n\endcsname{46}
\expandafter\gdef\csname svy@item/t11/e1/miss\endcsname{3}
\expandafter\gdef\csname svy@item/t11/e1/median\endcsname{1.0}
\expandafter\gdef\csname svy@item/t11/e1/cilow\endcsname{1.0}
\expandafter\gdef\csname svy@item/t11/e1/cihigh\endcsname{2.0}
\expandafter\gdef\csname svy@item/t11/e1/mean\endcsname{1.67}
\expandafter\gdef\csname svy@item/t11/e1/meanlow\endcsname{1.42}
\expandafter\gdef\csname svy@item/t11/e1/meanhigh\endcsname{1.92}
\expandafter\gdef\csname svy@item/t11/e1/se\endcsname{0.125}
\expandafter\gdef\csname svy@item/t11/e1/sd\endcsname{0.84}
\expandafter\gdef\csname svy@item/t11/e1/sdhigh\endcsname{0}
\expandafter\gdef\csname svy@item/t11/e1/verdict\endcsname{NO_PREDICTION}
\expandafter\gdef\csname svy@item/t11/e1/dip\endcsname{0}
\expandafter\gdef\csname svy@item/t11/e1/shapelow\endcsname{85}
\expandafter\gdef\csname svy@item/t11/e1/shapemid\endcsname{11}
\expandafter\gdef\csname svy@item/t11/e1/shapehigh\endcsname{4}
\expandafter\gdef\csname svy@item/t11/e2/n\endcsname{74}
\expandafter\gdef\csname svy@item/t11/e2/miss\endcsname{3}
\expandafter\gdef\csname svy@item/t11/e2/median\endcsname{2.0}
\expandafter\gdef\csname svy@item/t11/e2/cilow\endcsname{2.0}
\expandafter\gdef\csname svy@item/t11/e2/cihigh\endcsname{2.0}
\expandafter\gdef\csname svy@item/t11/e2/mean\endcsname{1.97}
\expandafter\gdef\csname svy@item/t11/e2/meanlow\endcsname{1.76}
\expandafter\gdef\csname svy@item/t11/e2/meanhigh\endcsname{2.19}
\expandafter\gdef\csname svy@item/t11/e2/se\endcsname{0.107}
\expandafter\gdef\csname svy@item/t11/e2/sd\endcsname{0.92}
\expandafter\gdef\csname svy@item/t11/e2/sdhigh\endcsname{0}
\expandafter\gdef\csname svy@item/t11/e2/verdict\endcsname{NO_PREDICTION}
\expandafter\gdef\csname svy@item/t11/e2/dip\endcsname{0}
\expandafter\gdef\csname svy@item/t11/e2/shapelow\endcsname{80}
\expandafter\gdef\csname svy@item/t11/e2/shapemid\endcsname{12}
\expandafter\gdef\csname svy@item/t11/e2/shapehigh\endcsname{8}
\expandafter\gdef\csname svy@item/t11/e3/n\endcsname{90}
\expandafter\gdef\csname svy@item/t11/e3/miss\endcsname{4}
\expandafter\gdef\csname svy@item/t11/e3/median\endcsname{2.0}
\expandafter\gdef\csname svy@item/t11/e3/cilow\endcsname{2.0}
\expandafter\gdef\csname svy@item/t11/e3/cihigh\endcsname{2.0}
\expandafter\gdef\csname svy@item/t11/e3/mean\endcsname{2.50}
\expandafter\gdef\csname svy@item/t11/e3/meanlow\endcsname{2.28}
\expandafter\gdef\csname svy@item/t11/e3/meanhigh\endcsname{2.72}
\expandafter\gdef\csname svy@item/t11/e3/se\endcsname{0.110}
\expandafter\gdef\csname svy@item/t11/e3/sd\endcsname{1.04}
\expandafter\gdef\csname svy@item/t11/e3/sdhigh\endcsname{0}
\expandafter\gdef\csname svy@item/t11/e3/verdict\endcsname{NO_PREDICTION}
\expandafter\gdef\csname svy@item/t11/e3/dip\endcsname{0}
\expandafter\gdef\csname svy@item/t11/e3/shapelow\endcsname{61}
\expandafter\gdef\csname svy@item/t11/e3/shapemid\endcsname{21}
\expandafter\gdef\csname svy@item/t11/e3/shapehigh\endcsname{18}
\expandafter\gdef\csname svy@item/t12/exposure/trend\endcsname{0.69}
\expandafter\gdef\csname svy@item/t12/exposure/trendlow\endcsname{0.53}
\expandafter\gdef\csname svy@item/t12/exposure/trendhigh\endcsname{0.84}
\expandafter\gdef\csname svy@item/t12/e0/n\endcsname{12}
\expandafter\gdef\csname svy@item/t12/e0/miss\endcsname{1}
\expandafter\gdef\csname svy@item/t12/e0/median\endcsname{1.0}
\expandafter\gdef\csname svy@item/t12/e0/cilow\endcsname{1.0}
\expandafter\gdef\csname svy@item/t12/e0/cihigh\endcsname{2.0}
\expandafter\gdef\csname svy@item/t12/e0/mean\endcsname{1.42}
\expandafter\gdef\csname svy@item/t12/e0/meanlow\endcsname{1.09}
\expandafter\gdef\csname svy@item/t12/e0/meanhigh\endcsname{1.74}
\expandafter\gdef\csname svy@item/t12/e0/se\endcsname{0.149}
\expandafter\gdef\csname svy@item/t12/e0/sd\endcsname{0.51}
\expandafter\gdef\csname svy@item/t12/e0/sdhigh\endcsname{0}
\expandafter\gdef\csname svy@item/t12/e0/verdict\endcsname{NO_PREDICTION}
\expandafter\gdef\csname svy@item/t12/e0/dip\endcsname{0}
\expandafter\gdef\csname svy@item/t12/e0/shapelow\endcsname{100}
\expandafter\gdef\csname svy@item/t12/e0/shapemid\endcsname{0}
\expandafter\gdef\csname svy@item/t12/e0/shapehigh\endcsname{0}
\expandafter\gdef\csname svy@item/t12/e1/n\endcsname{48}
\expandafter\gdef\csname svy@item/t12/e1/miss\endcsname{1}
\expandafter\gdef\csname svy@item/t12/e1/median\endcsname{2.5}
\expandafter\gdef\csname svy@item/t12/e1/cilow\endcsname{2.0}
\expandafter\gdef\csname svy@item/t12/e1/cihigh\endcsname{3.0}
\expandafter\gdef\csname svy@item/t12/e1/mean\endcsname{2.58}
\expandafter\gdef\csname svy@item/t12/e1/meanlow\endcsname{2.27}
\expandafter\gdef\csname svy@item/t12/e1/meanhigh\endcsname{2.90}
\expandafter\gdef\csname svy@item/t12/e1/se\endcsname{0.157}
\expandafter\gdef\csname svy@item/t12/e1/sd\endcsname{1.09}
\expandafter\gdef\csname svy@item/t12/e1/sdhigh\endcsname{0}
\expandafter\gdef\csname svy@item/t12/e1/verdict\endcsname{NO_PREDICTION}
\expandafter\gdef\csname svy@item/t12/e1/dip\endcsname{0}
\expandafter\gdef\csname svy@item/t12/e1/shapelow\endcsname{50}
\expandafter\gdef\csname svy@item/t12/e1/shapemid\endcsname{29}
\expandafter\gdef\csname svy@item/t12/e1/shapehigh\endcsname{21}
\expandafter\gdef\csname svy@item/t12/e2/n\endcsname{77}
\expandafter\gdef\csname svy@item/t12/e2/miss\endcsname{0}
\expandafter\gdef\csname svy@item/t12/e2/median\endcsname{3.0}
\expandafter\gdef\csname svy@item/t12/e2/cilow\endcsname{3.0}
\expandafter\gdef\csname svy@item/t12/e2/cihigh\endcsname{4.0}
\expandafter\gdef\csname svy@item/t12/e2/mean\endcsname{3.13}
\expandafter\gdef\csname svy@item/t12/e2/meanlow\endcsname{2.87}
\expandafter\gdef\csname svy@item/t12/e2/meanhigh\endcsname{3.39}
\expandafter\gdef\csname svy@item/t12/e2/se\endcsname{0.129}
\expandafter\gdef\csname svy@item/t12/e2/sd\endcsname{1.13}
\expandafter\gdef\csname svy@item/t12/e2/sdhigh\endcsname{0}
\expandafter\gdef\csname svy@item/t12/e2/verdict\endcsname{NO_PREDICTION}
\expandafter\gdef\csname svy@item/t12/e2/dip\endcsname{0}
\expandafter\gdef\csname svy@item/t12/e2/shapelow\endcsname{31}
\expandafter\gdef\csname svy@item/t12/e2/shapemid\endcsname{23}
\expandafter\gdef\csname svy@item/t12/e2/shapehigh\endcsname{46}
\expandafter\gdef\csname svy@item/t12/e3/n\endcsname{90}
\expandafter\gdef\csname svy@item/t12/e3/miss\endcsname{4}
\expandafter\gdef\csname svy@item/t12/e3/median\endcsname{4.0}
\expandafter\gdef\csname svy@item/t12/e3/cilow\endcsname{4.0}
\expandafter\gdef\csname svy@item/t12/e3/cihigh\endcsname{4.0}
\expandafter\gdef\csname svy@item/t12/e3/mean\endcsname{3.80}
\expandafter\gdef\csname svy@item/t12/e3/meanlow\endcsname{3.58}
\expandafter\gdef\csname svy@item/t12/e3/meanhigh\endcsname{4.02}
\expandafter\gdef\csname svy@item/t12/e3/se\endcsname{0.109}
\expandafter\gdef\csname svy@item/t12/e3/sd\endcsname{1.03}
\expandafter\gdef\csname svy@item/t12/e3/sdhigh\endcsname{0}
\expandafter\gdef\csname svy@item/t12/e3/verdict\endcsname{NO_PREDICTION}
\expandafter\gdef\csname svy@item/t12/e3/dip\endcsname{0}
\expandafter\gdef\csname svy@item/t12/e3/shapelow\endcsname{15}
\expandafter\gdef\csname svy@item/t12/e3/shapemid\endcsname{11}
\expandafter\gdef\csname svy@item/t12/e3/shapehigh\endcsname{74}
\expandafter\gdef\csname svy@item/t13/exposure/trend\endcsname{0.51}
\expandafter\gdef\csname svy@item/t13/exposure/trendlow\endcsname{0.34}
\expandafter\gdef\csname svy@item/t13/exposure/trendhigh\endcsname{0.68}
\expandafter\gdef\csname svy@item/t13/e0/n\endcsname{12}
\expandafter\gdef\csname svy@item/t13/e0/miss\endcsname{1}
\expandafter\gdef\csname svy@item/t13/e0/median\endcsname{2.0}
\expandafter\gdef\csname svy@item/t13/e0/cilow\endcsname{1.0}
\expandafter\gdef\csname svy@item/t13/e0/cihigh\endcsname{2.0}
\expandafter\gdef\csname svy@item/t13/e0/mean\endcsname{1.67}
\expandafter\gdef\csname svy@item/t13/e0/meanlow\endcsname{1.25}
\expandafter\gdef\csname svy@item/t13/e0/meanhigh\endcsname{2.08}
\expandafter\gdef\csname svy@item/t13/e0/se\endcsname{0.188}
\expandafter\gdef\csname svy@item/t13/e0/sd\endcsname{0.65}
\expandafter\gdef\csname svy@item/t13/e0/sdhigh\endcsname{0}
\expandafter\gdef\csname svy@item/t13/e0/verdict\endcsname{CONFIRMED}
\expandafter\gdef\csname svy@item/t13/e0/dip\endcsname{0}
\expandafter\gdef\csname svy@item/t13/e0/shapelow\endcsname{92}
\expandafter\gdef\csname svy@item/t13/e0/shapemid\endcsname{8}
\expandafter\gdef\csname svy@item/t13/e0/shapehigh\endcsname{0}
\expandafter\gdef\csname svy@item/t13/e1/n\endcsname{44}
\expandafter\gdef\csname svy@item/t13/e1/miss\endcsname{5}
\expandafter\gdef\csname svy@item/t13/e1/median\endcsname{3.0}
\expandafter\gdef\csname svy@item/t13/e1/cilow\endcsname{2.0}
\expandafter\gdef\csname svy@item/t13/e1/cihigh\endcsname{3.0}
\expandafter\gdef\csname svy@item/t13/e1/mean\endcsname{2.55}
\expandafter\gdef\csname svy@item/t13/e1/meanlow\endcsname{2.23}
\expandafter\gdef\csname svy@item/t13/e1/meanhigh\endcsname{2.86}
\expandafter\gdef\csname svy@item/t13/e1/se\endcsname{0.157}
\expandafter\gdef\csname svy@item/t13/e1/sd\endcsname{1.04}
\expandafter\gdef\csname svy@item/t13/e1/sdhigh\endcsname{0}
\expandafter\gdef\csname svy@item/t13/e1/verdict\endcsname{INCONCLUSIVE}
\expandafter\gdef\csname svy@item/t13/e1/dip\endcsname{0}
\expandafter\gdef\csname svy@item/t13/e1/shapelow\endcsname{43}
\expandafter\gdef\csname svy@item/t13/e1/shapemid\endcsname{41}
\expandafter\gdef\csname svy@item/t13/e1/shapehigh\endcsname{16}
\expandafter\gdef\csname svy@item/t13/e2/n\endcsname{74}
\expandafter\gdef\csname svy@item/t13/e2/miss\endcsname{3}
\expandafter\gdef\csname svy@item/t13/e2/median\endcsname{3.0}
\expandafter\gdef\csname svy@item/t13/e2/cilow\endcsname{3.0}
\expandafter\gdef\csname svy@item/t13/e2/cihigh\endcsname{3.0}
\expandafter\gdef\csname svy@item/t13/e2/mean\endcsname{2.93}
\expandafter\gdef\csname svy@item/t13/e2/meanlow\endcsname{2.65}
\expandafter\gdef\csname svy@item/t13/e2/meanhigh\endcsname{3.22}
\expandafter\gdef\csname svy@item/t13/e2/se\endcsname{0.143}
\expandafter\gdef\csname svy@item/t13/e2/sd\endcsname{1.23}
\expandafter\gdef\csname svy@item/t13/e2/sdhigh\endcsname{0}
\expandafter\gdef\csname svy@item/t13/e2/verdict\endcsname{REJECTED}
\expandafter\gdef\csname svy@item/t13/e2/dip\endcsname{0}
\expandafter\gdef\csname svy@item/t13/e2/shapelow\endcsname{35}
\expandafter\gdef\csname svy@item/t13/e2/shapemid\endcsname{35}
\expandafter\gdef\csname svy@item/t13/e2/shapehigh\endcsname{30}
\expandafter\gdef\csname svy@item/t13/e3/n\endcsname{89}
\expandafter\gdef\csname svy@item/t13/e3/miss\endcsname{5}
\expandafter\gdef\csname svy@item/t13/e3/median\endcsname{4.0}
\expandafter\gdef\csname svy@item/t13/e3/cilow\endcsname{3.0}
\expandafter\gdef\csname svy@item/t13/e3/cihigh\endcsname{4.0}
\expandafter\gdef\csname svy@item/t13/e3/mean\endcsname{3.44}
\expandafter\gdef\csname svy@item/t13/e3/meanlow\endcsname{3.19}
\expandafter\gdef\csname svy@item/t13/e3/meanhigh\endcsname{3.68}
\expandafter\gdef\csname svy@item/t13/e3/se\endcsname{0.123}
\expandafter\gdef\csname svy@item/t13/e3/sd\endcsname{1.16}
\expandafter\gdef\csname svy@item/t13/e3/sdhigh\endcsname{0}
\expandafter\gdef\csname svy@item/t13/e3/verdict\endcsname{INCONCLUSIVE}
\expandafter\gdef\csname svy@item/t13/e3/dip\endcsname{0}
\expandafter\gdef\csname svy@item/t13/e3/shapelow\endcsname{21}
\expandafter\gdef\csname svy@item/t13/e3/shapemid\endcsname{26}
\expandafter\gdef\csname svy@item/t13/e3/shapehigh\endcsname{53}
\expandafter\gdef\csname svy@item/t16/exposure/trend\endcsname{0.40}
\expandafter\gdef\csname svy@item/t16/exposure/trendlow\endcsname{0.27}
\expandafter\gdef\csname svy@item/t16/exposure/trendhigh\endcsname{0.53}
\expandafter\gdef\csname svy@item/t16/e0/n\endcsname{13}
\expandafter\gdef\csname svy@item/t16/e0/miss\endcsname{0}
\expandafter\gdef\csname svy@item/t16/e0/median\endcsname{1.0}
\expandafter\gdef\csname svy@item/t16/e0/cilow\endcsname{1.0}
\expandafter\gdef\csname svy@item/t16/e0/cihigh\endcsname{1.0}
\expandafter\gdef\csname svy@item/t16/e0/mean\endcsname{1.08}
\expandafter\gdef\csname svy@item/t16/e0/meanlow\endcsname{0.91}
\expandafter\gdef\csname svy@item/t16/e0/meanhigh\endcsname{1.24}
\expandafter\gdef\csname svy@item/t16/e0/se\endcsname{0.077}
\expandafter\gdef\csname svy@item/t16/e0/sd\endcsname{0.28}
\expandafter\gdef\csname svy@item/t16/e0/sdhigh\endcsname{0}
\expandafter\gdef\csname svy@item/t16/e0/verdict\endcsname{CONFIRMED}
\expandafter\gdef\csname svy@item/t16/e0/dip\endcsname{0}
\expandafter\gdef\csname svy@item/t16/e0/shapelow\endcsname{100}
\expandafter\gdef\csname svy@item/t16/e0/shapemid\endcsname{0}
\expandafter\gdef\csname svy@item/t16/e0/shapehigh\endcsname{0}
\expandafter\gdef\csname svy@item/t16/e1/n\endcsname{45}
\expandafter\gdef\csname svy@item/t16/e1/miss\endcsname{4}
\expandafter\gdef\csname svy@item/t16/e1/median\endcsname{1.0}
\expandafter\gdef\csname svy@item/t16/e1/cilow\endcsname{1.0}
\expandafter\gdef\csname svy@item/t16/e1/cihigh\endcsname{2.0}
\expandafter\gdef\csname svy@item/t16/e1/mean\endcsname{1.69}
\expandafter\gdef\csname svy@item/t16/e1/meanlow\endcsname{1.43}
\expandafter\gdef\csname svy@item/t16/e1/meanhigh\endcsname{1.95}
\expandafter\gdef\csname svy@item/t16/e1/se\endcsname{0.130}
\expandafter\gdef\csname svy@item/t16/e1/sd\endcsname{0.87}
\expandafter\gdef\csname svy@item/t16/e1/sdhigh\endcsname{0}
\expandafter\gdef\csname svy@item/t16/e1/verdict\endcsname{CONFIRMED}
\expandafter\gdef\csname svy@item/t16/e1/dip\endcsname{0}
\expandafter\gdef\csname svy@item/t16/e1/shapelow\endcsname{86}
\expandafter\gdef\csname svy@item/t16/e1/shapemid\endcsname{7}
\expandafter\gdef\csname svy@item/t16/e1/shapehigh\endcsname{7}
\expandafter\gdef\csname svy@item/t16/e2/n\endcsname{74}
\expandafter\gdef\csname svy@item/t16/e2/miss\endcsname{3}
\expandafter\gdef\csname svy@item/t16/e2/median\endcsname{2.0}
\expandafter\gdef\csname svy@item/t16/e2/cilow\endcsname{1.0}
\expandafter\gdef\csname svy@item/t16/e2/cihigh\endcsname{2.0}
\expandafter\gdef\csname svy@item/t16/e2/mean\endcsname{1.78}
\expandafter\gdef\csname svy@item/t16/e2/meanlow\endcsname{1.59}
\expandafter\gdef\csname svy@item/t16/e2/meanhigh\endcsname{1.98}
\expandafter\gdef\csname svy@item/t16/e2/se\endcsname{0.099}
\expandafter\gdef\csname svy@item/t16/e2/sd\endcsname{0.85}
\expandafter\gdef\csname svy@item/t16/e2/sdhigh\endcsname{0}
\expandafter\gdef\csname svy@item/t16/e2/verdict\endcsname{CONFIRMED}
\expandafter\gdef\csname svy@item/t16/e2/dip\endcsname{0}
\expandafter\gdef\csname svy@item/t16/e2/shapelow\endcsname{84}
\expandafter\gdef\csname svy@item/t16/e2/shapemid\endcsname{11}
\expandafter\gdef\csname svy@item/t16/e2/shapehigh\endcsname{5}
\expandafter\gdef\csname svy@item/t16/e3/n\endcsname{90}
\expandafter\gdef\csname svy@item/t16/e3/miss\endcsname{4}
\expandafter\gdef\csname svy@item/t16/e3/median\endcsname{2.0}
\expandafter\gdef\csname svy@item/t16/e3/cilow\endcsname{2.0}
\expandafter\gdef\csname svy@item/t16/e3/cihigh\endcsname{2.0}
\expandafter\gdef\csname svy@item/t16/e3/mean\endcsname{2.40}
\expandafter\gdef\csname svy@item/t16/e3/meanlow\endcsname{2.19}
\expandafter\gdef\csname svy@item/t16/e3/meanhigh\endcsname{2.61}
\expandafter\gdef\csname svy@item/t16/e3/se\endcsname{0.105}
\expandafter\gdef\csname svy@item/t16/e3/sd\endcsname{0.99}
\expandafter\gdef\csname svy@item/t16/e3/sdhigh\endcsname{0}
\expandafter\gdef\csname svy@item/t16/e3/verdict\endcsname{INCONCLUSIVE}
\expandafter\gdef\csname svy@item/t16/e3/dip\endcsname{0}
\expandafter\gdef\csname svy@item/t16/e3/shapelow\endcsname{64}
\expandafter\gdef\csname svy@item/t16/e3/shapemid\endcsname{20}
\expandafter\gdef\csname svy@item/t16/e3/shapehigh\endcsname{16}
\expandafter\gdef\csname svy@item/t17/exposure/trend\endcsname{0.75}
\expandafter\gdef\csname svy@item/t17/exposure/trendlow\endcsname{0.62}
\expandafter\gdef\csname svy@item/t17/exposure/trendhigh\endcsname{0.89}
\expandafter\gdef\csname svy@item/t17/e0/n\endcsname{12}
\expandafter\gdef\csname svy@item/t17/e0/miss\endcsname{1}
\expandafter\gdef\csname svy@item/t17/e0/median\endcsname{1.0}
\expandafter\gdef\csname svy@item/t17/e0/cilow\endcsname{1.0}
\expandafter\gdef\csname svy@item/t17/e0/cihigh\endcsname{2.0}
\expandafter\gdef\csname svy@item/t17/e0/mean\endcsname{1.50}
\expandafter\gdef\csname svy@item/t17/e0/meanlow\endcsname{1.07}
\expandafter\gdef\csname svy@item/t17/e0/meanhigh\endcsname{1.93}
\expandafter\gdef\csname svy@item/t17/e0/se\endcsname{0.195}
\expandafter\gdef\csname svy@item/t17/e0/sd\endcsname{0.67}
\expandafter\gdef\csname svy@item/t17/e0/sdhigh\endcsname{0}
\expandafter\gdef\csname svy@item/t17/e0/verdict\endcsname{REJECTED}
\expandafter\gdef\csname svy@item/t17/e0/dip\endcsname{0}
\expandafter\gdef\csname svy@item/t17/e0/shapelow\endcsname{92}
\expandafter\gdef\csname svy@item/t17/e0/shapemid\endcsname{8}
\expandafter\gdef\csname svy@item/t17/e0/shapehigh\endcsname{0}
\expandafter\gdef\csname svy@item/t17/e1/n\endcsname{47}
\expandafter\gdef\csname svy@item/t17/e1/miss\endcsname{2}
\expandafter\gdef\csname svy@item/t17/e1/median\endcsname{3.0}
\expandafter\gdef\csname svy@item/t17/e1/cilow\endcsname{2.0}
\expandafter\gdef\csname svy@item/t17/e1/cihigh\endcsname{3.0}
\expandafter\gdef\csname svy@item/t17/e1/mean\endcsname{2.70}
\expandafter\gdef\csname svy@item/t17/e1/meanlow\endcsname{2.39}
\expandafter\gdef\csname svy@item/t17/e1/meanhigh\endcsname{3.01}
\expandafter\gdef\csname svy@item/t17/e1/se\endcsname{0.155}
\expandafter\gdef\csname svy@item/t17/e1/sd\endcsname{1.06}
\expandafter\gdef\csname svy@item/t17/e1/sdhigh\endcsname{0}
\expandafter\gdef\csname svy@item/t17/e1/verdict\endcsname{INCONCLUSIVE}
\expandafter\gdef\csname svy@item/t17/e1/dip\endcsname{0}
\expandafter\gdef\csname svy@item/t17/e1/shapelow\endcsname{49}
\expandafter\gdef\csname svy@item/t17/e1/shapemid\endcsname{26}
\expandafter\gdef\csname svy@item/t17/e1/shapehigh\endcsname{25}
\expandafter\gdef\csname svy@item/t17/e2/n\endcsname{75}
\expandafter\gdef\csname svy@item/t17/e2/miss\endcsname{2}
\expandafter\gdef\csname svy@item/t17/e2/median\endcsname{3.0}
\expandafter\gdef\csname svy@item/t17/e2/cilow\endcsname{3.0}
\expandafter\gdef\csname svy@item/t17/e2/cihigh\endcsname{4.0}
\expandafter\gdef\csname svy@item/t17/e2/mean\endcsname{3.29}
\expandafter\gdef\csname svy@item/t17/e2/meanlow\endcsname{3.06}
\expandafter\gdef\csname svy@item/t17/e2/meanhigh\endcsname{3.53}
\expandafter\gdef\csname svy@item/t17/e2/se\endcsname{0.117}
\expandafter\gdef\csname svy@item/t17/e2/sd\endcsname{1.01}
\expandafter\gdef\csname svy@item/t17/e2/sdhigh\endcsname{0}
\expandafter\gdef\csname svy@item/t17/e2/verdict\endcsname{INCONCLUSIVE}
\expandafter\gdef\csname svy@item/t17/e2/dip\endcsname{0}
\expandafter\gdef\csname svy@item/t17/e2/shapelow\endcsname{23}
\expandafter\gdef\csname svy@item/t17/e2/shapemid\endcsname{31}
\expandafter\gdef\csname svy@item/t17/e2/shapehigh\endcsname{46}
\expandafter\gdef\csname svy@item/t17/e3/n\endcsname{94}
\expandafter\gdef\csname svy@item/t17/e3/miss\endcsname{0}
\expandafter\gdef\csname svy@item/t17/e3/median\endcsname{4.0}
\expandafter\gdef\csname svy@item/t17/e3/cilow\endcsname{4.0}
\expandafter\gdef\csname svy@item/t17/e3/cihigh\endcsname{4.0}
\expandafter\gdef\csname svy@item/t17/e3/mean\endcsname{4.06}
\expandafter\gdef\csname svy@item/t17/e3/meanlow\endcsname{3.88}
\expandafter\gdef\csname svy@item/t17/e3/meanhigh\endcsname{4.25}
\expandafter\gdef\csname svy@item/t17/e3/se\endcsname{0.092}
\expandafter\gdef\csname svy@item/t17/e3/sd\endcsname{0.89}
\expandafter\gdef\csname svy@item/t17/e3/sdhigh\endcsname{0}
\expandafter\gdef\csname svy@item/t17/e3/verdict\endcsname{CONFIRMED}
\expandafter\gdef\csname svy@item/t17/e3/dip\endcsname{0}
\expandafter\gdef\csname svy@item/t17/e3/shapelow\endcsname{8}
\expandafter\gdef\csname svy@item/t17/e3/shapemid\endcsname{11}
\expandafter\gdef\csname svy@item/t17/e3/shapehigh\endcsname{81}
\expandafter\gdef\csname svy@item/t18/exposure/trend\endcsname{0.34}
\expandafter\gdef\csname svy@item/t18/exposure/trendlow\endcsname{0.19}
\expandafter\gdef\csname svy@item/t18/exposure/trendhigh\endcsname{0.49}
\expandafter\gdef\csname svy@item/t18/e0/n\endcsname{12}
\expandafter\gdef\csname svy@item/t18/e0/miss\endcsname{1}
\expandafter\gdef\csname svy@item/t18/e0/median\endcsname{2.0}
\expandafter\gdef\csname svy@item/t18/e0/cilow\endcsname{1.0}
\expandafter\gdef\csname svy@item/t18/e0/cihigh\endcsname{2.0}
\expandafter\gdef\csname svy@item/t18/e0/mean\endcsname{1.92}
\expandafter\gdef\csname svy@item/t18/e0/meanlow\endcsname{1.49}
\expandafter\gdef\csname svy@item/t18/e0/meanhigh\endcsname{2.34}
\expandafter\gdef\csname svy@item/t18/e0/se\endcsname{0.193}
\expandafter\gdef\csname svy@item/t18/e0/sd\endcsname{0.67}
\expandafter\gdef\csname svy@item/t18/e0/sdhigh\endcsname{0}
\expandafter\gdef\csname svy@item/t18/e0/verdict\endcsname{REJECTED}
\expandafter\gdef\csname svy@item/t18/e0/dip\endcsname{0}
\expandafter\gdef\csname svy@item/t18/e0/shapelow\endcsname{83}
\expandafter\gdef\csname svy@item/t18/e0/shapemid\endcsname{17}
\expandafter\gdef\csname svy@item/t18/e0/shapehigh\endcsname{0}
\expandafter\gdef\csname svy@item/t18/e1/n\endcsname{48}
\expandafter\gdef\csname svy@item/t18/e1/miss\endcsname{1}
\expandafter\gdef\csname svy@item/t18/e1/median\endcsname{3.0}
\expandafter\gdef\csname svy@item/t18/e1/cilow\endcsname{2.0}
\expandafter\gdef\csname svy@item/t18/e1/cihigh\endcsname{3.0}
\expandafter\gdef\csname svy@item/t18/e1/mean\endcsname{2.77}
\expandafter\gdef\csname svy@item/t18/e1/meanlow\endcsname{2.46}
\expandafter\gdef\csname svy@item/t18/e1/meanhigh\endcsname{3.08}
\expandafter\gdef\csname svy@item/t18/e1/se\endcsname{0.153}
\expandafter\gdef\csname svy@item/t18/e1/sd\endcsname{1.06}
\expandafter\gdef\csname svy@item/t18/e1/sdhigh\endcsname{0}
\expandafter\gdef\csname svy@item/t18/e1/verdict\endcsname{INCONCLUSIVE}
\expandafter\gdef\csname svy@item/t18/e1/dip\endcsname{0}
\expandafter\gdef\csname svy@item/t18/e1/shapelow\endcsname{40}
\expandafter\gdef\csname svy@item/t18/e1/shapemid\endcsname{35}
\expandafter\gdef\csname svy@item/t18/e1/shapehigh\endcsname{25}
\expandafter\gdef\csname svy@item/t18/e2/n\endcsname{77}
\expandafter\gdef\csname svy@item/t18/e2/miss\endcsname{0}
\expandafter\gdef\csname svy@item/t18/e2/median\endcsname{3.0}
\expandafter\gdef\csname svy@item/t18/e2/cilow\endcsname{2.0}
\expandafter\gdef\csname svy@item/t18/e2/cihigh\endcsname{3.0}
\expandafter\gdef\csname svy@item/t18/e2/mean\endcsname{2.70}
\expandafter\gdef\csname svy@item/t18/e2/meanlow\endcsname{2.47}
\expandafter\gdef\csname svy@item/t18/e2/meanhigh\endcsname{2.93}
\expandafter\gdef\csname svy@item/t18/e2/se\endcsname{0.117}
\expandafter\gdef\csname svy@item/t18/e2/sd\endcsname{1.03}
\expandafter\gdef\csname svy@item/t18/e2/sdhigh\endcsname{0}
\expandafter\gdef\csname svy@item/t18/e2/verdict\endcsname{INCONCLUSIVE}
\expandafter\gdef\csname svy@item/t18/e2/dip\endcsname{0}
\expandafter\gdef\csname svy@item/t18/e2/shapelow\endcsname{45}
\expandafter\gdef\csname svy@item/t18/e2/shapemid\endcsname{34}
\expandafter\gdef\csname svy@item/t18/e2/shapehigh\endcsname{21}
\expandafter\gdef\csname svy@item/t18/e3/n\endcsname{94}
\expandafter\gdef\csname svy@item/t18/e3/miss\endcsname{0}
\expandafter\gdef\csname svy@item/t18/e3/median\endcsname{3.0}
\expandafter\gdef\csname svy@item/t18/e3/cilow\endcsname{3.0}
\expandafter\gdef\csname svy@item/t18/e3/cihigh\endcsname{4.0}
\expandafter\gdef\csname svy@item/t18/e3/mean\endcsname{3.26}
\expandafter\gdef\csname svy@item/t18/e3/meanlow\endcsname{3.04}
\expandafter\gdef\csname svy@item/t18/e3/meanhigh\endcsname{3.47}
\expandafter\gdef\csname svy@item/t18/e3/se\endcsname{0.110}
\expandafter\gdef\csname svy@item/t18/e3/sd\endcsname{1.07}
\expandafter\gdef\csname svy@item/t18/e3/sdhigh\endcsname{0}
\expandafter\gdef\csname svy@item/t18/e3/verdict\endcsname{INCONCLUSIVE}
\expandafter\gdef\csname svy@item/t18/e3/dip\endcsname{1}
\expandafter\gdef\csname svy@item/t18/e3/shapelow\endcsname{30}
\expandafter\gdef\csname svy@item/t18/e3/shapemid\endcsname{22}
\expandafter\gdef\csname svy@item/t18/e3/shapehigh\endcsname{48}
\expandafter\gdef\csname svy@item/t20/exposure/trend\endcsname{0.31}
\expandafter\gdef\csname svy@item/t20/exposure/trendlow\endcsname{0.19}
\expandafter\gdef\csname svy@item/t20/exposure/trendhigh\endcsname{0.44}
\expandafter\gdef\csname svy@item/t20/e0/n\endcsname{13}
\expandafter\gdef\csname svy@item/t20/e0/miss\endcsname{0}
\expandafter\gdef\csname svy@item/t20/e0/median\endcsname{1.0}
\expandafter\gdef\csname svy@item/t20/e0/cilow\endcsname{1.0}
\expandafter\gdef\csname svy@item/t20/e0/cihigh\endcsname{2.0}
\expandafter\gdef\csname svy@item/t20/e0/mean\endcsname{1.23}
\expandafter\gdef\csname svy@item/t20/e0/meanlow\endcsname{0.97}
\expandafter\gdef\csname svy@item/t20/e0/meanhigh\endcsname{1.50}
\expandafter\gdef\csname svy@item/t20/e0/se\endcsname{0.122}
\expandafter\gdef\csname svy@item/t20/e0/sd\endcsname{0.44}
\expandafter\gdef\csname svy@item/t20/e0/sdhigh\endcsname{0}
\expandafter\gdef\csname svy@item/t20/e0/verdict\endcsname{CONFIRMED}
\expandafter\gdef\csname svy@item/t20/e0/dip\endcsname{0}
\expandafter\gdef\csname svy@item/t20/e0/shapelow\endcsname{100}
\expandafter\gdef\csname svy@item/t20/e0/shapemid\endcsname{0}
\expandafter\gdef\csname svy@item/t20/e0/shapehigh\endcsname{0}
\expandafter\gdef\csname svy@item/t20/e1/n\endcsname{49}
\expandafter\gdef\csname svy@item/t20/e1/miss\endcsname{0}
\expandafter\gdef\csname svy@item/t20/e1/median\endcsname{1.0}
\expandafter\gdef\csname svy@item/t20/e1/cilow\endcsname{1.0}
\expandafter\gdef\csname svy@item/t20/e1/cihigh\endcsname{2.0}
\expandafter\gdef\csname svy@item/t20/e1/mean\endcsname{1.65}
\expandafter\gdef\csname svy@item/t20/e1/meanlow\endcsname{1.43}
\expandafter\gdef\csname svy@item/t20/e1/meanhigh\endcsname{1.88}
\expandafter\gdef\csname svy@item/t20/e1/se\endcsname{0.111}
\expandafter\gdef\csname svy@item/t20/e1/sd\endcsname{0.78}
\expandafter\gdef\csname svy@item/t20/e1/sdhigh\endcsname{0}
\expandafter\gdef\csname svy@item/t20/e1/verdict\endcsname{CONFIRMED}
\expandafter\gdef\csname svy@item/t20/e1/dip\endcsname{0}
\expandafter\gdef\csname svy@item/t20/e1/shapelow\endcsname{86}
\expandafter\gdef\csname svy@item/t20/e1/shapemid\endcsname{12}
\expandafter\gdef\csname svy@item/t20/e1/shapehigh\endcsname{2}
\expandafter\gdef\csname svy@item/t20/e2/n\endcsname{77}
\expandafter\gdef\csname svy@item/t20/e2/miss\endcsname{0}
\expandafter\gdef\csname svy@item/t20/e2/median\endcsname{2.0}
\expandafter\gdef\csname svy@item/t20/e2/cilow\endcsname{2.0}
\expandafter\gdef\csname svy@item/t20/e2/cihigh\endcsname{2.0}
\expandafter\gdef\csname svy@item/t20/e2/mean\endcsname{1.99}
\expandafter\gdef\csname svy@item/t20/e2/meanlow\endcsname{1.79}
\expandafter\gdef\csname svy@item/t20/e2/meanhigh\endcsname{2.19}
\expandafter\gdef\csname svy@item/t20/e2/se\endcsname{0.100}
\expandafter\gdef\csname svy@item/t20/e2/sd\endcsname{0.88}
\expandafter\gdef\csname svy@item/t20/e2/sdhigh\endcsname{0}
\expandafter\gdef\csname svy@item/t20/e2/verdict\endcsname{CONFIRMED}
\expandafter\gdef\csname svy@item/t20/e2/dip\endcsname{0}
\expandafter\gdef\csname svy@item/t20/e2/shapelow\endcsname{80}
\expandafter\gdef\csname svy@item/t20/e2/shapemid\endcsname{14}
\expandafter\gdef\csname svy@item/t20/e2/shapehigh\endcsname{6}
\expandafter\gdef\csname svy@item/t20/e3/n\endcsname{94}
\expandafter\gdef\csname svy@item/t20/e3/miss\endcsname{0}
\expandafter\gdef\csname svy@item/t20/e3/median\endcsname{2.0}
\expandafter\gdef\csname svy@item/t20/e3/cilow\endcsname{2.0}
\expandafter\gdef\csname svy@item/t20/e3/cihigh\endcsname{2.0}
\expandafter\gdef\csname svy@item/t20/e3/mean\endcsname{2.24}
\expandafter\gdef\csname svy@item/t20/e3/meanlow\endcsname{2.04}
\expandafter\gdef\csname svy@item/t20/e3/meanhigh\endcsname{2.45}
\expandafter\gdef\csname svy@item/t20/e3/se\endcsname{0.102}
\expandafter\gdef\csname svy@item/t20/e3/sd\endcsname{0.99}
\expandafter\gdef\csname svy@item/t20/e3/sdhigh\endcsname{0}
\expandafter\gdef\csname svy@item/t20/e3/verdict\endcsname{CONFIRMED}
\expandafter\gdef\csname svy@item/t20/e3/dip\endcsname{0}
\expandafter\gdef\csname svy@item/t20/e3/shapelow\endcsname{69}
\expandafter\gdef\csname svy@item/t20/e3/shapemid\endcsname{19}
\expandafter\gdef\csname svy@item/t20/e3/shapehigh\endcsname{12}
\expandafter\gdef\csname svy@item/t21/exposure/trend\endcsname{0.69}
\expandafter\gdef\csname svy@item/t21/exposure/trendlow\endcsname{0.56}
\expandafter\gdef\csname svy@item/t21/exposure/trendhigh\endcsname{0.83}
\expandafter\gdef\csname svy@item/t21/e0/n\endcsname{13}
\expandafter\gdef\csname svy@item/t21/e0/miss\endcsname{0}
\expandafter\gdef\csname svy@item/t21/e0/median\endcsname{1.0}
\expandafter\gdef\csname svy@item/t21/e0/cilow\endcsname{1.0}
\expandafter\gdef\csname svy@item/t21/e0/cihigh\endcsname{2.0}
\expandafter\gdef\csname svy@item/t21/e0/mean\endcsname{1.69}
\expandafter\gdef\csname svy@item/t21/e0/meanlow\endcsname{1.12}
\expandafter\gdef\csname svy@item/t21/e0/meanhigh\endcsname{2.26}
\expandafter\gdef\csname svy@item/t21/e0/se\endcsname{0.263}
\expandafter\gdef\csname svy@item/t21/e0/sd\endcsname{0.95}
\expandafter\gdef\csname svy@item/t21/e0/sdhigh\endcsname{0}
\expandafter\gdef\csname svy@item/t21/e0/verdict\endcsname{REJECTED}
\expandafter\gdef\csname svy@item/t21/e0/dip\endcsname{0}
\expandafter\gdef\csname svy@item/t21/e0/shapelow\endcsname{85}
\expandafter\gdef\csname svy@item/t21/e0/shapemid\endcsname{8}
\expandafter\gdef\csname svy@item/t21/e0/shapehigh\endcsname{7}
\expandafter\gdef\csname svy@item/t21/e1/n\endcsname{46}
\expandafter\gdef\csname svy@item/t21/e1/miss\endcsname{3}
\expandafter\gdef\csname svy@item/t21/e1/median\endcsname{4.0}
\expandafter\gdef\csname svy@item/t21/e1/cilow\endcsname{3.0}
\expandafter\gdef\csname svy@item/t21/e1/cihigh\endcsname{4.0}
\expandafter\gdef\csname svy@item/t21/e1/mean\endcsname{3.17}
\expandafter\gdef\csname svy@item/t21/e1/meanlow\endcsname{2.85}
\expandafter\gdef\csname svy@item/t21/e1/meanhigh\endcsname{3.50}
\expandafter\gdef\csname svy@item/t21/e1/se\endcsname{0.162}
\expandafter\gdef\csname svy@item/t21/e1/sd\endcsname{1.10}
\expandafter\gdef\csname svy@item/t21/e1/sdhigh\endcsname{0}
\expandafter\gdef\csname svy@item/t21/e1/verdict\endcsname{INCONCLUSIVE}
\expandafter\gdef\csname svy@item/t21/e1/dip\endcsname{1}
\expandafter\gdef\csname svy@item/t21/e1/shapelow\endcsname{31}
\expandafter\gdef\csname svy@item/t21/e1/shapemid\endcsname{17}
\expandafter\gdef\csname svy@item/t21/e1/shapehigh\endcsname{52}
\expandafter\gdef\csname svy@item/t21/e2/n\endcsname{76}
\expandafter\gdef\csname svy@item/t21/e2/miss\endcsname{1}
\expandafter\gdef\csname svy@item/t21/e2/median\endcsname{4.0}
\expandafter\gdef\csname svy@item/t21/e2/cilow\endcsname{4.0}
\expandafter\gdef\csname svy@item/t21/e2/cihigh\endcsname{4.0}
\expandafter\gdef\csname svy@item/t21/e2/mean\endcsname{3.67}
\expandafter\gdef\csname svy@item/t21/e2/meanlow\endcsname{3.47}
\expandafter\gdef\csname svy@item/t21/e2/meanhigh\endcsname{3.87}
\expandafter\gdef\csname svy@item/t21/e2/se\endcsname{0.102}
\expandafter\gdef\csname svy@item/t21/e2/sd\endcsname{0.89}
\expandafter\gdef\csname svy@item/t21/e2/sdhigh\endcsname{0}
\expandafter\gdef\csname svy@item/t21/e2/verdict\endcsname{INCONCLUSIVE}
\expandafter\gdef\csname svy@item/t21/e2/dip\endcsname{0}
\expandafter\gdef\csname svy@item/t21/e2/shapelow\endcsname{13}
\expandafter\gdef\csname svy@item/t21/e2/shapemid\endcsname{21}
\expandafter\gdef\csname svy@item/t21/e2/shapehigh\endcsname{66}
\expandafter\gdef\csname svy@item/t21/e3/n\endcsname{92}
\expandafter\gdef\csname svy@item/t21/e3/miss\endcsname{2}
\expandafter\gdef\csname svy@item/t21/e3/median\endcsname{4.0}
\expandafter\gdef\csname svy@item/t21/e3/cilow\endcsname{4.0}
\expandafter\gdef\csname svy@item/t21/e3/cihigh\endcsname{5.0}
\expandafter\gdef\csname svy@item/t21/e3/mean\endcsname{4.29}
\expandafter\gdef\csname svy@item/t21/e3/meanlow\endcsname{4.13}
\expandafter\gdef\csname svy@item/t21/e3/meanhigh\endcsname{4.46}
\expandafter\gdef\csname svy@item/t21/e3/se\endcsname{0.083}
\expandafter\gdef\csname svy@item/t21/e3/sd\endcsname{0.79}
\expandafter\gdef\csname svy@item/t21/e3/sdhigh\endcsname{0}
\expandafter\gdef\csname svy@item/t21/e3/verdict\endcsname{CONFIRMED}
\expandafter\gdef\csname svy@item/t21/e3/dip\endcsname{0}
\expandafter\gdef\csname svy@item/t21/e3/shapelow\endcsname{4}
\expandafter\gdef\csname svy@item/t21/e3/shapemid\endcsname{8}
\expandafter\gdef\csname svy@item/t21/e3/shapehigh\endcsname{88}
\expandafter\gdef\csname svy@item/t22/exposure/trend\endcsname{0.55}
\expandafter\gdef\csname svy@item/t22/exposure/trendlow\endcsname{0.37}
\expandafter\gdef\csname svy@item/t22/exposure/trendhigh\endcsname{0.72}
\expandafter\gdef\csname svy@item/t22/e0/n\endcsname{11}
\expandafter\gdef\csname svy@item/t22/e0/miss\endcsname{2}
\expandafter\gdef\csname svy@item/t22/e0/median\endcsname{1.0}
\expandafter\gdef\csname svy@item/t22/e0/cilow\endcsname{1.0}
\expandafter\gdef\csname svy@item/t22/e0/cihigh\endcsname{1.0}
\expandafter\gdef\csname svy@item/t22/e0/mean\endcsname{1.09}
\expandafter\gdef\csname svy@item/t22/e0/meanlow\endcsname{0.89}
\expandafter\gdef\csname svy@item/t22/e0/meanhigh\endcsname{1.29}
\expandafter\gdef\csname svy@item/t22/e0/se\endcsname{0.091}
\expandafter\gdef\csname svy@item/t22/e0/sd\endcsname{0.30}
\expandafter\gdef\csname svy@item/t22/e0/sdhigh\endcsname{0}
\expandafter\gdef\csname svy@item/t22/e0/verdict\endcsname{NO_PREDICTION}
\expandafter\gdef\csname svy@item/t22/e0/dip\endcsname{0}
\expandafter\gdef\csname svy@item/t22/e0/shapelow\endcsname{100}
\expandafter\gdef\csname svy@item/t22/e0/shapemid\endcsname{0}
\expandafter\gdef\csname svy@item/t22/e0/shapehigh\endcsname{0}
\expandafter\gdef\csname svy@item/t22/e1/n\endcsname{47}
\expandafter\gdef\csname svy@item/t22/e1/miss\endcsname{2}
\expandafter\gdef\csname svy@item/t22/e1/median\endcsname{2.0}
\expandafter\gdef\csname svy@item/t22/e1/cilow\endcsname{2.0}
\expandafter\gdef\csname svy@item/t22/e1/cihigh\endcsname{3.0}
\expandafter\gdef\csname svy@item/t22/e1/mean\endcsname{2.66}
\expandafter\gdef\csname svy@item/t22/e1/meanlow\endcsname{2.31}
\expandafter\gdef\csname svy@item/t22/e1/meanhigh\endcsname{3.01}
\expandafter\gdef\csname svy@item/t22/e1/se\endcsname{0.175}
\expandafter\gdef\csname svy@item/t22/e1/sd\endcsname{1.20}
\expandafter\gdef\csname svy@item/t22/e1/sdhigh\endcsname{0}
\expandafter\gdef\csname svy@item/t22/e1/verdict\endcsname{NO_PREDICTION}
\expandafter\gdef\csname svy@item/t22/e1/dip\endcsname{0}
\expandafter\gdef\csname svy@item/t22/e1/shapelow\endcsname{53}
\expandafter\gdef\csname svy@item/t22/e1/shapemid\endcsname{23}
\expandafter\gdef\csname svy@item/t22/e1/shapehigh\endcsname{24}
\expandafter\gdef\csname svy@item/t22/e2/n\endcsname{76}
\expandafter\gdef\csname svy@item/t22/e2/miss\endcsname{1}
\expandafter\gdef\csname svy@item/t22/e2/median\endcsname{3.0}
\expandafter\gdef\csname svy@item/t22/e2/cilow\endcsname{3.0}
\expandafter\gdef\csname svy@item/t22/e2/cihigh\endcsname{4.0}
\expandafter\gdef\csname svy@item/t22/e2/mean\endcsname{3.09}
\expandafter\gdef\csname svy@item/t22/e2/meanlow\endcsname{2.81}
\expandafter\gdef\csname svy@item/t22/e2/meanhigh\endcsname{3.38}
\expandafter\gdef\csname svy@item/t22/e2/se\endcsname{0.143}
\expandafter\gdef\csname svy@item/t22/e2/sd\endcsname{1.25}
\expandafter\gdef\csname svy@item/t22/e2/sdhigh\endcsname{0}
\expandafter\gdef\csname svy@item/t22/e2/verdict\endcsname{NO_PREDICTION}
\expandafter\gdef\csname svy@item/t22/e2/dip\endcsname{0}
\expandafter\gdef\csname svy@item/t22/e2/shapelow\endcsname{33}
\expandafter\gdef\csname svy@item/t22/e2/shapemid\endcsname{25}
\expandafter\gdef\csname svy@item/t22/e2/shapehigh\endcsname{42}
\expandafter\gdef\csname svy@item/t22/e3/n\endcsname{94}
\expandafter\gdef\csname svy@item/t22/e3/miss\endcsname{0}
\expandafter\gdef\csname svy@item/t22/e3/median\endcsname{4.0}
\expandafter\gdef\csname svy@item/t22/e3/cilow\endcsname{3.0}
\expandafter\gdef\csname svy@item/t22/e3/cihigh\endcsname{4.0}
\expandafter\gdef\csname svy@item/t22/e3/mean\endcsname{3.46}
\expandafter\gdef\csname svy@item/t22/e3/meanlow\endcsname{3.21}
\expandafter\gdef\csname svy@item/t22/e3/meanhigh\endcsname{3.70}
\expandafter\gdef\csname svy@item/t22/e3/se\endcsname{0.123}
\expandafter\gdef\csname svy@item/t22/e3/sd\endcsname{1.19}
\expandafter\gdef\csname svy@item/t22/e3/sdhigh\endcsname{0}
\expandafter\gdef\csname svy@item/t22/e3/verdict\endcsname{NO_PREDICTION}
\expandafter\gdef\csname svy@item/t22/e3/dip\endcsname{1}
\expandafter\gdef\csname svy@item/t22/e3/shapelow\endcsname{27}
\expandafter\gdef\csname svy@item/t22/e3/shapemid\endcsname{16}
\expandafter\gdef\csname svy@item/t22/e3/shapehigh\endcsname{57}
\expandafter\gdef\csname svy@item/t23/exposure/trend\endcsname{0.28}
\expandafter\gdef\csname svy@item/t23/exposure/trendlow\endcsname{0.16}
\expandafter\gdef\csname svy@item/t23/exposure/trendhigh\endcsname{0.39}
\expandafter\gdef\csname svy@item/t23/e0/n\endcsname{13}
\expandafter\gdef\csname svy@item/t23/e0/miss\endcsname{0}
\expandafter\gdef\csname svy@item/t23/e0/median\endcsname{1.0}
\expandafter\gdef\csname svy@item/t23/e0/cilow\endcsname{1.0}
\expandafter\gdef\csname svy@item/t23/e0/cihigh\endcsname{1.0}
\expandafter\gdef\csname svy@item/t23/e0/mean\endcsname{1.00}
\expandafter\gdef\csname svy@item/t23/e0/meanlow\endcsname{1.00}
\expandafter\gdef\csname svy@item/t23/e0/meanhigh\endcsname{1.00}
\expandafter\gdef\csname svy@item/t23/e0/se\endcsname{0.000}
\expandafter\gdef\csname svy@item/t23/e0/sd\endcsname{0.00}
\expandafter\gdef\csname svy@item/t23/e0/sdhigh\endcsname{0}
\expandafter\gdef\csname svy@item/t23/e0/verdict\endcsname{CONFIRMED}
\expandafter\gdef\csname svy@item/t23/e0/dip\endcsname{0}
\expandafter\gdef\csname svy@item/t23/e0/shapelow\endcsname{100}
\expandafter\gdef\csname svy@item/t23/e0/shapemid\endcsname{0}
\expandafter\gdef\csname svy@item/t23/e0/shapehigh\endcsname{0}
\expandafter\gdef\csname svy@item/t23/e1/n\endcsname{47}
\expandafter\gdef\csname svy@item/t23/e1/miss\endcsname{2}
\expandafter\gdef\csname svy@item/t23/e1/median\endcsname{1.0}
\expandafter\gdef\csname svy@item/t23/e1/cilow\endcsname{1.0}
\expandafter\gdef\csname svy@item/t23/e1/cihigh\endcsname{1.0}
\expandafter\gdef\csname svy@item/t23/e1/mean\endcsname{1.40}
\expandafter\gdef\csname svy@item/t23/e1/meanlow\endcsname{1.19}
\expandafter\gdef\csname svy@item/t23/e1/meanhigh\endcsname{1.62}
\expandafter\gdef\csname svy@item/t23/e1/se\endcsname{0.108}
\expandafter\gdef\csname svy@item/t23/e1/sd\endcsname{0.74}
\expandafter\gdef\csname svy@item/t23/e1/sdhigh\endcsname{0}
\expandafter\gdef\csname svy@item/t23/e1/verdict\endcsname{CONFIRMED}
\expandafter\gdef\csname svy@item/t23/e1/dip\endcsname{0}
\expandafter\gdef\csname svy@item/t23/e1/shapelow\endcsname{94}
\expandafter\gdef\csname svy@item/t23/e1/shapemid\endcsname{2}
\expandafter\gdef\csname svy@item/t23/e1/shapehigh\endcsname{4}
\expandafter\gdef\csname svy@item/t23/e2/n\endcsname{77}
\expandafter\gdef\csname svy@item/t23/e2/miss\endcsname{0}
\expandafter\gdef\csname svy@item/t23/e2/median\endcsname{1.0}
\expandafter\gdef\csname svy@item/t23/e2/cilow\endcsname{1.0}
\expandafter\gdef\csname svy@item/t23/e2/cihigh\endcsname{1.0}
\expandafter\gdef\csname svy@item/t23/e2/mean\endcsname{1.48}
\expandafter\gdef\csname svy@item/t23/e2/meanlow\endcsname{1.32}
\expandafter\gdef\csname svy@item/t23/e2/meanhigh\endcsname{1.64}
\expandafter\gdef\csname svy@item/t23/e2/se\endcsname{0.082}
\expandafter\gdef\csname svy@item/t23/e2/sd\endcsname{0.72}
\expandafter\gdef\csname svy@item/t23/e2/sdhigh\endcsname{0}
\expandafter\gdef\csname svy@item/t23/e2/verdict\endcsname{CONFIRMED}
\expandafter\gdef\csname svy@item/t23/e2/dip\endcsname{0}
\expandafter\gdef\csname svy@item/t23/e2/shapelow\endcsname{92}
\expandafter\gdef\csname svy@item/t23/e2/shapemid\endcsname{5}
\expandafter\gdef\csname svy@item/t23/e2/shapehigh\endcsname{3}
\expandafter\gdef\csname svy@item/t23/e3/n\endcsname{93}
\expandafter\gdef\csname svy@item/t23/e3/miss\endcsname{1}
\expandafter\gdef\csname svy@item/t23/e3/median\endcsname{2.0}
\expandafter\gdef\csname svy@item/t23/e3/cilow\endcsname{2.0}
\expandafter\gdef\csname svy@item/t23/e3/cihigh\endcsname{2.0}
\expandafter\gdef\csname svy@item/t23/e3/mean\endcsname{1.90}
\expandafter\gdef\csname svy@item/t23/e3/meanlow\endcsname{1.72}
\expandafter\gdef\csname svy@item/t23/e3/meanhigh\endcsname{2.09}
\expandafter\gdef\csname svy@item/t23/e3/se\endcsname{0.094}
\expandafter\gdef\csname svy@item/t23/e3/sd\endcsname{0.91}
\expandafter\gdef\csname svy@item/t23/e3/sdhigh\endcsname{0}
\expandafter\gdef\csname svy@item/t23/e3/verdict\endcsname{CONFIRMED}
\expandafter\gdef\csname svy@item/t23/e3/dip\endcsname{0}
\expandafter\gdef\csname svy@item/t23/e3/shapelow\endcsname{82}
\expandafter\gdef\csname svy@item/t23/e3/shapemid\endcsname{11}
\expandafter\gdef\csname svy@item/t23/e3/shapehigh\endcsname{7}
\expandafter\gdef\csname svy@item/t24/exposure/trend\endcsname{0.19}
\expandafter\gdef\csname svy@item/t24/exposure/trendlow\endcsname{0.09}
\expandafter\gdef\csname svy@item/t24/exposure/trendhigh\endcsname{0.28}
\expandafter\gdef\csname svy@item/t24/e0/n\endcsname{13}
\expandafter\gdef\csname svy@item/t24/e0/miss\endcsname{0}
\expandafter\gdef\csname svy@item/t24/e0/median\endcsname{1.0}
\expandafter\gdef\csname svy@item/t24/e0/cilow\endcsname{1.0}
\expandafter\gdef\csname svy@item/t24/e0/cihigh\endcsname{1.0}
\expandafter\gdef\csname svy@item/t24/e0/mean\endcsname{1.00}
\expandafter\gdef\csname svy@item/t24/e0/meanlow\endcsname{1.00}
\expandafter\gdef\csname svy@item/t24/e0/meanhigh\endcsname{1.00}
\expandafter\gdef\csname svy@item/t24/e0/se\endcsname{0.000}
\expandafter\gdef\csname svy@item/t24/e0/sd\endcsname{0.00}
\expandafter\gdef\csname svy@item/t24/e0/sdhigh\endcsname{0}
\expandafter\gdef\csname svy@item/t24/e0/verdict\endcsname{CONFIRMED}
\expandafter\gdef\csname svy@item/t24/e0/dip\endcsname{0}
\expandafter\gdef\csname svy@item/t24/e0/shapelow\endcsname{100}
\expandafter\gdef\csname svy@item/t24/e0/shapemid\endcsname{0}
\expandafter\gdef\csname svy@item/t24/e0/shapehigh\endcsname{0}
\expandafter\gdef\csname svy@item/t24/e1/n\endcsname{47}
\expandafter\gdef\csname svy@item/t24/e1/miss\endcsname{2}
\expandafter\gdef\csname svy@item/t24/e1/median\endcsname{1.0}
\expandafter\gdef\csname svy@item/t24/e1/cilow\endcsname{1.0}
\expandafter\gdef\csname svy@item/t24/e1/cihigh\endcsname{1.0}
\expandafter\gdef\csname svy@item/t24/e1/mean\endcsname{1.15}
\expandafter\gdef\csname svy@item/t24/e1/meanlow\endcsname{1.00}
\expandafter\gdef\csname svy@item/t24/e1/meanhigh\endcsname{1.30}
\expandafter\gdef\csname svy@item/t24/e1/se\endcsname{0.074}
\expandafter\gdef\csname svy@item/t24/e1/sd\endcsname{0.51}
\expandafter\gdef\csname svy@item/t24/e1/sdhigh\endcsname{0}
\expandafter\gdef\csname svy@item/t24/e1/verdict\endcsname{CONFIRMED}
\expandafter\gdef\csname svy@item/t24/e1/dip\endcsname{0}
\expandafter\gdef\csname svy@item/t24/e1/shapelow\endcsname{98}
\expandafter\gdef\csname svy@item/t24/e1/shapemid\endcsname{0}
\expandafter\gdef\csname svy@item/t24/e1/shapehigh\endcsname{2}
\expandafter\gdef\csname svy@item/t24/e2/n\endcsname{75}
\expandafter\gdef\csname svy@item/t24/e2/miss\endcsname{2}
\expandafter\gdef\csname svy@item/t24/e2/median\endcsname{1.0}
\expandafter\gdef\csname svy@item/t24/e2/cilow\endcsname{1.0}
\expandafter\gdef\csname svy@item/t24/e2/cihigh\endcsname{1.0}
\expandafter\gdef\csname svy@item/t24/e2/mean\endcsname{1.25}
\expandafter\gdef\csname svy@item/t24/e2/meanlow\endcsname{1.14}
\expandafter\gdef\csname svy@item/t24/e2/meanhigh\endcsname{1.37}
\expandafter\gdef\csname svy@item/t24/e2/se\endcsname{0.057}
\expandafter\gdef\csname svy@item/t24/e2/sd\endcsname{0.50}
\expandafter\gdef\csname svy@item/t24/e2/sdhigh\endcsname{0}
\expandafter\gdef\csname svy@item/t24/e2/verdict\endcsname{CONFIRMED}
\expandafter\gdef\csname svy@item/t24/e2/dip\endcsname{0}
\expandafter\gdef\csname svy@item/t24/e2/shapelow\endcsname{97}
\expandafter\gdef\csname svy@item/t24/e2/shapemid\endcsname{3}
\expandafter\gdef\csname svy@item/t24/e2/shapehigh\endcsname{0}
\expandafter\gdef\csname svy@item/t24/e3/n\endcsname{84}
\expandafter\gdef\csname svy@item/t24/e3/miss\endcsname{10}
\expandafter\gdef\csname svy@item/t24/e3/median\endcsname{1.0}
\expandafter\gdef\csname svy@item/t24/e3/cilow\endcsname{1.0}
\expandafter\gdef\csname svy@item/t24/e3/cihigh\endcsname{1.0}
\expandafter\gdef\csname svy@item/t24/e3/mean\endcsname{1.54}
\expandafter\gdef\csname svy@item/t24/e3/meanlow\endcsname{1.35}
\expandafter\gdef\csname svy@item/t24/e3/meanhigh\endcsname{1.72}
\expandafter\gdef\csname svy@item/t24/e3/se\endcsname{0.093}
\expandafter\gdef\csname svy@item/t24/e3/sd\endcsname{0.86}
\expandafter\gdef\csname svy@item/t24/e3/sdhigh\endcsname{0}
\expandafter\gdef\csname svy@item/t24/e3/verdict\endcsname{CONFIRMED}
\expandafter\gdef\csname svy@item/t24/e3/dip\endcsname{0}
\expandafter\gdef\csname svy@item/t24/e3/shapelow\endcsname{85}
\expandafter\gdef\csname svy@item/t24/e3/shapemid\endcsname{13}
\expandafter\gdef\csname svy@item/t24/e3/shapehigh\endcsname{2}
\expandafter\gdef\csname svy@item/accdev/exposure/trend\endcsname{-0.17}
\expandafter\gdef\csname svy@item/accdev/exposure/trendlow\endcsname{-0.28}
\expandafter\gdef\csname svy@item/accdev/exposure/trendhigh\endcsname{-0.06}
\expandafter\gdef\csname svy@item/accdev/e0/n\endcsname{13}
\expandafter\gdef\csname svy@item/accdev/e0/miss\endcsname{0}
\expandafter\gdef\csname svy@item/accdev/e0/median\endcsname{5.0}
\expandafter\gdef\csname svy@item/accdev/e0/cilow\endcsname{5.0}
\expandafter\gdef\csname svy@item/accdev/e0/cihigh\endcsname{5.0}
\expandafter\gdef\csname svy@item/accdev/e0/mean\endcsname{4.92}
\expandafter\gdef\csname svy@item/accdev/e0/meanlow\endcsname{4.76}
\expandafter\gdef\csname svy@item/accdev/e0/meanhigh\endcsname{5.09}
\expandafter\gdef\csname svy@item/accdev/e0/se\endcsname{0.077}
\expandafter\gdef\csname svy@item/accdev/e0/sd\endcsname{0.28}
\expandafter\gdef\csname svy@item/accdev/e0/sdhigh\endcsname{0}
\expandafter\gdef\csname svy@item/accdev/e0/verdict\endcsname{CONFIRMED}
\expandafter\gdef\csname svy@item/accdev/e0/dip\endcsname{0}
\expandafter\gdef\csname svy@item/accdev/e0/shapelow\endcsname{0}
\expandafter\gdef\csname svy@item/accdev/e0/shapemid\endcsname{0}
\expandafter\gdef\csname svy@item/accdev/e0/shapehigh\endcsname{100}
\expandafter\gdef\csname svy@item/accdev/e1/n\endcsname{47}
\expandafter\gdef\csname svy@item/accdev/e1/miss\endcsname{2}
\expandafter\gdef\csname svy@item/accdev/e1/median\endcsname{5.0}
\expandafter\gdef\csname svy@item/accdev/e1/cilow\endcsname{5.0}
\expandafter\gdef\csname svy@item/accdev/e1/cihigh\endcsname{5.0}
\expandafter\gdef\csname svy@item/accdev/e1/mean\endcsname{4.64}
\expandafter\gdef\csname svy@item/accdev/e1/meanlow\endcsname{4.41}
\expandafter\gdef\csname svy@item/accdev/e1/meanhigh\endcsname{4.86}
\expandafter\gdef\csname svy@item/accdev/e1/se\endcsname{0.111}
\expandafter\gdef\csname svy@item/accdev/e1/sd\endcsname{0.76}
\expandafter\gdef\csname svy@item/accdev/e1/sdhigh\endcsname{0}
\expandafter\gdef\csname svy@item/accdev/e1/verdict\endcsname{CONFIRMED}
\expandafter\gdef\csname svy@item/accdev/e1/dip\endcsname{0}
\expandafter\gdef\csname svy@item/accdev/e1/shapelow\endcsname{2}
\expandafter\gdef\csname svy@item/accdev/e1/shapemid\endcsname{4}
\expandafter\gdef\csname svy@item/accdev/e1/shapehigh\endcsname{94}
\expandafter\gdef\csname svy@item/accdev/e2/n\endcsname{73}
\expandafter\gdef\csname svy@item/accdev/e2/miss\endcsname{4}
\expandafter\gdef\csname svy@item/accdev/e2/median\endcsname{5.0}
\expandafter\gdef\csname svy@item/accdev/e2/cilow\endcsname{5.0}
\expandafter\gdef\csname svy@item/accdev/e2/cihigh\endcsname{5.0}
\expandafter\gdef\csname svy@item/accdev/e2/mean\endcsname{4.62}
\expandafter\gdef\csname svy@item/accdev/e2/meanlow\endcsname{4.48}
\expandafter\gdef\csname svy@item/accdev/e2/meanhigh\endcsname{4.75}
\expandafter\gdef\csname svy@item/accdev/e2/se\endcsname{0.069}
\expandafter\gdef\csname svy@item/accdev/e2/sd\endcsname{0.59}
\expandafter\gdef\csname svy@item/accdev/e2/sdhigh\endcsname{0}
\expandafter\gdef\csname svy@item/accdev/e2/verdict\endcsname{CONFIRMED}
\expandafter\gdef\csname svy@item/accdev/e2/dip\endcsname{0}
\expandafter\gdef\csname svy@item/accdev/e2/shapelow\endcsname{0}
\expandafter\gdef\csname svy@item/accdev/e2/shapemid\endcsname{6}
\expandafter\gdef\csname svy@item/accdev/e2/shapehigh\endcsname{94}
\expandafter\gdef\csname svy@item/accdev/e3/n\endcsname{89}
\expandafter\gdef\csname svy@item/accdev/e3/miss\endcsname{5}
\expandafter\gdef\csname svy@item/accdev/e3/median\endcsname{5.0}
\expandafter\gdef\csname svy@item/accdev/e3/cilow\endcsname{4.0}
\expandafter\gdef\csname svy@item/accdev/e3/cihigh\endcsname{5.0}
\expandafter\gdef\csname svy@item/accdev/e3/mean\endcsname{4.35}
\expandafter\gdef\csname svy@item/accdev/e3/meanlow\endcsname{4.15}
\expandafter\gdef\csname svy@item/accdev/e3/meanhigh\endcsname{4.54}
\expandafter\gdef\csname svy@item/accdev/e3/se\endcsname{0.097}
\expandafter\gdef\csname svy@item/accdev/e3/sd\endcsname{0.92}
\expandafter\gdef\csname svy@item/accdev/e3/sdhigh\endcsname{0}
\expandafter\gdef\csname svy@item/accdev/e3/verdict\endcsname{CONFIRMED}
\expandafter\gdef\csname svy@item/accdev/e3/dip\endcsname{0}
\expandafter\gdef\csname svy@item/accdev/e3/shapelow\endcsname{5}
\expandafter\gdef\csname svy@item/accdev/e3/shapemid\endcsname{10}
\expandafter\gdef\csname svy@item/accdev/e3/shapehigh\endcsname{85}
\expandafter\gdef\csname svy@item/accreason/exposure/trend\endcsname{0.15}
\expandafter\gdef\csname svy@item/accreason/exposure/trendlow\endcsname{0.01}
\expandafter\gdef\csname svy@item/accreason/exposure/trendhigh\endcsname{0.30}
\expandafter\gdef\csname svy@item/accreason/e0/n\endcsname{10}
\expandafter\gdef\csname svy@item/accreason/e0/miss\endcsname{3}
\expandafter\gdef\csname svy@item/accreason/e0/median\endcsname{3.5}
\expandafter\gdef\csname svy@item/accreason/e0/cilow\endcsname{1.0}
\expandafter\gdef\csname svy@item/accreason/e0/cihigh\endcsname{5.0}
\expandafter\gdef\csname svy@item/accreason/e0/mean\endcsname{3.50}
\expandafter\gdef\csname svy@item/accreason/e0/meanlow\endcsname{2.37}
\expandafter\gdef\csname svy@item/accreason/e0/meanhigh\endcsname{4.63}
\expandafter\gdef\csname svy@item/accreason/e0/se\endcsname{0.500}
\expandafter\gdef\csname svy@item/accreason/e0/sd\endcsname{1.58}
\expandafter\gdef\csname svy@item/accreason/e0/sdhigh\endcsname{0}
\expandafter\gdef\csname svy@item/accreason/e0/verdict\endcsname{INCONCLUSIVE}
\expandafter\gdef\csname svy@item/accreason/e0/dip\endcsname{0}
\expandafter\gdef\csname svy@item/accreason/e0/shapelow\endcsname{20}
\expandafter\gdef\csname svy@item/accreason/e0/shapemid\endcsname{30}
\expandafter\gdef\csname svy@item/accreason/e0/shapehigh\endcsname{50}
\expandafter\gdef\csname svy@item/accreason/e1/n\endcsname{46}
\expandafter\gdef\csname svy@item/accreason/e1/miss\endcsname{3}
\expandafter\gdef\csname svy@item/accreason/e1/median\endcsname{4.0}
\expandafter\gdef\csname svy@item/accreason/e1/cilow\endcsname{4.0}
\expandafter\gdef\csname svy@item/accreason/e1/cihigh\endcsname{5.0}
\expandafter\gdef\csname svy@item/accreason/e1/mean\endcsname{4.17}
\expandafter\gdef\csname svy@item/accreason/e1/meanlow\endcsname{3.87}
\expandafter\gdef\csname svy@item/accreason/e1/meanhigh\endcsname{4.48}
\expandafter\gdef\csname svy@item/accreason/e1/se\endcsname{0.153}
\expandafter\gdef\csname svy@item/accreason/e1/sd\endcsname{1.04}
\expandafter\gdef\csname svy@item/accreason/e1/sdhigh\endcsname{0}
\expandafter\gdef\csname svy@item/accreason/e1/verdict\endcsname{CONFIRMED}
\expandafter\gdef\csname svy@item/accreason/e1/dip\endcsname{0}
\expandafter\gdef\csname svy@item/accreason/e1/shapelow\endcsname{8}
\expandafter\gdef\csname svy@item/accreason/e1/shapemid\endcsname{7}
\expandafter\gdef\csname svy@item/accreason/e1/shapehigh\endcsname{85}
\expandafter\gdef\csname svy@item/accreason/e2/n\endcsname{72}
\expandafter\gdef\csname svy@item/accreason/e2/miss\endcsname{5}
\expandafter\gdef\csname svy@item/accreason/e2/median\endcsname{4.0}
\expandafter\gdef\csname svy@item/accreason/e2/cilow\endcsname{4.0}
\expandafter\gdef\csname svy@item/accreason/e2/cihigh\endcsname{5.0}
\expandafter\gdef\csname svy@item/accreason/e2/mean\endcsname{4.21}
\expandafter\gdef\csname svy@item/accreason/e2/meanlow\endcsname{3.97}
\expandafter\gdef\csname svy@item/accreason/e2/meanhigh\endcsname{4.44}
\expandafter\gdef\csname svy@item/accreason/e2/se\endcsname{0.119}
\expandafter\gdef\csname svy@item/accreason/e2/sd\endcsname{1.01}
\expandafter\gdef\csname svy@item/accreason/e2/sdhigh\endcsname{0}
\expandafter\gdef\csname svy@item/accreason/e2/verdict\endcsname{CONFIRMED}
\expandafter\gdef\csname svy@item/accreason/e2/dip\endcsname{0}
\expandafter\gdef\csname svy@item/accreason/e2/shapelow\endcsname{7}
\expandafter\gdef\csname svy@item/accreason/e2/shapemid\endcsname{8}
\expandafter\gdef\csname svy@item/accreason/e2/shapehigh\endcsname{85}
\expandafter\gdef\csname svy@item/accreason/e3/n\endcsname{90}
\expandafter\gdef\csname svy@item/accreason/e3/miss\endcsname{4}
\expandafter\gdef\csname svy@item/accreason/e3/median\endcsname{5.0}
\expandafter\gdef\csname svy@item/accreason/e3/cilow\endcsname{4.0}
\expandafter\gdef\csname svy@item/accreason/e3/cihigh\endcsname{5.0}
\expandafter\gdef\csname svy@item/accreason/e3/mean\endcsname{4.33}
\expandafter\gdef\csname svy@item/accreason/e3/meanlow\endcsname{4.15}
\expandafter\gdef\csname svy@item/accreason/e3/meanhigh\endcsname{4.52}
\expandafter\gdef\csname svy@item/accreason/e3/se\endcsname{0.092}
\expandafter\gdef\csname svy@item/accreason/e3/sd\endcsname{0.87}
\expandafter\gdef\csname svy@item/accreason/e3/sdhigh\endcsname{0}
\expandafter\gdef\csname svy@item/accreason/e3/verdict\endcsname{CONFIRMED}
\expandafter\gdef\csname svy@item/accreason/e3/dip\endcsname{0}
\expandafter\gdef\csname svy@item/accreason/e3/shapelow\endcsname{4}
\expandafter\gdef\csname svy@item/accreason/e3/shapemid\endcsname{10}
\expandafter\gdef\csname svy@item/accreason/e3/shapehigh\endcsname{86}
\expandafter\gdef\csname svy@item/g01/code\endcsname{G01}
\expandafter\gdef\csname svy@item/g01/pred\endcsname{}
\expandafter\gdef\csname svy@item/g01/label\endcsname{AI tools make me more productive}
\expandafter\gdef\csname svy@item/g01/usage/diff\endcsname{1.11}
\expandafter\gdef\csname svy@item/g01/usage/difflow\endcsname{0.74}
\expandafter\gdef\csname svy@item/g01/usage/diffhigh\endcsname{1.47}
\expandafter\gdef\csname svy@item/g01/usage/g\endcsname{1.07}
\expandafter\gdef\csname svy@item/g01/all/n\endcsname{233}
\expandafter\gdef\csname svy@item/g01/all/miss\endcsname{9}
\expandafter\gdef\csname svy@item/g01/all/median\endcsname{4.0}
\expandafter\gdef\csname svy@item/g01/all/cilow\endcsname{4.0}
\expandafter\gdef\csname svy@item/g01/all/cihigh\endcsname{5.0}
\expandafter\gdef\csname svy@item/g01/all/mean\endcsname{4.13}
\expandafter\gdef\csname svy@item/g01/all/meanlow\endcsname{3.99}
\expandafter\gdef\csname svy@item/g01/all/meanhigh\endcsname{4.28}
\expandafter\gdef\csname svy@item/g01/all/se\endcsname{0.075}
\expandafter\gdef\csname svy@item/g01/all/sd\endcsname{1.14}
\expandafter\gdef\csname svy@item/g01/all/sdhigh\endcsname{0}
\expandafter\gdef\csname svy@item/g01/all/verdict\endcsname{none}
\expandafter\gdef\csname svy@item/g01/all/dip\endcsname{0}
\expandafter\gdef\csname svy@item/g01/all/shapelow\endcsname{13}
\expandafter\gdef\csname svy@item/g01/all/shapemid\endcsname{6}
\expandafter\gdef\csname svy@item/g01/all/shapehigh\endcsname{81}
\expandafter\gdef\csname svy@item/g01/users/n\endcsname{167}
\expandafter\gdef\csname svy@item/g01/users/miss\endcsname{4}
\expandafter\gdef\csname svy@item/g01/users/median\endcsname{5.0}
\expandafter\gdef\csname svy@item/g01/users/cilow\endcsname{5.0}
\expandafter\gdef\csname svy@item/g01/users/cihigh\endcsname{5.0}
\expandafter\gdef\csname svy@item/g01/users/mean\endcsname{4.45}
\expandafter\gdef\csname svy@item/g01/users/meanlow\endcsname{4.32}
\expandafter\gdef\csname svy@item/g01/users/meanhigh\endcsname{4.58}
\expandafter\gdef\csname svy@item/g01/users/se\endcsname{0.067}
\expandafter\gdef\csname svy@item/g01/users/sd\endcsname{0.86}
\expandafter\gdef\csname svy@item/g01/users/sdhigh\endcsname{0}
\expandafter\gdef\csname svy@item/g01/users/verdict\endcsname{none}
\expandafter\gdef\csname svy@item/g01/users/dip\endcsname{0}
\expandafter\gdef\csname svy@item/g01/users/shapelow\endcsname{4}
\expandafter\gdef\csname svy@item/g01/users/shapemid\endcsname{5}
\expandafter\gdef\csname svy@item/g01/users/shapehigh\endcsname{91}
\expandafter\gdef\csname svy@item/g01/nonusers/n\endcsname{64}
\expandafter\gdef\csname svy@item/g01/nonusers/miss\endcsname{5}
\expandafter\gdef\csname svy@item/g01/nonusers/median\endcsname{4.0}
\expandafter\gdef\csname svy@item/g01/nonusers/cilow\endcsname{3.0}
\expandafter\gdef\csname svy@item/g01/nonusers/cihigh\endcsname{4.0}
\expandafter\gdef\csname svy@item/g01/nonusers/mean\endcsname{3.34}
\expandafter\gdef\csname svy@item/g01/nonusers/meanlow\endcsname{3.00}
\expandafter\gdef\csname svy@item/g01/nonusers/meanhigh\endcsname{3.69}
\expandafter\gdef\csname svy@item/g01/nonusers/se\endcsname{0.171}
\expandafter\gdef\csname svy@item/g01/nonusers/sd\endcsname{1.37}
\expandafter\gdef\csname svy@item/g01/nonusers/sdhigh\endcsname{0}
\expandafter\gdef\csname svy@item/g01/nonusers/verdict\endcsname{none}
\expandafter\gdef\csname svy@item/g01/nonusers/dip\endcsname{1}
\expandafter\gdef\csname svy@item/g01/nonusers/shapelow\endcsname{31}
\expandafter\gdef\csname svy@item/g01/nonusers/shapemid\endcsname{11}
\expandafter\gdef\csname svy@item/g01/nonusers/shapehigh\endcsname{58}
\expandafter\gdef\csname svy@item/g02/code\endcsname{G02}
\expandafter\gdef\csname svy@item/g02/pred\endcsname{}
\expandafter\gdef\csname svy@item/g02/label\endcsname{concerned about becoming too reliant on them}
\expandafter\gdef\csname svy@item/g02/usage/diff\endcsname{-0.29}
\expandafter\gdef\csname svy@item/g02/usage/difflow\endcsname{-0.69}
\expandafter\gdef\csname svy@item/g02/usage/diffhigh\endcsname{0.10}
\expandafter\gdef\csname svy@item/g02/usage/g\endcsname{-0.22}
\expandafter\gdef\csname svy@item/g02/all/n\endcsname{240}
\expandafter\gdef\csname svy@item/g02/all/miss\endcsname{2}
\expandafter\gdef\csname svy@item/g02/all/median\endcsname{4.0}
\expandafter\gdef\csname svy@item/g02/all/cilow\endcsname{4.0}
\expandafter\gdef\csname svy@item/g02/all/cihigh\endcsname{4.0}
\expandafter\gdef\csname svy@item/g02/all/mean\endcsname{3.50}
\expandafter\gdef\csname svy@item/g02/all/meanlow\endcsname{3.34}
\expandafter\gdef\csname svy@item/g02/all/meanhigh\endcsname{3.67}
\expandafter\gdef\csname svy@item/g02/all/se\endcsname{0.085}
\expandafter\gdef\csname svy@item/g02/all/sd\endcsname{1.31}
\expandafter\gdef\csname svy@item/g02/all/sdhigh\endcsname{0}
\expandafter\gdef\csname svy@item/g02/all/verdict\endcsname{none}
\expandafter\gdef\csname svy@item/g02/all/dip\endcsname{1}
\expandafter\gdef\csname svy@item/g02/all/shapelow\endcsname{29}
\expandafter\gdef\csname svy@item/g02/all/shapemid\endcsname{10}
\expandafter\gdef\csname svy@item/g02/all/shapehigh\endcsname{61}
\expandafter\gdef\csname svy@item/g02/users/n\endcsname{170}
\expandafter\gdef\csname svy@item/g02/users/miss\endcsname{1}
\expandafter\gdef\csname svy@item/g02/users/median\endcsname{4.0}
\expandafter\gdef\csname svy@item/g02/users/cilow\endcsname{3.0}
\expandafter\gdef\csname svy@item/g02/users/cihigh\endcsname{4.0}
\expandafter\gdef\csname svy@item/g02/users/mean\endcsname{3.41}
\expandafter\gdef\csname svy@item/g02/users/meanlow\endcsname{3.22}
\expandafter\gdef\csname svy@item/g02/users/meanhigh\endcsname{3.60}
\expandafter\gdef\csname svy@item/g02/users/se\endcsname{0.096}
\expandafter\gdef\csname svy@item/g02/users/sd\endcsname{1.26}
\expandafter\gdef\csname svy@item/g02/users/sdhigh\endcsname{0}
\expandafter\gdef\csname svy@item/g02/users/verdict\endcsname{none}
\expandafter\gdef\csname svy@item/g02/users/dip\endcsname{1}
\expandafter\gdef\csname svy@item/g02/users/shapelow\endcsname{30}
\expandafter\gdef\csname svy@item/g02/users/shapemid\endcsname{12}
\expandafter\gdef\csname svy@item/g02/users/shapehigh\endcsname{58}
\expandafter\gdef\csname svy@item/g02/nonusers/n\endcsname{68}
\expandafter\gdef\csname svy@item/g02/nonusers/miss\endcsname{1}
\expandafter\gdef\csname svy@item/g02/nonusers/median\endcsname{4.0}
\expandafter\gdef\csname svy@item/g02/nonusers/cilow\endcsname{4.0}
\expandafter\gdef\csname svy@item/g02/nonusers/cihigh\endcsname{5.0}
\expandafter\gdef\csname svy@item/g02/nonusers/mean\endcsname{3.71}
\expandafter\gdef\csname svy@item/g02/nonusers/meanlow\endcsname{3.36}
\expandafter\gdef\csname svy@item/g02/nonusers/meanhigh\endcsname{4.05}
\expandafter\gdef\csname svy@item/g02/nonusers/se\endcsname{0.173}
\expandafter\gdef\csname svy@item/g02/nonusers/sd\endcsname{1.43}
\expandafter\gdef\csname svy@item/g02/nonusers/sdhigh\endcsname{0}
\expandafter\gdef\csname svy@item/g02/nonusers/verdict\endcsname{none}
\expandafter\gdef\csname svy@item/g02/nonusers/dip\endcsname{1}
\expandafter\gdef\csname svy@item/g02/nonusers/shapelow\endcsname{27}
\expandafter\gdef\csname svy@item/g02/nonusers/shapemid\endcsname{4}
\expandafter\gdef\csname svy@item/g02/nonusers/shapehigh\endcsname{69}
\expandafter\gdef\csname svy@item/g03/code\endcsname{G03}
\expandafter\gdef\csname svy@item/g03/pred\endcsname{}
\expandafter\gdef\csname svy@item/g03/label\endcsname{AI coding tools are over-hyped}
\expandafter\gdef\csname svy@item/g03/usage/diff\endcsname{-0.85}
\expandafter\gdef\csname svy@item/g03/usage/difflow\endcsname{-1.22}
\expandafter\gdef\csname svy@item/g03/usage/diffhigh\endcsname{-0.47}
\expandafter\gdef\csname svy@item/g03/usage/g\endcsname{-0.65}
\expandafter\gdef\csname svy@item/g03/all/n\endcsname{236}
\expandafter\gdef\csname svy@item/g03/all/miss\endcsname{6}
\expandafter\gdef\csname svy@item/g03/all/median\endcsname{3.0}
\expandafter\gdef\csname svy@item/g03/all/cilow\endcsname{2.0}
\expandafter\gdef\csname svy@item/g03/all/cihigh\endcsname{3.0}
\expandafter\gdef\csname svy@item/g03/all/mean\endcsname{2.90}
\expandafter\gdef\csname svy@item/g03/all/meanlow\endcsname{2.73}
\expandafter\gdef\csname svy@item/g03/all/meanhigh\endcsname{3.08}
\expandafter\gdef\csname svy@item/g03/all/se\endcsname{0.088}
\expandafter\gdef\csname svy@item/g03/all/sd\endcsname{1.36}
\expandafter\gdef\csname svy@item/g03/all/sdhigh\endcsname{0}
\expandafter\gdef\csname svy@item/g03/all/verdict\endcsname{none}
\expandafter\gdef\csname svy@item/g03/all/dip\endcsname{1}
\expandafter\gdef\csname svy@item/g03/all/shapelow\endcsname{48}
\expandafter\gdef\csname svy@item/g03/all/shapemid\endcsname{14}
\expandafter\gdef\csname svy@item/g03/all/shapehigh\endcsname{38}
\expandafter\gdef\csname svy@item/g03/users/n\endcsname{170}
\expandafter\gdef\csname svy@item/g03/users/miss\endcsname{1}
\expandafter\gdef\csname svy@item/g03/users/median\endcsname{2.0}
\expandafter\gdef\csname svy@item/g03/users/cilow\endcsname{2.0}
\expandafter\gdef\csname svy@item/g03/users/cihigh\endcsname{3.0}
\expandafter\gdef\csname svy@item/g03/users/mean\endcsname{2.67}
\expandafter\gdef\csname svy@item/g03/users/meanlow\endcsname{2.47}
\expandafter\gdef\csname svy@item/g03/users/meanhigh\endcsname{2.87}
\expandafter\gdef\csname svy@item/g03/users/se\endcsname{0.099}
\expandafter\gdef\csname svy@item/g03/users/sd\endcsname{1.30}
\expandafter\gdef\csname svy@item/g03/users/sdhigh\endcsname{0}
\expandafter\gdef\csname svy@item/g03/users/verdict\endcsname{none}
\expandafter\gdef\csname svy@item/g03/users/dip\endcsname{1}
\expandafter\gdef\csname svy@item/g03/users/shapelow\endcsname{56}
\expandafter\gdef\csname svy@item/g03/users/shapemid\endcsname{14}
\expandafter\gdef\csname svy@item/g03/users/shapehigh\endcsname{30}
\expandafter\gdef\csname svy@item/g03/nonusers/n\endcsname{64}
\expandafter\gdef\csname svy@item/g03/nonusers/miss\endcsname{5}
\expandafter\gdef\csname svy@item/g03/nonusers/median\endcsname{4.0}
\expandafter\gdef\csname svy@item/g03/nonusers/cilow\endcsname{3.0}
\expandafter\gdef\csname svy@item/g03/nonusers/cihigh\endcsname{4.0}
\expandafter\gdef\csname svy@item/g03/nonusers/mean\endcsname{3.52}
\expandafter\gdef\csname svy@item/g03/nonusers/meanlow\endcsname{3.19}
\expandafter\gdef\csname svy@item/g03/nonusers/meanhigh\endcsname{3.84}
\expandafter\gdef\csname svy@item/g03/nonusers/se\endcsname{0.164}
\expandafter\gdef\csname svy@item/g03/nonusers/sd\endcsname{1.31}
\expandafter\gdef\csname svy@item/g03/nonusers/sdhigh\endcsname{0}
\expandafter\gdef\csname svy@item/g03/nonusers/verdict\endcsname{none}
\expandafter\gdef\csname svy@item/g03/nonusers/dip\endcsname{1}
\expandafter\gdef\csname svy@item/g03/nonusers/shapelow\endcsname{29}
\expandafter\gdef\csname svy@item/g03/nonusers/shapemid\endcsname{13}
\expandafter\gdef\csname svy@item/g03/nonusers/shapehigh\endcsname{58}
\expandafter\gdef\csname svy@item/g04/code\endcsname{G04}
\expandafter\gdef\csname svy@item/g04/pred\endcsname{}
\expandafter\gdef\csname svy@item/g04/label\endcsname{the more experience, the more I let them do}
\expandafter\gdef\csname svy@item/g04/usage/diff\endcsname{1.29}
\expandafter\gdef\csname svy@item/g04/usage/difflow\endcsname{0.93}
\expandafter\gdef\csname svy@item/g04/usage/diffhigh\endcsname{1.66}
\expandafter\gdef\csname svy@item/g04/usage/g\endcsname{1.08}
\expandafter\gdef\csname svy@item/g04/all/n\endcsname{233}
\expandafter\gdef\csname svy@item/g04/all/miss\endcsname{9}
\expandafter\gdef\csname svy@item/g04/all/median\endcsname{4.0}
\expandafter\gdef\csname svy@item/g04/all/cilow\endcsname{3.0}
\expandafter\gdef\csname svy@item/g04/all/cihigh\endcsname{4.0}
\expandafter\gdef\csname svy@item/g04/all/mean\endcsname{3.40}
\expandafter\gdef\csname svy@item/g04/all/meanlow\endcsname{3.23}
\expandafter\gdef\csname svy@item/g04/all/meanhigh\endcsname{3.57}
\expandafter\gdef\csname svy@item/g04/all/se\endcsname{0.087}
\expandafter\gdef\csname svy@item/g04/all/sd\endcsname{1.32}
\expandafter\gdef\csname svy@item/g04/all/sdhigh\endcsname{0}
\expandafter\gdef\csname svy@item/g04/all/verdict\endcsname{none}
\expandafter\gdef\csname svy@item/g04/all/dip\endcsname{1}
\expandafter\gdef\csname svy@item/g04/all/shapelow\endcsname{28}
\expandafter\gdef\csname svy@item/g04/all/shapemid\endcsname{15}
\expandafter\gdef\csname svy@item/g04/all/shapehigh\endcsname{57}
\expandafter\gdef\csname svy@item/g04/users/n\endcsname{169}
\expandafter\gdef\csname svy@item/g04/users/miss\endcsname{2}
\expandafter\gdef\csname svy@item/g04/users/median\endcsname{4.0}
\expandafter\gdef\csname svy@item/g04/users/cilow\endcsname{4.0}
\expandafter\gdef\csname svy@item/g04/users/cihigh\endcsname{4.0}
\expandafter\gdef\csname svy@item/g04/users/mean\endcsname{3.75}
\expandafter\gdef\csname svy@item/g04/users/meanlow\endcsname{3.57}
\expandafter\gdef\csname svy@item/g04/users/meanhigh\endcsname{3.92}
\expandafter\gdef\csname svy@item/g04/users/se\endcsname{0.091}
\expandafter\gdef\csname svy@item/g04/users/sd\endcsname{1.18}
\expandafter\gdef\csname svy@item/g04/users/sdhigh\endcsname{0}
\expandafter\gdef\csname svy@item/g04/users/verdict\endcsname{none}
\expandafter\gdef\csname svy@item/g04/users/dip\endcsname{0}
\expandafter\gdef\csname svy@item/g04/users/shapelow\endcsname{18}
\expandafter\gdef\csname svy@item/g04/users/shapemid\endcsname{14}
\expandafter\gdef\csname svy@item/g04/users/shapehigh\endcsname{68}
\expandafter\gdef\csname svy@item/g04/nonusers/n\endcsname{62}
\expandafter\gdef\csname svy@item/g04/nonusers/miss\endcsname{7}
\expandafter\gdef\csname svy@item/g04/nonusers/median\endcsname{2.0}
\expandafter\gdef\csname svy@item/g04/nonusers/cilow\endcsname{2.0}
\expandafter\gdef\csname svy@item/g04/nonusers/cihigh\endcsname{3.0}
\expandafter\gdef\csname svy@item/g04/nonusers/mean\endcsname{2.45}
\expandafter\gdef\csname svy@item/g04/nonusers/meanlow\endcsname{2.13}
\expandafter\gdef\csname svy@item/g04/nonusers/meanhigh\endcsname{2.77}
\expandafter\gdef\csname svy@item/g04/nonusers/se\endcsname{0.159}
\expandafter\gdef\csname svy@item/g04/nonusers/sd\endcsname{1.25}
\expandafter\gdef\csname svy@item/g04/nonusers/sdhigh\endcsname{0}
\expandafter\gdef\csname svy@item/g04/nonusers/verdict\endcsname{none}
\expandafter\gdef\csname svy@item/g04/nonusers/dip\endcsname{1}
\expandafter\gdef\csname svy@item/g04/nonusers/shapelow\endcsname{56}
\expandafter\gdef\csname svy@item/g04/nonusers/shapemid\endcsname{18}
\expandafter\gdef\csname svy@item/g04/nonusers/shapehigh\endcsname{26}
\expandafter\gdef\csname svy@item/g05/code\endcsname{G05}
\expandafter\gdef\csname svy@item/g05/pred\endcsname{}
\expandafter\gdef\csname svy@item/g05/label\endcsname{trust my AI tools more than myself}
\expandafter\gdef\csname svy@item/g05/usage/diff\endcsname{0.82}
\expandafter\gdef\csname svy@item/g05/usage/difflow\endcsname{0.47}
\expandafter\gdef\csname svy@item/g05/usage/diffhigh\endcsname{1.16}
\expandafter\gdef\csname svy@item/g05/usage/g\endcsname{0.63}
\expandafter\gdef\csname svy@item/g05/all/n\endcsname{238}
\expandafter\gdef\csname svy@item/g05/all/miss\endcsname{4}
\expandafter\gdef\csname svy@item/g05/all/median\endcsname{2.0}
\expandafter\gdef\csname svy@item/g05/all/cilow\endcsname{2.0}
\expandafter\gdef\csname svy@item/g05/all/cihigh\endcsname{3.0}
\expandafter\gdef\csname svy@item/g05/all/mean\endcsname{2.61}
\expandafter\gdef\csname svy@item/g05/all/meanlow\endcsname{2.44}
\expandafter\gdef\csname svy@item/g05/all/meanhigh\endcsname{2.78}
\expandafter\gdef\csname svy@item/g05/all/se\endcsname{0.087}
\expandafter\gdef\csname svy@item/g05/all/sd\endcsname{1.34}
\expandafter\gdef\csname svy@item/g05/all/sdhigh\endcsname{0}
\expandafter\gdef\csname svy@item/g05/all/verdict\endcsname{none}
\expandafter\gdef\csname svy@item/g05/all/dip\endcsname{1}
\expandafter\gdef\csname svy@item/g05/all/shapelow\endcsname{55}
\expandafter\gdef\csname svy@item/g05/all/shapemid\endcsname{15}
\expandafter\gdef\csname svy@item/g05/all/shapehigh\endcsname{30}
\expandafter\gdef\csname svy@item/g05/users/n\endcsname{170}
\expandafter\gdef\csname svy@item/g05/users/miss\endcsname{1}
\expandafter\gdef\csname svy@item/g05/users/median\endcsname{3.0}
\expandafter\gdef\csname svy@item/g05/users/cilow\endcsname{2.0}
\expandafter\gdef\csname svy@item/g05/users/cihigh\endcsname{3.0}
\expandafter\gdef\csname svy@item/g05/users/mean\endcsname{2.85}
\expandafter\gdef\csname svy@item/g05/users/meanlow\endcsname{2.64}
\expandafter\gdef\csname svy@item/g05/users/meanhigh\endcsname{3.05}
\expandafter\gdef\csname svy@item/g05/users/se\endcsname{0.103}
\expandafter\gdef\csname svy@item/g05/users/sd\endcsname{1.35}
\expandafter\gdef\csname svy@item/g05/users/sdhigh\endcsname{0}
\expandafter\gdef\csname svy@item/g05/users/verdict\endcsname{none}
\expandafter\gdef\csname svy@item/g05/users/dip\endcsname{1}
\expandafter\gdef\csname svy@item/g05/users/shapelow\endcsname{47}
\expandafter\gdef\csname svy@item/g05/users/shapemid\endcsname{17}
\expandafter\gdef\csname svy@item/g05/users/shapehigh\endcsname{36}
\expandafter\gdef\csname svy@item/g05/nonusers/n\endcsname{66}
\expandafter\gdef\csname svy@item/g05/nonusers/miss\endcsname{3}
\expandafter\gdef\csname svy@item/g05/nonusers/median\endcsname{2.0}
\expandafter\gdef\csname svy@item/g05/nonusers/cilow\endcsname{1.0}
\expandafter\gdef\csname svy@item/g05/nonusers/cihigh\endcsname{2.0}
\expandafter\gdef\csname svy@item/g05/nonusers/mean\endcsname{2.03}
\expandafter\gdef\csname svy@item/g05/nonusers/meanlow\endcsname{1.75}
\expandafter\gdef\csname svy@item/g05/nonusers/meanhigh\endcsname{2.31}
\expandafter\gdef\csname svy@item/g05/nonusers/se\endcsname{0.142}
\expandafter\gdef\csname svy@item/g05/nonusers/sd\endcsname{1.15}
\expandafter\gdef\csname svy@item/g05/nonusers/sdhigh\endcsname{0}
\expandafter\gdef\csname svy@item/g05/nonusers/verdict\endcsname{none}
\expandafter\gdef\csname svy@item/g05/nonusers/dip\endcsname{0}
\expandafter\gdef\csname svy@item/g05/nonusers/shapelow\endcsname{73}
\expandafter\gdef\csname svy@item/g05/nonusers/shapemid\endcsname{12}
\expandafter\gdef\csname svy@item/g05/nonusers/shapehigh\endcsname{15}
\expandafter\gdef\csname svy@item/g06/code\endcsname{G06}
\expandafter\gdef\csname svy@item/g06/pred\endcsname{}
\expandafter\gdef\csname svy@item/g06/label\endcsname{the tools I work with must be highly reliable}
\expandafter\gdef\csname svy@item/g06/usage/diff\endcsname{-0.21}
\expandafter\gdef\csname svy@item/g06/usage/difflow\endcsname{-0.48}
\expandafter\gdef\csname svy@item/g06/usage/diffhigh\endcsname{0.05}
\expandafter\gdef\csname svy@item/g06/usage/g\endcsname{-0.22}
\expandafter\gdef\csname svy@item/g06/all/n\endcsname{238}
\expandafter\gdef\csname svy@item/g06/all/miss\endcsname{4}
\expandafter\gdef\csname svy@item/g06/all/median\endcsname{4.0}
\expandafter\gdef\csname svy@item/g06/all/cilow\endcsname{4.0}
\expandafter\gdef\csname svy@item/g06/all/cihigh\endcsname{5.0}
\expandafter\gdef\csname svy@item/g06/all/mean\endcsname{4.15}
\expandafter\gdef\csname svy@item/g06/all/meanlow\endcsname{4.03}
\expandafter\gdef\csname svy@item/g06/all/meanhigh\endcsname{4.27}
\expandafter\gdef\csname svy@item/g06/all/se\endcsname{0.062}
\expandafter\gdef\csname svy@item/g06/all/sd\endcsname{0.95}
\expandafter\gdef\csname svy@item/g06/all/sdhigh\endcsname{0}
\expandafter\gdef\csname svy@item/g06/all/verdict\endcsname{none}
\expandafter\gdef\csname svy@item/g06/all/dip\endcsname{0}
\expandafter\gdef\csname svy@item/g06/all/shapelow\endcsname{6}
\expandafter\gdef\csname svy@item/g06/all/shapemid\endcsname{14}
\expandafter\gdef\csname svy@item/g06/all/shapehigh\endcsname{80}
\expandafter\gdef\csname svy@item/g06/users/n\endcsname{168}
\expandafter\gdef\csname svy@item/g06/users/miss\endcsname{3}
\expandafter\gdef\csname svy@item/g06/users/median\endcsname{4.0}
\expandafter\gdef\csname svy@item/g06/users/cilow\endcsname{4.0}
\expandafter\gdef\csname svy@item/g06/users/cihigh\endcsname{4.0}
\expandafter\gdef\csname svy@item/g06/users/mean\endcsname{4.08}
\expandafter\gdef\csname svy@item/g06/users/meanlow\endcsname{3.94}
\expandafter\gdef\csname svy@item/g06/users/meanhigh\endcsname{4.23}
\expandafter\gdef\csname svy@item/g06/users/se\endcsname{0.075}
\expandafter\gdef\csname svy@item/g06/users/sd\endcsname{0.97}
\expandafter\gdef\csname svy@item/g06/users/sdhigh\endcsname{0}
\expandafter\gdef\csname svy@item/g06/users/verdict\endcsname{none}
\expandafter\gdef\csname svy@item/g06/users/dip\endcsname{0}
\expandafter\gdef\csname svy@item/g06/users/shapelow\endcsname{7}
\expandafter\gdef\csname svy@item/g06/users/shapemid\endcsname{18}
\expandafter\gdef\csname svy@item/g06/users/shapehigh\endcsname{75}
\expandafter\gdef\csname svy@item/g06/nonusers/n\endcsname{68}
\expandafter\gdef\csname svy@item/g06/nonusers/miss\endcsname{1}
\expandafter\gdef\csname svy@item/g06/nonusers/median\endcsname{4.0}
\expandafter\gdef\csname svy@item/g06/nonusers/cilow\endcsname{4.0}
\expandafter\gdef\csname svy@item/g06/nonusers/cihigh\endcsname{5.0}
\expandafter\gdef\csname svy@item/g06/nonusers/mean\endcsname{4.29}
\expandafter\gdef\csname svy@item/g06/nonusers/meanlow\endcsname{4.07}
\expandafter\gdef\csname svy@item/g06/nonusers/meanhigh\endcsname{4.52}
\expandafter\gdef\csname svy@item/g06/nonusers/se\endcsname{0.111}
\expandafter\gdef\csname svy@item/g06/nonusers/sd\endcsname{0.92}
\expandafter\gdef\csname svy@item/g06/nonusers/sdhigh\endcsname{0}
\expandafter\gdef\csname svy@item/g06/nonusers/verdict\endcsname{none}
\expandafter\gdef\csname svy@item/g06/nonusers/dip\endcsname{0}
\expandafter\gdef\csname svy@item/g06/nonusers/shapelow\endcsname{6}
\expandafter\gdef\csname svy@item/g06/nonusers/shapemid\endcsname{4}
\expandafter\gdef\csname svy@item/g06/nonusers/shapehigh\endcsname{90}
\expandafter\gdef\csname svy@item/g07/code\endcsname{G07}
\expandafter\gdef\csname svy@item/g07/pred\endcsname{}
\expandafter\gdef\csname svy@item/g07/label\endcsname{willing to take moderate risks for high payoff}
\expandafter\gdef\csname svy@item/g07/usage/diff\endcsname{0.81}
\expandafter\gdef\csname svy@item/g07/usage/difflow\endcsname{0.48}
\expandafter\gdef\csname svy@item/g07/usage/diffhigh\endcsname{1.13}
\expandafter\gdef\csname svy@item/g07/usage/g\endcsname{0.74}
\expandafter\gdef\csname svy@item/g07/all/n\endcsname{234}
\expandafter\gdef\csname svy@item/g07/all/miss\endcsname{8}
\expandafter\gdef\csname svy@item/g07/all/median\endcsname{4.0}
\expandafter\gdef\csname svy@item/g07/all/cilow\endcsname{4.0}
\expandafter\gdef\csname svy@item/g07/all/cihigh\endcsname{4.0}
\expandafter\gdef\csname svy@item/g07/all/mean\endcsname{3.43}
\expandafter\gdef\csname svy@item/g07/all/meanlow\endcsname{3.28}
\expandafter\gdef\csname svy@item/g07/all/meanhigh\endcsname{3.58}
\expandafter\gdef\csname svy@item/g07/all/se\endcsname{0.075}
\expandafter\gdef\csname svy@item/g07/all/sd\endcsname{1.15}
\expandafter\gdef\csname svy@item/g07/all/sdhigh\endcsname{0}
\expandafter\gdef\csname svy@item/g07/all/verdict\endcsname{none}
\expandafter\gdef\csname svy@item/g07/all/dip\endcsname{1}
\expandafter\gdef\csname svy@item/g07/all/shapelow\endcsname{24}
\expandafter\gdef\csname svy@item/g07/all/shapemid\endcsname{17}
\expandafter\gdef\csname svy@item/g07/all/shapehigh\endcsname{59}
\expandafter\gdef\csname svy@item/g07/users/n\endcsname{164}
\expandafter\gdef\csname svy@item/g07/users/miss\endcsname{7}
\expandafter\gdef\csname svy@item/g07/users/median\endcsname{4.0}
\expandafter\gdef\csname svy@item/g07/users/cilow\endcsname{4.0}
\expandafter\gdef\csname svy@item/g07/users/cihigh\endcsname{4.0}
\expandafter\gdef\csname svy@item/g07/users/mean\endcsname{3.68}
\expandafter\gdef\csname svy@item/g07/users/meanlow\endcsname{3.51}
\expandafter\gdef\csname svy@item/g07/users/meanhigh\endcsname{3.84}
\expandafter\gdef\csname svy@item/g07/users/se\endcsname{0.082}
\expandafter\gdef\csname svy@item/g07/users/sd\endcsname{1.06}
\expandafter\gdef\csname svy@item/g07/users/sdhigh\endcsname{0}
\expandafter\gdef\csname svy@item/g07/users/verdict\endcsname{none}
\expandafter\gdef\csname svy@item/g07/users/dip\endcsname{0}
\expandafter\gdef\csname svy@item/g07/users/shapelow\endcsname{16}
\expandafter\gdef\csname svy@item/g07/users/shapemid\endcsname{15}
\expandafter\gdef\csname svy@item/g07/users/shapehigh\endcsname{69}
\expandafter\gdef\csname svy@item/g07/nonusers/n\endcsname{68}
\expandafter\gdef\csname svy@item/g07/nonusers/miss\endcsname{1}
\expandafter\gdef\csname svy@item/g07/nonusers/median\endcsname{3.0}
\expandafter\gdef\csname svy@item/g07/nonusers/cilow\endcsname{2.0}
\expandafter\gdef\csname svy@item/g07/nonusers/cihigh\endcsname{3.0}
\expandafter\gdef\csname svy@item/g07/nonusers/mean\endcsname{2.87}
\expandafter\gdef\csname svy@item/g07/nonusers/meanlow\endcsname{2.58}
\expandafter\gdef\csname svy@item/g07/nonusers/meanhigh\endcsname{3.15}
\expandafter\gdef\csname svy@item/g07/nonusers/se\endcsname{0.142}
\expandafter\gdef\csname svy@item/g07/nonusers/sd\endcsname{1.17}
\expandafter\gdef\csname svy@item/g07/nonusers/sdhigh\endcsname{0}
\expandafter\gdef\csname svy@item/g07/nonusers/verdict\endcsname{none}
\expandafter\gdef\csname svy@item/g07/nonusers/dip\endcsname{1}
\expandafter\gdef\csname svy@item/g07/nonusers/shapelow\endcsname{44}
\expandafter\gdef\csname svy@item/g07/nonusers/shapemid\endcsname{21}
\expandafter\gdef\csname svy@item/g07/nonusers/shapehigh\endcsname{35}
\expandafter\gdef\csname svy@item/g01/exposure/trend\endcsname{0.60}
\expandafter\gdef\csname svy@item/g01/exposure/trendlow\endcsname{0.45}
\expandafter\gdef\csname svy@item/g01/exposure/trendhigh\endcsname{0.74}
\expandafter\gdef\csname svy@item/g01/e0/n\endcsname{9}
\expandafter\gdef\csname svy@item/g01/e0/miss\endcsname{4}
\expandafter\gdef\csname svy@item/g01/e0/median\endcsname{1.0}
\expandafter\gdef\csname svy@item/g01/e0/cilow\endcsname{1.0}
\expandafter\gdef\csname svy@item/g01/e0/cihigh\endcsname{2.0}
\expandafter\gdef\csname svy@item/g01/e0/mean\endcsname{1.56}
\expandafter\gdef\csname svy@item/g01/e0/meanlow\endcsname{1.00}
\expandafter\gdef\csname svy@item/g01/e0/meanhigh\endcsname{2.11}
\expandafter\gdef\csname svy@item/g01/e0/se\endcsname{0.242}
\expandafter\gdef\csname svy@item/g01/e0/sd\endcsname{0.73}
\expandafter\gdef\csname svy@item/g01/e0/sdhigh\endcsname{0}
\expandafter\gdef\csname svy@item/g01/e0/verdict\endcsname{none}
\expandafter\gdef\csname svy@item/g01/e0/dip\endcsname{0}
\expandafter\gdef\csname svy@item/g01/e0/shapelow\endcsname{89}
\expandafter\gdef\csname svy@item/g01/e0/shapemid\endcsname{11}
\expandafter\gdef\csname svy@item/g01/e0/shapehigh\endcsname{0}
\expandafter\gdef\csname svy@item/g01/e1/n\endcsname{49}
\expandafter\gdef\csname svy@item/g01/e1/miss\endcsname{0}
\expandafter\gdef\csname svy@item/g01/e1/median\endcsname{4.0}
\expandafter\gdef\csname svy@item/g01/e1/cilow\endcsname{4.0}
\expandafter\gdef\csname svy@item/g01/e1/cihigh\endcsname{4.0}
\expandafter\gdef\csname svy@item/g01/e1/mean\endcsname{3.80}
\expandafter\gdef\csname svy@item/g01/e1/meanlow\endcsname{3.47}
\expandafter\gdef\csname svy@item/g01/e1/meanhigh\endcsname{4.12}
\expandafter\gdef\csname svy@item/g01/e1/se\endcsname{0.160}
\expandafter\gdef\csname svy@item/g01/e1/sd\endcsname{1.12}
\expandafter\gdef\csname svy@item/g01/e1/sdhigh\endcsname{0}
\expandafter\gdef\csname svy@item/g01/e1/verdict\endcsname{none}
\expandafter\gdef\csname svy@item/g01/e1/dip\endcsname{0}
\expandafter\gdef\csname svy@item/g01/e1/shapelow\endcsname{16}
\expandafter\gdef\csname svy@item/g01/e1/shapemid\endcsname{12}
\expandafter\gdef\csname svy@item/g01/e1/shapehigh\endcsname{72}
\expandafter\gdef\csname svy@item/g01/e2/n\endcsname{75}
\expandafter\gdef\csname svy@item/g01/e2/miss\endcsname{2}
\expandafter\gdef\csname svy@item/g01/e2/median\endcsname{4.0}
\expandafter\gdef\csname svy@item/g01/e2/cilow\endcsname{4.0}
\expandafter\gdef\csname svy@item/g01/e2/cihigh\endcsname{5.0}
\expandafter\gdef\csname svy@item/g01/e2/mean\endcsname{4.27}
\expandafter\gdef\csname svy@item/g01/e2/meanlow\endcsname{4.05}
\expandafter\gdef\csname svy@item/g01/e2/meanhigh\endcsname{4.48}
\expandafter\gdef\csname svy@item/g01/e2/se\endcsname{0.108}
\expandafter\gdef\csname svy@item/g01/e2/sd\endcsname{0.93}
\expandafter\gdef\csname svy@item/g01/e2/sdhigh\endcsname{0}
\expandafter\gdef\csname svy@item/g01/e2/verdict\endcsname{none}
\expandafter\gdef\csname svy@item/g01/e2/dip\endcsname{0}
\expandafter\gdef\csname svy@item/g01/e2/shapelow\endcsname{6}
\expandafter\gdef\csname svy@item/g01/e2/shapemid\endcsname{9}
\expandafter\gdef\csname svy@item/g01/e2/shapehigh\endcsname{85}
\expandafter\gdef\csname svy@item/g01/e3/n\endcsname{92}
\expandafter\gdef\csname svy@item/g01/e3/miss\endcsname{2}
\expandafter\gdef\csname svy@item/g01/e3/median\endcsname{5.0}
\expandafter\gdef\csname svy@item/g01/e3/cilow\endcsname{5.0}
\expandafter\gdef\csname svy@item/g01/e3/cihigh\endcsname{5.0}
\expandafter\gdef\csname svy@item/g01/e3/mean\endcsname{4.60}
\expandafter\gdef\csname svy@item/g01/e3/meanlow\endcsname{4.44}
\expandafter\gdef\csname svy@item/g01/e3/meanhigh\endcsname{4.76}
\expandafter\gdef\csname svy@item/g01/e3/se\endcsname{0.080}
\expandafter\gdef\csname svy@item/g01/e3/sd\endcsname{0.77}
\expandafter\gdef\csname svy@item/g01/e3/sdhigh\endcsname{0}
\expandafter\gdef\csname svy@item/g01/e3/verdict\endcsname{none}
\expandafter\gdef\csname svy@item/g01/e3/dip\endcsname{0}
\expandafter\gdef\csname svy@item/g01/e3/shapelow\endcsname{3}
\expandafter\gdef\csname svy@item/g01/e3/shapemid\endcsname{1}
\expandafter\gdef\csname svy@item/g01/e3/shapehigh\endcsname{96}
\expandafter\gdef\csname svy@item/g02/exposure/trend\endcsname{-0.18}
\expandafter\gdef\csname svy@item/g02/exposure/trendlow\endcsname{-0.37}
\expandafter\gdef\csname svy@item/g02/exposure/trendhigh\endcsname{0.01}
\expandafter\gdef\csname svy@item/g02/e0/n\endcsname{12}
\expandafter\gdef\csname svy@item/g02/e0/miss\endcsname{1}
\expandafter\gdef\csname svy@item/g02/e0/median\endcsname{4.0}
\expandafter\gdef\csname svy@item/g02/e0/cilow\endcsname{1.0}
\expandafter\gdef\csname svy@item/g02/e0/cihigh\endcsname{5.0}
\expandafter\gdef\csname svy@item/g02/e0/mean\endcsname{3.17}
\expandafter\gdef\csname svy@item/g02/e0/meanlow\endcsname{2.02}
\expandafter\gdef\csname svy@item/g02/e0/meanhigh\endcsname{4.31}
\expandafter\gdef\csname svy@item/g02/e0/se\endcsname{0.520}
\expandafter\gdef\csname svy@item/g02/e0/sd\endcsname{1.80}
\expandafter\gdef\csname svy@item/g02/e0/sdhigh\endcsname{0}
\expandafter\gdef\csname svy@item/g02/e0/verdict\endcsname{none}
\expandafter\gdef\csname svy@item/g02/e0/dip\endcsname{1}
\expandafter\gdef\csname svy@item/g02/e0/shapelow\endcsname{42}
\expandafter\gdef\csname svy@item/g02/e0/shapemid\endcsname{0}
\expandafter\gdef\csname svy@item/g02/e0/shapehigh\endcsname{58}
\expandafter\gdef\csname svy@item/g02/e1/n\endcsname{49}
\expandafter\gdef\csname svy@item/g02/e1/miss\endcsname{0}
\expandafter\gdef\csname svy@item/g02/e1/median\endcsname{4.0}
\expandafter\gdef\csname svy@item/g02/e1/cilow\endcsname{4.0}
\expandafter\gdef\csname svy@item/g02/e1/cihigh\endcsname{5.0}
\expandafter\gdef\csname svy@item/g02/e1/mean\endcsname{3.96}
\expandafter\gdef\csname svy@item/g02/e1/meanlow\endcsname{3.59}
\expandafter\gdef\csname svy@item/g02/e1/meanhigh\endcsname{4.33}
\expandafter\gdef\csname svy@item/g02/e1/se\endcsname{0.184}
\expandafter\gdef\csname svy@item/g02/e1/sd\endcsname{1.29}
\expandafter\gdef\csname svy@item/g02/e1/sdhigh\endcsname{0}
\expandafter\gdef\csname svy@item/g02/e1/verdict\endcsname{none}
\expandafter\gdef\csname svy@item/g02/e1/dip\endcsname{0}
\expandafter\gdef\csname svy@item/g02/e1/shapelow\endcsname{20}
\expandafter\gdef\csname svy@item/g02/e1/shapemid\endcsname{4}
\expandafter\gdef\csname svy@item/g02/e1/shapehigh\endcsname{76}
\expandafter\gdef\csname svy@item/g02/e2/n\endcsname{76}
\expandafter\gdef\csname svy@item/g02/e2/miss\endcsname{1}
\expandafter\gdef\csname svy@item/g02/e2/median\endcsname{4.0}
\expandafter\gdef\csname svy@item/g02/e2/cilow\endcsname{4.0}
\expandafter\gdef\csname svy@item/g02/e2/cihigh\endcsname{4.0}
\expandafter\gdef\csname svy@item/g02/e2/mean\endcsname{3.55}
\expandafter\gdef\csname svy@item/g02/e2/meanlow\endcsname{3.28}
\expandafter\gdef\csname svy@item/g02/e2/meanhigh\endcsname{3.82}
\expandafter\gdef\csname svy@item/g02/e2/se\endcsname{0.136}
\expandafter\gdef\csname svy@item/g02/e2/sd\endcsname{1.18}
\expandafter\gdef\csname svy@item/g02/e2/sdhigh\endcsname{0}
\expandafter\gdef\csname svy@item/g02/e2/verdict\endcsname{none}
\expandafter\gdef\csname svy@item/g02/e2/dip\endcsname{1}
\expandafter\gdef\csname svy@item/g02/e2/shapelow\endcsname{26}
\expandafter\gdef\csname svy@item/g02/e2/shapemid\endcsname{11}
\expandafter\gdef\csname svy@item/g02/e2/shapehigh\endcsname{63}
\expandafter\gdef\csname svy@item/g02/e3/n\endcsname{94}
\expandafter\gdef\csname svy@item/g02/e3/miss\endcsname{0}
\expandafter\gdef\csname svy@item/g02/e3/median\endcsname{4.0}
\expandafter\gdef\csname svy@item/g02/e3/cilow\endcsname{3.0}
\expandafter\gdef\csname svy@item/g02/e3/cihigh\endcsname{4.0}
\expandafter\gdef\csname svy@item/g02/e3/mean\endcsname{3.30}
\expandafter\gdef\csname svy@item/g02/e3/meanlow\endcsname{3.03}
\expandafter\gdef\csname svy@item/g02/e3/meanhigh\endcsname{3.57}
\expandafter\gdef\csname svy@item/g02/e3/se\endcsname{0.135}
\expandafter\gdef\csname svy@item/g02/e3/sd\endcsname{1.31}
\expandafter\gdef\csname svy@item/g02/e3/sdhigh\endcsname{0}
\expandafter\gdef\csname svy@item/g02/e3/verdict\endcsname{none}
\expandafter\gdef\csname svy@item/g02/e3/dip\endcsname{1}
\expandafter\gdef\csname svy@item/g02/e3/shapelow\endcsname{33}
\expandafter\gdef\csname svy@item/g02/e3/shapemid\endcsname{14}
\expandafter\gdef\csname svy@item/g02/e3/shapehigh\endcsname{53}
\expandafter\gdef\csname svy@item/g03/exposure/trend\endcsname{-0.54}
\expandafter\gdef\csname svy@item/g03/exposure/trendlow\endcsname{-0.73}
\expandafter\gdef\csname svy@item/g03/exposure/trendhigh\endcsname{-0.36}
\expandafter\gdef\csname svy@item/g03/e0/n\endcsname{11}
\expandafter\gdef\csname svy@item/g03/e0/miss\endcsname{2}
\expandafter\gdef\csname svy@item/g03/e0/median\endcsname{4.0}
\expandafter\gdef\csname svy@item/g03/e0/cilow\endcsname{3.0}
\expandafter\gdef\csname svy@item/g03/e0/cihigh\endcsname{5.0}
\expandafter\gdef\csname svy@item/g03/e0/mean\endcsname{4.27}
\expandafter\gdef\csname svy@item/g03/e0/meanlow\endcsname{3.74}
\expandafter\gdef\csname svy@item/g03/e0/meanhigh\endcsname{4.80}
\expandafter\gdef\csname svy@item/g03/e0/se\endcsname{0.237}
\expandafter\gdef\csname svy@item/g03/e0/sd\endcsname{0.79}
\expandafter\gdef\csname svy@item/g03/e0/sdhigh\endcsname{0}
\expandafter\gdef\csname svy@item/g03/e0/verdict\endcsname{none}
\expandafter\gdef\csname svy@item/g03/e0/dip\endcsname{0}
\expandafter\gdef\csname svy@item/g03/e0/shapelow\endcsname{0}
\expandafter\gdef\csname svy@item/g03/e0/shapemid\endcsname{18}
\expandafter\gdef\csname svy@item/g03/e0/shapehigh\endcsname{82}
\expandafter\gdef\csname svy@item/g03/e1/n\endcsname{46}
\expandafter\gdef\csname svy@item/g03/e1/miss\endcsname{3}
\expandafter\gdef\csname svy@item/g03/e1/median\endcsname{3.0}
\expandafter\gdef\csname svy@item/g03/e1/cilow\endcsname{2.0}
\expandafter\gdef\csname svy@item/g03/e1/cihigh\endcsname{4.0}
\expandafter\gdef\csname svy@item/g03/e1/mean\endcsname{3.26}
\expandafter\gdef\csname svy@item/g03/e1/meanlow\endcsname{2.87}
\expandafter\gdef\csname svy@item/g03/e1/meanhigh\endcsname{3.65}
\expandafter\gdef\csname svy@item/g03/e1/se\endcsname{0.195}
\expandafter\gdef\csname svy@item/g03/e1/sd\endcsname{1.32}
\expandafter\gdef\csname svy@item/g03/e1/sdhigh\endcsname{0}
\expandafter\gdef\csname svy@item/g03/e1/verdict\endcsname{none}
\expandafter\gdef\csname svy@item/g03/e1/dip\endcsname{1}
\expandafter\gdef\csname svy@item/g03/e1/shapelow\endcsname{39}
\expandafter\gdef\csname svy@item/g03/e1/shapemid\endcsname{13}
\expandafter\gdef\csname svy@item/g03/e1/shapehigh\endcsname{48}
\expandafter\gdef\csname svy@item/g03/e2/n\endcsname{76}
\expandafter\gdef\csname svy@item/g03/e2/miss\endcsname{1}
\expandafter\gdef\csname svy@item/g03/e2/median\endcsname{3.0}
\expandafter\gdef\csname svy@item/g03/e2/cilow\endcsname{2.0}
\expandafter\gdef\csname svy@item/g03/e2/cihigh\endcsname{4.0}
\expandafter\gdef\csname svy@item/g03/e2/mean\endcsname{3.08}
\expandafter\gdef\csname svy@item/g03/e2/meanlow\endcsname{2.79}
\expandafter\gdef\csname svy@item/g03/e2/meanhigh\endcsname{3.37}
\expandafter\gdef\csname svy@item/g03/e2/se\endcsname{0.146}
\expandafter\gdef\csname svy@item/g03/e2/sd\endcsname{1.27}
\expandafter\gdef\csname svy@item/g03/e2/sdhigh\endcsname{0}
\expandafter\gdef\csname svy@item/g03/e2/verdict\endcsname{none}
\expandafter\gdef\csname svy@item/g03/e2/dip\endcsname{1}
\expandafter\gdef\csname svy@item/g03/e2/shapelow\endcsname{45}
\expandafter\gdef\csname svy@item/g03/e2/shapemid\endcsname{15}
\expandafter\gdef\csname svy@item/g03/e2/shapehigh\endcsname{40}
\expandafter\gdef\csname svy@item/g03/e3/n\endcsname{94}
\expandafter\gdef\csname svy@item/g03/e3/miss\endcsname{0}
\expandafter\gdef\csname svy@item/g03/e3/median\endcsname{2.0}
\expandafter\gdef\csname svy@item/g03/e3/cilow\endcsname{2.0}
\expandafter\gdef\csname svy@item/g03/e3/cihigh\endcsname{2.0}
\expandafter\gdef\csname svy@item/g03/e3/mean\endcsname{2.34}
\expandafter\gdef\csname svy@item/g03/e3/meanlow\endcsname{2.09}
\expandafter\gdef\csname svy@item/g03/e3/meanhigh\endcsname{2.59}
\expandafter\gdef\csname svy@item/g03/e3/se\endcsname{0.126}
\expandafter\gdef\csname svy@item/g03/e3/sd\endcsname{1.22}
\expandafter\gdef\csname svy@item/g03/e3/sdhigh\endcsname{0}
\expandafter\gdef\csname svy@item/g03/e3/verdict\endcsname{none}
\expandafter\gdef\csname svy@item/g03/e3/dip\endcsname{1}
\expandafter\gdef\csname svy@item/g03/e3/shapelow\endcsname{64}
\expandafter\gdef\csname svy@item/g03/e3/shapemid\endcsname{14}
\expandafter\gdef\csname svy@item/g03/e3/shapehigh\endcsname{22}
\expandafter\gdef\csname svy@item/g04/exposure/trend\endcsname{0.76}
\expandafter\gdef\csname svy@item/g04/exposure/trendlow\endcsname{0.58}
\expandafter\gdef\csname svy@item/g04/exposure/trendhigh\endcsname{0.93}
\expandafter\gdef\csname svy@item/g04/e0/n\endcsname{8}
\expandafter\gdef\csname svy@item/g04/e0/miss\endcsname{5}
\expandafter\gdef\csname svy@item/g04/e0/median\endcsname{1.0}
\expandafter\gdef\csname svy@item/g04/e0/cilow\endcsname{1.0}
\expandafter\gdef\csname svy@item/g04/e0/cihigh\endcsname{3.0}
\expandafter\gdef\csname svy@item/g04/e0/mean\endcsname{1.62}
\expandafter\gdef\csname svy@item/g04/e0/meanlow\endcsname{0.86}
\expandafter\gdef\csname svy@item/g04/e0/meanhigh\endcsname{2.39}
\expandafter\gdef\csname svy@item/g04/e0/se\endcsname{0.324}
\expandafter\gdef\csname svy@item/g04/e0/sd\endcsname{0.92}
\expandafter\gdef\csname svy@item/g04/e0/sdhigh\endcsname{0}
\expandafter\gdef\csname svy@item/g04/e0/verdict\endcsname{none}
\expandafter\gdef\csname svy@item/g04/e0/dip\endcsname{0}
\expandafter\gdef\csname svy@item/g04/e0/shapelow\endcsname{75}
\expandafter\gdef\csname svy@item/g04/e0/shapemid\endcsname{25}
\expandafter\gdef\csname svy@item/g04/e0/shapehigh\endcsname{0}
\expandafter\gdef\csname svy@item/g04/e1/n\endcsname{47}
\expandafter\gdef\csname svy@item/g04/e1/miss\endcsname{2}
\expandafter\gdef\csname svy@item/g04/e1/median\endcsname{2.0}
\expandafter\gdef\csname svy@item/g04/e1/cilow\endcsname{2.0}
\expandafter\gdef\csname svy@item/g04/e1/cihigh\endcsname{3.0}
\expandafter\gdef\csname svy@item/g04/e1/mean\endcsname{2.62}
\expandafter\gdef\csname svy@item/g04/e1/meanlow\endcsname{2.25}
\expandafter\gdef\csname svy@item/g04/e1/meanhigh\endcsname{2.98}
\expandafter\gdef\csname svy@item/g04/e1/se\endcsname{0.181}
\expandafter\gdef\csname svy@item/g04/e1/sd\endcsname{1.24}
\expandafter\gdef\csname svy@item/g04/e1/sdhigh\endcsname{0}
\expandafter\gdef\csname svy@item/g04/e1/verdict\endcsname{none}
\expandafter\gdef\csname svy@item/g04/e1/dip\endcsname{1}
\expandafter\gdef\csname svy@item/g04/e1/shapelow\endcsname{53}
\expandafter\gdef\csname svy@item/g04/e1/shapemid\endcsname{17}
\expandafter\gdef\csname svy@item/g04/e1/shapehigh\endcsname{30}
\expandafter\gdef\csname svy@item/g04/e2/n\endcsname{76}
\expandafter\gdef\csname svy@item/g04/e2/miss\endcsname{1}
\expandafter\gdef\csname svy@item/g04/e2/median\endcsname{4.0}
\expandafter\gdef\csname svy@item/g04/e2/cilow\endcsname{3.0}
\expandafter\gdef\csname svy@item/g04/e2/cihigh\endcsname{4.0}
\expandafter\gdef\csname svy@item/g04/e2/mean\endcsname{3.34}
\expandafter\gdef\csname svy@item/g04/e2/meanlow\endcsname{3.07}
\expandafter\gdef\csname svy@item/g04/e2/meanhigh\endcsname{3.61}
\expandafter\gdef\csname svy@item/g04/e2/se\endcsname{0.134}
\expandafter\gdef\csname svy@item/g04/e2/sd\endcsname{1.17}
\expandafter\gdef\csname svy@item/g04/e2/sdhigh\endcsname{0}
\expandafter\gdef\csname svy@item/g04/e2/verdict\endcsname{none}
\expandafter\gdef\csname svy@item/g04/e2/dip\endcsname{0}
\expandafter\gdef\csname svy@item/g04/e2/shapelow\endcsname{26}
\expandafter\gdef\csname svy@item/g04/e2/shapemid\endcsname{20}
\expandafter\gdef\csname svy@item/g04/e2/shapehigh\endcsname{54}
\expandafter\gdef\csname svy@item/g04/e3/n\endcsname{93}
\expandafter\gdef\csname svy@item/g04/e3/miss\endcsname{1}
\expandafter\gdef\csname svy@item/g04/e3/median\endcsname{4.0}
\expandafter\gdef\csname svy@item/g04/e3/cilow\endcsname{4.0}
\expandafter\gdef\csname svy@item/g04/e3/cihigh\endcsname{5.0}
\expandafter\gdef\csname svy@item/g04/e3/mean\endcsname{4.08}
\expandafter\gdef\csname svy@item/g04/e3/meanlow\endcsname{3.85}
\expandafter\gdef\csname svy@item/g04/e3/meanhigh\endcsname{4.30}
\expandafter\gdef\csname svy@item/g04/e3/se\endcsname{0.113}
\expandafter\gdef\csname svy@item/g04/e3/sd\endcsname{1.09}
\expandafter\gdef\csname svy@item/g04/e3/sdhigh\endcsname{0}
\expandafter\gdef\csname svy@item/g04/e3/verdict\endcsname{none}
\expandafter\gdef\csname svy@item/g04/e3/dip\endcsname{0}
\expandafter\gdef\csname svy@item/g04/e3/shapelow\endcsname{12}
\expandafter\gdef\csname svy@item/g04/e3/shapemid\endcsname{10}
\expandafter\gdef\csname svy@item/g04/e3/shapehigh\endcsname{78}
\expandafter\gdef\csname svy@item/g05/exposure/trend\endcsname{0.59}
\expandafter\gdef\csname svy@item/g05/exposure/trendlow\endcsname{0.41}
\expandafter\gdef\csname svy@item/g05/exposure/trendhigh\endcsname{0.77}
\expandafter\gdef\csname svy@item/g05/e0/n\endcsname{10}
\expandafter\gdef\csname svy@item/g05/e0/miss\endcsname{3}
\expandafter\gdef\csname svy@item/g05/e0/median\endcsname{1.0}
\expandafter\gdef\csname svy@item/g05/e0/cilow\endcsname{1.0}
\expandafter\gdef\csname svy@item/g05/e0/cihigh\endcsname{1.0}
\expandafter\gdef\csname svy@item/g05/e0/mean\endcsname{1.00}
\expandafter\gdef\csname svy@item/g05/e0/meanlow\endcsname{1.00}
\expandafter\gdef\csname svy@item/g05/e0/meanhigh\endcsname{1.00}
\expandafter\gdef\csname svy@item/g05/e0/se\endcsname{0.000}
\expandafter\gdef\csname svy@item/g05/e0/sd\endcsname{0.00}
\expandafter\gdef\csname svy@item/g05/e0/sdhigh\endcsname{0}
\expandafter\gdef\csname svy@item/g05/e0/verdict\endcsname{none}
\expandafter\gdef\csname svy@item/g05/e0/dip\endcsname{0}
\expandafter\gdef\csname svy@item/g05/e0/shapelow\endcsname{100}
\expandafter\gdef\csname svy@item/g05/e0/shapemid\endcsname{0}
\expandafter\gdef\csname svy@item/g05/e0/shapehigh\endcsname{0}
\expandafter\gdef\csname svy@item/g05/e1/n\endcsname{49}
\expandafter\gdef\csname svy@item/g05/e1/miss\endcsname{0}
\expandafter\gdef\csname svy@item/g05/e1/median\endcsname{2.0}
\expandafter\gdef\csname svy@item/g05/e1/cilow\endcsname{2.0}
\expandafter\gdef\csname svy@item/g05/e1/cihigh\endcsname{2.0}
\expandafter\gdef\csname svy@item/g05/e1/mean\endcsname{2.29}
\expandafter\gdef\csname svy@item/g05/e1/meanlow\endcsname{1.95}
\expandafter\gdef\csname svy@item/g05/e1/meanhigh\endcsname{2.62}
\expandafter\gdef\csname svy@item/g05/e1/se\endcsname{0.165}
\expandafter\gdef\csname svy@item/g05/e1/sd\endcsname{1.15}
\expandafter\gdef\csname svy@item/g05/e1/sdhigh\endcsname{0}
\expandafter\gdef\csname svy@item/g05/e1/verdict\endcsname{none}
\expandafter\gdef\csname svy@item/g05/e1/dip\endcsname{0}
\expandafter\gdef\csname svy@item/g05/e1/shapelow\endcsname{66}
\expandafter\gdef\csname svy@item/g05/e1/shapemid\endcsname{16}
\expandafter\gdef\csname svy@item/g05/e1/shapehigh\endcsname{18}
\expandafter\gdef\csname svy@item/g05/e2/n\endcsname{76}
\expandafter\gdef\csname svy@item/g05/e2/miss\endcsname{1}
\expandafter\gdef\csname svy@item/g05/e2/median\endcsname{2.0}
\expandafter\gdef\csname svy@item/g05/e2/cilow\endcsname{2.0}
\expandafter\gdef\csname svy@item/g05/e2/cihigh\endcsname{2.0}
\expandafter\gdef\csname svy@item/g05/e2/mean\endcsname{2.36}
\expandafter\gdef\csname svy@item/g05/e2/meanlow\endcsname{2.06}
\expandafter\gdef\csname svy@item/g05/e2/meanhigh\endcsname{2.65}
\expandafter\gdef\csname svy@item/g05/e2/se\endcsname{0.150}
\expandafter\gdef\csname svy@item/g05/e2/sd\endcsname{1.30}
\expandafter\gdef\csname svy@item/g05/e2/sdhigh\endcsname{0}
\expandafter\gdef\csname svy@item/g05/e2/verdict\endcsname{none}
\expandafter\gdef\csname svy@item/g05/e2/dip\endcsname{1}
\expandafter\gdef\csname svy@item/g05/e2/shapelow\endcsname{63}
\expandafter\gdef\csname svy@item/g05/e2/shapemid\endcsname{13}
\expandafter\gdef\csname svy@item/g05/e2/shapehigh\endcsname{24}
\expandafter\gdef\csname svy@item/g05/e3/n\endcsname{94}
\expandafter\gdef\csname svy@item/g05/e3/miss\endcsname{0}
\expandafter\gdef\csname svy@item/g05/e3/median\endcsname{3.0}
\expandafter\gdef\csname svy@item/g05/e3/cilow\endcsname{3.0}
\expandafter\gdef\csname svy@item/g05/e3/cihigh\endcsname{4.0}
\expandafter\gdef\csname svy@item/g05/e3/mean\endcsname{3.24}
\expandafter\gdef\csname svy@item/g05/e3/meanlow\endcsname{2.99}
\expandafter\gdef\csname svy@item/g05/e3/meanhigh\endcsname{3.50}
\expandafter\gdef\csname svy@item/g05/e3/se\endcsname{0.129}
\expandafter\gdef\csname svy@item/g05/e3/sd\endcsname{1.25}
\expandafter\gdef\csname svy@item/g05/e3/sdhigh\endcsname{0}
\expandafter\gdef\csname svy@item/g05/e3/verdict\endcsname{none}
\expandafter\gdef\csname svy@item/g05/e3/dip\endcsname{1}
\expandafter\gdef\csname svy@item/g05/e3/shapelow\endcsname{35}
\expandafter\gdef\csname svy@item/g05/e3/shapemid\endcsname{19}
\expandafter\gdef\csname svy@item/g05/e3/shapehigh\endcsname{46}
\expandafter\gdef\csname svy@item/g06/exposure/trend\endcsname{-0.13}
\expandafter\gdef\csname svy@item/g06/exposure/trendlow\endcsname{-0.26}
\expandafter\gdef\csname svy@item/g06/exposure/trendhigh\endcsname{0.01}
\expandafter\gdef\csname svy@item/g06/e0/n\endcsname{13}
\expandafter\gdef\csname svy@item/g06/e0/miss\endcsname{0}
\expandafter\gdef\csname svy@item/g06/e0/median\endcsname{5.0}
\expandafter\gdef\csname svy@item/g06/e0/cilow\endcsname{4.0}
\expandafter\gdef\csname svy@item/g06/e0/cihigh\endcsname{5.0}
\expandafter\gdef\csname svy@item/g06/e0/mean\endcsname{4.31}
\expandafter\gdef\csname svy@item/g06/e0/meanlow\endcsname{3.64}
\expandafter\gdef\csname svy@item/g06/e0/meanhigh\endcsname{4.98}
\expandafter\gdef\csname svy@item/g06/e0/se\endcsname{0.308}
\expandafter\gdef\csname svy@item/g06/e0/sd\endcsname{1.11}
\expandafter\gdef\csname svy@item/g06/e0/sdhigh\endcsname{0}
\expandafter\gdef\csname svy@item/g06/e0/verdict\endcsname{none}
\expandafter\gdef\csname svy@item/g06/e0/dip\endcsname{0}
\expandafter\gdef\csname svy@item/g06/e0/shapelow\endcsname{8}
\expandafter\gdef\csname svy@item/g06/e0/shapemid\endcsname{0}
\expandafter\gdef\csname svy@item/g06/e0/shapehigh\endcsname{92}
\expandafter\gdef\csname svy@item/g06/e1/n\endcsname{48}
\expandafter\gdef\csname svy@item/g06/e1/miss\endcsname{1}
\expandafter\gdef\csname svy@item/g06/e1/median\endcsname{4.0}
\expandafter\gdef\csname svy@item/g06/e1/cilow\endcsname{4.0}
\expandafter\gdef\csname svy@item/g06/e1/cihigh\endcsname{5.0}
\expandafter\gdef\csname svy@item/g06/e1/mean\endcsname{4.25}
\expandafter\gdef\csname svy@item/g06/e1/meanlow\endcsname{3.99}
\expandafter\gdef\csname svy@item/g06/e1/meanhigh\endcsname{4.51}
\expandafter\gdef\csname svy@item/g06/e1/se\endcsname{0.131}
\expandafter\gdef\csname svy@item/g06/e1/sd\endcsname{0.91}
\expandafter\gdef\csname svy@item/g06/e1/sdhigh\endcsname{0}
\expandafter\gdef\csname svy@item/g06/e1/verdict\endcsname{none}
\expandafter\gdef\csname svy@item/g06/e1/dip\endcsname{0}
\expandafter\gdef\csname svy@item/g06/e1/shapelow\endcsname{6}
\expandafter\gdef\csname svy@item/g06/e1/shapemid\endcsname{6}
\expandafter\gdef\csname svy@item/g06/e1/shapehigh\endcsname{88}
\expandafter\gdef\csname svy@item/g06/e2/n\endcsname{77}
\expandafter\gdef\csname svy@item/g06/e2/miss\endcsname{0}
\expandafter\gdef\csname svy@item/g06/e2/median\endcsname{4.0}
\expandafter\gdef\csname svy@item/g06/e2/cilow\endcsname{4.0}
\expandafter\gdef\csname svy@item/g06/e2/cihigh\endcsname{5.0}
\expandafter\gdef\csname svy@item/g06/e2/mean\endcsname{4.21}
\expandafter\gdef\csname svy@item/g06/e2/meanlow\endcsname{4.01}
\expandafter\gdef\csname svy@item/g06/e2/meanhigh\endcsname{4.40}
\expandafter\gdef\csname svy@item/g06/e2/se\endcsname{0.098}
\expandafter\gdef\csname svy@item/g06/e2/sd\endcsname{0.86}
\expandafter\gdef\csname svy@item/g06/e2/sdhigh\endcsname{0}
\expandafter\gdef\csname svy@item/g06/e2/verdict\endcsname{none}
\expandafter\gdef\csname svy@item/g06/e2/dip\endcsname{0}
\expandafter\gdef\csname svy@item/g06/e2/shapelow\endcsname{4}
\expandafter\gdef\csname svy@item/g06/e2/shapemid\endcsname{17}
\expandafter\gdef\csname svy@item/g06/e2/shapehigh\endcsname{79}
\expandafter\gdef\csname svy@item/g06/e3/n\endcsname{91}
\expandafter\gdef\csname svy@item/g06/e3/miss\endcsname{3}
\expandafter\gdef\csname svy@item/g06/e3/median\endcsname{4.0}
\expandafter\gdef\csname svy@item/g06/e3/cilow\endcsname{4.0}
\expandafter\gdef\csname svy@item/g06/e3/cihigh\endcsname{4.0}
\expandafter\gdef\csname svy@item/g06/e3/mean\endcsname{3.98}
\expandafter\gdef\csname svy@item/g06/e3/meanlow\endcsname{3.76}
\expandafter\gdef\csname svy@item/g06/e3/meanhigh\endcsname{4.20}
\expandafter\gdef\csname svy@item/g06/e3/se\endcsname{0.109}
\expandafter\gdef\csname svy@item/g06/e3/sd\endcsname{1.04}
\expandafter\gdef\csname svy@item/g06/e3/sdhigh\endcsname{0}
\expandafter\gdef\csname svy@item/g06/e3/verdict\endcsname{none}
\expandafter\gdef\csname svy@item/g06/e3/dip\endcsname{0}
\expandafter\gdef\csname svy@item/g06/e3/shapelow\endcsname{10}
\expandafter\gdef\csname svy@item/g06/e3/shapemid\endcsname{19}
\expandafter\gdef\csname svy@item/g06/e3/shapehigh\endcsname{71}
\expandafter\gdef\csname svy@item/g07/exposure/trend\endcsname{0.46}
\expandafter\gdef\csname svy@item/g07/exposure/trendlow\endcsname{0.30}
\expandafter\gdef\csname svy@item/g07/exposure/trendhigh\endcsname{0.61}
\expandafter\gdef\csname svy@item/g07/e0/n\endcsname{13}
\expandafter\gdef\csname svy@item/g07/e0/miss\endcsname{0}
\expandafter\gdef\csname svy@item/g07/e0/median\endcsname{2.0}
\expandafter\gdef\csname svy@item/g07/e0/cilow\endcsname{2.0}
\expandafter\gdef\csname svy@item/g07/e0/cihigh\endcsname{3.0}
\expandafter\gdef\csname svy@item/g07/e0/mean\endcsname{2.54}
\expandafter\gdef\csname svy@item/g07/e0/meanlow\endcsname{2.01}
\expandafter\gdef\csname svy@item/g07/e0/meanhigh\endcsname{3.07}
\expandafter\gdef\csname svy@item/g07/e0/se\endcsname{0.243}
\expandafter\gdef\csname svy@item/g07/e0/sd\endcsname{0.88}
\expandafter\gdef\csname svy@item/g07/e0/sdhigh\endcsname{0}
\expandafter\gdef\csname svy@item/g07/e0/verdict\endcsname{none}
\expandafter\gdef\csname svy@item/g07/e0/dip\endcsname{0}
\expandafter\gdef\csname svy@item/g07/e0/shapelow\endcsname{54}
\expandafter\gdef\csname svy@item/g07/e0/shapemid\endcsname{31}
\expandafter\gdef\csname svy@item/g07/e0/shapehigh\endcsname{15}
\expandafter\gdef\csname svy@item/g07/e1/n\endcsname{48}
\expandafter\gdef\csname svy@item/g07/e1/miss\endcsname{1}
\expandafter\gdef\csname svy@item/g07/e1/median\endcsname{3.0}
\expandafter\gdef\csname svy@item/g07/e1/cilow\endcsname{2.0}
\expandafter\gdef\csname svy@item/g07/e1/cihigh\endcsname{4.0}
\expandafter\gdef\csname svy@item/g07/e1/mean\endcsname{2.88}
\expandafter\gdef\csname svy@item/g07/e1/meanlow\endcsname{2.52}
\expandafter\gdef\csname svy@item/g07/e1/meanhigh\endcsname{3.23}
\expandafter\gdef\csname svy@item/g07/e1/se\endcsname{0.178}
\expandafter\gdef\csname svy@item/g07/e1/sd\endcsname{1.23}
\expandafter\gdef\csname svy@item/g07/e1/sdhigh\endcsname{0}
\expandafter\gdef\csname svy@item/g07/e1/verdict\endcsname{none}
\expandafter\gdef\csname svy@item/g07/e1/dip\endcsname{1}
\expandafter\gdef\csname svy@item/g07/e1/shapelow\endcsname{44}
\expandafter\gdef\csname svy@item/g07/e1/shapemid\endcsname{19}
\expandafter\gdef\csname svy@item/g07/e1/shapehigh\endcsname{37}
\expandafter\gdef\csname svy@item/g07/e2/n\endcsname{74}
\expandafter\gdef\csname svy@item/g07/e2/miss\endcsname{3}
\expandafter\gdef\csname svy@item/g07/e2/median\endcsname{4.0}
\expandafter\gdef\csname svy@item/g07/e2/cilow\endcsname{3.0}
\expandafter\gdef\csname svy@item/g07/e2/cihigh\endcsname{4.0}
\expandafter\gdef\csname svy@item/g07/e2/mean\endcsname{3.49}
\expandafter\gdef\csname svy@item/g07/e2/meanlow\endcsname{3.24}
\expandafter\gdef\csname svy@item/g07/e2/meanhigh\endcsname{3.73}
\expandafter\gdef\csname svy@item/g07/e2/se\endcsname{0.122}
\expandafter\gdef\csname svy@item/g07/e2/sd\endcsname{1.05}
\expandafter\gdef\csname svy@item/g07/e2/sdhigh\endcsname{0}
\expandafter\gdef\csname svy@item/g07/e2/verdict\endcsname{none}
\expandafter\gdef\csname svy@item/g07/e2/dip\endcsname{0}
\expandafter\gdef\csname svy@item/g07/e2/shapelow\endcsname{19}
\expandafter\gdef\csname svy@item/g07/e2/shapemid\endcsname{20}
\expandafter\gdef\csname svy@item/g07/e2/shapehigh\endcsname{61}
\expandafter\gdef\csname svy@item/g07/e3/n\endcsname{90}
\expandafter\gdef\csname svy@item/g07/e3/miss\endcsname{4}
\expandafter\gdef\csname svy@item/g07/e3/median\endcsname{4.0}
\expandafter\gdef\csname svy@item/g07/e3/cilow\endcsname{4.0}
\expandafter\gdef\csname svy@item/g07/e3/cihigh\endcsname{4.0}
\expandafter\gdef\csname svy@item/g07/e3/mean\endcsname{3.83}
\expandafter\gdef\csname svy@item/g07/e3/meanlow\endcsname{3.62}
\expandafter\gdef\csname svy@item/g07/e3/meanhigh\endcsname{4.05}
\expandafter\gdef\csname svy@item/g07/e3/se\endcsname{0.110}
\expandafter\gdef\csname svy@item/g07/e3/sd\endcsname{1.04}
\expandafter\gdef\csname svy@item/g07/e3/sdhigh\endcsname{0}
\expandafter\gdef\csname svy@item/g07/e3/verdict\endcsname{none}
\expandafter\gdef\csname svy@item/g07/e3/dip\endcsname{0}
\expandafter\gdef\csname svy@item/g07/e3/shapelow\endcsname{14}
\expandafter\gdef\csname svy@item/g07/e3/shapemid\endcsname{10}
\expandafter\gdef\csname svy@item/g07/e3/shapehigh\endcsname{76}
\expandafter\gdef\csname svy@item/dceasy/code\endcsname{DC01}
\expandafter\gdef\csname svy@item/dceasy/pred\endcsname{}
\expandafter\gdef\csname svy@item/dceasy/label\endcsname{style for a familiar, easily checked task}
\expandafter\gdef\csname svy@item/dceasy/usage/diff\endcsname{0.07}
\expandafter\gdef\csname svy@item/dceasy/usage/difflow\endcsname{-0.35}
\expandafter\gdef\csname svy@item/dceasy/usage/diffhigh\endcsname{0.49}
\expandafter\gdef\csname svy@item/dceasy/usage/g\endcsname{0.05}
\expandafter\gdef\csname svy@item/dceasy/all/n\endcsname{220}
\expandafter\gdef\csname svy@item/dceasy/all/miss\endcsname{22}
\expandafter\gdef\csname svy@item/dceasy/all/median\endcsname{2.0}
\expandafter\gdef\csname svy@item/dceasy/all/cilow\endcsname{2.0}
\expandafter\gdef\csname svy@item/dceasy/all/cihigh\endcsname{2.0}
\expandafter\gdef\csname svy@item/dceasy/all/mean\endcsname{2.61}
\expandafter\gdef\csname svy@item/dceasy/all/meanlow\endcsname{2.42}
\expandafter\gdef\csname svy@item/dceasy/all/meanhigh\endcsname{2.81}
\expandafter\gdef\csname svy@item/dceasy/all/se\endcsname{0.098}
\expandafter\gdef\csname svy@item/dceasy/all/sd\endcsname{1.45}
\expandafter\gdef\csname svy@item/dceasy/all/sdhigh\endcsname{0}
\expandafter\gdef\csname svy@item/dceasy/all/verdict\endcsname{none}
\expandafter\gdef\csname svy@item/dceasy/all/dip\endcsname{1}
\expandafter\gdef\csname svy@item/dceasy/all/shapelow\endcsname{61}
\expandafter\gdef\csname svy@item/dceasy/all/shapemid\endcsname{6}
\expandafter\gdef\csname svy@item/dceasy/all/shapehigh\endcsname{33}
\expandafter\gdef\csname svy@item/dceasy/users/n\endcsname{160}
\expandafter\gdef\csname svy@item/dceasy/users/miss\endcsname{11}
\expandafter\gdef\csname svy@item/dceasy/users/median\endcsname{2.0}
\expandafter\gdef\csname svy@item/dceasy/users/cilow\endcsname{2.0}
\expandafter\gdef\csname svy@item/dceasy/users/cihigh\endcsname{2.0}
\expandafter\gdef\csname svy@item/dceasy/users/mean\endcsname{2.64}
\expandafter\gdef\csname svy@item/dceasy/users/meanlow\endcsname{2.40}
\expandafter\gdef\csname svy@item/dceasy/users/meanhigh\endcsname{2.87}
\expandafter\gdef\csname svy@item/dceasy/users/se\endcsname{0.119}
\expandafter\gdef\csname svy@item/dceasy/users/sd\endcsname{1.51}
\expandafter\gdef\csname svy@item/dceasy/users/sdhigh\endcsname{0}
\expandafter\gdef\csname svy@item/dceasy/users/verdict\endcsname{none}
\expandafter\gdef\csname svy@item/dceasy/users/dip\endcsname{1}
\expandafter\gdef\csname svy@item/dceasy/users/shapelow\endcsname{59}
\expandafter\gdef\csname svy@item/dceasy/users/shapemid\endcsname{7}
\expandafter\gdef\csname svy@item/dceasy/users/shapehigh\endcsname{34}
\expandafter\gdef\csname svy@item/dceasy/nonusers/n\endcsname{58}
\expandafter\gdef\csname svy@item/dceasy/nonusers/miss\endcsname{11}
\expandafter\gdef\csname svy@item/dceasy/nonusers/median\endcsname{2.0}
\expandafter\gdef\csname svy@item/dceasy/nonusers/cilow\endcsname{2.0}
\expandafter\gdef\csname svy@item/dceasy/nonusers/cihigh\endcsname{2.0}
\expandafter\gdef\csname svy@item/dceasy/nonusers/mean\endcsname{2.57}
\expandafter\gdef\csname svy@item/dceasy/nonusers/meanlow\endcsname{2.22}
\expandafter\gdef\csname svy@item/dceasy/nonusers/meanhigh\endcsname{2.92}
\expandafter\gdef\csname svy@item/dceasy/nonusers/se\endcsname{0.174}
\expandafter\gdef\csname svy@item/dceasy/nonusers/sd\endcsname{1.33}
\expandafter\gdef\csname svy@item/dceasy/nonusers/sdhigh\endcsname{0}
\expandafter\gdef\csname svy@item/dceasy/nonusers/verdict\endcsname{none}
\expandafter\gdef\csname svy@item/dceasy/nonusers/dip\endcsname{1}
\expandafter\gdef\csname svy@item/dceasy/nonusers/shapelow\endcsname{67}
\expandafter\gdef\csname svy@item/dceasy/nonusers/shapemid\endcsname{2}
\expandafter\gdef\csname svy@item/dceasy/nonusers/shapehigh\endcsname{31}
\expandafter\gdef\csname svy@item/dcsubtle/code\endcsname{DC02}
\expandafter\gdef\csname svy@item/dcsubtle/pred\endcsname{}
\expandafter\gdef\csname svy@item/dcsubtle/label\endcsname{style for an unfamiliar, subtle task}
\expandafter\gdef\csname svy@item/dcsubtle/usage/diff\endcsname{0.27}
\expandafter\gdef\csname svy@item/dcsubtle/usage/difflow\endcsname{-0.15}
\expandafter\gdef\csname svy@item/dcsubtle/usage/diffhigh\endcsname{0.69}
\expandafter\gdef\csname svy@item/dcsubtle/usage/g\endcsname{0.22}
\expandafter\gdef\csname svy@item/dcsubtle/all/n\endcsname{220}
\expandafter\gdef\csname svy@item/dcsubtle/all/miss\endcsname{22}
\expandafter\gdef\csname svy@item/dcsubtle/all/median\endcsname{4.0}
\expandafter\gdef\csname svy@item/dcsubtle/all/cilow\endcsname{4.0}
\expandafter\gdef\csname svy@item/dcsubtle/all/cihigh\endcsname{5.0}
\expandafter\gdef\csname svy@item/dcsubtle/all/mean\endcsname{3.98}
\expandafter\gdef\csname svy@item/dcsubtle/all/meanlow\endcsname{3.81}
\expandafter\gdef\csname svy@item/dcsubtle/all/meanhigh\endcsname{4.14}
\expandafter\gdef\csname svy@item/dcsubtle/all/se\endcsname{0.084}
\expandafter\gdef\csname svy@item/dcsubtle/all/sd\endcsname{1.25}
\expandafter\gdef\csname svy@item/dcsubtle/all/sdhigh\endcsname{0}
\expandafter\gdef\csname svy@item/dcsubtle/all/verdict\endcsname{none}
\expandafter\gdef\csname svy@item/dcsubtle/all/dip\endcsname{0}
\expandafter\gdef\csname svy@item/dcsubtle/all/shapelow\endcsname{16}
\expandafter\gdef\csname svy@item/dcsubtle/all/shapemid\endcsname{8}
\expandafter\gdef\csname svy@item/dcsubtle/all/shapehigh\endcsname{76}
\expandafter\gdef\csname svy@item/dcsubtle/users/n\endcsname{161}
\expandafter\gdef\csname svy@item/dcsubtle/users/miss\endcsname{10}
\expandafter\gdef\csname svy@item/dcsubtle/users/median\endcsname{4.0}
\expandafter\gdef\csname svy@item/dcsubtle/users/cilow\endcsname{4.0}
\expandafter\gdef\csname svy@item/dcsubtle/users/cihigh\endcsname{5.0}
\expandafter\gdef\csname svy@item/dcsubtle/users/mean\endcsname{4.04}
\expandafter\gdef\csname svy@item/dcsubtle/users/meanlow\endcsname{3.86}
\expandafter\gdef\csname svy@item/dcsubtle/users/meanhigh\endcsname{4.23}
\expandafter\gdef\csname svy@item/dcsubtle/users/se\endcsname{0.093}
\expandafter\gdef\csname svy@item/dcsubtle/users/sd\endcsname{1.18}
\expandafter\gdef\csname svy@item/dcsubtle/users/sdhigh\endcsname{0}
\expandafter\gdef\csname svy@item/dcsubtle/users/verdict\endcsname{none}
\expandafter\gdef\csname svy@item/dcsubtle/users/dip\endcsname{0}
\expandafter\gdef\csname svy@item/dcsubtle/users/shapelow\endcsname{15}
\expandafter\gdef\csname svy@item/dcsubtle/users/shapemid\endcsname{7}
\expandafter\gdef\csname svy@item/dcsubtle/users/shapehigh\endcsname{78}
\expandafter\gdef\csname svy@item/dcsubtle/nonusers/n\endcsname{57}
\expandafter\gdef\csname svy@item/dcsubtle/nonusers/miss\endcsname{12}
\expandafter\gdef\csname svy@item/dcsubtle/nonusers/median\endcsname{4.0}
\expandafter\gdef\csname svy@item/dcsubtle/nonusers/cilow\endcsname{4.0}
\expandafter\gdef\csname svy@item/dcsubtle/nonusers/cihigh\endcsname{5.0}
\expandafter\gdef\csname svy@item/dcsubtle/nonusers/mean\endcsname{3.77}
\expandafter\gdef\csname svy@item/dcsubtle/nonusers/meanlow\endcsname{3.39}
\expandafter\gdef\csname svy@item/dcsubtle/nonusers/meanhigh\endcsname{4.15}
\expandafter\gdef\csname svy@item/dcsubtle/nonusers/se\endcsname{0.191}
\expandafter\gdef\csname svy@item/dcsubtle/nonusers/sd\endcsname{1.44}
\expandafter\gdef\csname svy@item/dcsubtle/nonusers/sdhigh\endcsname{0}
\expandafter\gdef\csname svy@item/dcsubtle/nonusers/verdict\endcsname{none}
\expandafter\gdef\csname svy@item/dcsubtle/nonusers/dip\endcsname{1}
\expandafter\gdef\csname svy@item/dcsubtle/nonusers/shapelow\endcsname{23}
\expandafter\gdef\csname svy@item/dcsubtle/nonusers/shapemid\endcsname{10}
\expandafter\gdef\csname svy@item/dcsubtle/nonusers/shapehigh\endcsname{67}
\expandafter\gdef\csname svy@item/dceasy/attitude/diff\endcsname{0.19}
\expandafter\gdef\csname svy@item/dceasy/attitude/difflow\endcsname{-0.46}
\expandafter\gdef\csname svy@item/dceasy/attitude/diffhigh\endcsname{0.84}
\expandafter\gdef\csname svy@item/dceasy/attitude/g\endcsname{0.13}
\expandafter\gdef\csname svy@item/dceasy/trusters/n\endcsname{80}
\expandafter\gdef\csname svy@item/dceasy/trusters/miss\endcsname{10}
\expandafter\gdef\csname svy@item/dceasy/trusters/median\endcsname{2.0}
\expandafter\gdef\csname svy@item/dceasy/trusters/cilow\endcsname{2.0}
\expandafter\gdef\csname svy@item/dceasy/trusters/cihigh\endcsname{3.0}
\expandafter\gdef\csname svy@item/dceasy/trusters/mean\endcsname{2.62}
\expandafter\gdef\csname svy@item/dceasy/trusters/meanlow\endcsname{2.29}
\expandafter\gdef\csname svy@item/dceasy/trusters/meanhigh\endcsname{2.96}
\expandafter\gdef\csname svy@item/dceasy/trusters/se\endcsname{0.166}
\expandafter\gdef\csname svy@item/dceasy/trusters/sd\endcsname{1.49}
\expandafter\gdef\csname svy@item/dceasy/trusters/sdhigh\endcsname{0}
\expandafter\gdef\csname svy@item/dceasy/trusters/verdict\endcsname{none}
\expandafter\gdef\csname svy@item/dceasy/trusters/dip\endcsname{1}
\expandafter\gdef\csname svy@item/dceasy/trusters/shapelow\endcsname{60}
\expandafter\gdef\csname svy@item/dceasy/trusters/shapemid\endcsname{6}
\expandafter\gdef\csname svy@item/dceasy/trusters/shapehigh\endcsname{34}
\expandafter\gdef\csname svy@item/dceasy/skeptics/n\endcsname{30}
\expandafter\gdef\csname svy@item/dceasy/skeptics/miss\endcsname{5}
\expandafter\gdef\csname svy@item/dceasy/skeptics/median\endcsname{2.0}
\expandafter\gdef\csname svy@item/dceasy/skeptics/cilow\endcsname{1.0}
\expandafter\gdef\csname svy@item/dceasy/skeptics/cihigh\endcsname{3.0}
\expandafter\gdef\csname svy@item/dceasy/skeptics/mean\endcsname{2.43}
\expandafter\gdef\csname svy@item/dceasy/skeptics/meanlow\endcsname{1.86}
\expandafter\gdef\csname svy@item/dceasy/skeptics/meanhigh\endcsname{3.00}
\expandafter\gdef\csname svy@item/dceasy/skeptics/se\endcsname{0.278}
\expandafter\gdef\csname svy@item/dceasy/skeptics/sd\endcsname{1.52}
\expandafter\gdef\csname svy@item/dceasy/skeptics/sdhigh\endcsname{0}
\expandafter\gdef\csname svy@item/dceasy/skeptics/verdict\endcsname{none}
\expandafter\gdef\csname svy@item/dceasy/skeptics/dip\endcsname{1}
\expandafter\gdef\csname svy@item/dceasy/skeptics/shapelow\endcsname{67}
\expandafter\gdef\csname svy@item/dceasy/skeptics/shapemid\endcsname{3}
\expandafter\gdef\csname svy@item/dceasy/skeptics/shapehigh\endcsname{30}
\expandafter\gdef\csname svy@item/dcsubtle/attitude/diff\endcsname{0.39}
\expandafter\gdef\csname svy@item/dcsubtle/attitude/difflow\endcsname{-0.32}
\expandafter\gdef\csname svy@item/dcsubtle/attitude/diffhigh\endcsname{1.10}
\expandafter\gdef\csname svy@item/dcsubtle/attitude/g\endcsname{0.28}
\expandafter\gdef\csname svy@item/dcsubtle/trusters/n\endcsname{81}
\expandafter\gdef\csname svy@item/dcsubtle/trusters/miss\endcsname{9}
\expandafter\gdef\csname svy@item/dcsubtle/trusters/median\endcsname{4.0}
\expandafter\gdef\csname svy@item/dcsubtle/trusters/cilow\endcsname{4.0}
\expandafter\gdef\csname svy@item/dcsubtle/trusters/cihigh\endcsname{5.0}
\expandafter\gdef\csname svy@item/dcsubtle/trusters/mean\endcsname{3.84}
\expandafter\gdef\csname svy@item/dcsubtle/trusters/meanlow\endcsname{3.56}
\expandafter\gdef\csname svy@item/dcsubtle/trusters/meanhigh\endcsname{4.12}
\expandafter\gdef\csname svy@item/dcsubtle/trusters/se\endcsname{0.142}
\expandafter\gdef\csname svy@item/dcsubtle/trusters/sd\endcsname{1.28}
\expandafter\gdef\csname svy@item/dcsubtle/trusters/sdhigh\endcsname{0}
\expandafter\gdef\csname svy@item/dcsubtle/trusters/verdict\endcsname{none}
\expandafter\gdef\csname svy@item/dcsubtle/trusters/dip\endcsname{0}
\expandafter\gdef\csname svy@item/dcsubtle/trusters/shapelow\endcsname{21}
\expandafter\gdef\csname svy@item/dcsubtle/trusters/shapemid\endcsname{9}
\expandafter\gdef\csname svy@item/dcsubtle/trusters/shapehigh\endcsname{70}
\expandafter\gdef\csname svy@item/dcsubtle/skeptics/n\endcsname{29}
\expandafter\gdef\csname svy@item/dcsubtle/skeptics/miss\endcsname{6}
\expandafter\gdef\csname svy@item/dcsubtle/skeptics/median\endcsname{4.0}
\expandafter\gdef\csname svy@item/dcsubtle/skeptics/cilow\endcsname{2.0}
\expandafter\gdef\csname svy@item/dcsubtle/skeptics/cihigh\endcsname{5.0}
\expandafter\gdef\csname svy@item/dcsubtle/skeptics/mean\endcsname{3.45}
\expandafter\gdef\csname svy@item/dcsubtle/skeptics/meanlow\endcsname{2.79}
\expandafter\gdef\csname svy@item/dcsubtle/skeptics/meanhigh\endcsname{4.10}
\expandafter\gdef\csname svy@item/dcsubtle/skeptics/se\endcsname{0.320}
\expandafter\gdef\csname svy@item/dcsubtle/skeptics/sd\endcsname{1.72}
\expandafter\gdef\csname svy@item/dcsubtle/skeptics/sdhigh\endcsname{0}
\expandafter\gdef\csname svy@item/dcsubtle/skeptics/verdict\endcsname{none}
\expandafter\gdef\csname svy@item/dcsubtle/skeptics/dip\endcsname{1}
\expandafter\gdef\csname svy@item/dcsubtle/skeptics/shapelow\endcsname{38}
\expandafter\gdef\csname svy@item/dcsubtle/skeptics/shapemid\endcsname{0}
\expandafter\gdef\csname svy@item/dcsubtle/skeptics/shapehigh\endcsname{62}
\expandafter\gdef\csname svy@item/dcc/n\endcsname{154}
\expandafter\gdef\csname svy@item/dcc/length/q1\endcsname{260}
\expandafter\gdef\csname svy@item/dcc/length/q3\endcsname{667}
\expandafter\gdef\csname svy@sample/x01/label\endcsname{experience with AI tools of any kind}
\expandafter\gdef\csname svy@sample/x01/answered\endcsname{242}
\expandafter\gdef\csname svy@sample/x01/onscale\endcsname{242}
\expandafter\gdef\csname svy@sample/x01/offscale\endcsname{0}
\expandafter\gdef\csname svy@sample/x01/missing\endcsname{0}
\expandafter\gdef\csname svy@sample/x01/modal\endcsname{1-2 years}
\expandafter\gdef\csname svy@sample/x01/modalpct\endcsname{31}
\expandafter\gdef\csname svy@sample/x01/median\endcsname{5.0}
\expandafter\gdef\csname svy@sample/x01/medianlabel\endcsname{1-2 years}
\expandafter\gdef\csname svy@sample/x02/label\endcsname{how often coding agents are used}
\expandafter\gdef\csname svy@sample/x02/answered\endcsname{241}
\expandafter\gdef\csname svy@sample/x02/onscale\endcsname{241}
\expandafter\gdef\csname svy@sample/x02/offscale\endcsname{0}
\expandafter\gdef\csname svy@sample/x02/missing\endcsname{1}
\expandafter\gdef\csname svy@sample/x02/modal\endcsname{several per day}
\expandafter\gdef\csname svy@sample/x02/modalpct\endcsname{41}
\expandafter\gdef\csname svy@sample/x02/median\endcsname{6.0}
\expandafter\gdef\csname svy@sample/x02/medianlabel\endcsname{about daily}
\expandafter\gdef\csname svy@sample/drole/label\endcsname{primary role}
\expandafter\gdef\csname svy@sample/drole/answered\endcsname{228}
\expandafter\gdef\csname svy@sample/drole/onscale\endcsname{228}
\expandafter\gdef\csname svy@sample/drole/offscale\endcsname{0}
\expandafter\gdef\csname svy@sample/drole/missing\endcsname{14}
\expandafter\gdef\csname svy@sample/drole/modal\endcsname{senior developer}
\expandafter\gdef\csname svy@sample/drole/modalpct\endcsname{23}
\expandafter\gdef\csname svy@sample/dyearsexp/label\endcsname{years of professional experience}
\expandafter\gdef\csname svy@sample/dyearsexp/answered\endcsname{227}
\expandafter\gdef\csname svy@sample/dyearsexp/onscale\endcsname{227}
\expandafter\gdef\csname svy@sample/dyearsexp/offscale\endcsname{0}
\expandafter\gdef\csname svy@sample/dyearsexp/missing\endcsname{15}
\expandafter\gdef\csname svy@sample/dyearsexp/modal\endcsname{more than 20}
\expandafter\gdef\csname svy@sample/dyearsexp/modalpct\endcsname{23}
\expandafter\gdef\csname svy@sample/dyearsexp/median\endcsname{4.0}
\expandafter\gdef\csname svy@sample/dyearsexp/medianlabel\endcsname{6-10}
\expandafter\gdef\csname svy@sample/dorgtype/label\endcsname{type of organization}
\expandafter\gdef\csname svy@sample/dorgtype/answered\endcsname{228}
\expandafter\gdef\csname svy@sample/dorgtype/onscale\endcsname{228}
\expandafter\gdef\csname svy@sample/dorgtype/offscale\endcsname{0}
\expandafter\gdef\csname svy@sample/dorgtype/missing\endcsname{14}
\expandafter\gdef\csname svy@sample/dorgtype/modal\endcsname{software product company}
\expandafter\gdef\csname svy@sample/dorgtype/modalpct\endcsname{29}
\expandafter\gdef\csname svy@sample/dteamsizecat/label\endcsname{typical team size}
\expandafter\gdef\csname svy@sample/dteamsizecat/answered\endcsname{226}
\expandafter\gdef\csname svy@sample/dteamsizecat/onscale\endcsname{226}
\expandafter\gdef\csname svy@sample/dteamsizecat/offscale\endcsname{2}
\expandafter\gdef\csname svy@sample/dteamsizecat/missing\endcsname{14}
\expandafter\gdef\csname svy@sample/dteamsizecat/modal\endcsname{2-5}
\expandafter\gdef\csname svy@sample/dteamsizecat/modalpct\endcsname{36}
\expandafter\gdef\csname svy@sample/dteamsizecat/median\endcsname{2.0}
\expandafter\gdef\csname svy@sample/dteamsizecat/medianlabel\endcsname{2-5}
\expandafter\gdef\csname svy@sample/dtimecoding/label\endcsname{share of work time spent on code}
\expandafter\gdef\csname svy@sample/dtimecoding/answered\endcsname{228}
\expandafter\gdef\csname svy@sample/dtimecoding/onscale\endcsname{228}
\expandafter\gdef\csname svy@sample/dtimecoding/offscale\endcsname{0}
\expandafter\gdef\csname svy@sample/dtimecoding/missing\endcsname{14}
\expandafter\gdef\csname svy@sample/dtimecoding/modal\endcsname{51-75\%}
\expandafter\gdef\csname svy@sample/dtimecoding/modalpct\endcsname{32}
\expandafter\gdef\csname svy@sample/dtimecoding/median\endcsname{4.0}
\expandafter\gdef\csname svy@sample/dtimecoding/medianlabel\endcsname{51-75\%}
\expandafter\gdef\csname svy@sample/dgender/label\endcsname{gender}
\expandafter\gdef\csname svy@sample/dgender/answered\endcsname{216}
\expandafter\gdef\csname svy@sample/dgender/onscale\endcsname{216}
\expandafter\gdef\csname svy@sample/dgender/offscale\endcsname{11}
\expandafter\gdef\csname svy@sample/dgender/missing\endcsname{15}
\expandafter\gdef\csname svy@sample/dgender/modal\endcsname{male}
\expandafter\gdef\csname svy@sample/dgender/modalpct\endcsname{87}
\expandafter\gdef\csname svy@sample/tools/answered\endcsname{241}
\expandafter\gdef\csname svy@sample/tools/otherfilled\endcsname{13}
\expandafter\gdef\csname svy@sample/tools/otherclassified\endcsname{13}
\expandafter\gdef\csname svy@sample/tools/othernotool\endcsname{2}
\expandafter\gdef\csname svy@sample/tools/categories\endcsname{4}
\expandafter\gdef\csname svy@sample/tools/folded\endcsname{2}
\expandafter\gdef\csname svy@sample/tools/chat/n\endcsname{160}
\expandafter\gdef\csname svy@sample/tools/chat/pct\endcsname{66}
\expandafter\gdef\csname svy@sample/tools/chat/label\endcsname{chat assistant}
\expandafter\gdef\csname svy@sample/tools/autocomplete/n\endcsname{78}
\expandafter\gdef\csname svy@sample/tools/autocomplete/pct\endcsname{32}
\expandafter\gdef\csname svy@sample/tools/autocomplete/label\endcsname{IDE plugin autocomplete}
\expandafter\gdef\csname svy@sample/tools/pluginedit/n\endcsname{36}
\expandafter\gdef\csname svy@sample/tools/pluginedit/pct\endcsname{15}
\expandafter\gdef\csname svy@sample/tools/pluginedit/label\endcsname{IDE plugin autoedit mode}
\expandafter\gdef\csname svy@sample/tools/pluginagent/n\endcsname{67}
\expandafter\gdef\csname svy@sample/tools/pluginagent/pct\endcsname{28}
\expandafter\gdef\csname svy@sample/tools/pluginagent/label\endcsname{IDE plugin agent mode}
\expandafter\gdef\csname svy@sample/tools/ideagent/n\endcsname{30}
\expandafter\gdef\csname svy@sample/tools/ideagent/pct\endcsname{12}
\expandafter\gdef\csname svy@sample/tools/ideagent/label\endcsname{AI-native IDE agent mode}
\expandafter\gdef\csname svy@sample/tools/cliagent/n\endcsname{143}
\expandafter\gdef\csname svy@sample/tools/cliagent/pct\endcsname{59}
\expandafter\gdef\csname svy@sample/tools/cliagent/label\endcsname{terminal coding agent}
\expandafter\gdef\csname svy@sample/tools/github/n\endcsname{35}
\expandafter\gdef\csname svy@sample/tools/github/pct\endcsname{15}
\expandafter\gdef\csname svy@sample/tools/github/label\endcsname{GitHub Copilot on GitHub}
\expandafter\gdef\csname svy@sample/tools/selfhost/n\endcsname{65}
\expandafter\gdef\csname svy@sample/tools/selfhost/pct\endcsname{27}
\expandafter\gdef\csname svy@sample/tools/selfhost/label\endcsname{self-hosted LLM}
\expandafter\gdef\csname svy@sample/tools/other/n\endcsname{0}
\expandafter\gdef\csname svy@sample/tools/other/pct\endcsname{0}
\expandafter\gdef\csname svy@sample/tools/other/label\endcsname{other (not classified)}
\expandafter\gdef\csname svy@sample/tools/none/n\endcsname{15}
\expandafter\gdef\csname svy@sample/tools/none/pct\endcsname{6}
\expandafter\gdef\csname svy@sample/tools/none/label\endcsname{none of these}
\expandafter\gdef\csname svy@sample/tools/selfhostedagentharness/n\endcsname{4}
\expandafter\gdef\csname svy@sample/tools/selfhostedagentharness/pct\endcsname{2}
\expandafter\gdef\csname svy@sample/tools/selfhostedagentharness/label\endcsname{self-hosted agent harness}
\expandafter\gdef\csname svy@sample/tools/desktopagentapp/n\endcsname{2}
\expandafter\gdef\csname svy@sample/tools/desktopagentapp/pct\endcsname{1}
\expandafter\gdef\csname svy@sample/tools/desktopagentapp/label\endcsname{desktop agent app}
\expandafter\gdef\csname svy@sample/tools/ciagent/n\endcsname{1}
\expandafter\gdef\csname svy@sample/tools/ciagent/pct\endcsname{$<$1}
\expandafter\gdef\csname svy@sample/tools/ciagent/label\endcsname{CI agent}
\expandafter\gdef\csname svy@sample/tools/othernonagenttool/n\endcsname{2}
\expandafter\gdef\csname svy@sample/tools/othernonagenttool/pct\endcsname{1}
\expandafter\gdef\csname svy@sample/tools/othernonagenttool/label\endcsname{other non-agent tool}
\expandafter\gdef\csname svy@sample/tools/notoolnamed/n\endcsname{2}
\expandafter\gdef\csname svy@sample/tools/notoolnamed/pct\endcsname{1}
\expandafter\gdef\csname svy@sample/tools/notoolnamed/label\endcsname{no tool named}
\expandafter\gdef\csname svy@country/codes\endcsname{199}
\expandafter\gdef\csname svy@country/countries\endcsname{27}
\expandafter\gdef\csname svy@country/answered\endcsname{196}
\expandafter\gdef\csname svy@country/multiple\endcsname{3}
\expandafter\gdef\csname svy@country/germany\endcsname{124}
\expandafter\gdef\csname svy@country/germanypct\endcsname{63}
\expandafter\gdef\csname svy@country/shown\endcsname{6}
\expandafter\gdef\csname svy@country/textblock\endcsname{Germany (63\%), USA (14\%), Australia (3.1\%), UK (3.1\%), France (2.6\%), Denmark (2.0\%), Brazil (1.5\%), Argentina (1.0\%), Austria (1.0\%), Canada (1.0\%), Turkey (1.0\%), China (0.5\%), Colombia (0.5\%), Croatia (0.5\%), Ireland (0.5\%), Israel (0.5\%), Italy (0.5\%), Japan (0.5\%), Mexico (0.5\%), Netherlands (0.5\%), Norway (0.5\%), Poland (0.5\%), South Africa (0.5\%), Spain (0.5\%), Sweden (0.5\%), Switzerland (0.5\%), and Vietnam (0.5\%).}
\expandafter\gdef\csname svy@exposure/e0/n\endcsname{13}
\expandafter\gdef\csname svy@exposure/e0/pct\endcsname{5}
\expandafter\gdef\csname svy@exposure/e0/label\endcsname{no AI tools at all}
\expandafter\gdef\csname svy@exposure/e1/n\endcsname{49}
\expandafter\gdef\csname svy@exposure/e1/pct\endcsname{20}
\expandafter\gdef\csname svy@exposure/e1/label\endcsname{AI tools, but no agent practice}
\expandafter\gdef\csname svy@exposure/e2/n\endcsname{77}
\expandafter\gdef\csname svy@exposure/e2/pct\endcsname{32}
\expandafter\gdef\csname svy@exposure/e2/label\endcsname{agents, monthly to daily}
\expandafter\gdef\csname svy@exposure/e3/n\endcsname{94}
\expandafter\gdef\csname svy@exposure/e3/pct\endcsname{39}
\expandafter\gdef\csname svy@exposure/e3/label\endcsname{agents, several sessions a day}
\expandafter\gdef\csname svy@exposure/placed\endcsname{233}
\expandafter\gdef\csname svy@exposure/unknown\endcsname{9}
\expandafter\gdef\csname svy@exposure/unclassified\endcsname{2}
\expandafter\gdef\csname svy@exposure/contradicted\endcsname{7}
\expandafter\gdef\csname svy@exposure/contradict/high\endcsname{10}
\expandafter\gdef\csname svy@exposure/contradict/low\endcsname{13}
\expandafter\gdef\csname svy@trend/h/raw/n\endcsname{233}
\expandafter\gdef\csname svy@trend/h/raw/b\endcsname{0.34}
\expandafter\gdef\csname svy@trend/h/raw/blow\endcsname{0.25}
\expandafter\gdef\csname svy@trend/h/raw/bhigh\endcsname{0.42}
\expandafter\gdef\csname svy@trend/h/centred/n\endcsname{233}
\expandafter\gdef\csname svy@trend/h/centred/b\endcsname{-0.11}
\expandafter\gdef\csname svy@trend/h/centred/blow\endcsname{-0.14}
\expandafter\gdef\csname svy@trend/h/centred/bhigh\endcsname{-0.07}
\expandafter\gdef\csname svy@trend/e/raw/n\endcsname{230}
\expandafter\gdef\csname svy@trend/e/raw/b\endcsname{0.32}
\expandafter\gdef\csname svy@trend/e/raw/blow\endcsname{0.21}
\expandafter\gdef\csname svy@trend/e/raw/bhigh\endcsname{0.44}
\expandafter\gdef\csname svy@trend/e/centred/n\endcsname{230}
\expandafter\gdef\csname svy@trend/e/centred/b\endcsname{-0.12}
\expandafter\gdef\csname svy@trend/e/centred/blow\endcsname{-0.19}
\expandafter\gdef\csname svy@trend/e/centred/bhigh\endcsname{-0.05}
\expandafter\gdef\csname svy@trend/a/raw/n\endcsname{233}
\expandafter\gdef\csname svy@trend/a/raw/b\endcsname{0.66}
\expandafter\gdef\csname svy@trend/a/raw/blow\endcsname{0.56}
\expandafter\gdef\csname svy@trend/a/raw/bhigh\endcsname{0.76}
\expandafter\gdef\csname svy@trend/a/centred/n\endcsname{233}
\expandafter\gdef\csname svy@trend/a/centred/b\endcsname{0.22}
\expandafter\gdef\csname svy@trend/a/centred/blow\endcsname{0.15}
\expandafter\gdef\csname svy@trend/a/centred/bhigh\endcsname{0.28}
\expandafter\gdef\csname svy@trend/noe0/raw/n\endcsname{220}
\expandafter\gdef\csname svy@trend/noe0/raw/b\endcsname{0.31}
\expandafter\gdef\csname svy@trend/noe0/raw/blow\endcsname{0.21}
\expandafter\gdef\csname svy@trend/noe0/raw/bhigh\endcsname{0.41}
\expandafter\gdef\csname svy@trend/x02only/raw/n\endcsname{241}
\expandafter\gdef\csname svy@trend/x02only/raw/b\endcsname{0.15}
\expandafter\gdef\csname svy@trend/x02only/raw/blow\endcsname{0.11}
\expandafter\gdef\csname svy@trend/x02only/raw/bhigh\endcsname{0.18}
\expandafter\gdef\csname svy@trend/nogithub/raw/n\endcsname{233}
\expandafter\gdef\csname svy@trend/nogithub/raw/b\endcsname{0.34}
\expandafter\gdef\csname svy@trend/nogithub/raw/blow\endcsname{0.26}
\expandafter\gdef\csname svy@trend/nogithub/raw/bhigh\endcsname{0.42}
\expandafter\gdef\csname svy@attitude/trusters\endcsname{90}
\expandafter\gdef\csname svy@attitude/skeptics\endcsname{35}
\expandafter\gdef\csname svy@attitude/neither\endcsname{117}
\expandafter\gdef\csname svy@attitude/neitherpct\endcsname{48}
\expandafter\gdef\csname svy@attitude/trusters/userspct\endcsname{89}
\expandafter\gdef\csname svy@attitude/users/trusters\endcsname{80}
\expandafter\gdef\csname svy@attitude/users/trusterspct\endcsname{47}
\expandafter\gdef\csname svy@attitude/users/skeptics\endcsname{10}
\expandafter\gdef\csname svy@attitude/nonusers/skeptics\endcsname{24}
\expandafter\gdef\csname svy@attitude/skeptics/nonuserspct\endcsname{69}
\expandafter\gdef\csname svy@attitude/e0/trusters\endcsname{0}
\expandafter\gdef\csname svy@attitude/e0/skeptics\endcsname{8}
\expandafter\gdef\csname svy@attitude/e1/trusters\endcsname{9}
\expandafter\gdef\csname svy@attitude/e1/skeptics\endcsname{12}
\expandafter\gdef\csname svy@attitude/e2/trusters\endcsname{26}
\expandafter\gdef\csname svy@attitude/e2/skeptics\endcsname{8}
\expandafter\gdef\csname svy@attitude/e3/trusters\endcsname{54}
\expandafter\gdef\csname svy@attitude/e3/skeptics\endcsname{2}
\expandafter\gdef\csname svy@attitude/nonusers/trusters\endcsname{10}
\expandafter\gdef\csname svy@attitude/notier/trusters\endcsname{1}
\expandafter\gdef\csname svy@attitude/notier/skeptics\endcsname{5}
\expandafter\gdef\csname svy@attitude/moving\endcsname{5}
\expandafter\gdef\csname svy@attitude/flat\endcsname{2}
\expandafter\gdef\csname svy@attitude/reversed\endcsname{G03}
\expandafter\gdef\csname svy@curve/span\endcsname{0.75}
\expandafter\gdef\csname svy@curve/regions\endcsname{3}
\expandafter\gdef\csname svy@curve/bootstrap\endcsname{2000}
\expandafter\gdef\csname svy@curve/level\endcsname{95}
\expandafter\gdef\csname svy@curve/n\endcsname{232--239}
\expandafter\gdef\csname svy@curve/firstregion\endcsname{43}
\expandafter\gdef\csname svy@curve/lastregion\endcsname{157}
\expandafter\gdef\csname svy@deleg/n\endcsname{238}
\expandafter\gdef\csname svy@deleg/notsure\endcsname{4}
\expandafter\gdef\csname svy@deleg/untiered\endcsname{9}
\expandafter\gdef\csname svy@deleg/tiered\endcsname{229}
\expandafter\gdef\csname svy@deleg/rho\endcsname{0.76}
\expandafter\gdef\csname svy@deleg/cells\endcsname{25}
\expandafter\gdef\csname svy@deleg/cells/occupied\endcsname{17}
\expandafter\gdef\csname svy@deleg/kind/mean\endcsname{3.46}
\expandafter\gdef\csname svy@deleg/kind/code\endcsname{X04}
\expandafter\gdef\csname svy@deleg/scope/mean\endcsname{2.91}
\expandafter\gdef\csname svy@deleg/scope/code\endcsname{X05}
\expandafter\gdef\csname svy@deleg/gap/above\endcsname{122}
\expandafter\gdef\csname svy@deleg/gap/level\endcsname{95}
\expandafter\gdef\csname svy@deleg/gap/below\endcsname{21}
\expandafter\gdef\csname svy@deleg/gap/mean\endcsname{0.53}
\expandafter\gdef\csname svy@deleg/e0/n\endcsname{13}
\expandafter\gdef\csname svy@deleg/e0/kindmean\endcsname{1.00}
\expandafter\gdef\csname svy@deleg/e0/scopemean\endcsname{1.15}
\expandafter\gdef\csname svy@deleg/e1/n\endcsname{47}
\expandafter\gdef\csname svy@deleg/e1/kindmean\endcsname{2.85}
\expandafter\gdef\csname svy@deleg/e1/scopemean\endcsname{2.21}
\expandafter\gdef\csname svy@deleg/e2/n\endcsname{75}
\expandafter\gdef\csname svy@deleg/e2/kindmean\endcsname{3.61}
\expandafter\gdef\csname svy@deleg/e2/scopemean\endcsname{2.88}
\expandafter\gdef\csname svy@deleg/e3/n\endcsname{94}
\expandafter\gdef\csname svy@deleg/e3/kindmean\endcsname{4.15}
\expandafter\gdef\csname svy@deleg/e3/scopemean\endcsname{3.68}
\expandafter\gdef\csname svy@deleg/kind/b\endcsname{0.82}
\expandafter\gdef\csname svy@deleg/kind/blow\endcsname{0.70}
\expandafter\gdef\csname svy@deleg/kind/bhigh\endcsname{0.93}
\expandafter\gdef\csname svy@deleg/scope/b\endcsname{0.78}
\expandafter\gdef\csname svy@deleg/scope/blow\endcsname{0.65}
\expandafter\gdef\csname svy@deleg/scope/bhigh\endcsname{0.91}
\expandafter\gdef\csname svy@deleg/gap/b\endcsname{0.03}
\expandafter\gdef\csname svy@deleg/gap/blow\endcsname{-0.08}
\expandafter\gdef\csname svy@deleg/gap/bhigh\endcsname{0.15}
\expandafter\gdef\csname svy@togetherness/n\endcsname{140}
\expandafter\gdef\csname svy@togetherness/a1/label\endcsname{tried a few times}
\expandafter\gdef\csname svy@togetherness/a2/label\endcsname{occasionally}
\expandafter\gdef\csname svy@togetherness/a3/label\endcsname{regularly, months}
\expandafter\gdef\csname svy@togetherness/a4/label\endcsname{regularly, years}
\expandafter\gdef\csname svy@usage/slots\endcsname{9}
\expandafter\gdef\csname svy@usage/index/mean\endcsname{0.68}
\expandafter\gdef\csname svy@usage/breadth/mean\endcsname{0.29}
\expandafter\gdef\csname svy@usage/rho/breadth\endcsname{0.34}
\expandafter\gdef\csname svy@usage/rho/exposure\endcsname{0.86}
\expandafter\gdef\csname svy@usage/agree\endcsname{158}
\expandafter\gdef\csname svy@usage/agree/base\endcsname{233}
\expandafter\gdef\csname svy@usage/unbanded\endcsname{1}
\expandafter\gdef\csname svy@usage/deleg/unbanded\endcsname{1}
\expandafter\gdef\csname svy@usage/deleg/banded\endcsname{237}
\expandafter\gdef\csname svy@usage/u0/n\endcsname{34}
\expandafter\gdef\csname svy@usage/u0/pct\endcsname{14}
\expandafter\gdef\csname svy@usage/u0/label\endcsname{index below 0.25}
\expandafter\gdef\csname svy@usage/u1/n\endcsname{34}
\expandafter\gdef\csname svy@usage/u1/pct\endcsname{14}
\expandafter\gdef\csname svy@usage/u1/label\endcsname{index 0.25 to 0.50}
\expandafter\gdef\csname svy@usage/u2/n\endcsname{35}
\expandafter\gdef\csname svy@usage/u2/pct\endcsname{15}
\expandafter\gdef\csname svy@usage/u2/label\endcsname{index 0.50 to 0.75}
\expandafter\gdef\csname svy@usage/u3/n\endcsname{138}
\expandafter\gdef\csname svy@usage/u3/pct\endcsname{57}
\expandafter\gdef\csname svy@usage/u3/label\endcsname{index 0.75 and above}
\expandafter\gdef\csname svy@usage/u0/kindmean\endcsname{1.47}
\expandafter\gdef\csname svy@usage/u0/scopemean\endcsname{1.53}
\expandafter\gdef\csname svy@usage/u0/dn\endcsname{34}
\expandafter\gdef\csname svy@usage/u1/kindmean\endcsname{2.97}
\expandafter\gdef\csname svy@usage/u1/scopemean\endcsname{2.06}
\expandafter\gdef\csname svy@usage/u1/dn\endcsname{31}
\expandafter\gdef\csname svy@usage/u2/kindmean\endcsname{3.51}
\expandafter\gdef\csname svy@usage/u2/scopemean\endcsname{2.71}
\expandafter\gdef\csname svy@usage/u2/dn\endcsname{35}
\expandafter\gdef\csname svy@usage/u3/kindmean\endcsname{4.04}
\expandafter\gdef\csname svy@usage/u3/scopemean\endcsname{3.52}
\expandafter\gdef\csname svy@usage/u3/dn\endcsname{137}
\expandafter\gdef\csname svy@usage/kind/b\endcsname{0.77}
\expandafter\gdef\csname svy@usage/kind/blow\endcsname{0.69}
\expandafter\gdef\csname svy@usage/kind/bhigh\endcsname{0.86}
\expandafter\gdef\csname svy@usage/scope/b\endcsname{0.68}
\expandafter\gdef\csname svy@usage/scope/blow\endcsname{0.58}
\expandafter\gdef\csname svy@usage/scope/bhigh\endcsname{0.78}
\expandafter\gdef\csname svy@usage/gap/b\endcsname{0.09}
\expandafter\gdef\csname svy@usage/gap/blow\endcsname{0.00}
\expandafter\gdef\csname svy@usage/gap/bhigh\endcsname{0.19}
\expandafter\gdef\csname svy@usage/scatter/users\endcsname{171}
\expandafter\gdef\csname svy@usage/scatter/nonusers\endcsname{69}
\expandafter\gdef\csname svy@usage/scatter/unclassified\endcsname{0}
\expandafter\gdef\csname svy@usage/scatter/n\endcsname{240}
\expandafter\gdef\csname svy@trust/items\endcsname{4}
\expandafter\gdef\csname svy@trust/items/all\endcsname{7}
\expandafter\gdef\csname svy@trust/items/list\endcsname{G01, G03, G04, G05}
\expandafter\gdef\csname svy@trust/items/out\endcsname{G02, G06, G07}
\expandafter\gdef\csname svy@trust/minanswers\endcsname{3}
\expandafter\gdef\csname svy@trust/mean\endcsname{3.31}
\expandafter\gdef\csname svy@trust/sd\endcsname{1.05}
\expandafter\gdef\csname svy@trust/n\endcsname{238}
\expandafter\gdef\csname svy@trust/missing\endcsname{4}
\expandafter\gdef\csname svy@trust/alpha\endcsname{0.82}
\expandafter\gdef\csname svy@trust/alpha/n\endcsname{223}
\expandafter\gdef\csname svy@trust/alpha/all\endcsname{0.66}
\expandafter\gdef\csname svy@trust/rho/usage\endcsname{0.54}
\expandafter\gdef\csname svy@trust/rho/exposure\endcsname{0.49}
\expandafter\gdef\csname svy@trust/scatter/r\endcsname{0.58}
\expandafter\gdef\csname svy@trust/scatter/users\endcsname{171}
\expandafter\gdef\csname svy@trust/scatter/nonusers\endcsname{65}
\expandafter\gdef\csname svy@trust/scatter/unclassified\endcsname{1}
\expandafter\gdef\csname svy@trust/scatter/n\endcsname{237}
\expandafter\gdef\csname svy@trust/scatter/jitterx\endcsname{0.04}
\expandafter\gdef\csname svy@trust/scatter/jittery\endcsname{0.16}
\expandafter\gdef\csname svy@usage/ridge/bins\endcsname{20}
\expandafter\gdef\csname svy@usage/ridge/binwidth\endcsname{0.05}
\expandafter\gdef\csname svy@uptake/n\endcsname{241}
\expandafter\gdef\csname svy@uptake/untiered\endcsname{8}
\expandafter\gdef\csname svy@uptake/tiered\endcsname{233}
\expandafter\gdef\csname svy@uptake/cells\endcsname{42}
\expandafter\gdef\csname svy@uptake/cells/occupied\endcsname{31}
\expandafter\gdef\csname svy@uptake/cells/empty\endcsname{11}
\expandafter\gdef\csname svy@uptake/cells/pure\endcsname{19}
\expandafter\gdef\csname svy@uptake/cells/mixed\endcsname{12}
\expandafter\gdef\csname svy@uptake/rho\endcsname{0.41}
\expandafter\gdef\csname svy@uptake/rho/x01\endcsname{0.36}
\expandafter\gdef\csname svy@uptake/rho/x02\endcsname{0.91}
\expandafter\gdef\csname svy@uptake/notier/label\endcsname{no tier}
\expandafter\gdef\csname svy@verdict/margin\endcsname{0.50}
\expandafter\gdef\csname svy@verdict/level\endcsname{95}
\expandafter\gdef\csname svy@verdict/dipbalance\endcsname{30}
\expandafter\gdef\csname svy@agg/items\endcsname{15}
\expandafter\gdef\csname svy@agg/allitems\endcsname{19}
\expandafter\gdef\csname svy@agg/unpredicted\endcsname{4}
\expandafter\gdef\csname svy@agg/all/confirmed\endcsname{10}
\expandafter\gdef\csname svy@agg/all/inconclusive\endcsname{4}
\expandafter\gdef\csname svy@agg/all/rejected\endcsname{1}
\expandafter\gdef\csname svy@agg/all/baseline\endcsname{1.04}
\expandafter\gdef\csname svy@agg/all/sdconfirmed\endcsname{0.97}
\expandafter\gdef\csname svy@agg/all/sdinconclusive\endcsname{1.09}
\expandafter\gdef\csname svy@agg/all/sdrejected\endcsname{1.24}
\expandafter\gdef\csname svy@agg/users/confirmed\endcsname{11}
\expandafter\gdef\csname svy@agg/users/inconclusive\endcsname{2}
\expandafter\gdef\csname svy@agg/users/rejected\endcsname{2}
\expandafter\gdef\csname svy@agg/users/baseline\endcsname{0.98}
\expandafter\gdef\csname svy@agg/users/sdconfirmed\endcsname{0.91}
\expandafter\gdef\csname svy@agg/users/sdinconclusive\endcsname{0.98}
\expandafter\gdef\csname svy@agg/users/sdrejected\endcsname{1.16}
\expandafter\gdef\csname svy@agg/nonusers/confirmed\endcsname{8}
\expandafter\gdef\csname svy@agg/nonusers/inconclusive\endcsname{6}
\expandafter\gdef\csname svy@agg/nonusers/rejected\endcsname{1}
\expandafter\gdef\csname svy@agg/nonusers/baseline\endcsname{0.97}
\expandafter\gdef\csname svy@agg/nonusers/sdconfirmed\endcsname{0.82}
\expandafter\gdef\csname svy@agg/nonusers/sdinconclusive\endcsname{1.13}
\expandafter\gdef\csname svy@agg/nonusers/sdrejected\endcsname{0.93}
\expandafter\gdef\csname svy@cat/c/items\endcsname{2}
\expandafter\gdef\csname svy@cat/c/name\endcsname{Specification, Requirements, and Architectural Direction}
\expandafter\gdef\csname svy@cat/c/cluster\endcsname{Intent formation}
\expandafter\gdef\csname svy@cat/c/all/confirmed\endcsname{1}
\expandafter\gdef\csname svy@cat/c/users/confirmed\endcsname{1}
\expandafter\gdef\csname svy@cat/c/nonusers/confirmed\endcsname{1}
\expandafter\gdef\csname svy@cat/i/items\endcsname{2}
\expandafter\gdef\csname svy@cat/i/name\endcsname{Knowledge Work, Learning, and Discovery}
\expandafter\gdef\csname svy@cat/i/cluster\endcsname{Intent formation}
\expandafter\gdef\csname svy@cat/i/all/confirmed\endcsname{1}
\expandafter\gdef\csname svy@cat/i/users/confirmed\endcsname{2}
\expandafter\gdef\csname svy@cat/i/nonusers/confirmed\endcsname{0}
\expandafter\gdef\csname svy@cat/e/items\endcsname{2}
\expandafter\gdef\csname svy@cat/e/name\endcsname{Context Engineering and Prompting Discipline}
\expandafter\gdef\csname svy@cat/e/cluster\endcsname{Design operationalization}
\expandafter\gdef\csname svy@cat/e/all/confirmed\endcsname{1}
\expandafter\gdef\csname svy@cat/e/users/confirmed\endcsname{0}
\expandafter\gdef\csname svy@cat/e/nonusers/confirmed\endcsname{2}
\expandafter\gdef\csname svy@cat/h/items\endcsname{1}
\expandafter\gdef\csname svy@cat/h/name\endcsname{Tooling, Infrastructure, and Environment Design}
\expandafter\gdef\csname svy@cat/h/cluster\endcsname{Design operationalization}
\expandafter\gdef\csname svy@cat/h/all/confirmed\endcsname{1}
\expandafter\gdef\csname svy@cat/h/users/confirmed\endcsname{1}
\expandafter\gdef\csname svy@cat/h/nonusers/confirmed\endcsname{1}
\expandafter\gdef\csname svy@cat/b/items\endcsname{1}
\expandafter\gdef\csname svy@cat/b/name\endcsname{Workflow Orchestration}
\expandafter\gdef\csname svy@cat/b/cluster\endcsname{Design operationalization}
\expandafter\gdef\csname svy@cat/b/all/confirmed\endcsname{0}
\expandafter\gdef\csname svy@cat/b/users/confirmed\endcsname{0}
\expandafter\gdef\csname svy@cat/b/nonusers/confirmed\endcsname{0}
\expandafter\gdef\csname svy@cat/d/items\endcsname{3}
\expandafter\gdef\csname svy@cat/d/name\endcsname{Human Judgment and Quality Calibration}
\expandafter\gdef\csname svy@cat/d/cluster\endcsname{Evaluation}
\expandafter\gdef\csname svy@cat/d/all/confirmed\endcsname{2}
\expandafter\gdef\csname svy@cat/d/users/confirmed\endcsname{3}
\expandafter\gdef\csname svy@cat/d/nonusers/confirmed\endcsname{1}
\expandafter\gdef\csname svy@cat/f/items\endcsname{2}
\expandafter\gdef\csname svy@cat/f/name\endcsname{Iteration, Refactoring, and Cleanup}
\expandafter\gdef\csname svy@cat/f/cluster\endcsname{Evaluation}
\expandafter\gdef\csname svy@cat/f/all/confirmed\endcsname{2}
\expandafter\gdef\csname svy@cat/f/users/confirmed\endcsname{2}
\expandafter\gdef\csname svy@cat/f/nonusers/confirmed\endcsname{1}
\expandafter\gdef\csname svy@cat/a/items\endcsname{1}
\expandafter\gdef\csname svy@cat/a/name\endcsname{Accountability and Ownership of Outcomes}
\expandafter\gdef\csname svy@cat/a/cluster\endcsname{Evaluation}
\expandafter\gdef\csname svy@cat/a/all/confirmed\endcsname{1}
\expandafter\gdef\csname svy@cat/a/users/confirmed\endcsname{1}
\expandafter\gdef\csname svy@cat/a/nonusers/confirmed\endcsname{1}
\expandafter\gdef\csname svy@cat/j/items\endcsname{1}
\expandafter\gdef\csname svy@cat/j/name\endcsname{Cultural Framing and Legitimacy Claims}
\expandafter\gdef\csname svy@cat/j/cluster\endcsname{cross-cutting}
\expandafter\gdef\csname svy@cat/j/all/confirmed\endcsname{1}
\expandafter\gdef\csname svy@cat/j/users/confirmed\endcsname{1}
\expandafter\gdef\csname svy@cat/j/nonusers/confirmed\endcsname{1}

\DeclareRobustCommand{\svy}[1]{%
  \ifcsname svy@#1\endcsname
    \csname svy@#1\endcsname
  \else
    \GenericError{}{Survey macro '#1' is not defined in generated/survey-numbers.tex}%
      {See article/generated/survey-numbers.tex for the available keys.}{}%
  \fi}

\newcommand{\tablefont}{\footnotesize}

\definecolor{percentbarcol}{HTML}{4E8256}  
\newlength{\percentbarunit}
\newcommand{\percentagebar}[1]{%
  \begin{tikzpicture}[baseline=-0.55ex, x=\percentbarunit, y=1ex]
    \useasboundingbox (0,-0.30) rectangle (100,1.15);
    \fill[percentbarcol] (0,0) rectangle (#1,0.75);
    \draw[percentbarcol, line width=0.15pt] (100,-0.20) -- (100,0.95);
  \end{tikzpicture}}

\definecolor{likertA}{HTML}{2C5985}  
\definecolor{likertB}{HTML}{7FA6C9}  
\definecolor{likertC}{HTML}{D9D9D9}  
\definecolor{likertD}{HTML}{E8A25C}  
\definecolor{likertE}{HTML}{B25E17}  
\newlength{\likertunit}
\newcommand{\likertbar}[6]{%
  \begin{tikzpicture}[baseline=-0.55ex, x=\likertunit, y=1ex]
    \useasboundingbox (0,-0.30) rectangle (200,1.15);
    \pgfmathsetmacro{\lxa}{100-(#1+#2+0.5*#3)}%
    \pgfmathsetmacro{\lxb}{\lxa+#1}%
    \pgfmathsetmacro{\lxc}{\lxb+#2}%
    \pgfmathsetmacro{\lxd}{\lxc+#3}%
    \pgfmathsetmacro{\lxe}{\lxd+#4}%
    \pgfmathsetmacro{\lxf}{\lxe+#5}%
    \fill[likertA] (\lxa,0) rectangle (\lxb,0.75);
    \fill[likertB] (\lxb,0) rectangle (\lxc,0.75);
    \fill[likertC] (\lxc,0) rectangle (\lxd,0.75);
    \fill[likertD] (\lxd,0) rectangle (\lxe,0.75);
    \fill[likertE] (\lxe,0) rectangle (\lxf,0.75);
    \draw[black!55, line width=0.15pt] (100,-0.20) -- (100,0.95);
    \ifnum#6=1\relax
      \node[anchor=south, inner sep=0pt, font=\tiny] at (100,0.80) {$\vee$};
    \fi
  \end{tikzpicture}}

\DeclareRobustCommand{\likertswatch}[1]{\textcolor{#1}{\rule[-0.1ex]{1.4ex}{1.1ex}}}
\DeclareRobustCommand{\likertlegend}{%
  \likertswatch{likertA}\,1 the human alone\quad
  \likertswatch{likertB}\,2 mainly the human\quad
  \likertswatch{likertC}\,3 about equally\quad
  \likertswatch{likertD}\,4 mainly the agent\quad
  \likertswatch{likertE}\,5 the agent alone}
\DeclareRobustCommand{\likertlegendagree}{%
  \likertswatch{likertA}\,1 strongly disagree\quad
  \likertswatch{likertB}\,2 disagree\quad
  \likertswatch{likertC}\,3 neither\quad
  \likertswatch{likertD}\,4 agree\quad
  \likertswatch{likertE}\,5 strongly agree}
\DeclareRobustCommand{\likertlegenddividecollab}{%
  \likertswatch{likertA}\,1 clearly DIVIDE\quad
  \likertswatch{likertB}\,2 rather DIVIDE\quad
  \likertswatch{likertC}\,3 equally both\quad
  \likertswatch{likertD}\,4 rather COLLAB\quad
  \likertswatch{likertE}\,5 clearly COLLAB}

\expandafter\def\csname pred@H\endcsname{\textcolor{likertA}{H}}
\expandafter\def\csname pred@A\endcsname{\textcolor{likertE}{A}}
\expandafter\def\csname pred@E\endcsname{E}
\expandafter\def\csname pred@X\endcsname{}
\newcommand*{\pred}[1]{%
  \ifcsname pred@#1\endcsname\csname pred@#1\endcsname
  \else\PackageError{hairoles}{Unknown prediction letter '#1' in \string\pred}{}\fi}

\expandafter\def\csname rankmove@up\endcsname#1{#1$\uparrow$}
\expandafter\def\csname rankmove@down\endcsname#1{#1$\downarrow$}
\expandafter\def\csname rankmove@same\endcsname#1{\textcolor{black!45}{$\bullet$}}
\DeclareRobustCommand*{\rankmove}[2]{%
  \ifcsname rankmove@#2\endcsname\csname rankmove@#2\endcsname{#1}%
  \else\PackageError{hairoles}{Unknown rank-move mark '#2' in \string\rankmove}{}\fi}

\newcommand{\attitudechartwidth}{0.44\linewidth}  
\pgfplotsset{attitudechart/.style={
  width=\attitudechartwidth, height=56mm,
  ytick={1,2,3,4,5},
  scaled ticks=false, tick align=outside, tick pos=left,
  axis line style={black!45}, tick style={black!45},
  xtick style={draw=none},
  title style={font=\scriptsize\itshape, yshift=-2pt},
  label style={font=\scriptsize}, tick label style={font=\scriptsize},
  xticklabel style={align=center, font=\tiny},
  ylabel={mean}, ylabel style={yshift=-4pt}}}
\pgfplotsset{attitudemid/.style={no marks, draw=black!30, dotted, line width=0.5pt}}
\tikzset{attitudelabel/.style={anchor=west, font=\tiny, inner sep=1pt, xshift=2pt}}
\definecolor{attcurveA}{HTML}{2F5E8C}  
\definecolor{attcurveB}{HTML}{A8443F}  
\definecolor{attcurveC}{HTML}{4E8256}  
\definecolor{attcurveD}{HTML}{8A6234}  
\definecolor{attcurveE}{HTML}{6B4B8F}  
\pgfplotsset{attcurvechart/.style={attitudechart,
  xlabel={AI usage index}, xlabel style={font=\scriptsize, yshift=3pt}}}
\pgfplotsset{attcurveline/.style={no marks, thick, forget plot}}
\pgfplotsset{attcurveends/.style={only marks, mark size=1.7pt, forget plot}}
\pgfplotsset{attcurvebar/.style={mark=-, mark size=1.8pt, line width=0.5pt, forget plot}}

\newcommand{\delegchartwidth}{0.66\linewidth}
\newcommand{\delegpiescale}{3.5}
\colorlet{delegtier0}{likertA}
\colorlet{delegtier1}{likertB}
\colorlet{delegtier2}{likertD}
\colorlet{delegtier3}{likertE}
\newcommand{\delegdisc}[4]{%
  \fill[delegtier#4] (axis cs:#1,#2) circle[radius={sqrt(#3)*\delegpiescale pt}];}
\newcommand{\delegwedge}[6]{%
  \fill[delegtier#6, draw=white, line width=0.3pt]
    (axis cs:#1,#2)
    -- ++({-90+360*(#4/#3)}:{sqrt(#3)*\delegpiescale pt})
    arc[start angle={-90+360*(#4/#3)}, end angle={-90+360*(#5/#3)},
        radius={sqrt(#3)*\delegpiescale pt}] -- cycle;}
\pgfplotsset{delegdiagonal/.style={no marks, draw=black!30, dashed, line width=0.5pt}}
\pgfplotsset{delegchart/.style={
  width=\delegchartwidth, height=\delegchartwidth,  
  xmin=0.4, xmax=5.6, ymin=0.4, ymax=5.6,
  grid=both, grid style={black!12},
  axis line style={black!45}, tick style={black!45},
  tick align=outside, tick pos=left, scaled ticks=false,
  xticklabel style={rotate=30, anchor=north east, align=right, font=\tiny, inner sep=1.5pt},
  yticklabel style={font=\tiny}, label style={font=\scriptsize},
  xlabel={X04: kinds of work delegated fully},
  ylabel={X05: scope and decision-making delegated}}}

\newcommand{\togchartwidth}{0.66\linewidth}
\newcommand{\togpiescale}{3.5}  
\definecolor{togtier0}{HTML}{BFDCC0}  
\definecolor{togtier1}{HTML}{8FC08F}  
\definecolor{togtier2}{HTML}{5A9459}  
\definecolor{togtier3}{HTML}{2F6B33}  
\newcommand{\togdisc}[4]{%
  \fill[togtier#4] (axis cs:#1,#2) circle[radius={sqrt(#3)*\togpiescale pt}];}
\newcommand{\togwedge}[6]{%
  \fill[togtier#6, draw=white, line width=0.3pt]
    (axis cs:#1,#2)
    -- ++({-90+360*(#4/#3)}:{sqrt(#3)*\togpiescale pt})
    arc[start angle={-90+360*(#4/#3)}, end angle={-90+360*(#5/#3)},
        radius={sqrt(#3)*\togpiescale pt}] -- cycle;}
\pgfplotsset{togchart/.style={
  width=\togchartwidth, height=\togchartwidth,  
  xmin=1.4, xmax=6.6, ymin=1.4, ymax=6.6,
  grid=both, grid style={black!12},
  axis line style={black!45}, tick style={black!45},
  tick align=outside, tick pos=left, scaled ticks=false,
  xticklabel style={rotate=30, anchor=north east, align=right, font=\tiny, inner sep=1.5pt},
  yticklabel style={font=\tiny}, label style={font=\scriptsize},
  xlabel={P02: togetherness reached in the best human pairing},
  ylabel={P03: togetherness reached with a coding agent}}}

\pgfplotsset{usagemarkuser/.style={only marks, mark=*, mark size=1.1pt,
  draw=likertE, fill=likertE}}
\pgfplotsset{usagemarknonuser/.style={only marks, mark=diamond*, mark size=1.5pt,
  draw=likertA, fill=likertA}}
\pgfplotsset{usagemarkunknown/.style={only marks, mark=o, mark size=1.2pt,
  draw=black!45}}
\pgfplotsset{trustregress/.style={no marks, thick, draw=black!50, forget plot}}
\newcommand{\trustscatterwidth}{0.66\linewidth}
\pgfplotsset{trustscatter/.style={
  width=\trustscatterwidth, height=\trustscatterwidth,  
  xmin=-0.08, xmax=1.08, ymin=0.80, ymax=5.20,
  xtick={0,0.25,0.5,0.75,1}, ytick={1,2,3,4,5},
  grid=both, grid style={black!12},
  axis line style={black!45}, tick style={black!45},
  tick align=outside, tick pos=left, scaled ticks=false,
  tick label style={font=\tiny}, label style={font=\scriptsize},
  xlabel={usage index (X01 and X02 weighted response ranks)},
  ylabel={trust index (G01, G03-mirrored, G04, G05)}}}

\newcommand{\practquote}[2]{``#2'' (#1)}

\newcounter{responsei}
\renewcommand{\theresponsei}{Q\arabic{responsei}}
\newenvironment{response}[2]%
{\begin{quotation}\emergencystretch=2em\refstepcounter{responsei}{\bfseries \theresponsei}(Block #2): \itshape\label{block#2}}%
{\end{quotation}}
\newcommand{\gap}{[\ldots]}
\definecolor{coded}{gray}{0.35}  
\definecolor{codedhigh}{rgb}{0.0,0.0,0.85} 
\newcommand*{\coded}[2]{\allowbreak\textcolor{coded}{[{\scriptsize\Code{#1}}}\,\penalty 200\hskip 0pt\relax #2\textcolor{coded}{]}}
\newcommand*{\codedhigh}[2]{\allowbreak\textcolor{coded}{[{\scriptsize\Code{#1}}}\,\penalty 200\hskip 0pt\relax\textcolor{codedhigh}{#2}\textcolor{coded}{]}}
\newlength{\qindent}
\newcommand{\qitem}[2]{\par\addvspace{\medskipamount}\noindent\hangindent=\qindent\hangafter=1
  \makebox[\qindent][l]{\textbf{#1}}#2\par}
\newcommand{\qopts}[1]{\par\nobreak
  {\leftskip=\qindent\relax\small\itshape\noindent #1\par}}
\newcommand{\qscale}[1]{\par\addvspace{\medskipamount}\noindent{\small\textit{Answer options:}\ \textit{#1}\par}}
\newcommand{\qsep}{\,\textperiodcentered\,}   
\newcommand{\qfree}{\emph{[free-text box]}}

\begin{document}

\parbox{80mm}{\hfil\textbf{\textcolor{red}{This article ends on page \pageref{end-of-article}, the rest is appendices.}\\~}}

\title{How developers split roles among them and a coding agent and what ``collaboration'' means}

\author{Linus Ververs}
\orcid{0009-0005-6170-310X}
\affiliation{%
  \institution{Freie Universit\"at Berlin}
  \department{Institut f\"ur Informatik}
  \city{Berlin}
  \country{Germany}}
\email{linus.ververs@fu-berlin.de}

\author{Helya Dadkhah}
\orcid{0009-0003-0164-9207}
\affiliation{%
  \institution{Freie Universit\"at Berlin}
  \department{Institut f\"ur Informatik}
  \city{Berlin}
  \country{Germany}}
\email{helyadadkhah.g@gmail.com}

\author{Lutz Prechelt}
\orcid{0000-0001-5580-4426}
\affiliation{%
  \institution{Freie Universit\"at Berlin}
  \department{Institut f\"ur Informatik}
  \city{Berlin}
  \country{Germany}}
\email{prechelt@inf.fu-berlin.de}

\renewcommand{\shortauthors}{Ververs, Dadkhah, Prechelt}

\begin{abstract}
\emph{Background:} 
Coding agents such as Claude Code become more and more competent,
so that their role in the development process can grow. \\
\emph{Objective:} 
How do developers characterize the role-split between them and their agent?
How would they ideally want to collaborate with their agent?\\
\emph{Methods:} 
We perform a 3-step, mixed-methods study.
We apply qualitative open coding to 2025 early adopters' blog articles;
perform a quantitative survey of 2026 practitioners; 
develop and apply a method for coding-agent-supported open coding of survey freetext responses.\\
\emph{Results:} 
The 2025 practitioners have the agent work in supporting roles only. 
Accountability was firmly human.
In 2026, both are fundamentally still true.
Gradually, however, the role-split is now different, with much more agent involvement.
The developers' notion of collaboration has three facets:
Predominantly a separation into human-only, agent-only, and collaborative work steps,
with defined conditions and goals.
Second, accommodating personal preferences without engineering justification.
Third, a holistic view of collaboration covering the entire development process.\\
\emph{Conclusions:} 
Defining one's role relative to the agent will need to be a recurring activity.
The different views on collaboration are likely a source of friction and conflict for teams;
future work should find out how to cope with that.\\
\end{abstract}

\ccsdesc[300]{Software and its engineering~Software development techniques}
\ccsdesc[500]{Software and its engineering~Collaboration in software development}

\keywords{agentic coding, human-AI collaboration, software engineering practice,
  semi-automated open coding}

\maketitle

\section{Introduction}

We define agentic coding as the use of a coding agent for any code-related activity in software development.
We define a coding agent as the combination of an LLM (large-language-model)
and a generic coding harness, as exemplified by tools such as
Claude Code \cite{Anthropic26code},
GitHub Copilot CLI \cite{GitHub26cli},
OpenAI Codex \cite{OpenAI26codex},
OpenCode \cite{OpenCode26},
and many others~\cite{MSV26}.

After early research prototypes~\cite{YanJimWet24}, these tools became mainstream with the
introduction of Claude Code in 2025-02~\cite{Anthropic25ClaudeCodeRelease, Karpathy25} and have 
since seen an
unprecedented growth in usage by developers around the world~\cite{LiZhaHas26, RobMatDeg26, 
HasLiLin26, PopGroBot26, Willison26llms}. 

Compared to previous software development tools, they are extremely versatile,
supporting activities as diverse as
vibe coding~\cite{Hashimoto25ghostty, Willison25vibe, Steinberger25arena, Woolf26},
technology and architecture discussions~\cite{Ronacher25ninety, Steinberger25talk, 
Steinberger25workflow, Pocock26, HitMasLyu26}, 
complex debugging episodes~\cite{Hashimoto25ghostty, Valsorda25},
and anything in between.

Not everybody is yet able to turn these capabilities into productivity improvements.
An August 2026 McKinsey study reports 25\% of organizations seeing twofold or higher
improvements, but also 30\% seeing reductions in productivity~\cite{RelLunPal26}.

Large improvements require drastic changes to the development style, 
which raises several questions:
How does this change the role (and the self-image) of developers?
What is the role of the coding agent?
How do developer and coding agent play together? 
Does the developer merely \emph{delegate} the execution of tasks to the coding agent?
Or should we speak of a \emph{collaboration} between the two?

The present work tackles these questions by a 3-step, mixed methods study.
In Step 1, we study blog articles of top early adopters of coding agents in order to
describe what role-split successful coding agent users of 2025 used.
In Step 2, we perform a quantitative survey in order to
find out to what extent this role-split still applies and applies to a broader developer population in 2026.
In Step 3, we analyze a freetext response from that same survey in order to
characterize the notions of collaboration these developers hold.

\subsection{Research Questions}

\begin{itemize}
  \item Step 1: How have early adopters of agentic coding split roles between them and a coding agent in 2025?
  \item Step 2: How do developers do it in 2026?
  \item Step 3 research method: How can we employ coding agents to support exploratory analysis of 
    survey freetext responses?
  \item Step 3: What is developers' notion of ``collaboration'' with a coding agent?
\end{itemize}

\subsection{Research Contributions}

We consider our research purely exploratory.
Although the survey is quantitative in principle, we assume no statistical generalization and
will describe the results in a mostly qualitative manner.
Our focus is on explaining, on telling a story, not on measuring.

From this perspective, we consider our research contributions to be the following:
\begin{itemize}
  \item Step 1: We explain what role-split worked for balancing productivity with 
    human responsibility and quality assurance in 2025
    (Section~\ref{sec:results1}).
  \item Step 2: We explain how this has shifted towards more joint activity in 2026
    (Section~\ref{sec:results2}).
  \item Step 3 research method: We describe our method for an open coding process where the human injects the major ideas
    and maintains control throughout and the agent adds minor ideas and performs all of the grunt work
    (Section~\ref{sec:methods3}).
  \item Step 3: We explain three facets of the notion of ``collaboration'' that developers hold in 2026:
    activity-centered collaboration, holistic collaboration, and personal preferences
    (Section~\ref{sec:results3}).
\end{itemize}

\subsection{Structure of the Article}

We next discuss related work (Section~\ref{sec:relatedwork}).
This is followed by three pairs of a methods section and a results section,
one for each of the three steps (Sections~\ref{sec:methods1} to \ref{sec:results3}),
then a discussion of threats to validity (Section~\ref{sec:threats-validity}).
We end with conclusions (Section~\ref{sec:conclusions}).

In order to keep the main body of the article short, readable, and interesting,
we push large parts of the details into appendices.
Appendices~\ref{app:result1}, \ref{app:step2}, \ref{app:step3}
correspond to steps 1, 2, 3.

\subsection{Data Availability}

We publish a data package\footnote{\url{https://figshare.com/articles/journal_contribution/34031112?file=69493860}}
that contains 
the Step 1 input data,
Step 2 questionnaire,
raw result data,
and evaluation scripts for the quantitative evaluation,
Step 3 infrastructure for the agent-supported open coding process and
exact results of the open coding.

\section{Related Work}\label{sec:relatedwork}

\subsection{Roles in Software Development}

In classical software development processes, roles are a key mechanism for
conceptualizing how work is organized:
what kind of person (the role) performs which activities,
and which activities form a bundle that describes a role.
Plan-driven processes such as V-Modell~XT define dozens of roles
to create order among their many activities~\cite{VModellXT}.
Responsibility assignment matrices further gradate participation in an
activity into different kinds, such as primary result responsibility,
execution, support, and review \cite{PMI2017}.

Agile processes approach the problem differently.
Their focus on flexibility \cite{BecBeeBen01}
and self-organization \cite{HodNobMar13, MoeDinDyb10}
means that detailed sets of roles make no sense; roles must become much coarser
\cite{SchSut20}.
Even then, however, clearly assigning certain key responsibilities is important:
in Scrum, for instance, solely the Product Owner is responsible for deciding
the requirements to be fulfilled \cite{SchSut20}.

Role clarity is important for psychological safety
\cite{BunBou10, LenFel18}.
Role conflicts can lead to members leaving a team or organization and
can even occasionally rip apart an entire team~\cite{BarPre19}.
Human beings need sufficient time for absorbing changes \cite{HodNob17, WeiQui99},
so given the fast pace of progress of coding agents, clarifying the human/agent role-split
in agentic coding is a problem the software development community will need to solve urgently.

There is initial evidence that turning agentic coding capabilities into large productivity
improvements at the team level requires substantial role changes~\cite{RelLunPal26}.

\subsection{Coding Agent Usage Patterns}

In this topic area, scientific studies tend to lag far behind current practice, because publication cycles
cannot keep up with the speed of technical progress and corresponding developer behavior changes --
in particular since the November 2025 improvements in model reliability 
\cite{Willison26llms}
that have made agentic coding workflows considerably easier to adopt.
Recent large-scale industrial experience reports 
\cite{DuaZhaJia26}
represent some of the more current empirical work, yet even these
describe a state of practice that has since evolved substantially.
We therefore rely on practitioner writings to a large degree here.

Observed usage of coding agents spans a broad spectrum.
At one end, developers treat agents as a particularly powerful editor,
delegating only mechanical, precisely defined operations;
at the other end lies \emph{vibe coding}: assigning full authority over
the code and internal design to the agent
\cite{ChoBirBad26}.

\citeauthor{XiCheGuo25} propose the theoretical framework of differentiating two different levels 
of collaboration between developers and AI agents: Either an instructor-executor role-split, i.e., 
where 
\say{humans serve as issuers of instructions, while agents act as executors, participating as 
assistants to humans in collaboration} 
or an equal partnership, where 
\say{agents [participate] on an equal footing with humans}~\cite[p. 18]{XiCheGuo25}.

Adoption varies considerably across individuals and organizations.
Individual readiness for AI-assisted development differs along dimensions such as
prior experience, perceived risk, and organizational support
\cite{DobGreBor26}.
At the team level, both individual disposition and team culture shape
whether and how developers adopt generative-AI tools
\cite{MilChoUll26}.
Organizations may need to provide more empathy than usual to get the more
anxious developers on board
\cite{GunGruHod26}.

Comprehensive surveys have traced the technical evolution
of LLM-based code generation agents, categorizing their architectures,
applications across development phases, and evaluation benchmarks
\cite{DonJiaQia25}.
Large-scale empirical studies of agent-authored contributions in open-source
development show that while agent-generated pull requests are often completed
faster, they are accepted at lower rates and tend to exhibit lower structural
complexity~\cite{LiZhaHas26}.
A complementary study of 567~pull requests generated using Claude Code across
157~open-source projects confirms that agents are commonly used for refactoring,
documentation, and testing, but that human modification remains frequently
necessary~\cite{WatLiKas25}.

Research roadmaps have articulated strategic visions for how generative~AI
may transform software engineering processes and products
\cite{AmaMetAut26}.
These works operate at a conceptual level, identifying what generative~AI
enhances, obsolesces, or otherwise changes within software engineering practice.
However, they do not examine how individual engineers operationalize their
roles within concrete agentic workflows, nor do they differentiate human
and agent responsibilities at the level of day-to-day practice.
In an industrial context, those engineering organizations that are seeing the biggest
productivity improvements from coding agents most often report 
workflow and process redesign as a top driver of these improvements
and this usually involves redesigning developer roles~\cite{RelLunPal26}.

Taken together, the related work establishes that coding agents are technically
capable, increasingly adopted, and actively contributing code in real-world settings.
What remains underexplored is the human role as a distinct analytic object
within these workflows -- which responsibilities developers retain,
which they delegate, and how they conceptualize the boundary.
This is the gap addressed by the present study.
The only study with a similar mission that we know of is \citeauthor{ChoBirBad26},
but their stance emphasizes the split between human and agent where we emphasize the collaboration.

\subsection{Notions of Collaboration}

We will look here at notions from several different fields outside of software engineering
and, in the spirit of strengthening our theoretical sensitivity~\cite{StrCor90},
suggest how they might apply to our setting of agentic coding.

In doing so, we constantly encounter the question of whether or how anthropomorphizing LLMs
is acceptable.
Three major stances on this in the literature are 
the anti-mentalistic view that rejects using anthropomorphic terms \cite{BenKol20} or 
advises great caution \cite{Shanahan24},
the pragmatic view that human terms and ideas are a (and possibly the only) practical way
of conversing about how LLMs behave \cite{ShaMcDRey23,MarLinOla26}, and
the engineering view (by which one can \emph{avoid} the philosophical issue) of 
defining these properties as measurable constructs
and then measuring them on LLMs \cite{JiaXuZhu23,ParZouKam26}.
We share the pragmatic view and will formulate accordingly; no philosophical position is implied.
That said, back to notions of collaboration:

In political science, the highly influential definition of Gray~\cite{Gray89} suggests 
that collaboration should be viewed as a means for coping with conflict.
For applying this to our domain of agentic coding, we ignore the underlying notion
that collaboration is for pursuing self-interest and 
instead focus on the human and agent's joint handling of the many \emph{tradeoffs} involved in 
building software which opens many collaboration opportunities.


Malone and Crowston~\cite{MalCro94} draw on work in computer science, organization theory, 
operations research, economics, linguistics, and psychology for formulating a broad framework of
coordination; in our frame of collaboration here, coordination is a subtopic.
The framework characterizes coordination as 
\Quote{the process of managing dependencies among activities}.
Some of these activities, such as goal selection, task decomposition, and task assignment \cite[Table 1]{MalCro94},
are highly pertinent to the present work.
Managing task/subtask dependencies can be based on top-down goal decomposition or 
bottom-up goal identification \cite[Section 2.2.4]{MalCro94}.
Group decisions can be made by authority, voting, or consensus \cite[Section 2.4.1]{MalCro94}.
One important aspect of such decisions is goal conflicts \cite[Section 3.1.2]{MalCro94}.
All of these ideas can be applied to human/agent collaboration and some of them will indeed
recur further down.

Singh~\cite{Singh92} defines coordination as the integration and harmonious adjustment of 
individual work efforts towards the accomplishment of a larger goal.
This sounds a lot like what human and agent together will want to achieve.

In 1995, Roschelle and Teasley suggested that collaborative problem solving in their given computerized
setting could be described by \Quote{an emergent, socially-negotiated set of knowledge elements, such as 
goals, problem state descriptions and problem solving actions} and that these were constructed
through shared language, situation, and activity \cite{RosTea95}.
At this level of abstraction, their conceptualization is applicable to a human/agent setting today
and the interesting statement is then that the knowledge elements are emerging and
socially negotiated.

Andrews et al.~\cite{AndLilSri23} review the notion of shared mental models in terms of the rich
literature that exists on the topic.
It suggests concepts such as shared cognition, distributed cognition, 
team knowledge, transactive (i.e. team) memory, and team situation awareness.
Many of these are applicable to our perspective here, but we do not use these particular terms
in order to arrive at a representation that is more coherent and more precisely aligned with our data
than would be possible otherwise.
(The article, which appeared in 2023, also discusses the application to human/AI teams, 
but these parts seem barely relevant today.)

\subsection{Using Coding Agents for Supporting Open Coding}

In Step 3 of the present work we will perform open coding on freetext answers of the survey
from Step 2 and Section~\ref{sec:methods3} explains how we used a coding agent for accelerating 
and improving this process.

Barros et al. \cite{BarAzeNet25} provide a systematic mapping study of LLM use for qualitative research
(in software engineering or elsewhere), but their study predates agentic coding and does not even
contain the word ``agent''.

Alves et al. \cite{AlvAraAra26} report on researcher discussion of LLM use for thematic analysis:
researchers emphasized a need for human oversight. 
Gomes et al. \cite{GomFraSoa25} found LLMs to be of little help for defining codes automatically 
from scratch in open coding. 
Our process will be strongly human-guided.

Few studies appear to exist so far that use coding agents for qualitative analysis.
In software engineering, we found only one: Baltes et al. \cite{BalCheTre26} analyze 
web posts of how developers complain about AI slop.
They use Claude Code for both developing the codebook and applying it to their data.
The authors provide a data package that documents 11 versions of the codebook
and the exact prompt for applying it, but do not document the procedure used for the codebook development.
They appear to consider the codebook to be the \emph{input} for the coding process 
(a content analysis (quantitative) perspective, although they do not use the term);
we consider it the \emph{output} from the coding process (an exploratory analysis perspective).

\section{Methods 1: Open Coding of 2025 Practitioner Blogs for Understanding the Role-Split}\label{sec:methods1}

\subsection{Data Sources: Hashimoto, Ronacher, Steinberger, Willison}

The qualitative analysis uses as primary data: practitioner-authored blog posts, long-form
technical write-ups, and one recorded live-coding session.
These sources were written by early adopters who have been actively using coding agents
since shortly after this new class of tool had appeared and in particular
before the \Quote{November inflection}~\cite{Willison26llms}
(the introduction of Claude Opus 4.5 in 2025-11~\cite{Anthropic25opus45}), 
which increased the reliability of coding agents to a level
that accelerated their subsequent mainstream adoption. 

Our sources' authors describe their workflows in detail: how tasks are framed,
how agents are instructed, which tools are invoked, how feedback is gathered,
how failures are handled, and how outputs are reviewed and integrated.
These narratives were produced independently of our research;
we consider them to capture authentic practice better than interviews could.

Four practitioners were selected using the following criteria:
\begin{itemize}
  \item High-quality contribution: the posts describe real software
    outcomes that can often be inspected directly (e.g., open-source  repositories).
  \item Neutral standpoint: the narratives include downsides, failure
    modes, or explicit trade-offs, and a clear statement of non-affiliation
    with tool vendors.
  \item Established reputation: the authors are highly capable software
    engineers with a documented track record predating coding agents.
\end{itemize}

We treat the four identified practitioners, their writings, and records as four cases.
They cover production infrastructure
(Armin Ronacher~\cite{Ronacher25ninety,Ronacher25vibes}),
highly parallel CLI-native solo workflows
(Peter Steinberger~\cite{Steinberger25speed,Steinberger25talk,Steinberger25workflow,Steinberger25arena}),
a single shipped feature documented at session level
(Mitchell Hashimoto~\cite{Hashimoto25ghostty}),
and practitioner-oriented methodological writing on agentic practice
(Simon Willison~\cite{Willison25loops,Willison25parallel,Willison25dangerous,Willison25tips,Willison25vibe,Willison25proven}).

\subsection{Analysis Method}

The analysis uses elements of Straussian Grounded Theory Methodology (GTM)~\cite{StrCor90}:
open coding, constant comparison, memoing, theoretical coding, and theoretical sensitivity.
In the context of open coding as we use it, the term ``code'' is synonymous with ``concept'', 
not limited to a mere label.

In the first round of open coding, a substantial portion of implementation
work was observed to be delegated to coding agents.
We noticed the role-split between software engineer and agent to be an important issue,
so the second round of open coding focused on aspects of software development that remained 
distinctly human.
Each paragraph of source material was coded by the primary action or responsibility.

Related codes were then grouped into higher-level categories
along two analytic dimensions: \emph{locus of agency} (human judgment versus
delegated agent execution) and \emph{functional role} within the development
workflow.
This yielded nine categories (A--J, excluding G), each capturing a
distinct aspect of the human/agent role-split.\footnote{These code characters express 
analysis history, not meaning, but we keep them here
to simplify matching article statements to research history.}
The nine categories were further integrated into three
\emph{clusters}.

\section{Results 1: 2025 Early Adopters' Role-Split}
\label{sec:results1}

The qualitative analysis identified nine categories (labeled A--J,
excluding~G) that capture aspects of how practitioners split
work between themselves and coding agents.
Each category contains facets of the human role, the agent role, or both.
The categories group into three higher-level \emph{human role clusters}:
intent formation, design operationalization, and evaluation.

We present each category with its human, agent, and shared facets,
followed by the cross-cutting findings that emerged from selective coding.

\subsection{Cluster: Intent Formation}

The first cluster comprises the categories that establish what the
system is for and which constraints must be obeyed.

\subsubsection{Specification, Requirements, and Architectural Direction (C, Human/Agent)}

All practitioners begin by establishing intent (defining purpose,
constraints, and success criteria) before any delegation to the agent.
Architecture, technology-stack, and boundary decisions remain human.
Agents may draft specification artifacts or brainstorm alternatives,
but hold no decision authority.
For more information, see Appendix~\ref{app:result1-category-C}.

\subsubsection{Knowledge Work, Learning, and Discovery (I, Human/Agent)}

Engineers use agents to accelerate research, expand exploratory capacity,
and enable fast iteration.
For instance, agents can trace codepaths when building code understanding
and allow for prototyping multiple approaches quickly
and selecting the most promising ones.
Both help the engineer to build personal understanding.
For more information, see Appendix~\ref{app:result1-category-I}.

\subsection{Cluster: Design Operationalization}

The second cluster translates intent into structures, constraints, and process controls
that constrain and inform implementation work. 

\subsubsection{Context Engineering and Prompting Discipline (E, Human)}

Once intent is sufficiently specified, practitioners translate it into
agent-actionable inputs through context and prompt engineering.
Context engineering in our sense concerns selecting, structuring, and constraining
what the agent is allowed (and sometimes required) to see.
Prompt engineering concerns how tasks, constraints, and intentions are
framed so the agent acts on them as intended.
Both are human responsibilities with little delegation.
For more information, see Appendix~\ref{app:result1-category-E}.

\subsubsection{Tooling, Infrastructure, and Environment Design (H, Human/Agent)}

Engineers select and combine agent tools, design execution environments
with appropriate sandboxing and safety boundaries, build infrastructure
to scale agentic workflows (parallelism, escalation, queueing), and
establish version-control and recovery affordances.
Decision-making in this category is not delegated;
agents may execute setup and configuration work once decisions are made.
For more information, see Appendix~\ref{app:result1-category-H}.

\subsubsection{Workflow Orchestration (B, Human)}

Practitioners design the structure of agent-driven work:
breaking tasks into reviewable, PR-sized chunks;
designing iteration loops (exploratory vs.\ lockstep refinement);
and coordinating multiple parallel agent sessions while retaining
control over prioritization and integration.
No delegation was observed in this category.
For more information, see Appendix~\ref{app:result1-category-B}.

\subsection{Cluster: Evaluation}

The third cluster governs accept/reject decisions and corrective action
after agent output exists.

\subsubsection{Implementation and Quality Calibration (D, Human/Agent)}

Agents contribute by generating candidate solutions, 
evaluating them with self-checking
(tests, linters, type checkers) and by supporting code review.
Engineers hold authority over quality standards, using domain
expertise to accept, discard, or modify agent output.
They calibrate review depth to risk, define what counts as evidence
that a change works, and intervene decisively when output repeatedly
fails to meet expectations.
For more information, see Appendix~\ref{app:result1-category-D}.

\subsubsection{Iteration, Refactoring, and Cleanup (F, Human/Agent)}

Decisions about \emph{when and why} to refactor or clean up remain human;
agents execute the mechanical work (moving functions, updating documentation,
fixing build errors) under human direction.
Production readiness is achieved through repeated human refinement:
engineers prompt until output is close and may perform finishing touches
and cleanup before merge.
For more information, see Appendix~\ref{app:result1-category-F}.

\subsubsection{Accountability and Ownership of Outcomes (A, Human)}

All four practitioners frame responsibility as a professional obligation
that cannot be delegated to the agent, regardless of who wrote the code.
Four complementary framings emerged:\\
Ronacher: ownership is non-transferable regardless of AI authorship;\\
Willison: ownership means delivering code proven to work;\\
Hashimoto: ownership is exercised through final human judgment before shipping;\\
Steinberger: ownership is maintained through architectural stewardship with selective
verification.\\
For more information, see Appendix~\ref{app:result1-category-A}.

\subsection{Cross-Cutting Category}

\subsubsection{Cultural Framing and Legitimacy Claims (J, Human)}

Practitioners draw boundaries between acceptable and unacceptable use
of AI in development, anchoring legitimacy in demonstrable correctness,
accountability, and transparency rather than in the degree of automation itself.
For more information, see Appendix~\ref{app:result1-category-J}.

\subsection{Summary}
\label{sec:results1-summary}

\begin{itemize}
\item The human role is not limited to a single function but is distributed
  across complementary activities in all three clusters.

\item {Intent formation is an explicit, ongoing human responsibility.}
  Engineers do not merely ``know what they want'' and pass it to agents.
  They actively construct intent through iterative specification,
  architectural reasoning, exploratory prototyping, and interpretation
  of failures.
  Agents can surface options and constraints but do not decide intent.

\item {Design operationalization is a distinct engineering competence.}
  Translating intent into agent-suitable structures is 
  neither traditional coding nor passive oversight.
  It constitutes a new core engineering competence: designing systems
  in which agents can act reliably.

\item {Accountability for outcomes is invariant under delegation.}
  Across all four cases, the human holds himself fully accountable no matter how much work
  is delegated.

\item {Implementation expertise shifts but does not disappear from execution.}
  The shift in implementation labor does not eliminate the need for expertise in that area.
  Architecting the system and evaluating the agent outputs still require a deep understanding.
\end{itemize}

\subsection{RACI Mapping}

The concepts of the RACI matrix~\cite{PMI2017} form a useful theoretical lens for taking stock
of our findings:

\begin{itemize}
  \item R (Responsible: Who does it?): Our four practitioners take a definite stance
    regarding responsibility for nearly all tasks.
    It is the human for nearly all tasks (especially the foundation-creating ones), 
    the agent for a few,
    and both together for a few others, such as research and code understanding.
  \item A (Accountable): An agent can never be accountable for any task.
  \item C (Consulted): The agent can be consulted, especially for gathering knowledge about an unknown code base.
  \item I (Informed): Successful work requires the agent to be informed about all important
    matters relevant for the current task, which is a result of successful context engineering.
\end{itemize}

\section{Methods 2: Quantitative Survey}
\label{sec:methods2-survey}\label{sec:methods2}

\subsection{Goals and Design Decisions}

The Step 2 survey was designed based on the results of Step 1.
It pursues a number of goals:
\begin{itemize}
  \item Validate the role-split found in Step 1 with a broader developer population
    and determine the amount of variance among developers (see Section~\ref{sec:results2-validate} for results).
  \item Check to what extent a trend towards more agent or more human responsibility appears to exist
     (Section~\ref{sec:results2-rolessplit-shift})
  \item Compare coding agent users to non-users (Section~\ref{sec:results2-rolessplit-users-nonusers}).
  \item Characterize some aspects of the general attitude towards coding agents
    (Section~\ref{sec:results2-general-attitude}).
  \item Characterize the notion of collaboration between a developer and a coding agent
    held by our respondents (Step 3, Section~\ref{sec:results3}).
\end{itemize}

Design decisions: We strived to keep the survey short (15 minutes target time).
We strived for neutral wording and for avoiding nudging respondents in any direction.
We made all questions optional to avoid forcing responses in cases where the respondent
had nothing to say.
The latter decision has resulted in lots of extra cases because almost no two questions get responses 
from the same set of respondents.
Fortunately, the numbers involved in the edge cases are always small and so we can ignore them
or push the discussion to the appendix in order to keep the main text readable.

For instance we split our respondents into coding agent users and non-users,
but \svy{sample/undefinedusers} respondents end up in neither group, because they have not answered 
all of the questions used by the splitting criterion.

Furthermore, we offered a number of freetext fields in otherwise categorical questions
in order to avoid overconstraining our respondents.
This, too, made the evaluation more complicated but reduced the bias in our data.

Two examples:
In the demographical data, ``country'' was a freetext field and \svy{country/multiple} respondents
chose to provide more than one country in their response.
In the list of AI tools used, ``other'' was accompanied by a freetext field,
which we evaluated and categorized manually,
which resulted in four additional categories beyond what our categorical question had originally contained.

\subsection{Questionnaire}

This section provides a short overview of the questionnaire.
Find the complete questionnaire in Appendix~\ref{app:method2-questionnaire}.
Where needed, we will also explain each question in the context of presenting its results.
The survey had 6 pages.

Page 0: States topic, voluntariness, ``there are no right or wrong answers'', anonymity, privacy (EU-GDPR),
authors' names.

Page 1: AI tool exposure and attitude.
\begin{itemize}
  \item 3 AI tool exposure items X01 to X03\footnote{These are 
    the technical identifiers of each questionnaire item. We use these throughout the article to make
    tracing to and from the data package easier.}.
  \item 7 AI tool attitude Likert items G01 to G07 with topics such as ``makes me more productive'', ``concerned about becoming too reliant'',
    ``trust tool more than myself'', shown in randomized order.
  \item 2 items X04 and X05 regarding what respondents are comfortable delegating to a coding agent 
    (X04: single-choice ordinal scale from mechanical work to judgment work;
    X05: single-choice ordinal scale from small and precisely defined to open-ended with minimal specification).
\end{itemize}
Discussion: The tool attitude and delegation questions are also meant to get the respondents thinking about
how they are using coding agents, which serves as preparation to obtain more accurate responses
on page 2, the centerpiece of the survey.

Page 2: Role-split inventory. 
\begin{itemize}
  \item 19 task-allocation task-type items T01 to T24 (with gaps in numbering, shown in randomized order)
    about ``Who should do each of the following?'' 
    on a 5-point scale from ``Human'' over ``Both about equally'' to ``Agent'';
    results Table~\ref{tab:items-users} shows the concrete list.
  \item 2 situation-type items DC01 and DC02 about delegation to agent vs.~collaboration with agent on a 5-point scale.
  \item Freetext response DCc ``How I would like to work together with an agent''.
\end{itemize}

Page 3: Accountability attitude. 
\begin{itemize}
  \item 2 items R01 and R02 on an agree/disagree scale.
\end{itemize}
Discussion: Asking about accountability only \emph{after} asking about task-allocation avoids biasing respondents
by perceived social desirability, which would presumably have meant more human-leaning responses.

Page 4: Pair programming experience and togetherness attitude. 
\begin{itemize}
  \item 3 items P01 to P03 on ordinal scales: pair programming exposure,
    best togetherness in pair programming, best togetherness with coding agent.
  \item 3 items W01 to W03 about the togetherness attribute on an agree/disagree scale.
\end{itemize}
We will explain the notion of togetherness in the results section.

Page 5: Demographics.
\begin{itemize}
  \item 7 items D01 to D07 (role, years of experience, type of organization, typical team size, 
    fraction of time spent on coding-related tasks, gender, country).
  \item multi-line freetext Z01 for general comments on working with coding agents.
  \item multi-line freetext Z02 for general comments on the survey.
  \item freetext email-address field Z03 for getting notified about results (split off from the data before analysis).
\end{itemize}

\subsection{Advertising and Data Collection}

The survey was advertised by email through the authors' personal networks and advertised 
on LinkedIn in 21 international groups in the topic areas of AI and Software Engineering. 
Find the full text of the LinkedIn post in Appendix~\ref{app:step2-linkedin-msg}. 
As a result of this approach, our survey has no fixed base population.
This has two consequences: First, a response rate cannot be determined. 
Second, our respondent set is not representative of anything in particular and 
we must hence keep quantitative interpretation very rough.

Our first and last data records are from \svy{sample/firstdate} and \svy{sample/lastdate},
respectively.

\subsection{Data Filtering}

Our data contains \svy{sample/nonempty} records of people who answered at least one question.
We removed three types of record we deemed invalid or unusable:
\begin{enumerate}
	\item Filtering out respondents who answered too few of the central questions. 
	Here, we 
	removed \svy{funnel/fewtasks} respondents (\svy{funnel/fewtaskspct}\,\%) who answered fewer than
	10 of the 19 task-allocation items.
	\item Filtering out straight-liner respondents. 
	Here, we removed \svy{funnel/straightliners} (\svy{funnel/straightlinerspct}\,\%) respondents 
	who gave the same answer to each of the task-allocation items.
	\item  Filtering out respondents who gave the wrong response to the attention-check item
    which was embedded into the 19 page 2 task-allocation items. 
	Here, we removed \svy{funnel/attcheck} (\svy{funnel/attcheckpct}\,\%) respondents who failed 
	the attention-check.
\end{enumerate}

This leaves \svy{sample/eligible}  (\svy{sample/eligiblepct}\,\%) eligible respondents.

\subsection{Respondent Demographics}

We only provide a rough overview here;
see Appendix~\ref{app:method2-demographics} for more detail.
Of the \svy{sample/eligible} respondents, those \svy{country/answered} who answered 
\Quote{In which country do you currently work?} gave the following responses:
\svy{country/textblock}

The respondents are overwhelmingly male, come from a broad mix of different types of organization,
and work in a broad variety of roles.
A big majority of them work alone or in small teams up to 10 people and spend a large part of their
time on code-related tasks. 
Half of them work in senior roles and most have long professional experience.
See Table~\ref{tab:sample-background} in Appendix~\ref{app:method2-demographics} for details.

\subsection{``user'', ``non-user''}

We split our respondents into two groups we will sometimes want to compare: 
users of agentic coding versus non-users.
We use the page 1 questions X01 to X03 for the classification.
\emph{Users} are all respondents who 
state they use AI tools for software development (X01) and 
state they use coding agents at least monthly (X02) and 
state they use at least one AI tool we consider an agentic coding tool 
(X03: a terminal coding agent such as Claude Code, Codex CLI, GitHub Copilot CLI, etc.,
an IDE AI plugin in agent mode, GitHub Copilot on GitHub, or 
an AI-native IDE such as Cursor or Windsurf in agent mode).

Everybody else who answered all three questions is a \emph{non-user}.
This leads to \svy{sample/users} users (\svy{sample/userspct}\%)
and \svy{sample/nonusers} non-users (\svy{sample/nonuserspct}\%).
\svy{sample/undefinedusers} have not answered all three questions and
land in neither group.

\subsection{Assumptions, Scale Embedding, ``usage index'', ``trust index''}

A quantitative survey yields so much descriptive information that it is difficult
to comprehend it as a whole.
Ours being exploratory, however, means we need to perform useful sensemaking.
This requires aggregating many response items into understandable, larger-grain units of meaning.
For that, purely logical combinations such as for user/non-user above do not always suffice.

In order to facilitate such advanced aggregation, we make a few mild assumptions.
Our individual response items are mostly on binary scales (items of type ``check all that apply'')
or on ordinal scales (e.g. a 5-point agree-disagree scale).

Aggregation 1: we may embed a 5-point ordinal scale into a 1-to-5 numerical interval scale 
in order to compute means in order to simplify the comparison of distributions etc.
This assumes that the distance between any two neighboring points on the 5-point scale 
is the same.
For example, we will do this for the task allocation items T01 and up: ``The human alone'' becomes 1,
``Both about equally'' becomes 3, ``The agent alone'' becomes 5.
We treat ``not applicable'' and ``not sure'' responses like no responses.

Aggregation 2: we may embed a group of binary items into an absolute scale (count scale).
This assumes that it is sensible to assign the same weight to each item.
For example, we will count how many of the AI tools of item X03 respondents have checked.

Aggregation 3: we define the agent \textbf{usage index} as follows:
For X01 (how long have you used AI tools for SW development), map the 6-point ordinal scale
to range 0\ldots 1. Let us call the result $L$.
For X02 (how often do you use coding agents), map the 7-point ordinal scale 
to range 0\ldots 1. Let us call the result $I$, for intensity.
The usage index is then the weighted average $(L+2I)/3$ of both numbers.
This is qualitatively sensible, but quantitatively questionable, as it assumes that
the ordinal steps are equidistant and that the aspects should be weighted in this manner.
Hence, the usage index should be considered a rough approximation only.

Aggregation 4: we define the agent \textbf{trust index} as follows:
Consider the four attitude items 
G01 (\Quote{AI tools make me more productive in my development work}), 
G03 (\Quote{In my experience, AI coding tools are over-hyped relative to what they actually deliver}), 
G04 (\Quote{The more experience I gain with AI tools, the more of my work I am willing to let them do}), and
G05 (\Quote{For many tasks, I trust my AI tools more than I trust myself.}).
Map the 5-point agree/disagree scales of each to 0\ldots 1 each, but invert the polarity
of G03.
Average the result.
These four Likert items have high agreement 
(Cronbach’s $\alpha = \svy{trust/alpha}$ over the \svy{trust/alpha/n} respondents who answered all four items),
which indicates it may be sensible to integrate them into a Likert scale by plain averaging: 
The trust index.
This operation assumes equidistance within each agree/disagree scale and equal relevance
of the four factors.

Given the exploratory nature of our survey where small differences are hardly of interest,
we deem these assumptions harmless if one keeps in mind that the usage index needs to be taken
with a grain of salt.

\subsection{Comparing Step 2 Task-Allocation Responses to Step 1 Expectations}\label{sec:methods2-verdictrule}

Most of our 19 task-allocation items can be mapped to some statement in one of 
the 9 qualitative categories from Section~\ref{sec:results1} 
(see Appendix~\ref{app:result1} for specific statements).
That statement would then predict what our four 2025 practitioners would have answered
to our survey item back in 2025: 
Whether they would prefer that task to be done by the human (H),
the agent (A), or both together (E, for ``equally'' in our phrasing).

For 4 of our task-allocation items (T04, T11, T12 and T22) the data is not sufficient to ground a 
prediction, either because our practitioners disagree 
(like for T22) 
or there is too little evidence to conclude the practitioners' opinions on that matter 
(like for T04).

For checking where and to what extent the views of our 2026 respondents agree with those
practitioners' 2025 views, we use that prediction as the reference and compare
the response distribution to it.
Comparing a distribution over a 5-point ordinal scale with a single fixed prediction, however,
is not a canonical operation.
We use the following approach:
\begin{itemize}
  \item Embed the responses into a 1-to-5 interval scale.
  \item Compute the mean M and its 95\% confidence interval CI.
  \item Define the E region to range from 2.5 to 3.5.
    Anything below is the H region (human), anything above is the A region (agent).
  \item Consider the response distribution to be an H overall response if the entire CI lies \emph{below} the E region.
    Likewise it is an E overall response if the entire CI lies \emph{within} the E region and
    an A overall response if the entire CI lies \emph{above} the E region.
  \item Otherwise, the overall response is undefined.
  \item Deviating from the definition above, we also consider the overall response to be undefined
    if the distribution is bimodal.
    We consider it bimodal if ``there is little in the middle and much on both sides'':
    The ``3'' responses are a local minimum in the distribution (there are more H region responses
    and also more A region responses) and there are at least \svy{verdict/dipbalance}\,\% as many responses 
    in the smaller outer region as in the larger one.
  \item The prediction is considered confirmed if it agrees with the overall response,
    considered rejected if it disagrees, and 
    its status is inconclusive if the overall response is undefined.
\end{itemize}
The results of this approach will be discussed in Section~\ref{sec:results2-validate}.

\section{Results 2: Quantitative Survey Results}\label{sec:results2}

\subsection{Agent and Other AI Tool Use}

\begin{table}
  \caption{\gentag{tab-sample-agentuse.tex}AI tool experience and coding-agent use. 
    Percentages are of the respondents who
    answered that question ($n$ in the block heading), so each block sums to 100.}
  \label{tab:sample-agentuse}
  \centering\tablefont\setlength{\tabcolsep}{4pt}
\begin{tabular}{@{}lrrc@{\qquad}lrrc@{}}
\toprule
Characteristic & $n$ & \% & percent & Characteristic & $n$ & \% & percent \\
\midrule
\multicolumn{4}{c}{\textit{experience with AI tools of any kind (n=242)}} & \multicolumn{4}{c}{\textit{how often coding agents are used (n=241)}} \\
does not use them & 18 & 7 & \percentagebar{7} & never & 29 & 12 & \percentagebar{12} \\
less than 3 months & 3 & 1 & \percentagebar{1} & tried once or twice & 27 & 11 & \percentagebar{11} \\
3-6 months & 29 & 12 & \percentagebar{12} & monthly & 11 & 5 & \percentagebar{5} \\
6-12 months & 52 & 21 & \percentagebar{21} & weekly & 10 & 4 & \percentagebar{4} \\
1-2 years & 76 & 31 & \percentagebar{31} & several per week & 35 & 15 & \percentagebar{15} \\
more than 2 years & 64 & 26 & \percentagebar{26} & about daily & 30 & 12 & \percentagebar{12} \\
 &  &  &  & several per day & 99 & 41 & \percentagebar{41} \\
\bottomrule
\end{tabular}

\end{table}

As you can see in Table~\ref{tab:sample-agentuse},
a big majority of our respondents has been using AI tools for half a year or more
and a majority is using coding agents daily.
There are, however, also quite a few non-users of coding agents, so we will be able to compare
the two resulting worldviews meaningfully.
There are only few respondents who use coding agents, but do so only rarely.

Table~\ref{tab:sample-tools} in the appendix shows how frequently each specific type of
AI tool was used for software development by our respondents.
A clear majority (\svy{sample/tools/chat/pct}\,\%) uses chat assistants, 
a clear majority (\svy{sample/tools/cliagent/pct}\,\%) uses terminal coding agents,
all other tool types are far less common.
We find the frequency of self-hosted LLMs (\svy{sample/tools/selfhost/pct}\,\%) surprisingly high,
which underscores this is a sample containing many power users.

\subsection{How Developers Split Roles Among Them and a Coding Agent}\label{sec:results2-validate}\label{sec:results2-rolessplit-shift}\label{sec:results2-rolessplit-users-nonusers}

\subsubsection{Users}

Table~\ref{tab:items-users} shows the role-split data for the users group.
Please study that table now: 
The set of items, 
the resulting ``who does what'' response distribution for each,
how that relates to your own style and expectation, and 
how the predicted polarity P
relates to the overall response O in the sense of Section~\ref{sec:methods2-verdictrule}.

\begin{table}[p]
  \caption{\gentag{tab-items-users.tex}How the \svy{sample/users} \textbf{users} respondents split roles with their agent.
    Item: The Appendix~\ref{app:method2-questionnaire} item number and corresponding
    Section~\ref{sec:results1} category letter.
    Activity: abbreviated text of the item.
    P: predicted polarity (H, E, or A according to Section~\ref{sec:methods2-verdictrule}), if any.
    O: observed overall response polarity.
    V: verdict on P ($+$~prediction confirmed, $-$~prediction rejected, else inconclusive).
    Distribution: The bar is the answer distribution, always 100 units long, with the center 
    of the ``3'' bar at a fixed position:
    \likertlegend.
    A~$\vee$ over the bar marks bimodal distributions according to Section~\ref{sec:methods2-verdictrule}.
    M: arithmetic mean and its CI (based on the standard error).
    $n$: number of respondents who gave an answer.
    Region (\%): size of the H, E, and A regions.
    Rows are sorted by the position of the bar, most human-attributed first.}
  \label{tab:items-users}
  \centering\tablefont\setlength{\tabcolsep}{3.5pt}
\begin{tabular}{@{}llcclclrrrr@{}}
\toprule
Item & Activity & P & O & V & Distribution & $M$ [95\,\% CI] & $n$ & \multicolumn{3}{c}{Region (\%)} \\
\cmidrule(lr){9-11}
 & & & & & {\scriptsize human\,$\leftrightarrow$\,agent} & & & H & E & A \\
\midrule
T24-J & what counts as acceptable use of AI & \pred{H} & \pred{H} & $+$ & \likertbar{71}{19}{8}{1}{1}{0} & 1.40 [1.29, 1.52] & 159 & 90 & 8 & 2 \\
T23-A & final judgment that a change is ready to ship & \pred{H} & \pred{H} & $+$ & \likertbar{48}{38}{8}{5}{1}{0} & 1.71 [1.58, 1.84] & 170 & 86 & 8 & 6 \\
T01-C & what a task must achieve to be acceptable & \pred{H} & \pred{H} & $+$ & \likertbar{37}{46}{9}{7}{1}{0} & 1.88 [1.74, 2.01] & 171 & 83 & 9 & 8 \\
T10-H & choosing which AI tools and agents to use & \pred{H} & \pred{H} & $+$ & \likertbar{39}{43}{10}{7}{1}{0} & 1.89 [1.75, 2.04] & 169 & 82 & 10 & 8 \\
T20-F & deciding when and why to refactor or clean up & \pred{H} & \pred{H} & $+$ & \likertbar{25}{49}{17}{7}{2}{0} & 2.13 [1.99, 2.27] & 171 & 74 & 17 & 9 \\
T16-D & defining what counts as evidence it works & \pred{H} & \pred{H} & $+$ & \likertbar{28}{45}{16}{10}{1}{0} & 2.12 [1.97, 2.27] & 164 & 73 & 16 & 11 \\
T11-H & designing environments, sandboxing, safety boundaries & \pred{X} & \pred{H} & & \likertbar{21}{48}{17}{10}{4}{0} & 2.26 [2.10, 2.42] & 164 & 69 & 17 & 14 \\
T09-E & writing and improving prompts / task descriptions & \pred{H} & & & \likertbar{21}{43}{20}{12}{4}{0} & 2.35 [2.19, 2.52] & 167 & 64 & 20 & 16 \\
T03-C & developing/critiquing ideas in a spec or plan & \pred{E} & & & \likertbar{7}{47}{31}{13}{2}{0} & 2.57 [2.44, 2.71] & 170 & 54 & 31 & 15 \\
T04-I & understanding domain / unfamiliar codebase & \pred{X} & \pred{E} & & \likertbar{6}{43}{32}{15}{4}{0} & 2.67 [2.53, 2.82] & 166 & 49 & 32 & 19 \\
T07-E & judging which information matters for a task & \pred{H} & \pred{E} & $-$ & \likertbar{13}{33}{26}{23}{5}{0} & 2.75 [2.58, 2.92] & 168 & 46 & 26 & 28 \\
T18-D & reviewing a change for gaps, risks, problems & \pred{E} & \pred{E} & $+$ & \likertbar{6}{30}{28}{28}{8}{0} & 3.01 [2.84, 3.17] & 171 & 36 & 28 & 36 \\
T13-B & splitting work into human-reviewable chunks & \pred{H} & \pred{E} & $-$ & \likertbar{10}{17}{30}{26}{17}{0} & 3.21 [3.02, 3.40] & 163 & 27 & 30 & 43 \\
T22-F & removing small imperfections near the end & \pred{X} & & & \likertbar{9}{21}{20}{33}{17}{1} & 3.29 [3.11, 3.48] & 170 & 30 & 20 & 50 \\
T12-H & carrying out setup and configuration once chosen & \pred{X} & & & \likertbar{5}{17}{17}{44}{17}{1} & 3.49 [3.32, 3.66] & 167 & 22 & 17 & 61 \\
T17-D & implementing and checking with tests/linters & \pred{A} & \pred{A} & $+$ & \likertbar{2}{13}{19}{43}{23}{0} & 3.72 [3.57, 3.88] & 169 & 15 & 19 & 66 \\
T05-I & tracing code paths, summarizing how parts work & \pred{A} & \pred{A} & $+$ & \likertbar{0}{7}{23}{46}{24}{0} & 3.87 [3.74, 4.00] & 169 & 7 & 23 & 70 \\
T21-F & carrying out mechanical refactors/cleanup/fixes & \pred{A} & \pred{A} & $+$ & \likertbar{0}{8}{14}{46}{32}{0} & 4.01 [3.88, 4.15] & 168 & 8 & 14 & 78 \\
T06-I & throwaway prototypes for alternative approaches & \pred{A} & \pred{A} & $+$ & \likertbar{1}{3}{12}{44}{40}{0} & 4.17 [4.04, 4.30] & 169 & 4 & 12 & 84 \\
\bottomrule
\end{tabular}

\end{table}

\begin{table}[p]
  \caption{\gentag{tab-items-nonusers.tex}Like Table~\ref{tab:items-users}, but for the \svy{sample/nonusers} non-users.
    Rows are sorted by \emph{this} table's bars, so an item's row can differ from
    Table~\ref{tab:items-users}.
    The final column Move is additional. It shows how many ranks a row has moved up ($\uparrow$)
    or down ($\downarrow$) compared to Table~\ref{tab:items-users}; a dot marks no change.
    The smaller $n$ makes the CIs wider, so a tendency towards more inconclusive verdicts is expected.}
  \label{tab:items-nonusers}
  \centering\tablefont\setlength{\tabcolsep}{3.5pt}
\begin{tabular}{@{}llcclclrrrrr@{}}
\toprule
Item & Activity & P & O & V & Distribution & $M$ [95\,\% CI] & $n$ & \multicolumn{3}{c}{Region (\%)} & Move \\
\cmidrule(lr){9-11}
 & & & & & {\scriptsize human\,$\leftrightarrow$\,agent} & & & H & E & A & \\
\midrule
T24-J & what counts as acceptable use of AI & \pred{H} & \pred{H} & $+$ & \likertbar{91}{6}{0}{2}{1}{0} & 1.16 [1.01, 1.32] & 67 & 97 & 0 & 3 & \rankmove{0}{same} \\
T23-A & final judgment that a change is ready to ship & \pred{H} & \pred{H} & $+$ & \likertbar{78}{18}{1}{3}{0}{0} & 1.30 [1.14, 1.46] & 67 & 96 & 1 & 3 & \rankmove{0}{same} \\
T01-C & what a task must achieve to be acceptable & \pred{H} & \pred{H} & $+$ & \likertbar{66}{29}{0}{3}{2}{0} & 1.44 [1.25, 1.63] & 68 & 95 & 0 & 5 & \rankmove{0}{same} \\
T20-F & deciding when and why to refactor or clean up & \pred{H} & \pred{H} & $+$ & \likertbar{57}{32}{10}{1}{0}{0} & 1.57 [1.39, 1.74] & 69 & 89 & 10 & 1 & \rankmove{1}{up} \\
T11-H & designing environments, sandboxing, safety boundaries & \pred{X} & \pred{H} & & \likertbar{61}{27}{8}{4}{0}{0} & 1.56 [1.36, 1.76] & 66 & 88 & 8 & 4 & \rankmove{2}{up} \\
T16-D & defining what counts as evidence it works & \pred{H} & \pred{H} & $+$ & \likertbar{63}{26}{5}{6}{0}{0} & 1.54 [1.33, 1.75] & 65 & 89 & 5 & 6 & \rankmove{0}{same} \\
T10-H & choosing which AI tools and agents to use & \pred{H} & \pred{H} & $+$ & \likertbar{61}{23}{11}{5}{0}{0} & 1.61 [1.38, 1.83] & 61 & 84 & 11 & 5 & \rankmove{3}{down} \\
T07-E & judging which information matters for a task & \pred{H} & \pred{H} & $+$ & \likertbar{45}{33}{15}{7}{0}{0} & 1.85 [1.62, 2.08] & 67 & 78 & 15 & 7 & \rankmove{3}{up} \\
T04-I & understanding domain / unfamiliar codebase & \pred{X} & \pred{H} & & \likertbar{29}{46}{18}{7}{0}{0} & 2.03 [1.82, 2.24] & 69 & 75 & 18 & 7 & \rankmove{1}{up} \\
T03-C & developing/critiquing ideas in a spec or plan & \pred{E} & \pred{H} & $-$ & \likertbar{33}{39}{22}{4}{2}{0} & 2.01 [1.79, 2.24] & 69 & 72 & 22 & 6 & \rankmove{1}{down} \\
T09-E & writing and improving prompts / task descriptions & \pred{H} & \pred{H} & $+$ & \likertbar{47}{25}{17}{9}{2}{0} & 1.92 [1.64, 2.19] & 59 & 72 & 17 & 11 & \rankmove{3}{down} \\
T12-H & carrying out setup and configuration once chosen & \pred{X} & & & \likertbar{30}{33}{21}{13}{3}{0} & 2.27 [1.99, 2.54] & 67 & 63 & 21 & 16 & \rankmove{3}{up} \\
T22-F & removing small imperfections near the end & \pred{X} & & & \likertbar{35}{29}{17}{11}{8}{0} & 2.26 [1.95, 2.58] & 65 & 64 & 17 & 19 & \rankmove{1}{up} \\
T13-B & splitting work into human-reviewable chunks & \pred{H} & & & \likertbar{28}{27}{31}{11}{3}{0} & 2.35 [2.07, 2.63] & 62 & 55 & 31 & 14 & \rankmove{1}{down} \\
T17-D & implementing and checking with tests/linters & \pred{A} & & & \likertbar{24}{35}{20}{18}{3}{0} & 2.41 [2.13, 2.69] & 66 & 59 & 20 & 21 & \rankmove{1}{up} \\
T18-D & reviewing a change for gaps, risks, problems & \pred{E} & & & \likertbar{17}{34}{31}{15}{3}{0} & 2.54 [2.28, 2.79] & 67 & 51 & 31 & 18 & \rankmove{4}{down} \\
T05-I & tracing code paths, summarizing how parts work & \pred{A} & & & \likertbar{13}{29}{32}{20}{6}{0} & 2.76 [2.48, 3.03] & 66 & 42 & 32 & 26 & \rankmove{0}{same} \\
T21-F & carrying out mechanical refactors/cleanup/fixes & \pred{A} & & & \likertbar{20}{26}{14}{37}{3}{1} & 2.77 [2.46, 3.08] & 65 & 46 & 14 & 40 & \rankmove{0}{same} \\
T06-I & throwaway prototypes for alternative approaches & \pred{A} & & & \likertbar{12}{29}{21}{30}{8}{1} & 2.92 [2.63, 3.21] & 66 & 41 & 21 & 38 & \rankmove{0}{same} \\
\bottomrule
\end{tabular}

\end{table}

All of our predictions in the top and bottom parts of the table are confirmed.
In the middle, many are inconclusive, either because the 2025 practitioners' statements did
not support a prediction in the first place or because the response distribution was too
ambiguous for a verdict.

Only \svy{agg/users/rejected} predictions come out rejected, and 
\svy{agg/users/inconclusive} inconclusive.
In both rejected cases, what had been a purely human task (H) in 2025 was a shared task (E) now;
one relying on better automated context management of recent coding agents (T07),
the other relying on improved planning capabilities (T13).

\subsubsection{Non-users}

Table~\ref{tab:items-nonusers} shows the corresponding data for the non-users group.
As we can see, the general trend is the same: a big majority of the items have the same
position in the ranking or one nearby (column Move).
\svy{agg/nonusers/rejected} of our predictions was rejected and
\svy{agg/nonusers/inconclusive} came out inconclusive.
The number of inconclusive comparisons is larger than for the users,
partly due to the smaller number of respondents and hence wider confidence intervals and
partly because the variance in the responses themselves is larger as well.

Non-users lean much more heavily to the human side for nearly all task-allocation items. 
With only one exception (T13, also not confirmed for the users), 
all our H predictions are confirmed by them.
One E prediction comes out as H as well.
None of the A predictions is confirmed.

\subsubsection{Accountability}\label{sec:results2-accountability}

The followup questions regarding accountability were answered nearly unanimously
(see Table~\ref{tab:acc} in Appendix~\ref{app:result2-acc} for details):
Users and non-users overwhelmingly agree that developers need to be held accountable
for the agent's mistakes (question R01).
Users also overwhelmingly agree that mistakes in agent-generated code can be due to 
agent error, human error, or both.
Non-users have slightly more disagreement here -- which our data cannot explain but is presumably 
a sign of naivete.

\subsubsection{Assessing Generalizability}\label{sec:results2-german-nongerman}

\begin{table}
	\caption{\gentag{tab-items-compare-country.tex}
		Comparing the answers of German and non-German respondents regarding our role-split questions. 
		Most columns match Table~\ref{tab:items-users}.
		O, $M$ and $n$ are reported per group.
		$\Delta M$ is the difference of the two means (Germany minus elsewhere) with its CI.}
	\label{tab:items-compare-country}
	\centering\tablefont\setlength{\tabcolsep}{3.5pt}
\begin{tabular}{@{}llccrrcrrl@{}}
\toprule
Item & Activity & P & \multicolumn{3}{c}{Germany} & \multicolumn{3}{c}{elsewhere} & $\Delta M$ [95\,\% CI] \\
\cmidrule(lr){4-6}\cmidrule(lr){7-9}
 & & & O & $M$ & $n$ & O & $M$ & $n$ & \\
\midrule
T12-H & carrying out setup and configuration once chosen & \pred{X} & \pred{E} & 3.08 & 120 & & 3.36 & 70 & -0.27 [-0.66, 0.12] \\
T23-A & final judgment that a change is ready to ship & \pred{H} & \pred{H} & 1.52 & 122 & \pred{H} & 1.75 & 72 & -0.23 [-0.49, 0.03] \\
T18-D & reviewing a change for gaps, risks, problems & \pred{E} & \pred{E} & 2.81 & 122 & & 3.01 & 72 & -0.20 [-0.54, 0.14] \\
T11-H & designing environments, sandboxing, safety boundaries & \pred{X} & \pred{H} & 2.03 & 117 & \pred{H} & 2.17 & 70 & -0.14 [-0.46, 0.19] \\
T09-E & writing and improving prompts / task descriptions & \pred{H} & \pred{H} & 2.21 & 121 & & 2.33 & 64 & -0.12 [-0.49, 0.25] \\
T24-J & what counts as acceptable use of AI & \pred{H} & \pred{H} & 1.27 & 119 & \pred{H} & 1.36 & 66 & -0.09 [-0.31, 0.12] \\
T17-D & implementing and checking with tests/linters & \pred{A} & & 3.32 & 120 & & 3.40 & 72 & -0.09 [-0.46, 0.29] \\
T16-D & defining what counts as evidence it works & \pred{H} & \pred{H} & 1.90 & 118 & \pred{H} & 1.96 & 70 & -0.06 [-0.37, 0.25] \\
T01-C & what a task must achieve to be acceptable & \pred{H} & \pred{H} & 1.69 & 124 & \pred{H} & 1.73 & 71 & -0.04 [-0.31, 0.23] \\
T13-B & splitting work into human-reviewable chunks & \pred{H} & \pred{E} & 2.96 & 114 & & 3.00 & 67 & -0.04 [-0.45, 0.38] \\
T04-I & understanding domain / unfamiliar codebase & \pred{X} & & 2.46 & 122 & & 2.46 & 70 & 0.00 [-0.31, 0.32] \\
T07-E & judging which information matters for a task & \pred{H} & & 2.52 & 122 & & 2.51 & 70 & 0.00 [-0.37, 0.37] \\
T20-F & deciding when and why to refactor or clean up & \pred{H} & \pred{H} & 1.97 & 124 & \pred{H} & 1.94 & 72 & 0.02 [-0.26, 0.31] \\
T05-I & tracing code paths, summarizing how parts work & \pred{A} & & 3.61 & 120 & & 3.56 & 71 & 0.04 [-0.29, 0.38] \\
T21-F & carrying out mechanical refactors/cleanup/fixes & \pred{A} & \pred{A} & 3.70 & 119 & & 3.65 & 71 & 0.05 [-0.31, 0.41] \\
T03-C & developing/critiquing ideas in a spec or plan & \pred{E} & & 2.45 & 124 & & 2.38 & 71 & 0.07 [-0.24, 0.38] \\
T22-F & removing small imperfections near the end & \pred{X} & & 3.00 & 120 & & 2.87 & 71 & 0.13 [-0.28, 0.53] \\
T10-H & choosing which AI tools and agents to use & \pred{H} & \pred{H} & 1.95 & 122 & \pred{H} & 1.68 & 65 & 0.27 [-0.00, 0.55] \\
T06-I & throwaway prototypes for alternative approaches & \pred{A} & \pred{A} & 3.97 & 121 & & 3.65 & 71 & 0.32 [-0.03, 0.67] \\
\bottomrule
\end{tabular}

\end{table}

Our non-representative and German-heavy respondent sample raises the question whether our results
contain a bias that is so large as to distort the conclusions drawn.
We use a simple technique for convincing ourselves that this is not the case.

The idea is comparing the \svy{country/germanypct}\% German respondents as one group
to the remainder (all respondents from the rest of the world) as the other group.
If the detailed results for these two groups are similar, at least the geographical bias
cannot be very large.
We apply this idea to the role-split data just discussed above.

See Table~\ref{tab:items-compare-country} for the outcome.
Rough comparison: The overall responses agree in all rows where they are defined for both groups.
Finer comparison: The differences of the means are never larger than about a third of a scale point and often even
smaller than 0.1. 
We have very little reason to be concerned.

A similar comparison for the accountability questions is shown in 
Table~\ref{tab:items-compare-country-acc} in Appendix~\ref{app:result2-acc}.

\subsection{General Attitudes Toward AI Tools}\label{sec:results2-general-attitude}

\begin{table}
  \caption{\gentag{tab-items-attitude2.tex}Attitudes towards AI tools in software development 
    as held by agent users only. The distribution notation is like in Table~\ref{tab:items-users},
    including the v-shape mark that indicates polarization (few responses in the middle).}
  \label{tab:items-attitude2}
  \centering\tablefont\setlength{\tabcolsep}{3.5pt}
\begin{tabular}{@{}llclr@{}}
\toprule
Item & Statement & Distribution & $M$ [95\,\% CI] & $n$ \\
 & & {\scriptsize disagree\,$\leftrightarrow$\,agree} & & \\
\midrule
G03 & AI coding tools are over-hyped & \likertbar{20}{36}{14}{19}{11}{1} & 2.67 [2.47, 2.87] & 170 \\
G05 & trust my AI tools more than myself & \likertbar{18}{29}{17}{21}{15}{1} & 2.85 [2.64, 3.05] & 170 \\
G02 & concerned about becoming too reliant on them & \likertbar{8}{22}{12}{37}{21}{1} & 3.41 [3.22, 3.60] & 170 \\
G04 & the more experience, the more I let them do & \likertbar{5}{13}{14}{37}{31}{0} & 3.75 [3.57, 3.92] & 169 \\
G07 & willing to take moderate risks for high payoff & \likertbar{4}{12}{15}{49}{20}{0} & 3.68 [3.51, 3.84] & 164 \\
G06 & the tools I work with must be highly reliable & \likertbar{1}{6}{18}{33}{42}{0} & 4.08 [3.94, 4.23] & 168 \\
G01 & AI tools make me more productive & \likertbar{2}{2}{5}{31}{60}{0} & 4.45 [4.32, 4.58] & 167 \\
\bottomrule
\end{tabular}

\end{table}

As we see in Table~\ref{tab:items-attitude2}, our agent users overwhelmingly agree that their
AI tools make them more productive 
(G01: \svy{item/g01/users/shapehigh}\,\% agree or strongly agree).
A big majority is willing to take some risks to maximize the benefits 
(G07: \svy{item/g07/users/shapehigh}\,\%)
and reports a tendency to use them for more and more things the more experience they gain 
(G04: \svy{item/g04/users/shapehigh}\,\%).
The top three items show polarization, i.e., nearly everybody has a clear opinion:
Most respondents are concerned about becoming overly reliant on AI tools 
(G02: \svy{item/g02/users/shapehigh}\,\% agree or strongly agree), 
but others disagree
(G02: \svy{item/g02/users/shapelow}\,\% disagree or strongly disagree).
Most do not find AI coding tools to be over-hyped 
(G03: \svy{item/g03/users/shapelow}\,\%), but others do (G03: \svy{item/g03/users/shapehigh}\,\%).
Many state they often trust their tools more than themselves 
(G05: \svy{item/g05/users/shapehigh}\,\%), but more disagree (G05: \svy{item/g05/users/shapelow}\,\%).

\begin{figure}
  \caption{\gentag{fig-trust-scatter.tex}Trust index versus usage index, one mark per
    respondent, jittered to make all points visible. 
    \likertswatch{likertE}~circles are the \svy{trust/scatter/users} agent users,
    \likertswatch{likertA}~diamonds the \svy{trust/scatter/nonusers} non-users, and the open
    grey circles the \svy{trust/scatter/unclassified} respondents the split leaves
    undetermined.
    The line is a simple linear regression. $r = \svy{trust/scatter/r}$.}
  \label{fig:trust-scatter}
  \centering
\begin{tikzpicture}
\begin{axis}[trustscatter]
\addplot[trustregress] coordinates {(0,1.89705) (1,3.94328)};
\addplot[usagemarkuser] coordinates {(0.8607,4.5891) (0.9608,4.9431) (0.9979,4.8788) (0.813,1.9798) (0.5595,3.3892) (1.0317,3.6564) (0.9626,4.9189) (0.7729,3.853) (0.9883,4.358) (0.7758,2.6042) (0.7894,4.8492) (0.8401,1.0275) (0.8631,3.9067) (0.6042,4.1484) (0.9332,3.636) (1.0243,4.8078) (0.7369,4.5667) (0.7274,4.3618) (0.9517,4.3732) (0.7859,4.0948) (0.7771,2.6539) (0.8955,3.1062) (0.7047,3.2388) (0.6624,2.8489) (0.7189,2.9505) (0.8235,3.4686) (0.508,2.2077) (0.8952,4.4653) (0.9967,4.4924) (0.8963,4.141) (0.8576,4.0131) (1.0076,3.4663) (0.7799,3.6554) (1.0141,3.9424) (0.8378,4.9958) (0.6342,2.2546) (0.7238,3.0994) (0.7109,2.894) (0.6152,3.8588) (0.3213,3.2549) (1.0257,3.1632) (0.8947,3.3762) (0.7782,2.1454) (0.951,3.1868) (0.7793,3.4426) (1.013,4.079) (0.9198,3.4881) (0.8878,2.672) (0.8641,3.1169) (0.6733,3.1565) (0.9878,3.2424) (0.8461,2.2115) (0.5641,1.6609) (0.7796,2.6519) (0.5191,0.8957) (0.9145,4.8758) (0.6236,2.1084) (0.6208,2.0876) (0.7787,1.2923) (1.0069,4.3048) (0.7693,3.0972) (0.857,3.5739) (0.7673,2.7966) (0.7977,4.2288) (0.831,4.3595) (0.4945,1.9004) (0.991,2.9229) (0.7405,3.865) (0.7832,3.3443) (0.9595,2.6723) (0.8951,3.458) (0.9521,3.2825) (0.9578,2.9932) (0.6081,4.6265) (0.8641,4.0895) (0.8602,3.3233) (0.7291,2.488) (0.9346,1.591) (0.3038,3.9476) (0.9918,3.0123) (0.6821,2.5091) (0.8199,4.5093) (0.5083,1.8751) (0.9353,4.1299) (0.8365,4.3654) (0.9467,4.7755) (0.9594,4.7388) (0.9619,4.2068) (0.8474,4.6312) (0.8328,5.1071) (0.933,3.858) (1.0293,4.7952) (0.9388,3.1422) (0.9617,4.0896) (1.021,4.1328) (0.8391,4.8512) (1.0019,5.144) (0.97,4.0101) (1.0389,3.9357) (0.8941,3.817) (0.7748,4.5319) (0.9608,3.6579) (0.9128,4.3048) (0.8668,3.5918) (0.9584,4.8565) (1.0247,4.1515) (0.9836,3.8386) (0.8318,4.378) (0.8519,4.9057) (0.9692,1.6727) (0.9677,4.6812) (0.7828,4.3828) (0.8633,4.3868) (0.8128,4.2844) (0.8118,3.9639) (1.0372,5.1482) (0.785,4.388) (0.9507,3.8271) (0.9228,3.3573) (0.7141,2.1563) (0.4771,3.1688) (0.8144,3.8455) (0.7035,3.8938) (0.8464,3.4816) (0.7403,2.8789) (0.7029,2.8536) (0.5113,4.6172) (0.9277,3.3457) (0.784,4.0918) (1.0081,3.3855) (0.8981,4.4595) (0.9076,3.3879) (0.7973,2.6786) (0.8323,4.8707) (0.9866,3.8782) (0.9859,3.4582) (0.9633,3.1948) (0.5195,1.852) (0.9526,2.6512) (0.838,2.9061) (0.9682,5.0679) (0.9773,5.0252) (0.4493,2.6721) (1.0199,3.3521) (0.7594,3.7311) (0.4285,3.9256) (0.7124,3.0646) (0.7864,3.9028) (0.9797,4.0321) (0.8846,3.7929) (0.7368,2.9412) (0.9602,4.5387) (0.332,1.6249) (0.8589,2.5905) (0.857,3.5304) (0.8812,3.4018) (0.5948,3.2494) (0.8324,2.875) (1.0003,3.3257) (1.0319,3.7141) (1.0387,3.2519) (0.9342,3.6725) (0.9225,2.7855) (0.6728,4.2421) (0.9318,3.5396) (0.8493,1.4328) (0.5506,4.1168) (0.976,4.3872) (0.8104,3.683) (0.6895,2.4389) (0.4373,4.1716)};
\addplot[usagemarknonuser] coordinates {(0.2717,2.6298) (0.2229,1.8563) (0.2727,2.9842) (0.3502,2.6602) (0.2949,3.5826) (0.9628,4.0913) (0.4781,2.2574) (0.8318,3.1459) (0.1029,1.9187) (0.0025,1.7596) (0.1648,3.4876) (0.2359,3.8931) (0.0267,0.9478) (0.1516,1.0914) (0.073,1.0458) (0.1337,1.1252) (0.3157,4.146) (0.3502,1.3776) (-0.0007,1.6852) (0.2231,1.302) (0.596,1.909) (-0.0256,1.0885) (0.346,3.0687) (0.3042,2.5128) (0.2519,1.4186) (0.1119,1.4762) (0.4698,3.0943) (0.2284,3.3775) (0.1016,4.0565) (0.0279,1.2086) (0.477,3.1507) (0.6807,3.0404) (0.0433,1.7842) (0.4089,3.8777) (0.6787,3.1319) (0.9543,4.5887) (0.756,2.4952) (0.4946,3.3968) (0.1568,1.4938) (-0.0116,2.0002) (0.0167,1.7071) (0.1726,2.6176) (0.3236,3.3511) (0.0988,3.1983) (0.0307,1.1973) (0.1497,1.7891) (0.6594,2.3981) (0.0969,1.372) (0.2298,3.5262) (0.9311,4.176) (0.2547,2.2718) (0.2394,2.2974) (0.2638,1.2904) (0.2237,3.2149) (0.4424,3.7056) (0.3665,3.6951) (0.7863,4.1056) (0.0771,1.1331) (0.2774,3.41) (0.7037,4.1791) (0.6362,1.8757) (0.356,4.2253) (0.4087,1.8706) (0.0353,2.6292) (0.2316,3.7435)};
\addplot[usagemarkunknown] coordinates {(1.0257,2.0039)};
\end{axis}
\end{tikzpicture}
\end{figure}
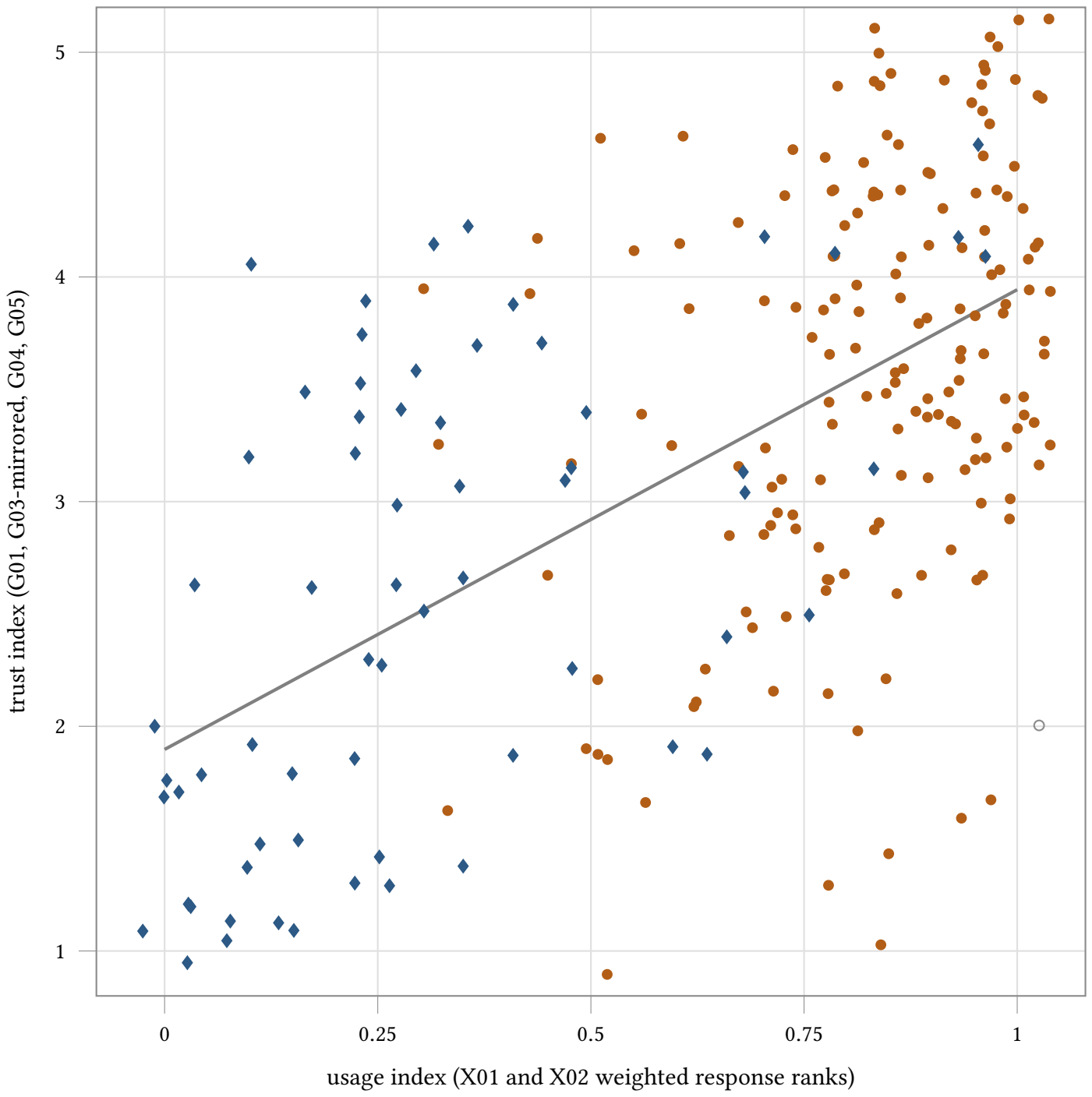

What about non-users?
A more granular question than a mere user/non-user split would be this: 
How do these attitudes depend on actual experience with agents?
G01, G03, G04, G05 together form our trust index.
If we plot this against the usage index at an individual level, we get 
Figure~\ref{fig:trust-scatter}.
We see a trend that trust increases with intensity of usage,
but the trend is not particularly strong.

\begin{figure}
  \caption{\gentag{fig-items-attitude-index.tex}
    AI usage index versus AI attitude. Each line is a loess local linear regression smoother fit 
    with a span of \svy{curve/span}, i.e. each point of the line is computed from that fraction
    of the data, giving data the more weight, the closer it lies to that x-value.
    The three error bars per item are that fit's own
    \svy{curve/level}\,\% interval, computed from \svy{curve/bootstrap} bootstrap resamples of the
    respondents, one for each third of the range and each drawn at that subgroup's mean index.
    Lower usage has fewer respondents, so the bars are wider there.}
  \label{fig:items-attitude-index}
  \centering
  \input{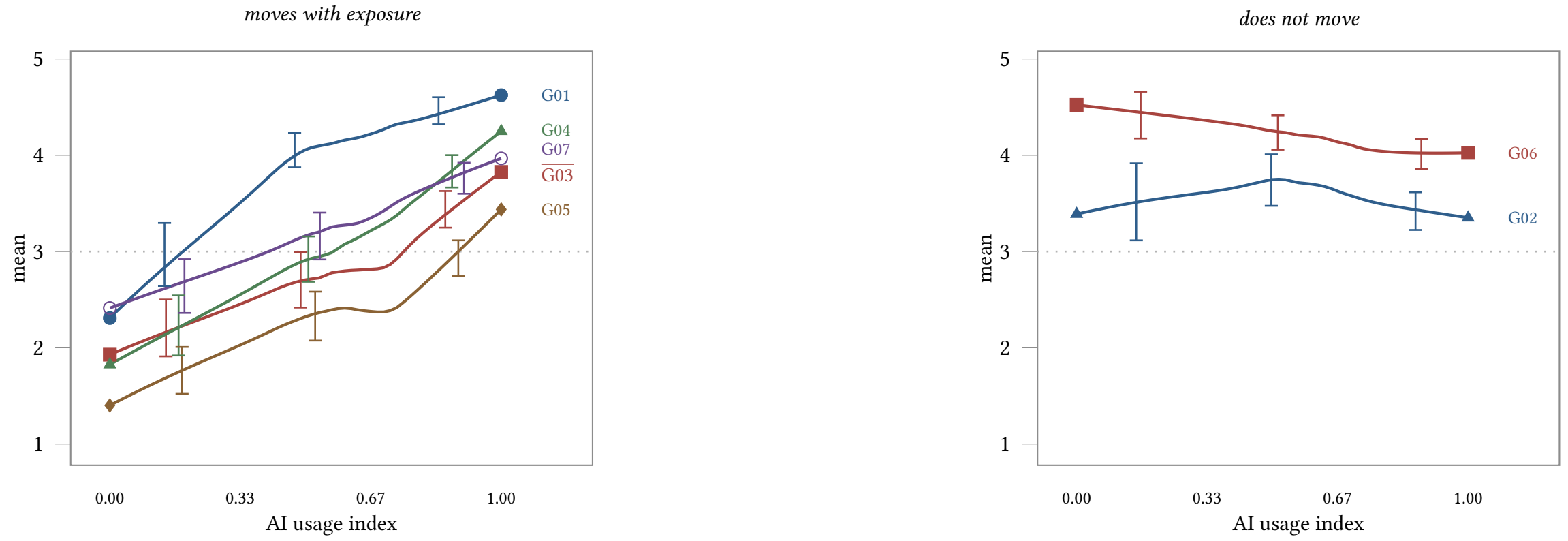}
\end{figure}

If we take away individual variation, however, and plot the trend for each attitude variable
separately, the trend becomes impressively clear; see Figure~\ref{fig:items-attitude-index}:
Five of the seven items strongly increase with increasing AI and agent usage, only two do not.

Of those five, G01 and G04 do so almost by definition, but for the others, the resulting
statements are interesting:
With more experience, respondents become more willing to take risks for more payoff (G07),
less and less find the AI enthusiasm (in software development!) to be a hype (G03),
and increasingly tend to trust the agent more than themselves (G05).

Discussion: There are certainly different ways to read these results, but we believe they typically mean
our respondents become less afraid of making the most of agentic tools (G07),
see more and more reasons why the enthusiasm about these tools is warranted (G03),
and learn to trust them and rely on them (G05).

The concern of becoming overly reliant on AI tools (G02) does \emph{not} change with
more usage.
We conjecture this reflects different views of the profession, where it is heading,
and how different people feel about that, but this is beyond the current work
and left to future research.  

\subsection{Delegation Comfort}

Our questions X04 and X05 specifically ask about working with agents.
What are you comfortable delegating to an agent completely with respect to 
the kind of work (X04) and the decision-making involved (X05)?

We can again evaluate this relative to the respondents' intensity of agent-use.
The outcomes are unremarkable: agent non-users (would) mostly delegate only small,
well-defined tasks that are purely mechanical or involve only modest coding.
Heavy agent users almost all delegate substantial coding tasks as well and also
much more of the decision-making.
To view the full granularity of these results, see
Figure~\ref{fig:deleg-tiers} in Appendix~\ref{app:result2-delegation}.

\subsection{Togetherness}\label{sec:results2-togetherness}

In pair programming, \emph{togetherness} describes how well the pair has achieved a joint
understanding of the product, goal, and process, i.e., how well each member
understands where the partner is with their understanding of those things \cite{ZiePre21,Zieris20}.

Low togetherness implies that explaining something to the partner will be cumbersome, misunderstandings abound,
and many proposals during the process will never get a proper response because 
it is too hard for the partner to understand them.
High togetherness, in contrast, may look almost like mind-reading to an outside observer:
Partners can often complete the other's sentences and understand even garbled utterances
effortlessly; the technical problem to be solved may be difficult, but the collaboration towards
solving it looks as if it posed no problem whatsoever.
(See also the description given to the respondents in the questionnaire as printed in
Appendix~\ref{app:method2-questionnaire-p4}.)

\begin{figure}
  \caption{\gentag{fig-togetherness.tex}How togetherness in human pair programming (P02)
    relates to togetherness in human/agent pair programming (P03) for the
    \svy{togetherness/n} agent users who answered all of P01--P03 and gave neither item's
    lowest-rank answer (``None'', ``I have not pair programmed'', or ``I hardly use coding
    agents'', respectively). 
    Bubble area is the number of respondents in that cell. 
    Bubble wedges are how much pair programming they have done
    (P01):
    \likertswatch{togtier0}~\svy{togetherness/a1/label},
    \likertswatch{togtier1}~\svy{togetherness/a2/label},
    \likertswatch{togtier2}~\svy{togetherness/a3/label},
    \likertswatch{togtier3}~\svy{togetherness/a4/label}.}
  \label{fig:togetherness}
  \centering
\begin{tikzpicture}
\begin{axis}[togchart,
  xtick={2,3,4,5,6},
  xticklabels={{low},{fairly low},{moderate},{high},{very high}},
  ytick={2,3,4,5,6},
  yticklabels={{low},{fairly low},{moderate},{high},{very high}}]
\addplot[delegdiagonal, forget plot] coordinates {(2,2) (6,6)};
\togdisc{2}{2}{2}{0}
\togdisc{2}{5}{1}{0}
\togdisc{3}{3}{1}{0}
\togwedge{3}{4}{4}{0}{3}{0}
\togwedge{3}{4}{4}{3}{4}{1}
\togwedge{3}{5}{5}{0}{4}{0}
\togwedge{3}{5}{5}{4}{5}{3}
\togwedge{4}{2}{3}{0}{1}{0}
\togwedge{4}{2}{3}{1}{2}{1}
\togwedge{4}{2}{3}{2}{3}{2}
\togwedge{4}{3}{6}{0}{3}{0}
\togwedge{4}{3}{6}{3}{4}{2}
\togwedge{4}{3}{6}{4}{6}{3}
\togwedge{4}{4}{15}{0}{9}{0}
\togwedge{4}{4}{15}{9}{14}{1}
\togwedge{4}{4}{15}{14}{15}{3}
\togwedge{4}{5}{27}{0}{11}{0}
\togwedge{4}{5}{27}{11}{19}{1}
\togwedge{4}{5}{27}{19}{24}{2}
\togwedge{4}{5}{27}{24}{27}{3}
\togwedge{4}{6}{4}{0}{2}{0}
\togwedge{4}{6}{4}{2}{4}{1}
\togdisc{5}{2}{1}{3}
\togwedge{5}{3}{11}{0}{2}{0}
\togwedge{5}{3}{11}{2}{3}{1}
\togwedge{5}{3}{11}{3}{6}{2}
\togwedge{5}{3}{11}{6}{11}{3}
\togwedge{5}{4}{25}{0}{5}{0}
\togwedge{5}{4}{25}{5}{16}{1}
\togwedge{5}{4}{25}{16}{20}{2}
\togwedge{5}{4}{25}{20}{25}{3}
\togwedge{5}{5}{17}{0}{6}{0}
\togwedge{5}{5}{17}{6}{11}{1}
\togwedge{5}{5}{17}{11}{12}{2}
\togwedge{5}{5}{17}{12}{17}{3}
\togwedge{5}{6}{4}{0}{2}{0}
\togwedge{5}{6}{4}{2}{3}{1}
\togwedge{5}{6}{4}{3}{4}{2}
\togdisc{6}{2}{1}{1}
\togwedge{6}{3}{2}{0}{1}{1}
\togwedge{6}{3}{2}{1}{2}{3}
\togwedge{6}{4}{6}{0}{2}{2}
\togwedge{6}{4}{6}{2}{6}{3}
\togwedge{6}{5}{3}{0}{1}{1}
\togwedge{6}{5}{3}{1}{2}{2}
\togwedge{6}{5}{3}{2}{3}{3}
\togdisc{6}{6}{2}{1}
\end{axis}
\end{tikzpicture}
\end{figure}
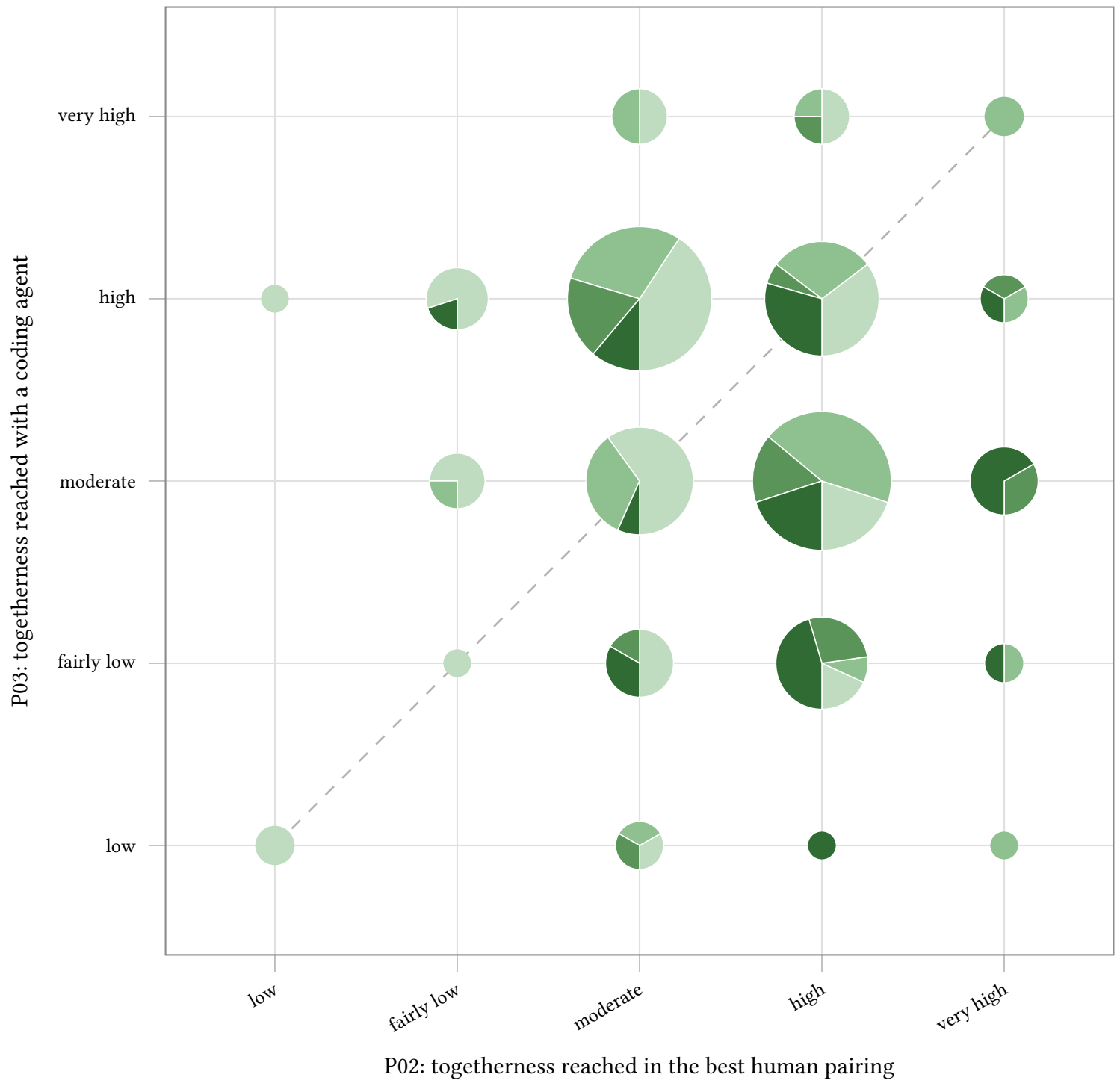

The three items on pair programming and togetherness in our questionnaire regard how
much pair programming experience the respondent has (P01)
and how good the best togetherness was they have reached with a fellow human (P02)
or with a coding agent (P03).
For the responses, see Figure~\ref{fig:togetherness}.
The number of data points in each piechart is not large, so we must be careful not to interpret
too much into what we see.

Therefore, let us focus on the four largest groups: 
Those with moderate or high human/human togetherness (columns 3 and 4 up the horizontal scale)
combined with moderate or high human/agent togetherness.
Two observations: 
First, there are more respondents for whom the best togetherness was different
between human/human and human/agent pair programming 
than there are respondents for whom they are the same (smaller bubbles on the diagonal).
Second, there are about as many for whom human/human pair programming went better
as there are others for whom human/agent pair programming went better.
The former have higher human/human pair programming experience on average.

Discussion: The interpretation of these data is not obvious, but it appears that collaboration
and togetherness are aspects of agentic coding that deserve research attention.
Our Step 3 investigation will shed some light on this, but does not use a togetherness
perspective specifically.

\section{Methods 3: Open Coding of Freetext Survey Responses}\label{sec:methods3}

A quantitative survey with fixed answer categories works well when the 
researcher can be reasonably sure of the categories in which the respondents
will be able and willing to think.

For the question ``What does collaboration with a coding agent mean for our respondents?'',
this was not the case for us, so we decided to capture this one via a freetext question,
hoping for sufficient number and quality of responses to feed a proper qualitative analysis.

\subsection{Question and Responses}

In our questionnaire, the question was situated just below the task-allocation inventory,
so our respondents had just thought about how they handled various individual types of work steps.
It was part of a block consisting of two definitions, two simple scenario questions,
and the freetext question. The block went like this:
\begin{quotation}
  Consider these work styles:\\
  DIVIDE: ``I want a clean division of labor: clear interfaces, each side doing what it’s best at, 
  hand-offs I can inspect.
  That leaves me free while the agent works, I have to judge the result.''\\
  COLLAB: ``I and the agent collaborate, also when the agent performs the larger amount of work. 
  We are solving the task hand-in-hand in a manner none of us could do alone. 
  That costs me continuous attention, but leaves me with few misunderstandings and a good grasp of the result.''
  
  Which style would you aim at for the following two tasks?\\
  Assume both are serious but not highly critical, you are using a coding agent for each, 
  and you will own the outcome.\\
  Answer options: 
  1 Clearly DIVIDE · 2 Rather DIVIDE · 3 Equally both · 4 Rather COLLAB · 5 Clearly COLLAB · N/A Not sure\\
  DC01 Clear what is wanted and easy to see when it is right; a change of a familiar kind in code you know well.\\
  DC02 A part of the system you do not know well or subtleties in the requirements; mistakes would not be obvious,
  and you will be maintaining the result.
  
  When would you want which, and why? 
  Please explain your view: sketch your desired work style, even if you do not yet achieve it in practice often; 
  what stands in the way, if anything.\\
  (A few sentences here would help us a lot—this is where we’ll learn the most.)\\
  DCc How I would like to work together with an agent:
\end{quotation}

The question resulted in \svy{item/dcc/n} responses.
The middle half of these (from first to third quartile) was
from \svy{item/dcc/length/q1} to \svy{item/dcc/length/q3} characters long and sufficiently rich for a valuable analysis.

\subsection{Analysis Approach: Agent-assisted Open Coding}\label{sec:methods3-agent-assistance}

As in Step 1, our approach was open coding in Strauss/Corbin GTM tradition~\cite{StrCor90}:
using open coding, constant comparison, memoing,
theoretical coding, and theoretical sensitivity.

Given the granularity of our data (short, focused, self-contained individual responses),
we decided to use our coding agent to help with the coding process:
The human was to provide initial coding families (as a codebook), conventions (in a memo),
and an initial list of ideas what to look for (explicated theoretical sensitivity, also in a memo)
and the agent would contribute most of the grunt work of constant comparison.
We report on our approach according to the guidelines of~\cite{BalAngAro25}.

We implemented agent support in the form of a small commandline-driven database application
\emph{codebook.py} and a Claude Code skill \emph{open-coding}, both available in the data package.
The actual coding as well as the implementation of \emph{codebook.py} was performed by Claude Opus 5.

\emph{codebook.py} provides CRUD (create, read, update, delete) functionality for codes
(the codebook), memos, and codings (applying a code to a contiguous stretch of text within a single
response).
It also supports switching between a representation of the data in YAML files,
which is useful for initial import, versioning, and manual inspection,
and another representation as a SQL database, allowing \emph{codebook.py} to perform
queries for responses (called ``blocks'' here) containing certain words or having certain
combinations of codes attached or not attached.
This makes the coding work traceable (via a log of write calls done with \emph{codebook.py})
and allows orderly work with more data than would fit into the LLM's context window.
The latter aspect is not crucial for our data here, but presumably allows leaner sessions
and hence improved LLM focus.
Find the list of commands supported by \emph{codebook.py} in Appendix~\ref{app:method3-codebook-py}.

The \emph{open-coding} skill is 450 lines long and explains
the nature and representation of the data and some corresponding ideas regarding theoretical sensitivity,
a three-step procedure the agent had to follow in each application of the skill,
including constraints (what not to do) and practical tips,
and a format convention for interacting with the human user.

A central idea of the skill was a four-step dialog process:
(1)~Each use of the skill would come with a task formulated by the user, the character of which was variable.
(2)~Based on the format convention, the agent would then prepare a text document with 
observations and suggestions, usually concerning code definitions as well as 
individual applications of codes to blocks.
(3)~The user would analyze this document
(called the ``batched review document'')
and write his reactions into it: accepting some suggestions,
rejecting or modifying others, and sometimes making some of his own.
(4)~Finally, the agent would execute the resulting suggestions on the database.

These Markdown documents were around 300 to 700 lines long and typically took
the user one to three hours to process.
A lot of work could be done in this manner in any single round:
it took only eight calls of the skill to get to the end results presented below.
The resulting process was exhausting for the user, but proceeded several times as fast
as manual coding would have, and resulted in a presumably higher consistency in applying the codes,
in particular around code modifications.

\subsection{Analysis Process}\label{sec:methods3-analysis-process}

We (specifically: Prechelt) started the analysis manually and only defined the above approach once we had 
immersed ourselves in the data and gained initial insights.

By the time the agent-supported main part of the work started, we had defined the main six
coding families (presented in Sections \ref{sec:results3-collab-codes} and \ref{sec:results3-divide-codes}) 
and a few helper codes and written the initial content
of memos 1 to 3 that concerned work status, conventions, and goals (to-dos).
During the first few rounds of analysis, the skill was improved and partially generalized,
resulting in two more memos of a more technical-procedural type.
The list of titles of the memos, seven overall, is given in Appendix~\ref{app:method3-memos-list},
the memos themselves are available in the data package.

The eight rounds of agent-supported coding had the following topics:
\begin{enumerate}
  \item Review and consolidate the three \Code{collab-*} code families that resulted from manual coding.
  \item Consolidating the various provisional codes for odd cases into a cohesive family.
  \item Review and consolidate the three \Code{divide-*} code families that resulted from manual coding.
  \item Applying the codes to the remaining blocks for completeness and validation. 
    Manual coding had previously worked on only about one third of the data available by now.
  \item Conceptualizing the personal preferences category the human user had noted in the data recently.
  \item Finally tackle the hypothesis about holistic notions of collaboration that had been
    waiting to be checked since round 1 (but required the concepts to be clearer than they were then).
  \item Applying the codes to 18 blocks of new data for completeness and validation. 
    This encountered no need for modifying the codebook, which can be considered evidence of 
    theoretical saturation~\cite{StrCor90}.
  \item Cleanup: While writing the article, we detected overly frequent uses of
    \Code{preference-*} codes and performed a coding round in which we tightened their use.
\end{enumerate}
The content of these steps will become more concrete in the results section.
Two examples of complete prompts used for the steps are given in Appendix~\ref{app:method3-example-prompts}.
The prompts and batched review documents of all eight rounds are available in the data package.

\subsection{Outcome Quality}\label{sec:methods3-outcome-quality}

In the steps the agent performed alone (constant comparison and applying the existing codes)
we relied on it fully. 
Our active quality management was restricted to general decisions (concept definitions)
and left specific decisions (applying a concept definition to an individual case) to the agent.

The results of this would not have been good enough for a quantitative content analysis:
For instance, we found the agent sometimes did not mark the proper stretch (rather too long or too short)
where the chosen code was appropriate.
It was sometimes more eager to make annotations than appropriate, which resulted in some
incorrect additional codings.
Neither of these appears to be frequent, though.

But for our exploratory analysis, these problems did not matter:
We only wanted to understand phenomena that existed in the responses;
frequencies are of almost no interest.
If a few of the instances the agent claimed to have found were questionable,
this was not a problem as long as there were enough others that were convincing.

\section{Results 3: Notions of Collaboration}\label{sec:results3}

\subsection{Quantitative Answers}

\begin{table}
  \caption{\gentag{tab-items-dc.tex}The two collaboration-style items.
  Many columns are as in Table~\ref{tab:items-users}, but distribution shows the
    DIVIDE\,/\,COLLAB scale: \likertlegenddividecollab.}
  \label{tab:items-dc}
  \centering\tablefont\setlength{\tabcolsep}{3.5pt}
\begin{tabular}{@{}llclrrrr@{}}
\toprule
Item & Situation & Distribution & $M$ [95\,\% CI] & $n$ & \multicolumn{3}{c}{Region (\%)} \\
\cmidrule(lr){6-8}
 & & {\scriptsize DIVIDE\,$\leftrightarrow$\,COLLAB} & & & D & E & C \\
\midrule
\addlinespace\multicolumn{8}{@{}l}{\textbf{Users (n = 171)}} \\
DC01 & style for a familiar, easily checked task & \likertbar{31}{28}{7}{16}{18}{1} & 2.64 [2.40, 2.87] & 160 & 59 & 7 & 34 \\
DC02 & style for an unfamiliar, subtle task & \likertbar{6}{9}{7}{32}{46}{0} & 4.04 [3.86, 4.23] & 161 & 15 & 7 & 78 \\
\addlinespace\multicolumn{8}{@{}l}{\textbf{Non-users (n = 69)}} \\
DC01 & style for a familiar, easily checked task & \likertbar{19}{48}{2}{19}{12}{1} & 2.57 [2.22, 2.92] & 58 & 67 & 2 & 31 \\
DC02 & style for an unfamiliar, subtle task & \likertbar{12}{11}{10}{21}{46}{1} & 3.77 [3.39, 4.15] & 57 & 23 & 10 & 67 \\
\bottomrule
\end{tabular}

\end{table}

As seen in Table~\ref{tab:items-dc},
even for the simple, non-risky task (DC01), roughly a third of users prefers to collaborate
instead of just delegating it to their agent; slightly less for the non-users.
And even for the difficult task (DC02) some users (and many more non-users) would \emph{not} collaborate
and rather (presumably) solve the task without much agent help.

Discussion: These answers suggest that our respondents may hold fundamentally (rather than only gradually) 
different views about collaboration with coding agents.
There may be a lot to find in our qualitative analysis.

\subsection{Engineering considerations 1: The \Code{collab-*} code 
families}\label{sec:results3-collab-codes}

The first three families of codes were found immediately during the initial
manual analysis steps because the theoretical sensitivity for them was high:
We firmly expected that respondents would be talking about \emph{conditions} for collaboration,
\emph{means} of collaboration, and \emph{goals} for the collaboration,
leading to the code families 
\Code{collab-if-*} (5 codes at the end of the analysis), 
\Code{collab-by-*} (8 codes), and 
\Code{collab-for-*} (9 codes).

The full list of codes is presented in Appendix~\ref{app:result3-collab-codes}
along with the number of instances and length of definition for each.
The actual codings and concept definitions are available in the data package.
One example concept definition is provided in Appendix~\ref{app:method3-codedef-example}.

In the main article text, we will illustrate the code families by examples
of DCc responses to which they apply, preferring frequent codes over rarer ones.

So what does collaboration look like according to our respondents?
The most frequent response is question-and-answer (\Code{collab-by-interview}):

\begin{response}{2}{15}
\coded{collab-for-codeunderstand}{I often collaborate by first letting the agent propose or implement
a prototype and then
\codedhigh{collab-by-interview}{asking many questions about it to the agent until I understand the
implementation and processes}.} \gap
\end{response}

We will use this notation throughout.
The block number provides a coordinate in the data package,
the highlighted coded stretch is the one in focus,
other coded stretches provide context and may be referred to later in the article.

The second-most common one amounts to ``I am the boss''  (\Code{collab-by-humanauthority}):

\begin{response}{54}{3111}
\gap \codedhigh{collab-by-humanauthority}{I use it as an assistant and advisor
\coded{collab-if-unfamiliar}{for more complex or unfamiliar tasks}, but I always have the final
say.} \gap
\end{response}

\Code{collab-by-approval} is on third rank:

\begin{response}{117}{4665} 
\coded{collab-for-codeunderstand}{I would like to understand the whole code-base},
\codedhigh{collab-by-approval}{using the agent as making suggestions that i then can choose to
accept}.
Ideally I would have an idea of what i want to do and
\coded{collab-by-partitioning}{then figure out how to do it in small steps}. \gap
\\
\coded{collab-if-agentunreliable}{I am still a bit distrustful of AI} and are
\coded{constraint-agentobstacle}{hesitant to pay money for it}.
\coded{preference-keepskill}{I also have the fear that one becomes too dependent.}
\end{response}

These examples also partially illustrate the \Code{collab-if-*} and \Code{collab-for-*} families:
Conditions under which people want to COLLAB are for example 
if they are unfamiliar with the codebase or the technology (\ref{block3111}) or
if they expect the agent's work is not going to be reliable (\Code{collab-if-agentunreliable}).
Goals for which they want to COLLAB are for example
for making sure they understand the code that was produced (\ref{block15}),
for learning about the technology, the domain, or good practice (\Code{collab-for-learning}), or
for making sure the work stays aimed at the intended goal (\Code{collab-for-steering}).

Togetherness is often a goal of collaboration, but rarely named as such explicitly.
More often, it is implicit, e.g. in
\Code{collab-for-requnderstand} (see in \ref{block4998}, \ref{block228}), 
\Code{collab-for-codeunderstand} (\ref{block15}, \ref{block4665}, \ref{block228}), and
\Code{collab-for-ownership} or 
is achieved as a natural side-effect of the chosen procedure as in
\Code{collab-by-interview}  (\ref{block15}) and
\Code{collab-by-tradeoffdialog} (\ref{block3599}).

Only sometimes is the togetherness idea mentioned explicitly:
\begin{response}{57}{3243}
\gap 
\coded{collab-if-unfamiliar}{In exploratory scenarios where I am trying to gain a better
understanding of an unfamiliar codebase or requirements are vague I genereally take a more
collaborative approach with the agent} and \coded{collab-for-codeunderstand}{will let the agent do
the heavy lifting of providing a summary of the code and only review the evidence it produces} and
then, \codedhigh{collab-for-togetherness}{based on this mutual understanding of the code},
\coded{collab-for-split}{we will refine and groom the task in order to
\coded{divide-by-phase}{derive an implementation plan that I can confidently hand off to the
agent}}.
\gap
\end{response}
See also \ref{block4998} below.
Generally, examples often contain codes from families discussed in earlier or later sections.

\subsection{Engineering considerations 2: The \Code{divide-*} code families}\label{sec:results3-divide-codes}

Many responses would also involve statements regarding when, how, or what-for to use DIVIDE mode.
We think there are two reasons for this.
One is that this often makes it easier to characterize when, how, or what-for to use COLLAB mode.
The other will be discussed in Section~\ref{sec:results3-holistic-codes}.

The conditions are straightforward:
DIVIDE if there is nearly no ambiguity (\Code{divide-if-clear}),
if the result can easily be validated (\Code{divide-if-checkable}), etc.

The goals are also mostly obvious:
DIVIDE for sheer development speed (\Code{divide-for-speed}),
for being able to do something else in the meantime (\Code{divide-for-parallel}),
for prototyping and trying out things (\Code{divide-for-probe}), or
for avoiding uninteresting busywork (\Code{divide-for-nondrudgery}).

How respondents phrase the means, however, is highly informative for our research goal.
The most frequent entry is a clearly phased development (\Code{divide-by-phase}) 
as seen in \ref{block3243} or, in a rather tidy formulation, here:

\begin{response}{143}{4998}
\coded{collab-for-togetherness}{Working together with the AI to reach
\coded{collab-for-requnderstand}{the same understanding of the problem and the desired solution}}.
\codedhigh{divide-by-phase}{Then the Agent implements it on its own and I review it}
\end{response}

A very different, almost opposite perspective is next: The agent as a code reviewer
(\Code{divide-by-agentreview}).
See this very rich response of somebody who wants to retain a high amount of control:

\begin{response}{16}{228}  
\coded{divide-plain}{I want to use an agent to do simple, repetitive work} as well as
\coded{collab-for-codeunderstand}{help me understand an existing codebase}.
\coded{collab-for-requnderstand}{It could also help in understanding tricky tasks and what might be
important to consider.}
\codedhigh{divide-by-agentreview}{I also like to let an agent do the first code review before
handing a PR over to a human colleague for review}, this way I can eliminate typos, small mistakes
or missing test cases.
But despite all the help, improvements and efficiency an agent can provide,
\coded{solo-for-exercise}{
\coded{solo-work}{I still want to be able to work on something myself
\coded{constraint-socialpressure}{without being judged for working slower than with an agent}.}
I want this because it is the only way to make sure I improve my
knowledge and skills about coding and will be able to judge what an agent produces in the future.}
\end{response}

Ranked third (and everything beyond is far less frequent than these three) are statements
in the general direction of spec-driven development (\Code{divide-by-spec}):

\begin{response}{69}{3599}
I prefer an asynchronous DIVIDE model:
\codedhigh{divide-by-spec}{I state the outcome and non-negotiable constraints},
\coded{divide-for-parallel}{then the agent researches, plans, implements, and verifies without
continuous supervision}.
\coded{collab-if-unfamiliar}{For unfamiliar or subtle work},
\coded{collab-by-tradeoffdialog}{I want the agent to surface evidence, assumptions, and a small
number of material decisions at checkpoints rather than pair step-by-step}.
\coded{collab-if-agentunreliable}{The main blockers are reliability, context loss, and agents that
stop at partial scaffolds or ask questions the repository could answer};
\coded{divide-by-guardrails}{persistent project instructions, broad tool access, and behavioral
verification make autonomy practical}.
\end{response}

Find the full list of these codes in Appendix~\ref{app:result3-divide-codes}.

\subsection{The standing of the answer: The \Code{constraint-*} family}

Our question had asked for an ideal, desired work style.
Some respondents remark that their response is provisional
because they have not yet settled on a work style (\Code{constraint-unsettledstyle})
or because the agents' capabilities keep improving (\Code{constraint-agentsgetbetter}). 
Some remark their ideal cannot be attained due to other influences:
limited agent accessibility (\Code{constraint-agentobstacle}, see~\ref{block4665}, \ref{block5144}),
limited human discipline or attention (\Code{constraint-selfdiscipline}),
or expectations of managers or colleagues (\Code{constraint-socialpressure}, see~\ref{block228}).
Find the full list of these codes in Appendix~\ref{app:result3-constraints}.

\subsection{Odd Bits: Solo Work, Question Refusal, Non-answers}

20 responses did not quite fit into the frame of our analysis.
Of these, 7 talked about work the respondent wanted to do alone, in neither a COLLAB nor a DIVIDE manner
(\Code{solo-work}, see~\ref{block228}).
5 talked partially or completely about something other than what we had asked.
8 non-users refused the question.
The responses involving \Code{solo-work} were used in our analysis, the others were not.

\subsection{Non-Engineering considerations: The \Code{preference-*} family}\label{sec:results3-preference-codes}

Many of the statements you have already seen above and generally of the
\Code{collab-*} and \Code{divide-*} kinds are worded as involving the respondent's wishes.
Semantically, however, they all argue on rational grounds, hence the term engineering considerations.
There are, however, exceptions where no such justification of the wish is given,
making the statement a pure personal preference.

\ref{block5363} shows two such codes:
\Code{preference-making} is about wanting to be the one who creates the software,
\linebreak\Code{preference-worksatisfaction} is about what agent use does to the enjoyment of the work.

\begin{response}{153}{5363}
\coded{constraint-unsettledstyle}{Right now I am unsure about the whole development lifecycle.
While I see benefits in using agents and delegating whole tasks to it,
\coded{collab-for-ownership}{it limits my own understanding of the existing solution and I fear to
loose required knowledge}.
\\
\codedhigh{preference-making}{In addition it interferes with my personal taste, I like to be a
developer, and I see software as craftsmanship.
What is my value by only defining some tasks.}
\coded{constraint-selfdiscipline}{While still being responsible I have to motivate me to read code
and concepts of another identity (the agent) and judge.
\codedhigh{preference-worksatisfaction}{This quickly gets boring, repetitive} and leads to a
situation where I just sign a contract where the machine says it works.}}
\end{response}

There are five such codes;
all are rare: together they occur in only 14 of the 154 responses.

Some of the preference responses sound as if the respondent finds the personal preference enough of a justification:
\begin{response}{80}{4405}
\codedhigh{preference-disengagement}{I want to be as high up the org chart as possible. 
I've been coding for almost 30 years and I never want to write a single line again.}
\end{response}

For the authors of others, there may be an engineering argument behind, just not stated explicitly:
\begin{response}{149}{5144}
\codedhigh{preference-worksatisfaction}{I find it more satisfying when I can articulate the end goal
completely on my own, without any assistance from the agent.
\coded{collab-by-humanauthorship}{In a sense, I feel like I use my agent to type up what I think.}
This gives me the satisfaction of coding too.}
\gap
\end{response}

Find the full list of these codes in Appendix~\ref{app:result3-preferences}.

\subsection{Holistic versus activity-centered notions of collaboration: The \Code{holistic-*} family}\label{sec:results3-holistic-codes}

Our question only asked about working \emph{together} with an agent, 
so why do \Code{divide-*} statements even appear?
The answer to this question so far had been that they usually serve as a rhetorical device 
that makes it easier to explain the collaboration by contrasting it with something else:
``I collaborate on X, but delegate Y.''

Some statements, however, are not like this.
They mention delegation for \emph{completeness}, not contrast, because they state a comprehensive idea
of a development process.
Concise example:

\begin{response}{86}{4467}
\coded{divide-by-phase}{
\codedhigh{holistic-process}{
\coded{collab-for-split}{Chat, plan, establish execution steps}, define prompt, let 'er rip, review
outcome, human test, wash rinse repeat}}
\end{response}

This respondent appears to have a very different notion of collaboration:
It is not the individual activity done together, it is the entire role-split over
all activities that are required.

Here is another example of this perspective, spelling it out in much more detail:

\begin{response}{90}{4496}
\coded{divide-by-phase}{
\codedhigh{holistic-process}{My desired work style on larger projects is structured in several
phases:\\
1) Setup and environment to work together with the agent.\\
\coded{divide-by-guardrails}{That involves setting guidelines for the interactions with the agent and
the style of edits, and making a very crude summary of what the project is about. Write this into
markdown files, accessible at any time. Then setup a coding environment with linters etc.} (COLLAB)\\
2) \coded{collab-for-split}{Sketch out the general structure.\\
What are the goals, what are we going to
do to achieve them, what is the overall structure of the code or project.} Write this into markdown
files, accessible at any time. (COLLAB)\\
3) Start coding.\\
\coded{collab-for-codeunderstand}{Write a minimal version of the code that can run at least part of
my application and that I can understand easily as a human.} (COLLAB)\\
4) Iterate.\\
\coded{divide-by-spec}{Define small tasks and feature requests that are
\coded{divide-if-checkable}{systematically checkable} and hand them to sub-agents.} (DIVIDE)\\
5) Crosscheck and document.\\
\coded{divide-by-agentreview}{Use the agent to flag any inconsistencies or risks that may come up in
the code base and test unexplored paths.} Document the features and how they work together. If I'm
implementing algorithms from papers, go back and crosscheck that the algorithms actually match the
math, after the full implementation. (COLLAB)}}
\end{response}
Note how steps 1 and 5 talk about a \Code{divide-*} concept, but the respondent views them
as COLLAB. That is a sign of their holistic perspective: 
Our code for step 1 focuses on the DIVIDE enabled by the guardrails, but the respondent's attention
is on the collaboration in \emph{defining} those guardrails in the first place.
Our code for step 5 focuses on the DIVIDE represented by assigning the review task to the agent,
but the respondent emphasizes the larger context in which this is happening.

This holistic type of perspective can also take different forms than a process model.
\Code{holistic-fraction} are short statements that focus on \emph{how much} of the work the agent is doing:

\begin{response}{84}{4434}
\coded{preference-disengagement}{
\codedhigh{holistic-fraction}{The dream is 0\% collab 100\% divide.} I do not want to have more
conversations with more people to get work done. I want the agent to capture my intent and do the job
so I can be freed from the shackles of my screens}
\end{response}

\begin{response}{110}{4624}
\gap
\codedhigh{holistic-fraction}{I am spending most of my time as Product Owner/QA while being able to
outsource my full stack knowledge through agents.}
\end{response}

\Code{holistic-roleshift} statements focus on how the developer's role has changed:

\begin{response}{64}{3484}
\coded{divide-by-orchestration}{
\codedhigh{holistic-roleshift}{In the future i will probably be more of a orchestrator that has
enough knowledge to build the agent pipeline and
\coded{divide-by-spec}{enough domain knowledge to work out clear requirements that the agent can then
implement by itself}. I will work more on the process than in the process.}}
\end{response}

\begin{response}{11}{83}
In genereal,
\coded{divide-if-clear}{if the requirements of a task or the desired outcome is clear for me} and
\coded{divide-if-checkable}{i clearly know how to test/see the outcome}, i would fully DIVIDE,
\coded{divide-for-parallel}{most of the times even in multiple projects at the same time}.
\codedhigh{holistic-roleshift}{That shifts my day to day work from only implementing features to a
broader understanding of the multiple projects i work on}, and
\coded{divide-by-spec}{it requires me to clearly say what i want the agent to achieve}.
(multiple projects here can also be one project with multiple worktrees to develop at both)
\\
\coded{collab-if-unfamiliar}{For systems i dont know, i like to use it for exploration, but there
its more COLLAB.} \coded{collab-for-codeunderstand}{I let the agent explore the codebase, but
together with me, so I get a better understanding to be able to work in the way above.}
\end{response}

Find the codes list in Appendix~\ref{app:result3-holistic} and note how much longer the definitions
are compared to other families.
We do not claim to understand the border well between the local view of collaboration
and the holistic view, but we are sure it is an important distinction.

\section{Threats to Validity}\label{sec:threats-validity}

Step 1 data: The set of practitioners and sources we picked may have been a misleading one.
This appears unlikely, given how well the results of Step 1 and Step 2 fit together.

Step 1 analysis: The manner in which we have sliced and diced the data into our categories
is not canonical.
The problem is only mild, because the categories themselves do not bear much load.
Bad categories would mean that the observations within each are too inhomogeneous for
interpretation or that our survey respondents have difficulty making sense of our items.
Neither danger has materialized.

Step 2 data: There is no well-defined base population from which we have gathered
our data and hence no representativeness.
This would be a problem if we attempted to make precise quantitative statements
that are only useful when they are generalizable.
However, we make no such statements.
Rough statements (which we do make) are likely robust, as the comparison between
the German versus non-German respondents in Section~\ref{sec:results2-german-nongerman} shows.

Step 2 analysis: There are many more ways in which these data could have been analyzed.
But our conclusions do not rely on souped-up statistical analysis for which the consequences
of using different approaches might be difficult to assess. 
Rather, our analyses are all straightforward and stay close to the immediate questions.

Step 3 data: Respondents who have not provided freetext may hold a very different view.
Yes, our work only explains views we have found, it does not claim these are the only
ones in existence.

Step 3 data: These responses lack detail, compared to an interview.
Yes. There is surely more to say about collaboration than we do here.
Our research is only the first step.
Still, we do not expect more detailed research to invalidate any of the statements
made herein.

Step 3 analysis: There are many more ways in which these data could be conceptualized.
This is correct, but not a threat.
The purpose of such a qualitative analysis is not ruling out other analyses.
Rather, it must be meaningful and intersubjectively credible.
In the terms of Tracy \cite{Tracy10}, it should have a Worthy Topic, make a
Significant Contribution, and create Resonance and it should possess Rich Rigor and Credibility.
We present enough detail that our readers can judge meaningfulness and credibility themselves.
Employing a coding agent for the open coding process greatly accelerated our work
and allowed it to be more comprehensive than it might have been otherwise.
It also resulted in codings that are not always precise.
This, however, is unproblematic: Our results do not depend on the individual codings and their
quantitative distribution, but only on discovering and correctly characterizing the 
relevant phenomena.

\section{Conclusions}\label{sec:conclusions}

\subsection{How have early adopters of agentic coding split roles between them and a coding agent in 2025?}\label{sec:conclusion-step1}

As seen in Section~\ref{sec:results1}, especially~\ref{sec:results1-summary},
they draw a sharp line of separation between them and the agent. 
Assigning full accountability to themselves, they perform the entire intent formation (requirements definition),
design operationalization (defining the ground on which an agent can then act),
and critical evaluation of outcomes.
The agent performs what may be the larger volume of work, but in a supporting role only.

\subsection{How do developers split roles between them and a coding agent in 2026?}\label{sec:conclusion-step2}

As seen in Section~\ref{sec:results2},
the attitude towards accountability, a few dissenters notwithstanding, 
is the same (Section~\ref{sec:results2-accountability}).
The role-split is fundamentally the same, but gradually quite different:
In Table~\ref{tab:items-users}, many of the items on the strongly ``human'' side of the scale
have large fractions of responses on levels 2 and 3 (mostly the human or human and agent about equally);
we are seeing a lot more agent involvement.
This is easily explainable by two items from Table~\ref{tab:items-attitude2}:
Respondents overwhelmingly report their agents make them more productive and
a huge majority reports the tendency that the more experience with agents they gain,
the more work they hand over.
This attitude, however, is dependent on actual use of agentic coding;
non-users are more reserved (Figure~\ref{fig:items-attitude-index}).

We expect this trend to continue: As agent capabilities continue to increase,
the human developers will likely enlist more and more agent support (and hence involvement) 
in the steps the core of which will have to remain human (for some time at least).

The general order of more-humanish vs. more-agentish tasks is roughly the same for the non-users
(Table~\ref{tab:items-nonusers}), especially near top and bottom, but, unsurprisingly, 
non-users of agents are much less inclined to hand over work to agents (bottom items).  
Teams wanting to move to agent-centric work styles may have a lot of convincing to do.

\subsection{Research method: How can we employ coding agents to support exploratory analysis of 
  survey freetext responses?}\label{sec:conclusion-methods3}

Our attempt at building agent assistance resulted in a relatively simple setup
using one skill and one modest Python script using a local SQLite DB
(Section~\ref{sec:methods3-agent-assistance}).
This allowed an analysis process in which
the human remained in complete control of the general direction,
could easily inject ideas about new code families, and 
would supervise the agent's work throughout,
yet the agent performed the constant comparison busywork, 
flagged needs for additional codes within a family,
made (mostly good) suggestions for such codes,
and applied the existing codes (Section~\ref{sec:methods3-analysis-process}).
We doubt the setup is precise enough for a quantitative content analysis
(Section~\ref{sec:methods3-outcome-quality}),
but for our purposes of exploratory analysis here, it was highly productive and helpful.

\subsection{What is developers' notion of ``collaboration'' with a coding agent?}\label{sec:conclusion-step3}

Regarding how developers think about what collaboration with a coding agent
should be like, we found three facets.
The mainstream manner is thinking in terms of individual work steps, which are characterized by
the means applied (mode of collaboration, nature of work step), 
the conditions when to collaborate in this way,
and the engineering reasons for and goals of doing it 
(Sections \ref{sec:results3-collab-codes} and \ref{sec:results3-divide-codes}).

Second, a few respondents (14 of 154) complement these engineering considerations with personal ones and formulate
personal preferences they also want to see fulfilled (Section~\ref{sec:results3-preference-codes}).

Third, some respondents also or primarily hold a \emph{holistic} view of collaboration:
For describing how they want to collaborate, they sketch an entire development process
or characterize their role as the owner and controller of that process (Section~\ref{sec:results3-holistic-codes}).

\subsection{Outlook}

The early adopters in Step 1 of our work largely shared the same view of agentic coding:
This is a powerful tool that should be utilized intensely, but needs careful control for good results.

Just one year later, there is a lot more diversity of views.
The Step 2 data show a lot of individual variation in the what-for and how-much of adoption, 
as well as a chasm between users and non-users with respect to their general attitude.
The semantic analysis of Step 3 shows there are different views of what collaboration with
a coding agent even is.
In particular, the holistic view implies that agentic coding may dominate the team process instead
of being discretionary developer tool use and
some of the personal styles may contradict what this new team process is going to demand.

We expect both phenomena to become a source of tension and possible conflict for teams.
We conjecture that ``roles'' and ``collaboration'' will be useful theoretical lenses
for future research into the resulting team dynamics.
We furthermore conjecture that ``togetherness'' between developer and agent may be 
a goal that most participants agree to be worth pursuing.

\subsection{``collaboration'' and ``togetherness'' instead of ``ceding autonomy''}

Choudhuri et al. \cite{ChoBirBad26} pursue the same mission as our work:
Characterizing the human/agent role-split (they do not use that term).
Where we proceed bottom-up, from practitioner writings,
they proceed top-down, from existing psychological models.
They formulate the issue as one of ceding autonomy to AI (or not).
Two remarks on that.

First, we believe ``collaboration with AI'' is a more useful
theoretical lens for teams in practice,
because ceding autonomy sounds like only a single degree of freedom,
whereas, as we have seen above, collaboration offers a multitude of
dimensions.

Second, we believe the very notion of AI autonomy is a dead end.
Agents are tools. They follow the developer's prompt.
But they are also complex adaptive systems \cite{Lansing03}
and hence their behavior is difficult to predict.
The central engineering issue in agentic coding is
not granting or denying agent autonomy, it is whether the agent understands the
developer's prompt as intended and whether the developer
understands the consequences in the resulting code.
The approach for solving both of these is striving for
higher togetherness between developer and agent.

\clearpage  
\begin{acks}
We thank our many survey respondents.
\end{acks}

\bibliographystyle{ACM-Reference-Format}
\bibliography{special.bib}
\label{end-of-article}

\clearpage
\appendix
\section*{Appendix}

\section{Step 1: Practitioner Blog Analysis Details}\label{app:result1}

Here, we provide the most pertinent quotes from the four studied practitioners that 
support the findings of our first qualitative analysis (Step 1).
The structure is aligned to the clusters reported in Section~\ref{sec:results1}. 
Where we used our findings to create a survey question for Step 2 (see Section~\ref{sec:methods2})
we directly refer to the survey item (T01--T25).
Those usually come with a prediction (A, H, E) that is supported by our qualitative findings. 
If the qualitative findings are inconclusive and therefore do not support a prediction we briefly 
discuss why. 
Where our qualitative findings did not serve as a basis for a task-allocation item, we only list 
the quotes without any further discussion. 

\subsection{Cluster: Intent Formation}

\subsubsection{Specification, Requirements, and Architectural Direction (C, Human/Agent)}\label{app:result1-category-C}


\paragraph{T01 \pred{H}: Intent and requirement definition.}

Ronacher and Willison explicitly discuss setting a goal and defining requirements as an activity 
they perform (a human activity). 
Hashimoto does so more implicitly as they discuss a past project where they started off with making 
a plan. 
Steinberger discusses how they come up with initial ideas but that they are \say{woefully under-spec}.

All of them either directly or indirectly show that the initial idea / intent comes from them. 
We therefore predict that our practitioners would lean to the human side for this task-allocation 
item. 

\begin{itemize}
  \item \practquote{AR90-P5}{I set a high bar, especially that I can operate it reliably.
    I've run similar systems before and knew what I wanted.}~\cite{Ronacher25ninety}
  \item \practquote{AR90-P7}{I pay close attention to the fundamentals of system
    architecture, code layout, and database interaction. I'm incredibly opinionated. As a
    result, there are certain things I don't let the AI do. I know it won't reach the point
    where I could sign off on a commit.}~\cite{Ronacher25ninety}
  \item \practquote{SIMDL-P5}{If you can reduce your problem to a clear goal and
    a set of tools that can iterate towards that goal a coding agent can often brute force
    its way to an effective solution.}~\cite{Willison25loops}
  \item \practquote{SIMDL-P49}{Not every problem responds well to this pattern of working.
  	The thing to look out for here are problems with clear success criteria where finding a
  	good solution is likely to involve (potentially slightly tedious) trial and
  	error.}~\cite{Willison25loops}
  \item \practquote{SIMV-P9}{Planning in advance. Sitting down to hack something together goes much 
  better if you start with a high level plan.}~\cite{Willison25vibe}
  \item \practquote{SIMV-P10}{You're not just responsible for writing the code---you're
    researching approaches, deciding on high-level architecture, writing specifications,
    defining success criteria, designing agentic loops, planning QA, [\dots] and spending
    so much time on code review.}~\cite{Willison25vibe}
  \item \practquote{SIMP-P23}{Code that started from your own specification is a lot less
    effort to review. If you already decided what to solve, picked the approach and worked
    out a detailed specification for the work itself, confirming it was built to your needs
    can take a lot less time.}~\cite{Willison25parallel}
  \item \practquote{MIT-P15}{I began by coming up with a rough plan of how I wanted this to 
  work.}~\cite{Hashimoto25ghostty}
  \item \practquote{PETJ-P76}{The far more fun approach is when I do UI-based work. I often start 
  with sth simple and woefully under-spec my requests, and watch the model build and see the 
  browser update in real time. Then I queue in additional changes and iterate on the 
  feature.}~\cite{Steinberger25talk}
  \item \practquote{PETJ-P77}{Often I get ideas for related interactions and iterate on other parts 
  as well while I build it, that work I do in a different agent.}~\cite{Steinberger25talk}
\end{itemize}

\paragraph{Architecture, stack, and boundary decisions.}
\begin{itemize}
  \item \practquote{PETM-P19}{The hardest part is distributed system design, picking the
    right dependencies, platforms and a forward-thinking database schema.}~\cite{Steinberger25workflow}
  \item \practquote{PETS-P9}{The important decisions these days are language/ecosystem and
    dependencies. [\dots] Go wasn't something I gave even the slightest thought even a few
    months ago, but eventually I played around and found that agents are really great at
    writing it, and its simple type system makes linting fast.}~\cite{Steinberger25speed}
  \item \practquote{PETJ-P101}{Just because I don't write the code anymore doesn't
   	mean I don't think hard about architecture, system design, dependencies, features or how
   	to delight users.}~\cite{Steinberger25talk}
\end{itemize}

\paragraph{T03 \pred{E}: Drafting and critiquing specifications.}

Three practitioners explicitly discuss drafting and critiquing a specification as a shared activity:
Ronacher describes a \say{back-and-forth}, Hashimoto writes about creating a plan 
\say{interactively with an agent}, Steinberger writes about a \say{discussion with codex}.
Willison arguably leans more towards the human side than the other three, as his statements imply that 
the specification should come from the human as he talks about the benefits of code that started 
from \say{your own specification}.

We do not see this as a complete contradiction to the other three practitioners and therefore  
predict that our practitioners in sum would lean towards the center (E) for this 
task-allocation item.

\begin{itemize}
  \item \practquote{AR90-P10}{I began in the traditional way: system design, schema, architecture. 
  At this state I don't let the AI write, but I loop it in AI as a kind of rubber duck. The 
  back-and-forth helps me see mistakes, even if I don't need or trust the 
  answers.}~\cite{Ronacher25ninety}
  \item \practquote{MIT-P25--26}{Another thing to notice is that I ask it to only create a plan and 
  not to write any code. Since this is a relatively vague request, it's important that I review a 
  plan before it goes off and does a ton of work (and spends a ton of tokens doing it). Tip: 
  Creating a comprehensive plan interactively with an agent is a really important first-step for 
  anything non-trivial.}~\cite{Hashimoto25ghostty}
  \item \practquote{PETJ-P74}{My current approach is usually that I start a discussion with
    codex, I paste in some websites, some ideas, ask it to read code, and we flesh out a new
    feature together. If it's something tricky, I ask it to write everything into a spec,
    give that to GPT-5-Pro for review (via chatgpt.com) to see if it has better ideas
    (surprisingly often, this greatly improves my plan!) and then paste back what I think is
    useful into the main context to update the file.}~\cite{Steinberger25talk}
  \item \practquote{SIMV-P9}{Planning in advance. Sitting down to hack something together
    goes much better if you start with a high level plan. Working with an agent makes this
    even more important---you can iterate on the plan first, then hand it off to the agent to
    write the code.}~\cite{Willison25vibe}
  \item \practquote{SIMP-P23}{Code that started from your own specification is a lot less effort to 
  review. If you already decided what to solve, picked the approach and worked out a detailed 
  specification for the work itself, confirming it was built to your needs can take a lot less 
  time.}~\cite{Willison25parallel}
  \item \practquote{SIMP-P24}{If I tell them exactly how to build something the work needed to 
  review the resulting changes is a whole lot less taxing.}~\cite{Willison25parallel}
\end{itemize}

\paragraph{Specification as ongoing, adaptive work.}
\begin{itemize}
  \item \practquote{MIT-P60}{I don't think I can figure this out on my own, so I
    decide to pivot. I decided that for these problematic titlebar styles, let's put the
    update notices in the bottom-right corner of the window overlaid on top of the content
    view rather than in the titlebar.}~\cite{Hashimoto25ghostty}
  \item \practquote{AR90-P11}{I did get the foundation wrong once. I initially argued myself
    into a more complex setup than I wanted. That's a part where I later used the LLM to
    redo a larger part early and clean it up.}~\cite{Ronacher25ninety}
\end{itemize}

\subsubsection{Knowledge Work, Learning, and Discovery (I, Human/Agent)}\label{app:result1-category-I}

\paragraph{T04 \pred{X}: Building one's own understanding.}

Building one's own understanding is not discussed very explicitly by the four practitioners:
Only Hashimoto is explicit about creating and maintaining his own understanding. 
Steinberger, on the other hand, explains that they do not read the code, but only know about the 
architecture and overall design of the system, which we interpret as leaning more towards the agent 
side.  
Ronacher once discusses this matter, but more in a normative fashion leaning toward human 
understanding. 
Willison only talks indirectly about that matter by explaining that coding agents are great for 
research tasks and generating explanations, but not discussing the importance of human 
understanding. 
He also describes using those generated explanations as context for his agents, which frames the 
building of understanding more as an agent's responsibility. 
We find our qualitative evidence to be lacking and therefore do not predict what our practitioners 
would be answering for this task-allocation item.

\begin{itemize}
  
  \item \practquote{SIMP-P9}{Research tasks answer questions or provide recommendations
  	without making modifications to a project that you plan to keep.}~\cite{Willison25parallel}
  \item \practquote{SIMP-P13}{If you need a reminder about how a portion of your existing system 
  	works, modern reasoning LLMs can provide a detailed, actionable answer in just a minute or 
  	two.}~\cite{Willison25parallel}
  \item \practquote{SIMP-P15--16}{Ask them to make notes on where your signed cookies are set and
    read, or how your application uses subprocesses and threads, or which aspects of your
    JSON API aren't yet covered by your documentation. These LLM-generated explanations are
    worth stashing away somewhere, because they can make excellent context to paste into
    further prompts in the future.}~\cite{Willison25parallel}
  \item \practquote{MIT-P32}{I'll often make these few hail mary attempts to fix a bug. If the 
  agent can figure it out, I can study it and learn myself. If it doesn't, it costs me very little. 
  If the agent figures it out and I don't understand it, I back it out. I'm not shipping code I 
  don't understand. While it's failing, I'm also tabbed out searching the issue and trying to 
  figure it out myself.}~\cite{Hashimoto25ghostty}  
  \item \practquote{MIT-P15}{I began by coming up with a rough plan of how I wanted this to work. 
  Ghostty uses Sparkle, an extremely popular macOS update framework. I poked around their docs and 
  found that they support custom UI through Obj-C protocols.}~\cite{Hashimoto25ghostty} 
  \item \practquote{MIT-P41}{Adding documentation is a really important step because it helps 
  reaffirm your own understanding of the code as well as educate future agents that may read and 
  modify this code.}~\cite{Hashimoto25ghostty} 
  \item \practquote{AR90-P21}{The most important piece of working with an agent is the same as 
  regular software engineering. You need to understand your state machines, how the system behaves 
  at any point in time, your database.}~\cite{Ronacher25ninety}  
  \item \practquote{PETS-P8}{These days I don't read much code anymore. I watch the stream and 
  sometimes look at key parts, but I gotta be honest - most code I don't read. I do know where 
  which components are and how things are structured and how the overall system is designed, and 
  that's usually all that's needed.}~\cite{Steinberger25speed}  
\end{itemize}

\paragraph{T05 \pred{A}: Tracing code paths.}

Tracing code paths is not discussed very explicitly by the four practitioners:
Only Willison is explicit about using agents to trace code paths and how they employ
them.  
Steinberger does so more implicitly. 
The other two do not talk about tracing code paths as a human activity, but also do not explicitly 
lean toward the agent side.  
We therefore predict that the four practitioners likely would lean toward the agent side for this 
task-allocation item, but acknowledge that the support for this prediction is weak. 

\begin{itemize}
  \item \practquote{SIMP-P14}{It doesn't matter how large your codebase is: coding agents are
    extremely effective with tools like grep and can follow codepaths through dozens of
    different files if they need to.}~\cite{Willison25parallel}
  \item \practquote{SIMP-P13}{If you need a reminder about how a portion of your existing system 
	works, modern reasoning LLMs can provide a detailed, actionable answer in just a minute or 
	two.}~\cite{Willison25parallel}
  \item \practquote{SIMS-P2}{I actually don't think documentation is too important: LLMs can
    read the code a lot faster than you to figure out how to use it.}~\cite{Willison25tips}
  \item \practquote{PETS-P14}{Instead of ``plan mode'', I simply start a conversation with the
    model, ask a question, let it google, explore code, create a plan together, and when I'm
    happy with what I see, I write ``build''}~\cite{Steinberger25speed}
  \item \practquote{PETJ-P39}{codex is far FAR more careful and reads much more files in your repo 
  before deciding what to do.}~\cite{Steinberger25talk}
\end{itemize}

\paragraph{T06 \pred{A}: Rapid exploratory prototyping.}

Three practitioners explicitly write about using coding agents for prototyping.
Steinberger does so implicitly when they describe their \say{UI-based work}.
We therefore predict that our practitioners would lean towards the agent side for this 
task-allocation item.   

\begin{itemize}
  \item \practquote{SIMP-P11}{Today's coding agents can build a proof of concept with
    new libraries and resolve those kinds of basic questions. Libraries too new to be in the
    training data? Doesn't matter: tell them to checkout the repos for those new dependencies
    and read the code to figure out how to use them.}~\cite{Willison25parallel}
  \item \practquote{AR90-P26}{For me, this has reached the point where I can't imagine
    working any other way. [\dots] This way of working unlocks paths I'd normally skip or
    defer.}~\cite{Ronacher25ninety}
  \item \practquote{AR90-P28}{It allows me to directly play with one or two implementations of a 
  problem. [\dots] I tried three different OpenAPI implementations and approaches in a 
  day.}~\cite{Ronacher25ninety}
  \item \practquote{MIT-P17}{AI is a very good prototyper, so even knowing what you don't know
    is useful enough to get started.}~\cite{Hashimoto25ghostty}
  \item \practquote{PETJ-P76}{The far more fun approach is when I do UI-based work. I often start 
	with sth simple and woefully under-spec my requests, and watch the model build and see the 
	browser update in real time. Then I queue in additional changes and iterate on the 
	feature. Often I don't fully know how something should look like, and that way I can play with 
	the idea and iterate and see it slowly come to life. I often saw codex build something 
	interesting I didn't even think of. I don't reset, I simply iterate and morph the chaos into 
	the shape that feels right.}~\cite{Steinberger25talk}
  \item \practquote{PETS-P26}{Rarely do I have a complete picture of what I want in my head. Sure, 
  I have a rough idea, but often that drastically changes as I explore the problem domain. So 
  systems that take the complete idea as input and then deliver output wouldn't work well for me. I 
  need to play with it, touch it, feel it, see it, that’s how I evolve 
  it.}~\cite{Steinberger25speed}
\end{itemize}

\subsection{Cluster: Design Operationalization}

The second cluster translates intent into agent-executable structures,
constraints, and process controls.

\subsubsection{Context Engineering and Prompting Discipline (E, Human)}\label{app:result1-category-E}


\paragraph{T07 \pred{H}: Curating access to relevant sources.}

Three practitioners explicitly describe handing the relevant sources to the agent:
Ronacher calls it to \say{bring the right information to the context}, 
Willison pastes \say{URLs to issues directly into Claude	Code}, and
Steinberger describes pointing to the adequate sources in their AGENTS file, but also mentioned not 
precisely pointing to the correct sources. 
Hashimoto does so implicitly by giving a prompt he used. 
We therefore predict that our practitioners would lean towards the human side for this 
task-allocation item. 

\begin{itemize}
  \item \practquote{AR90-P17}{You constantly need to learn how to bring the right information
    to the context. For me, this means pointing the AI to existing implementations and giving
    it very specific instructions on how to follow along.}~\cite{Ronacher25ninety}
  \item \practquote{SIMS-P2}{I'm having great results from maintaining a GitHub issues
    collection for projects and pasting URLs to issues directly into Claude
    Code.}~\cite{Willison25tips}
  \item \practquote{PETS-P26}{I maintain docs for subsystems and features in a docs folder in
    each project, and use a script + some instructions in my global AGENTS file to force the
    model to read docs on certain topics. This pays off more the larger the project is, so I
    don't use it everywhere}~\cite{Steinberger25speed}
	\item \practquote{MIT-P71}{[Prompt]: Read the Sparkle documentaiton [sic!] as necessary to 
	understand the functionality.}~\cite{Hashimoto25ghostty}
	\item \practquote{PETS-P26}{I cross-reference projects all the time, esp if I know that I 
	already solved sth somewhere else, I ask codex to look in../project-folder and that’s usually 
	enough for it to infer from context where to look. This is extremely useful to save on 
	prompts.}~\cite{Steinberger25speed}
\end{itemize}

\paragraph{Engineering context through codebase and tooling affordances.}
\begin{itemize}
  \item \practquote{PETS-P26}{After all, I don't design codebases to be easy to navigate for
    me, I engineer them so agents can work in it efficiently.}~\cite{Steinberger25speed}
  \item \practquote{MIT-P70}{I manually created a file with a scaffold of incomplete
    functions and various TODO comments. I then started a session to complete that work for
    me:}~\cite{Hashimoto25ghostty}
  \item \practquote{MIT-P72}{Tip: AI is very good at fill-in-the-blank or
    draw-the-rest-of-the-owl. My pattern here of creating scaffolding with descriptive
    function names, parameters, todo comments, etc. is a really common one for me and it
    works very well.}~\cite{Hashimoto25ghostty}
\end{itemize}

\paragraph{T09 \pred{H}: Writing and improving prompts.}

Three practitioners explicitly mention prompting as a human activity.
Steinberger describes how their prompting improved and how they do so, which makes it a human 
activity. 
Ronacher describes prompting, i.e., providing the right context, as an important (human) skill.
Hashimoto and Willison describe how they prompt, which again makes it a human activity. 
We therefore predict that our practitioners would lean towards the human side for this 
task-allocation item. 

\begin{itemize}
  \item \practquote{PETS-P26}{With codex, my prompts gotten much shorter, I often type again,
    and many times I add images, especially when iterating on UI (or text copies with CLIs).
    If you show the model what's wrong, just a few words are enough to make it do what you
    want.}~\cite{Steinberger25speed}
  \item \practquote{PETJ-P49}{Back when using claude, I used to write (ofc not, I speak) very 
  extensive prompts, since this model ``gets me'' the more context I supply. While this is true 
  with any model, I noticed that my prompts became significantly shorter with codex. Often it's 
  just 1-2 sentences + an image.}~\cite{Steinberger25talk}
  \item \practquote{AR90-P17}{You constantly need to learn how to bring the right information
	to the context. For me, this means pointing the AI to existing implementations and giving
	it very specific instructions on how to follow along.}~\cite{Ronacher25ninety}
  \item \practquote{MIT-P49}{I start vague, and slowly get more specific with how I'd
    approach it.}~\cite{Hashimoto25ghostty}
  \item \practquote{SIMP-P24}{I described my more authoritarian approach to prompting models for 
  code back in March. If I tell them exactly how to build something the work needed to review the 
  resulting changes is a whole lot less taxing.}~\cite{Willison25parallel}  
  \item \practquote{SIMV-P9}{You need to provide clear instructions, ensure they have the necessary 
  context and provide actionable feedback on what they produce.}~\cite{Willison25vibe}
\end{itemize}

\paragraph{Persistent instruction artifacts.}
\begin{itemize}
  \item \practquote{PETJ-P11}{My agents do git atomic commits themselves. In order to
    maintain a mostly clean commit history, I iterated a lot on my agent
    file.}~\cite{Steinberger25talk}
  \item \practquote{PETJ-P91}{My Agent file is currently \textasciitilde{}800 lines long and feels 
  like a collection of organizational scar tissue. I didn't write it, codex did, and anytime sth 
  happens I ask it to make a concise note in there. I should clean this up at some point, but 
  despite it being large it works incredibly well, and gpt really mostly honors entries 
  there.}~\cite{Steinberger25talk}
  \item \practquote{PETM-P25}{Agents can use them, one line in CLAUDE.md is enough `logs:
    axiom or vercel cli'.}~\cite{Steinberger25workflow}
\end{itemize}

\subsubsection{Tooling, Infrastructure, and Environment Design (H, Human/Agent)}\label{app:result1-category-H}


\paragraph{T10 \pred{H}: Tool selection and multi-tool orchestration.}

All four practitioners explicitly discuss how they chose their agent tools or how they orchestrate 
them framing it as a human activity.
We therefore predict that our practitioners would lean towards the human side for this 
task-allocation item. 

\begin{itemize}
  \item \practquote{AR90-P15}{Each has strengths, but the constant is Codex for code review
    after PRs. [\dots] Claude is indispensable still when debugging and needing a lot of tool
    access}~\cite{Ronacher25ninety}
  \item \practquote{MIT-P21}{I want to enable custom, unobtrusive update notifications and
	installs by customizing SPUUserDriver. [\dots] Create a plan to put it
	there. Consult the oracle.}~\cite{Hashimoto25ghostty}
  \item \practquote{MIT-P22}{Frequent question: `What is the oracle?' It is an
    Amp-specific read-only subagent that uses a slower, higher-cost model that is generally
    better at thinking about things. I consult the oracle for all
    planning.}~\cite{Hashimoto25ghostty}
  \item \practquote{PETJ-P10}{I’ve completely moved to codex cli as daily driver. I run 
  between 3-8 in parallel in a 3x3 terminal grid, most of them in the same folder, some experiments 
  go in separate folders.}~\cite{Steinberger25talk}
  \item \practquote{PETJ-P23}{Codex is more like the introverted engineer that chugs along and
    just gets stuff done. It reads much more files before starting work so even small prompts
    usually do exactly what I want.}~\cite{Steinberger25talk}
  \item \practquote{SIMP-P26}{My daily drivers are currently Claude Code (on Sonnet 4.5), Codex CLI 
  (on GPT-5-Codex), and Codex Cloud (for asynchronous tasks, frequently launched from my 
  phone.)}~\cite{Willison25parallel}  
  \item \practquote{SIMP-P28}{I’m still settling into patterns that work for me. I imagine I’ll be 
  iterating on my processes for a long time to come, especially as the landscape of coding agents 
  continues to evolve.}~\cite{Willison25parallel}
  \item \practquote{SIMP-P29}{I frequently have multiple terminal windows open running different 
  coding agents in different directories. These are currently a mixture of Claude Code and Codex 
  CLI, running in YOLO mode (no approvals) for tasks where I’m confident malicious instructions 
  can’t sneak into the context.}~\cite{Willison25parallel}
    
\end{itemize}

\paragraph{T11 \pred{X}: Environment design and sandboxing for safety.}

Here our analysis shows that the practitioners do not have a shared opinion on sandbox design. 
Willison explicitly discusses setting up a sandbox and designing the agent's boundaries, 
which we interpret as him leaning more toward the human side. 
Ronacher describes their normative stance, that coding agents should not \say{take over without 
[human] judgment}, but does not discuss sandboxing explicitly.
Steinberger shows that they give their coding agent access to the whole projects folder and does 
not discuss setting any further boundaries. 
They also describe a non-coding agent assistant that has full access to everything on all their 
computers. 
We interpret this that Steinberger would lean more towards the agent side of this task-allocation 
item while Willison and Ronacher would arguably lean more toward the human side. 
As the practitioners seemingly disagree on that matter we do not predict anything here. 

\begin{itemize}
  \item \practquote{SIMD-P65--66}{Some people will try to convince you that prompt injection
    attacks can be solved using more AI to detect the attacks. This does not work 100\%
    reliably, which means it's not a useful security defense at all. The only solution that's
    credible is to run coding agents in a sandbox.}~\cite{Willison25dangerous}
  \item \practquote{SIMD-P58}{Any time an LLM system combines access to private data with
    exposure to untrusted content and the ability to externally communicate, there's an
    opportunity for attackers to trick the system into leaking that private data back to
    them.}~\cite{Willison25dangerous}
  \item \practquote{SIMDL-P19}{Run your agent in a secure sandbox that restricts the files
    and secrets it can access and the network connections it can make.}~\cite{Willison25loops}
  \item \practquote{SIMDL-P21}{Despite the existence of container escapes I think option 1
    using Docker or the new Apple container tool is a reasonable risk to accept for most
    people.}~\cite{Willison25loops}
  \item \practquote{SIMDL-P24}{Coding agents themselves implement various levels of
    sandboxing, but so far I've not seen convincing enough documentation of these to trust
    them.}~\cite{Willison25loops}
  \item \practquote{SIMD-P91}{Those policies can control which files are visible but can
    also allow-list network connections. Anthropic run an HTTP proxy and allow the Claude
    Code environment to talk to that, then use the proxy to control which domains it can
    communicate with.}~\cite{Willison25dangerous}
  \item \practquote{SIMDL-P42}{Try to provide credentials to test or
    staging environments where any damage can be well contained. If a credential can spend
    money, set a tight budget limit.}~\cite{Willison25loops}
  \item \practquote{AR90-P33}{If you let the AI take over without judgment, you'll end up
    with brittle systems and painful surprises (data loss, security holes, unscalable
    software). The tools are powerful, but they don't absolve you of
    responsibility.}~\cite{Ronacher25ninety}
  \item \practquote{PETS-P21}{My current focus is Clawdis, an AI assistant that has full
    access to everything on all my computers, messages, emails, home automation, cameras,
    lights, music, [\dots] heck it can even control the temperature of my
    bed.}~\cite{Steinberger25speed}
  \item {\emergencystretch=3em \practquote{PETS-P30--46}{This is my \texttt{\textasciitilde/.codex/config.toml}:
    [\dots] \texttt{[projects."/Users/steipete/}\allowbreak\texttt{Projects"]}
    \texttt{trust\_level = "trusted"}}~\cite{Steinberger25speed}\par}
\end{itemize}

\paragraph{T12 \pred{X}: Setup and configuration once the environment is chosen.}

Here our analysis is conflicting on who sets up an environment.
Ronacher describes how he sets up AWS, Pulumi and a Postgres test infrastructure via an agent, and 
Willison hands the PyTorch/CUDA setup of an unfamiliar device to Claude Code. 
Steinberger describes automating domain registration and DNS through agent skills. 
The environments the practitioners configure themselves are almost always the agent's own 
operating environment: 
Willison installs a tool the agent will call and documents it in an \texttt{AGENTS.md}, and 
Steinberger sets up a web-based agent environment by hand. 
 
Since the evidence points in different directions for the two kinds of environment, 
deciding on a prediction for this task-allocation item would require declaring one of them 
inferior to the other. We therefore opt for making no prediction.

\begin{itemize}
  \item \practquote{AR90-P28}{Claude got me through AWS and Pulumi. Work I generally dislike became 
  a few days instead of weeks. It also debugged the setup issues as it was going through
  them.}~\cite{Ronacher25ninety}
  \item \practquote{AR90-P28}{I got a recommendation on Twitter to use testcontainers for
    testing against Postgres. [\dots] It would have been quite an involved project to migrate
    to. Claude did it in an hour for all tests.}~\cite{Ronacher25ninety}
  \item \practquote{SIMD-P24--25}{I wanted to try the newly released DeepSeek-OCR model on an
    NVIDIA Spark, but doing so requires figuring out how to run a model using PyTorch and CUDA,
    which is never easy and is a whole lot harder on an ARM64 device. [\dots] I SSHd into the
    Spark, started a fresh Docker container and told Claude Code to figure it out. It took 40
    minutes and three additional prompts but it solved the problem, and I got to have
    breakfast and tinker with some other projects while it was working.}~\cite{Willison25dangerous}
  \item \practquote{SIMDL-P33}{For a project that involved taking screenshots of various
    websites I installed my own shot-scraper CLI tool and dropped the following in
    \texttt{AGENTS.md}:}~\cite{Willison25loops}
  \item \practquote{PETJ-P53}{Google's Jules looks nice but was really annoying to set up [\dots]
    The only one that stuck is codex web. It also is annoying to setup and broken [\dots] but I
    had an older version of my environment and made it work, with the price of slower wramp-up
    times.}~\cite{Steinberger25talk}
  \item \practquote{PETS-P28}{There's a skill to register domains and change
    DNS.}~\cite{Steinberger25speed}
  \item \practquote{PETJ-P93}{If you don't know or don't use ast-grep as codebase linter, stop
    here and ask your model to set this up as a git hook to block commits}~\cite{Steinberger25talk}
\end{itemize}

\paragraph{Infrastructure to scale agentic workflows.}
\begin{itemize}
  \item \practquote{PETS-P16}{I built oracle [\dots] about a month ago - it's a CLI that
    allows the agent to run GPT 5 Pro and upload files + a prompt and manages sessions so
    answers can be retrieved later. I did this because many times when agents were stuck, I
    asked it to write everything into a markdown file and then did the query
    myself}~\cite{Steinberger25speed}
  \item \practquote{PETM-P25}{Pick services that have CLIs: vercel, psql, gh,
    axiom.}~\cite{Steinberger25workflow}
  \item \practquote{PETM-P24}{Even removed my last MCP, since Claude sometimes would go off
    spinning up Playwright unasked when it could simply read the code - which is faster and
    pollutes the context less.}~\cite{Steinberger25workflow}
  \item \practquote{PETJ-P54}{I use codex web as my short-term issue tracker. Whenever I'm on
    the go and have an idea, I do a one-liner via the iOS app and later review this on my
    Mac.}~\cite{Steinberger25talk}
  \item \practquote{PETJ-P26}{\textasciitilde{}230k usable context vs claude's
    156k.}~\cite{Steinberger25talk}
  \item \practquote{PETJ-P26}{Codex allows to queue messages. [\dots] I often
    queue related feature tasks and it just reliably works them off.}~\cite{Steinberger25talk}
\end{itemize}

\paragraph{Version-control and recovery affordances.}
\begin{itemize}
  \item \practquote{ARAY-P28}{Some agentic coding tools have begun spinning up worktrees or
    creating checkpoints in git for restore, in-conversation branch and undo
    features.}~\cite{Ronacher25vibes}
  \item \practquote{ARAY-P29}{I increasingly want to better understand genuine human input
    and tell it apart from machine output. I want to see the prompts and the attempts that
    failed along the way.}~\cite{Ronacher25vibes}
  \item \practquote{ARAY-P33}{Code review to me feels like it needs to become part of the
    VCS.}~\cite{Ronacher25vibes}
  \item \practquote{ARAY-P26}{The pull request model on GitHub doesn't
    carry enough information to review AI generated code properly --- I wish I could see the
    prompts that led to changes.}~\cite{Ronacher25vibes}
\end{itemize}

\subsubsection{Workflow Orchestration (B, Human)}\label{app:result1-category-B}


\paragraph{T13 \pred{H}: Task chunking for reviewable units.}

Ronacher and Hashimoto explicitly describe creating reviewable chunks as tasks for their agents. 
Steinberger does so implicitly by using \say{blast radius} as a dimension to think about when 
creating tasks for agents.
Willison does not discuss this aspect explicitly but puts an emphasis on code review, which implies 
that the tasks handed out to agents need to be in reviewable chunks. He is not explicit about who 
creates these chunks. 
We therefore predict that our practitioners would lean toward the human side for this 
task-allocation item. 

\begin{itemize}
  \item \practquote{AR90-P18}{I generally create PR-sized chunks that I can
    review.}~\cite{Ronacher25ninety}
  \item \practquote{MIT-P24}{Notice here that I don't send the agent off to build the full
    feature. [\dots] Second, smaller chunks of work are easier to review, understand, and
    iterate on.}~\cite{Hashimoto25ghostty}
   \item \practquote{PETJ-P17}{Whenever I work, I think about the `blast radius'. I didn’t come up 
   with that term, I do love it tho. When I think of a change I have a pretty good feeling about 
   how long it’ll take and how many files it will touch. I can throw many small bombs at my 
   codebase or a one `Fat Man' and a few small ones. If you throw multiple large bombs, it’ll be 
   impossible to do isolated commits, much harder to reset if sth goes 
   wrong.}~\cite{Steinberger25talk}
   \item \practquote{SIMV-P9}{A culture of code review. This one explains itself. If you’re fast 
   and productive at code review you’re going to have a much better time working with LLMs than if 
   you’d rather write code yourself than review the same thing written by someone (or something) 
   else.}~\cite{Willison25vibe}
\end{itemize}

\paragraph{Loop design.}
\begin{itemize}
  \item \practquote{SIMDL-P14}{Each of these tools provides its own version of what I like
    to call YOLO mode, where everything gets approved by default.}~\cite{Willison25loops}
  \item \practquote{SIMDL-P6}{My preferred
    definition of an LLM agent is something that runs tools in a loop to achieve a goal. The
    art of using them well is to carefully design the tools and loop for them to
    use.}~\cite{Willison25loops}
  \item \practquote{AR90-P19}{It requires intuition to know when each approach is more likely
    to lead to the right results. Familiarity with the agent also helps understanding when a
    task will not go anywhere, avoiding wasted cycles.}~\cite{Ronacher25ninety}
\end{itemize}

\paragraph{Parallel coordination.}
\begin{itemize}
  \item \practquote{PETS-P26}{I usually work on multiple projects at the same time.
    Depending on complexity that can be between 3-8. [\ldots] Bigger tasks I keep for moments where 
    I'm distracted - for example while writing this, I run refactors on 4 projects here that will 
    take around 1-2h each to complete.}~\cite{Steinberger25speed}
  \item \practquote{PETS-P28}{I usually work on two Macs. My MacBook Pro on the big screen, and a 
  Jump Desktop session to my Mac Studio on another screen. Some projects are cooking there, some 
  here. Sometimes I edit different parts of the same project on each machine and sync via 
  git.}~\cite{Steinberger25speed}
  \item \practquote{PETJ-P10}{I run between 3-8 in parallel in a 3x3 terminal grid,
    most of them in the same folder, some experiments go in separate folders. I experimented
    with worktrees, PRs but always revert back to this setup as it gets stuff done the
    fastest.}~\cite{Steinberger25talk}
\end{itemize}

\subsection{Cluster: Evaluation}


\subsubsection{Implementation and Quality Calibration (D, Human/Agent)}\label{app:result1-category-D}


\paragraph{Explicit quality standards.}
\begin{itemize}
  \item \practquote{MIT-P73}{It actually did a really bad job here and I ended up throwing all
    of this code away. The code it produced worked, but it was obviously the wrong
    approach.}~\cite{Hashimoto25ghostty}
  \item \practquote{AR90-P23}{Here's an example: I asked it to build a rate limiter. It
    `worked' but lacked jitter and used poor storage decisions. Easy to fix if you know rate
    limiters, dangerous if you don't.}~\cite{Ronacher25ninety}
\end{itemize}

\paragraph{T16 \pred{H}: Defining what counts as evidence that a change works.}

Two practitioners explicitly discuss what to review and test as a human activity.  
Willison describes manual testing and automated testing to confirm a change by the agent works as 
intended. 
Steinberger describes instructing their agents to create automated tests, i.e., the human decides 
that automated tests are required.  
Ronacher discusses upholding strong invariants to \say{let code crash loudly when they fail}, which 
we also interpret as leaning toward the human side on that matter.
Hashimoto is not explicit about who is responsible for the decision. We found no basis that they 
would lean toward the agent side.
We therefore predict that our practitioners would lean toward the human side for this 
task-allocation item. 

\begin{itemize}
  \item \practquote{SIMY-P8}{The first is manual testing. If you haven't seen the code do
    the right thing yourself, that code doesn't work.}~\cite{Willison25proven}
  \item \practquote{SIMY-P10}{If possible I like to reduce these steps to a sequence of
    terminal commands which I can paste, along with their output, into a comment in the code
    review.}~\cite{Willison25proven}
  \item \practquote{SIMY-P13}{The second step in proving a change works is automated
    testing. This is so much easier now that we have LLM tooling, which means there's no
    excuse at all for skipping this step.}~\cite{Willison25proven}
  \item \practquote{SIMY-P14}{Your contribution should bundle the change with an automated test 
  that proves the change works. That test should fail if you revert the 
  implementation.}~\cite{Willison25proven}
  \item \practquote{SIMV-P9}{Without tests? Your agent might claim something works without
  	having actually tested it at all [\ldots] Beyond automated tests, you need to be really good
  	at manually testing software, including predicting and digging into
  	edge-cases.}~\cite{Willison25vibe}  
  \item \practquote{PETJ-P85}{Ask the model to write tests after each feature/fix is done. Use
    the same context. This will lead to far better tests, and likely uncover a bug in your
    implementation. If it's purely a UI tweak, tests likely make less sense, but for anything
    else, do it. AI generally is bad at writing good tests, it's still helpful tho, and let's be
    honest - are you writing tests for every fix you make?}~\cite{Steinberger25talk}
  \item \practquote{AR90-P24}{Agents also operate on conventional wisdom from the internet and in 
  tern do things I would never do myself. It loves to use dependencies (particularly outdated 
  ones). It loves to swallow errors and take away all tracebacks. I’d rather uphold strong 
  invariants and let code crash loudly when they fail, than hide problems. If you don’t fight this, 
  you end up with opaque, unobservable systems.}~\cite{Ronacher25ninety}
\end{itemize}

\paragraph{T17 \pred{A}: Candidate solution generation with self-checking.}

Willison, Hashimoto, and Steinberger all explicitly describe how they task their agents to fulfill 
a task and how they instruct the agent to verify/test their implementation at the end.  
Ronacher uses agents to write code, but finds their ability to write tests lacking. 

We therefore predict that our practitioners would lean toward the agent side for this 
task-allocation item.

\begin{itemize}
  \item \practquote{SIMY-P19}{To master these [agentic] tools you need to learn how to get them to 
  prove their changes work as well.}~\cite{Willison25proven}
  \item \practquote{SIMS-P2}{I love pytest for this - one of my projects has 1500 tests and Claude 
  	Code is really good at selectively executing just tests relevant to the change it is making, 
  	and then running the whole suite at the end. Give them the ability to interactively test the 
  	code they are writing too. Notes on how to start a development server (for web projects) are 
  	useful, then you can have them use Playwright or curl to try things out. [\ldots] Linters, type 
  	checkers, auto-formatters - give coding agents helpful tools to run and they'll use 
  	them.}~\cite{Willison25tips}
  \item \practquote{MIT-P70}{I manually created a file with a scaffold of incomplete functions
    and various TODO comments. I then started a session to complete that work for
    me:}~\cite{Hashimoto25ghostty}
  \item \practquote{MIT-P72}{Tip: AI is very good at fill-in-the-blank or
    draw-the-rest-of-the-owl. My pattern here of creating scaffolding with descriptive
    function names, parameters, todo comments, etc. is a really common one for me and it
    works very well.}~\cite{Hashimoto25ghostty}
  \item \practquote{PETS-P6}{The simplest form is text, so by default, whatever I wanna
    build, it starts as CLI. Agents can call it directly and verify output - closing the
    loop.}~\cite{Steinberger25speed}
  \item \practquote{MIT-P83--85}{I had a meeting to hop onto so I decided to let the agent
    fix it while I was busy: [\ldots] Run the build and fix the errors [\ldots] ``I broke a
    bunch of things, please fix my mess.'' is another frequent use case I have for agents.
    I'd say generally this fits within the same fill-in-the-blank pattern as
    before.}~\cite{Hashimoto25ghostty}
  \item \practquote{MIT-P95}{Tip: Agents are great at generating tests and simulations. The
  	simulation code specifically that it generated here is honestly pretty gross but it
  	works and it's not part of the release binary so the quality doesn't matter to
  	me.}~\cite{Hashimoto25ghostty}
  \item \practquote{PETM-P22}{Bigger changes always get tests. Automated ones usually
    aren't great, but the model almost always finds issues when you ask it to write tests
    IN THE SAME CONTEXT.}~\cite{Steinberger25workflow}
  \item \practquote{AR90-P28}{While they suck at writing tests, they turned out great at setting up 
  test infrastructure I didn’t know I needed.}~\cite{Ronacher25ninety}
	
\end{itemize}

\paragraph{T18 \pred{E}: Reviewing a change for missing areas, risks, and problems.}

Ronacher uses Codex for code reviews, but also discusses setting explicit quality controls in the 
form of invariants, and reviewing code himself. We interpret this as a shared responsibility between 
agent and human.    
Hashimoto shows a prompt where they task their agent with identifying areas of improvement as the 
last step of a task, but also does manual reviews. 
Steinberger does not mention doing manual reviews, instead he briefly mentions agents doing 
reviews, being engaged in manual testing himself and explains that he does not read a lot of code 
himself. Instead he observes what the agent is doing and interferes when necessary. 
We interpret this as a shared responsibility.

We therefore predict that our practitioners would lean toward shared for this task-allocation item. 

\begin{itemize}
  \item \practquote{AR90-P15}{Each has strengths, but the constant is Codex for code review
    after PRs. It's very good at that.}~\cite{Ronacher25ninety}
  \item \practquote{AR90-P24}{It loves to swallow errors and take away all
  	tracebacks. I'd rather uphold strong invariants and let code crash loudly when they fail,
  	than hide problems. If you don't fight this, you end up with opaque, unobservable
  	systems.}~\cite{Ronacher25ninety}
  \item \practquote{ARAY-P38}{I still treat it like regular software engineering and review a 
  	lot.}~\cite{Ronacher25vibes}
  \item \practquote{AR90-P31}{At the same time, for me, AI doesn’t own the code. I still review 
  every line, shape the architecture, and carry the responsibility for how it runs in production.}~\cite{Ronacher25ninety}
  \item \practquote{MIT-P108}{My last prompt to an agent is always to ask what else I might be
  	missing.}~\cite{Hashimoto25ghostty}
  \item \practquote{MIT-P109}{[Prompt:] Are there any other improvements you can see to be made 
  with the @macos/Sources/Features/Update feature? Don't write any code. Consult the oracle. 
  Consider parts of the code that can also get more unit tests added.}~\cite{Hashimoto25ghostty}
  \item \practquote{MIT-P126}{I think the feature is beautiful, works great, and after a final 
  manual review I merged it}~\cite{Hashimoto25ghostty}
  \item \practquote{PETM-P6}{After going all-in on VS Code, I went fully back to Ghostty for my 
  main setup, with VS Code on the side to look up code and Cursor/GPT-5 for reviews (sometimes also 
  the CLI).}~\cite{Steinberger25workflow}
  \item \practquote{PETS-P8}{These days I don’t read much code anymore. I watch the stream and 
  sometimes look at key parts, but I gotta be honest - most code I don’t 
  read.}~\cite{Steinberger25speed}
  \item \practquote{PETS-P28}{Whenever prompts start taking too long or I see sth ugly flying by in 
  the code stream, I’ll deal with it right away.}~\cite{Steinberger25speed}
  \item \practquote{PETJ-P21}{I run one dev server, as I evolve my project I click through it and 
  test multiple changes at once}~\cite{Steinberger25talk}
\end{itemize}

\paragraph{Intervention and pivot.}
\begin{itemize}
  \item \practquote{MIT-P59--60}{This whole time, I've also spent time trying to solve this
    myself through manual research and human effort. My more specific prompts are based on
    things I've learned through that process. But overall, it was clearly not
    working. I don't think I can figure this out on my own, so I decide to 
    pivot.}~\cite{Hashimoto25ghostty}
  \item \practquote{MIT-P31}{You can see in chats 11 to 14 that we're entering the
    slop zone. The code the agent created has a critical bug, and it's absolutely failing to
    fix it.}~\cite{Hashimoto25ghostty}
  \item \practquote{MIT-P95}{Tip: Agents are great at generating tests and simulations. The
    simulation code specifically that it generated here is honestly pretty gross but it
    works and it's not part of the release binary so the quality doesn't matter to
    me.}~\cite{Hashimoto25ghostty}
  \item \practquote{MIT-P96}{I then ran the various simulations and found a number of UX
    improvements.}~\cite{Hashimoto25ghostty}
\end{itemize}

\subsubsection{Iteration, Refactoring, and Cleanup (F, Human/Agent)}\label{app:result1-category-F}

\paragraph{T20 \pred{H}: Deciding when and why to refactor.}

Three of our practitioners write about deciding when and what to refactor themselves. 
Ronacher and Steinberger explicitly discuss when and why they decided to task their agent with 
refactoring. 
Hashimoto describes how they identified an area for refactoring and then doing the refactoring 
themselves. 
Willison does not explicitly discuss this. 
We therefore predict that our practitioners would lean to the human side for this task-allocation 
item. 

\begin{itemize}
  \item \practquote{AR90-P28}{The code looks more organized than it would otherwise have been
    because the cost of refactoring is quite low. You need to know what you do, but if set up
    well, refactoring becomes easy.}~\cite{Ronacher25ninety}
  \item \practquote{AR90-P11}{I did get the foundation wrong once. I initially argued myself
    into a more complex setup than I wanted. That's a part where I later used the LLM to
    redo a larger part early and clean it up.}~\cite{Ronacher25ninety}
	\item \practquote{MIT-P74}{When I studied what it did, I realized it was because the view
		model was structured in a suboptimal way, so I switched gears into cleanup mode to give
		AI (and humans, if I chose to write it myself) the better framework to work
		with.}~\cite{Hashimoto25ghostty}
	\item \practquote{MIT-P77}{Experience has taught me that the cleanliness of a UI frontend
		and business logic backend is often dictated by the quality of the view model in
		between. So I spent some time manually restructured the view model. [\dots] I knew
		from experience that this small bit of manual work in the middle would set the agents
		up for success in future sessions for both the frontend and backend.}~\cite{Hashimoto25ghostty}
	\item \practquote{PETS-P28}{In the past I often took dedicated days to refactor and clean up
		projects, I do this much more ad-hoc now. Whenever prompts start taking too long or I see
		sth ugly flying by in the code stream, I'll deal with it right away.}~\cite{Steinberger25speed}
\end{itemize}

\paragraph{T21 \pred{A}: Mechanical refactors, cleanup, and build fixes.}

Three practitioners explicitly hand mechanical refactoring and cleanup to their agents.
Steinberger states that all of their refactoring is done by agents and lists the typical work,
e.g., removing duplicated and dead code, updating dependencies and restructuring files.
Hashimoto lets the agent fix the build while in a meeting and, after restructuring code by hand,
has the agent update the dependent code.
Willison hands deprecation warnings to an agent to fix.
Ronacher does not discuss this explicitly. We found no basis that they would lean toward the
human side.
We therefore predict that our practitioners would lean toward the agent side for this
task-allocation item.

\begin{itemize}
  \item \practquote{PETJ-P69}{I spend about 20\% of my time on refactoring. Ofc all of that is
    done by agents, I don't waste my time doing that manually.}~\cite{Steinberger25talk}
  \item \practquote{PETJ-P70}{Typical refactor work is using jscpd for code duplication,
  	knip for dead code, [\dots] updating dependencies, tool upgrades, file restructuring,
  	[\dots] There's always something to do.}~\cite{Steinberger25talk}
  \item \practquote{MIT-P83--84}{I had a meeting to hop onto so I decided to let the agent fix it 
  while I was busy: Run the build and fix the errors}~\cite{Hashimoto25ghostty}
  \item \practquote{MIT-P79}{After doing the restructuring, the first thing I did was
    ask the agent to once again complete the owl, this time inspecting my changes and
    updating dependent code to the new style and removing the old:}~\cite{Hashimoto25ghostty}
  \item \practquote{SIMP-P18--19}{Now we're moving on to code edits that we intend to keep,
    albeit with very low-stakes. It turns out there are a lot of problems that really just
    require a little bit of extra cognitive overhead which can be outsourced to a bot.
    Warnings are a great example. Is your test suite spitting out a warning that something
    you are using is deprecated? Chuck that at a bot---tell it to run the test suite and
    figure out how to fix the warning. No need to take a break from what you're doing to
    resolve minor irritations like that.}~\cite{Willison25parallel}
\end{itemize}

\paragraph{T22 \pred{X}: Manual polish to reach production quality.}

Only Hashimoto explicitly states he polishes agent results by hand.
Ronacher prompts until the result is close and then cleans up, but does not say whether he
or the agent does the cleanup.
Steinberger asks the model to change whatever he does not like.
Willison does not discuss polishing.
The practitioners do not agree on this matter, so we make no prediction.

\begin{itemize}
  \item \practquote{MIT-P103}{This required some back and forth and manual polish, but it
    got there.}~\cite{Hashimoto25ghostty}
  \item \practquote{MIT-P64}{It did a really good job with this. I did a lot of
    manual polish after (moving stuff around, renaming things, etc.) but the core work was
    solid.}~\cite{Hashimoto25ghostty}
  \item \practquote{AR90-P18}{Agent loop with finishing touches: Prompt until the
    result is close, then clean up.}~\cite{Ronacher25ninety}
  \item \practquote{MIT-P5}{I almost always go in after an AI does work and iterate myself
    for awhile, too. [\dots] Therefore, you may see some discrepancies between what the AI
    produced and what ended up in the final code.}~\cite{Hashimoto25ghostty}
  \item \practquote{MIT-P110}{This highlighted some real issues, so I went ahead and asked
    it to implement them. I find its easier to tell the agent ``okay just do it all'' rather
    than ask it specific things to do, since I can always easily clean it up in selective
    commits later.}~\cite{Hashimoto25ghostty}
  \item \practquote{PETS-P26}{I basically never revert or use checkpointing. If something isn't
    how I like it, I ask the model to change it.}~\cite{Steinberger25speed}
  \item \practquote{SIMP-P5}{I can only focus on reviewing and landing one significant
    change at a time, but I'm finding an increasing number of tasks that can still be fired
    off in parallel without adding too much cognitive overhead to my primary
    work.}~\cite{Willison25parallel}
\end{itemize}

\subsubsection{Accountability and Ownership of Outcomes (A, Human)}\label{app:result1-category-A}

\paragraph{T23 \pred{H}: Ownership of outcomes.}

All four practitioners keep the ownership of outcomes with the human.
Ronacher treats every line as their own responsibility and states that AI does not own the code.
Willison calls delivering code without proof that it works a dereliction of duty.
Hashimoto insists on a thorough final manual review before shipping AI-written code.
Steinberger reads little of the code, but still thinks hard about architecture and system design
and keeps track of how the overall system is structured, which we interpret as owning the outcome
on the system level.
We therefore predict that our practitioners would lean toward the human side for this
task-allocation item.

\begin{itemize}
  \item \practquote{AR90-P6}{I still treat every line as my responsibility, judged
    as if I wrote it myself. AI doesn't change that.}~\cite{Ronacher25ninety}
  \item \practquote{AR90-P31}{At the same time, for me, AI doesn't own the code. I still
    review every line, shape the architecture, and carry the responsibility for how it
    runs in production.}~\cite{Ronacher25ninety}
  \item \practquote{AR90-P33}{If you let the AI take over without judgment, you'll end up
    with brittle systems and painful surprises (data loss, security holes, unscalable
    software). The tools are powerful, but they don't absolve you of
    responsibility.}~\cite{Ronacher25ninety}
  \item \practquote{ARAY-P38}{This may be a little more controversial, but what I haven't
    managed this year is to give in to the machine. I still treat it like regular software
    engineering and review a lot.}~\cite{Ronacher25vibes}
  \item \practquote{SIMY-P5}{As software engineers we don't just crank out
    code---in fact these days you could argue that's what the LLMs are for. We need to
    deliver code that works---and we need to include proof that it works as
    well.}~\cite{Willison25proven}
  \item \practquote{SIMY-P2}{In all of the debates about the value of AI-assistance in
    software development there's one depressing anecdote that I keep on seeing:
    the junior engineer, empowered by some class of LLM tool, who deposits giant, untested
    PRs on their coworkers---or open source maintainers---and expects the ``code review''
    process to handle the rest.}~\cite{Willison25proven}
  \item \practquote{SIMY-P3}{This is rude, a waste of other people's time, and is
    honestly a dereliction of duty as a software developer.}~\cite{Willison25proven}
  \item \practquote{MIT-P132}{`Final manual review' is super super super important. [\dots]
    Please don't ever ship AI-written code without a thorough manual
    review.}~\cite{Hashimoto25ghostty}
  \item \practquote{PETS-P8}{These days I don't read much code anymore. I watch the stream and
    sometimes look at key parts, but I gotta be honest - most code I don't read. I do know
    where which components are and how things are structured and how the overall system is
    designed, and that's usually all that's needed.}~\cite{Steinberger25speed}
  \item \practquote{PETJ-P101}{Just because I don't write the code anymore doesn't
    mean I don't think hard about architecture, system design, dependencies, features or how
    to delight users.}~\cite{Steinberger25talk}
  \item \practquote{PETS-P8}{The real unlock into building like a factory was GPT 5.
    It took me a few weeks after the release to see it [\dots] but then I started trusting
    the model more and more.}~\cite{Steinberger25speed}
\end{itemize}

\subsection{Cross-cutting category}

\subsubsection{Cultural Framing and Legitimacy Claims (J, Human)}\label{app:result1-category-J}


\paragraph{T24 \pred{H}: Acceptable and unacceptable use of AI.}

Three practitioners explicitly draw a line between acceptable and unacceptable use of AI
themselves.
Willison separates \emph{vibe coding} from \emph{vibe engineering} and calls handing untested
AI-generated pull requests to others rude and a dereliction of duty.
Ronacher suggests that new social contracts are needed, tries to set them through contribution
guidelines and pull request templates, and argues that agency and responsibility should remain
with humans.
Hashimoto urges others never to ship AI-written code without a thorough manual review.
Steinberger does not discuss this explicitly. We found no basis that they would lean toward the
agent side.
We therefore predict that our practitioners would lean toward the human side for this
task-allocation item.

\begin{itemize}
  \item \practquote{SIMV-P2}{I feel like vibe coding is pretty well
    established now as covering the fast, loose and irresponsible way of building software
    with AI---entirely prompt-driven, and with no attention paid to how the code actually
    works. This leaves us with a terminology gap: what should we call the other end of the
    spectrum, where seasoned professionals accelerate their work with LLMs while staying
    proudly and confidently accountable for the software they produce?}~\cite{Willison25vibe}
  \item \practquote{SIMV-P15}{I've never really liked the artificial distinction between
    ``coders'' and ``engineers''---that's always smelled to me a bit like gatekeeping. But
    in this case a bit of gatekeeping is exactly what we need!}~\cite{Willison25vibe}
  \item \practquote{SIMV-P7}{This feels very different from classic vibe coding, where I
    outsource a simple, low-stakes task to an LLM and accept the result if it appears to work.
    [\dots] Iterating with coding agents to produce production-quality code that I'm confident
    I can maintain in the future feels like a different process entirely.}~\cite{Willison25vibe}
  \item \practquote{ARAY-P17}{An unexpected aspect of using AI so much is that we talk far more
    about vibes than anything else. This way of working is less than a year old, yet it
    challenges half a century of software engineering experience.}~\cite{Ronacher25vibes}
  \item \practquote{ARAY-P38}{I also recognize that an increasing number of people are not
    working with this model of engineering but instead completely given in to the machine. As
    crazy as that sounds, I have seen some people be quite successful with this. I don't yet
    know how to reason about this}~\cite{Ronacher25vibes}
  \item \practquote{ARAY-P18}{I found a lot of conventional wisdom I don't agree with, but I
    have nothing to back up my opinions.}~\cite{Ronacher25vibes}
  \item \practquote{ARAY-P19}{It's also important to know that some of the vibes come with
    intentional signalling. Plenty of people whose views you can find online have a financial 
    interest in one product over another, for instance because they are investors in it or they are 
    paid influencers.}~\cite{Ronacher25vibes}
  \item \practquote{ARAY-P38}{I still treat it like regular software engineering and review a lot. 
  I also recognize that an increasing number of people are not working with this model of 
  engineering but instead completely given in to the machine. [\ldots] the way of working in that 
  new world is very different from the world that I’m comfortable with. And my suspicion is that 
  because that world is here to stay, we might need some new social contracts to separate these 
  out.}~\cite{Ronacher25vibes}
  \item \practquote{ARAY-P40}{Personally, I've tried to attack this problem with contribution
    guidelines and pull request templates. But this seems a little like a fight against
    windmills. This might be something where the solution will not come from changing what
    we're doing. Instead, it might come from vocal people who are also pro-AI engineering
    speaking out on what good behavior in an agentic codebase looks like. And it is not just
    to throw up unreviewed code [\dots]}~\cite{Ronacher25vibes}
  \item \practquote{ARAY-P14}{These systems we now create have human tendencies, but
    elevating them to a human level would be a mistake. I increasingly take issue with
    calling these machines ``agents,'' yet I have no better word for it. I take issue with
    ``agent'' as a term because agency and responsibility should remain with
    humans.}~\cite{Ronacher25vibes}
  \item \practquote{SIMY-P2}{In all of the debates about the value of AI-assistance in
    software development there's one depressing anecdote that I keep on seeing:
    the junior engineer, empowered by some class of LLM tool, who deposits giant, untested
    PRs on their coworkers---or open source maintainers---and expects the ``code review''
    process to handle the rest.}~\cite{Willison25proven}
  \item \practquote{SIMY-P3}{This is rude, a waste of other people's time, and is
    honestly a dereliction of duty as a software developer.}~\cite{Willison25proven}
  \item \practquote{MIT-P132}{``Final manual review'' is super super super important. [\dots]
    Please don't ever ship AI-written code without a thorough manual
    review.}~\cite{Hashimoto25ghostty}
  \item \practquote{MIT-P4}{This post will share every single agentic coding session I had
    on the path to shipping this feature, unedited and in full. Alongside it, I'll provide
    some additional context about my process and reasoning.}~\cite{Hashimoto25ghostty}
  \item \practquote{MIT-P3}{I'm regularly asked to share non-trivial examples of how I use AI
    and agentic coding tools and this felt like a golden opportunity to walk through my
    process with a well-scoped, non-trivial, real-world, shipping
    feature}~\cite{Hashimoto25ghostty}
\end{itemize}

\section{Step 2: Quantitative Survey Details}\label{app:step2}

\subsection{Questionnaire}\label{app:method2-questionnaire}

This appendix reproduces the survey instrument as it was fielded. 
The attitude items G01--G07 and the activity items of survey page~2 were each shown in randomized
order.
The item identifiers set in bold (\textbf{T01}, \textbf{G01}, \dots) were \emph{not}
shown to respondents; they are reproduced here so that every item can be matched to
the prose and tables in Sections~\ref{sec:methods2} and \ref{sec:results2}.

\subsubsection{Page 0: Developers and AI Coding Agents: Who Does What?}

You are invited to take part in a research study on \textbf{how software developers
work with AI coding tools and coding agents}. The survey takes about
\textbf{15 minutes}. You are welcome to take part whether or not you use AI tools in
your work.

Participation is \textbf{voluntary}---you may skip questions or stop at any time.
Your responses will be handled \textbf{anonymously} and are used only in aggregate for
academic research. There are \textbf{no right or wrong answers}; we are interested in
your attitude and experience. The results are intended to be published as a research
article. By continuing, you consent to take part under these conditions. This survey
fulfills the research ethics rules of Freie Universit\"at Berlin.

EU-GDPR privacy statement: The Data Controller for this survey is Prof.\ Dr.\ Lutz
Prechelt, \nolinkurl{prechelt@inf.fu-berlin.de}. Your data will be submitted to Processor
\url{1blu.de} who run the server you are currently using. If you provide your email address
in the final step, you have the right to demand deletion of your data at any time. The
survey will store only your answers plus timestamps.

Thank you for participating!

Helia Dadkhah, Linus Ververs, Lutz Prechelt

\subsubsection{Page 1: Attitude Toward AI and Prior Experience}

In this survey, \textbf{AI tools} and \textbf{AI assistance} mean any tool
\textbf{powered by a large language model (LLM)} that supports software
development---the common denominator, covering chat assistants, LLM-based IDE
autocomplete, and coding agents alike.

A \textbf{coding agent} is a specific kind of AI tool that can read and modify a
codebase across multiple steps on its own---planning, editing files, and running
commands---such as Claude Code, OpenAI Codex CLI, Cursor's agent mode, or similar.
This differs from single-suggestion autocomplete (e.g.\ classic IntelliSense) or a
one-off AI chat answer.

\qitem{X01}{How long have you been using AI tools of any kind for software development?}
\qopts{I don't use them \qsep Less than 3 months \qsep 3-6 months \qsep 6--12 months
  \qsep 1--2 years \qsep More than 2 years}

\qitem{X02}{How often do you typically use \textbf{coding agents} (as defined above)
  that act across a codebase over multiple steps?}
\qopts{Never \qsep Tried once or twice \qsep Monthly \qsep Weekly \qsep Several
  sessions per week \qsep About daily \qsep Several sessions per day}

\qitem{X03}{Which AI tools do you use for software development? (select all that apply)}
\qopts{Chat assistant (ChatGPT, Claude, Gemini, \dots) \qsep IDE AI plugin autocomplete
  (GitHub Copilot, \dots) \qsep IDE AI plugin autoedit mode \qsep IDE AI plugin agent
  mode \qsep AI-native IDE agent mode (Cursor, Windsurf, \dots) \qsep Terminal coding
  agent (Claude Code, Codex CLI, GitHub Copilot CLI, \dots) \qsep GitHub Copilot on
  GitHub \qsep Self-hosted LLMs \qsep Other: \qsep I don't use AI tools for development}

\addvspace{\medskipamount}\noindent
How much do you agree with the following statements?

\qscale{1 Strongly disagree \qsep 2 Disagree \qsep 3 Neither agree nor disagree \qsep
  4 Agree \qsep 5 Strongly agree \qsep N/A Not sure}

\qitem{G01}{AI tools make me more productive in my development work.}
\qitem{G02}{I am concerned about becoming too reliant on AI tools.}
\qitem{G03}{In my experience, AI coding tools are over-hyped relative to what they
  actually deliver.}
\qitem{G04}{The more experience I gain with AI tools, the more of my work I am willing
  to let them do.}
\qitem{G05}{For many tasks, I trust my AI tools more than I trust myself.}
\qitem{G06}{The tools I work with must be highly reliable.}
\qitem{G07}{I am willing to take moderate risks with a tool if it promises high payoffs.}

\qitem{X04}{Which kinds of development work (not just their preparation) are you
  typically comfortable delegating to a coding agent fully? (select one)}
\qopts{None---I prefer to do all of it myself \qsep Only mechanical work---boilerplate,
  formatting, renames, repetitive edits \qsep Also modest coding work---tests, small
  features, straightforward bug fixes \qsep Also substantial coding work---larger
  features/refactorings, tricky bugs \qsep Also judgment work---making a major design
  decision, accepting a pull request, etc. \qsep Not sure}

\qitem{X04c}{Optional comment}
\qopts{\qfree}

\qitem{X05}{How much scope and decision-making are you comfortable delegating to a
  coding agent for a serious but not highly critical task? (select one)}
\qopts{None---I am not comfortable delegating tasks to a coding agent \qsep Small,
  precisely defined tasks only---I keep all the decisions \qsep Moderate tasks---the
  agent decides minor details \qsep Large tasks---the agent makes substantial
  implementation decisions \qsep Open-ended tasks---the agent works with broad autonomy
  and minimal specification}

\qitem{X05c}{Optional comment}
\qopts{\qfree}

\subsubsection{Page 2: Task Allocation}

When developing software with the help of a coding agent, who should do each of the
following?

Many activities have a deciding part and a carrying-out part. Where an item names only
one of them, rate only that part; the other part is covered by other items.

\qscale{1 The human alone \qsep 2 Mainly the human, with help from the agent \qsep
  3 The human and the agent about equally \qsep 4 Mainly the agent, with help from the
  human \qsep 5 The agent alone \qsep N/A Not sure / not applicable}

\qitem{T01}{Deciding what a development task must achieve for its result to be acceptable.}
\qitem{T03}{Developing and critiquing the ideas in a specification document or
  implementation plan (not writing the document).}
\qitem{T04}{Building your own understanding of a problem domain and of what an
  unfamiliar codebase is for.}
\qitem{T05}{Tracing code paths and summarizing how parts of a codebase work.}
\qitem{T06}{Building throwaway prototypes to try out alternative approaches.}
\qitem{T07}{Judging which information (code, docs, tests, logs) matters for a task
  (not gathering it).}
\qitem{T09}{Writing and iteratively improving the prompts and task descriptions given
  to the agent.}
\qitem{T10}{Choosing which AI tools and agents to use for your work and how to combine
  them.}
\qitem{T11}{Designing (as opposed to setting up) execution environments, sandboxing,
  permissions, and safety boundaries.}
\qitem{T12}{Carrying out setup and configuration once the tools and environment have
  been chosen.}
\qitem{T13}{Splitting work into chunks suitable for a human to review.}
\qitem{T16}{Defining what counts as evidence that a change actually works (rather than
  producing that evidence).}
\qitem{T17}{Producing an implementation and checking it with tests, linters, or type
  checkers.}
\qitem{T18}{Reviewing a code change to flag missing areas, risks, or problems.}
\qitem{T20}{Deciding when and why a large refactoring or cleanup is needed.}
\qitem{T21}{Carrying out mechanical refactors/cleanup/updates/build-fixes.}
\qitem{T22}{Removing the small remaining imperfections in an almost-finished result.}
\qitem{T23}{Making the final judgment that a change is correct and ready to ship
  (beyond what tests show).}
\qitem{T24}{Deciding what counts as acceptable vs.\ unacceptable use of AI in software
  development.}
\qitem{T25}{select ``Mainly the agent, with help from the human'' for this item
  (attention check)}

\addvspace{\medskipamount}\noindent
Whatever mix of answers you gave above, there is a question of how, not only how much.
Consider these work styles:

DIVIDE: ``I want a clean division of labor: clear interfaces, each side doing what it's
best at, hand-offs I can inspect. That leaves me free while the agent works, I have to
judge the result.''

COLLAB: ``I and the agent collaborate, also when the agent performs the larger amount of
work. We are solving the task hand-in-hand in a manner none of us could do alone. That
costs me continuous attention, but leaves me with few misunderstandings and a good grasp
of the result.''

Which style would you aim at for the following two tasks?
Assume both are serious but not highly critical, you are using a coding agent for each,
and you will own the outcome.

\qscale{1 Clearly DIVIDE \qsep 2 Rather DIVIDE \qsep 3 Equally both \qsep 4 Rather
  COLLAB \qsep 5 Clearly COLLAB \qsep N/A Not sure}

\qitem{DC01}{Clear what is wanted and easy to see when it is right; a change of a
  familiar kind in code you know well.}
\qitem{DC02}{A part of the system you do not know well or subtleties in the
  requirements; mistakes would not be obvious, and you will be maintaining the result.}

\addvspace{\medskipamount}\noindent
When would you want which, and why?
Please explain your view: sketch your desired work style,
even if you do not yet achieve it in practice often; what stands in the way, if anything.

(A few sentences here would help us a lot---this is where we'll learn the most.)

\qitem{DCc}{How I would like to work together with an agent}
\qopts{\qfree}

\subsubsection{Page 3: Accountability}

How much do you agree with the following statements?

\qscale{the five-point agree/disagree scale of G01--G07 above}

\qitem{R01}{If code written by a coding agent turns out to be defective, the developer
  who used the agent should be held just as accountable as if they had written the code
  themselves.}
\qitem{R02}{If code written by a coding agent turns out to be defective, the reason can
  be AI tool error, human error, or both.}

\subsubsection{Page 4: Working Closely Together}
\label{app:method2-questionnaire-p4}

When two developers work on one task together (pair programming), the quality of their
exchange varies. Where it is high, each has a good grasp of what the other is currently
thinking, knows, and is aiming at: explanations stay short, misunderstandings are rare,
little has to be redone, and both contribute their best. Where it is low, the two talk
past each other, much has to be re-explained, more mistakes happen, and one of them may
participate much less.

We call these outcomes the \textbf{togetherness} of the pair.
Good pairings achieve high togetherness throughout most of the session.
Difficult pairings experience low togetherness frequently.

\qitem{P01}{How much pair programming (two developers on one task at the same time,
  sharing a screen in person or remotely) have you done?}
\qopts{None \qsep Tried it a few times \qsep Occasionally \qsep Regularly for some
  months \qsep Regularly for years}

\qitem{P02}{Thinking of the pair programming that worked best for you: how much
  togetherness did the two of you reach?}
\qopts{I have not pair programmed \qsep Low---we often talked past each other \qsep
  Fairly low---much had to be re-explained \qsep Moderate---it worked, with some
  friction \qsep High---we usually followed each other easily \qsep Very high---brief
  hints were enough, we rarely misunderstood each other}

\qitem{P03}{And at its best, how much togetherness do you reach with a coding agent?}
\qopts{I hardly use coding agents \qsep Low---we often talk past each other \qsep
  Fairly low---much has to be re-explained \qsep Moderate---it works, with some
  friction \qsep High---we usually follow each other easily \qsep Very high---brief
  hints are enough, we rarely misunderstand each other}

\addvspace{\medskipamount}\noindent
How much do you agree with the following statements about togetherness?

\qscale{the five-point agree/disagree scale of G01--G07 above}

\qitem{W01}{More togetherness with a coding agent would make my work substantially better.}
\qitem{W02}{Togetherness is something only people can have with each other; an AI cannot
  take part in it.}
\qitem{W03}{Staying that closely engaged with an agent would cost me more attention than
  it is worth.}

\subsubsection{Page 5: Demographics and General Remarks}

Who are you?

\qitem{D01}{What best describes your current primary role?}
\qopts{Software developer / engineer \qsep Senior developer / engineer \qsep Tech lead /
  staff engineer \qsep Engineering manager \qsep Architect \qsep DevOps / platform /
  infrastructure \qsep QA / test engineer \qsep Data / ML engineer \qsep Researcher /
  educator \qsep Student \qsep Other:}

\qitem{D02}{How many years of professional software development experience do you have?}
\qopts{Less than 1 \qsep 1--2 \qsep 3--5 \qsep 6--10 \qsep 11--15 \qsep 16--20 \qsep
  More than 20}

\qitem{D03}{What type of organization do you primarily work in?}
\qopts{Software product company \qsep IT / software consultancy \qsep Non-software
  company with in-house development \qsep Startup \qsep Public sector / government \qsep
  University / research institution \qsep Freelance / self-employed \qsep Not currently
  working in software \qsep Other:}

\qitem{D04}{What is the typical size of the software teams you work in?}
\qopts{I mostly work alone \qsep 2--5 \qsep 6--10 \qsep 11--25 \qsep 26--50 \qsep More
  than 50 \qsep Not applicable}

\qitem{D05}{Roughly what share of your work time is spent on directly code-related tasks?}
\qopts{0--10\% \qsep 11--25\% \qsep 26--50\% \qsep 51--75\% \qsep 76--90\% \qsep More
  than 90\%}

\qitem{D06}{Which gender do you most identify with?}
\qopts{male \qsep female \qsep Non-binary \qsep Prefer not to say}

\qitem{D07}{In which country do you currently work?}
\qopts{\qfree}

\qitem{Z01}{Anything not captured above you'd like to tell us about working with coding
  agents}
\qopts{\qfree}

\qitem{Z02}{Optional comments on the survey itself}
\qopts{\qfree}

\addvspace{\medskipamount}\noindent
Receiving results:

Enter your email address below if you want to receive the results of this survey as soon
as we have worked them out.
The address will be stored along with your responses, but we will remove it before we
process the responses.
Or drop us an email to \nolinkurl{prechelt@inf.fu-berlin.de} with subject ``Human/AI
survey: Please send me the results''.

\qitem{Z03}{Email address}
\qopts{\qfree}

\subsection{LinkedIn Posting Text}\label{app:step2-linkedin-msg}

\begin{quotation}
Is your way of working with coding agents mainstream or \ldots\ special?

I don't know either. That is why we are running a scientific survey:
``Software developers and coding agents: who does what?'' 15 minutes, anonymous.
Which parts of the work should the developer do and which the coding agent?
If you do not use coding agents at all, your answer is just as valuable.

Academic research from my group at Freie Universität Berlin; the results are headed into a journal article.

Leave your email address at the end and we will send you the results as soon as we have worked them out.

Do you like the survey? Then please repost! Recommendations like yours are the only way we can reach enough people.

((survey link here))
\end{quotation}

\subsection{Respondent Demographics}\label{app:method2-demographics}

See Table~\ref{tab:sample-background}.

\begin{table}
  \caption{\gentag{tab-sample-background.tex}Demographics data of the \svy{sample/eligible} respondents:
    Primary role, gender, years of professional experience, typical team size,
    type of organization, and fraction of work time typically spent on code-related tasks.
    Answers that are not a point on the scale
    (``not applicable'', ``prefer not to say'') are listed withour percentage.}
  \label{tab:sample-background}
  \centering\tablefont\setlength{\tabcolsep}{4pt}
\begin{tabular}{@{}lrrc@{\qquad}lrrc@{\qquad}lrrc@{}}
\toprule
Characteristic & $n$ & \% & percent & Characteristic & $n$ & \% & percent & Characteristic & $n$ & \% & percent \\
\midrule
\multicolumn{4}{c}{\textit{primary role (n=228)}} & \multicolumn{4}{c}{\textit{years of professional experience (n=227)}} & \multicolumn{4}{c}{\textit{type of organization (n=228)}} \\
developer / engineer & 37 & 16 & \percentagebar{16} & less than 1 & 25 & 11 & \percentagebar{11} & software product company & 67 & 29 & \percentagebar{29} \\
senior developer & 53 & 23 & \percentagebar{23} & 1-2 & 25 & 11 & \percentagebar{11} & IT / software consultancy & 29 & 13 & \percentagebar{13} \\
tech lead / staff & 33 & 14 & \percentagebar{14} & 3-5 & 29 & 13 & \percentagebar{13} & non-software, in-house & 34 & 15 & \percentagebar{15} \\
engineering manager & 11 & 5 & \percentagebar{5} & 6-10 & 51 & 22 & \percentagebar{22} & startup & 19 & 8 & \percentagebar{8} \\
architect & 14 & 6 & \percentagebar{6} & 11-15 & 30 & 13 & \percentagebar{13} & public sector & 8 & 4 & \percentagebar{4} \\
DevOps / platform & 3 & 1 & \percentagebar{1} & 16-20 & 14 & 6 & \percentagebar{6} & university / research & 41 & 18 & \percentagebar{18} \\
QA / test & 0 & 0 & \percentagebar{0} & more than 20 & 53 & 23 & \percentagebar{23} & freelance / self-employed & 15 & 7 & \percentagebar{7} \\
data / ML & 3 & 1 & \percentagebar{1} & \multicolumn{4}{c}{\textit{typical team size (n=226)}} & not working in software & 6 & 3 & \percentagebar{3} \\
researcher / educator & 23 & 10 & \percentagebar{10} & works alone & 48 & 21 & \percentagebar{21} & other & 9 & 4 & \percentagebar{4} \\
student & 25 & 11 & \percentagebar{11} & 2-5 & 82 & 36 & \percentagebar{36} & \multicolumn{4}{c}{\textit{share of work time spent on code (n=228)}} \\
other & 26 & 11 & \percentagebar{11} & 6-10 & 66 & 29 & \percentagebar{29} & 0-10\% & 12 & 5 & \percentagebar{5} \\
\multicolumn{4}{c}{\textit{gender (n=216)}} & 11-25 & 13 & 6 & \percentagebar{6} & 11-25\% & 31 & 14 & \percentagebar{14} \\
male & 187 & 87 & \percentagebar{87} & 26-50 & 5 & 2 & \percentagebar{2} & 26-50\% & 53 & 23 & \percentagebar{23} \\
female & 21 & 10 & \percentagebar{10} & more than 50 & 12 & 5 & \percentagebar{5} & 51-75\% & 72 & 32 & \percentagebar{32} \\
non-binary & 8 & 4 & \percentagebar{4} & not applicable & 2 &  &  & 76-90\% & 45 & 20 & \percentagebar{20} \\
prefer not to say & 11 &  &  &  &  &  &  & more than 90\% & 15 & 7 & \percentagebar{7} \\
\bottomrule
\end{tabular}

\end{table}

\subsection{Agent and Other AI Tool Use}

\begin{table}
  \caption{\gentag{tab-sample-tools.tex}Which types of AI tools respondents use. Multiple answers were allowed.
    \textsuperscript{\dag}~marks the entries required for being considered an agent user (versus non-user).
    The bottom \svy{sample/tools/categories} entries were formed by
    grouping free-text \emph{Other} responses.}
  \label{tab:sample-tools}
  \centering\tablefont\setlength{\tabcolsep}{4pt}
\begin{tabular}{@{}lrrc@{}}
\toprule
Characteristic & $n$ & \% & percent \\
\midrule
\multicolumn{4}{c}{\textit{AI tools used (n=241, multiple answers)}} \\
chat assistant & 160 & 66 & \percentagebar{66} \\
IDE plugin autocomplete & 78 & 32 & \percentagebar{32} \\
IDE plugin autoedit mode & 36 & 15 & \percentagebar{15} \\
IDE plugin agent mode\textsuperscript{\dag} & 67 & 28 & \percentagebar{28} \\
AI-native IDE agent mode\textsuperscript{\dag} & 30 & 12 & \percentagebar{12} \\
terminal coding agent\textsuperscript{\dag} & 143 & 59 & \percentagebar{59} \\
GitHub Copilot on GitHub\textsuperscript{\dag} & 35 & 15 & \percentagebar{15} \\
self-hosted LLM & 65 & 27 & \percentagebar{27} \\
\multicolumn{4}{@{}l}{\quad\textit{includes free-text answers}} \\
self-hosted agent harness\textsuperscript{\dag} & 4 & 2 & \percentagebar{2} \\
desktop agent app\textsuperscript{\dag} & 2 & 1 & \percentagebar{1} \\
other non-agent tool & 2 & 1 & \percentagebar{1} \\
CI agent\textsuperscript{\dag} & 1 & $<$1 & \percentagebar{0.5} \\
\midrule
none of these & 15 & 6 & \percentagebar{6} \\
\bottomrule
\end{tabular}

\end{table}

With respect to Table~\ref{tab:sample-agentuse} in the main text,
for \svy{exposure/contradict/high} respondents their two answers reflected in the table
cannot both be right:
they report using agents at least several times a week in X02, but ticked no agent tool in X03.
We think the likeliest explanation is that they read X02 as a continuation of X01, which asks
about AI tools in general (not just coding agents) and sits directly above it,
so we count these as non-users.
Conversely, \svy{exposure/contradict/low} respondents tick an agent tool in X03,
but reporting that they have never used one or tried it only once or twice in X02.
We count these as non-users as well.
Effectively, we treat X03 as the reliable signal of \emph{what kind} of delegation a
respondent practices and X02 as the reliable signal of \emph{how much}, and combine them
by taking, for each respondent, the lower of what the two answers support.

\subsection{Accountability}\label{app:result2-acc}

See Table~\ref{tab:acc} for the precise data for users and non-users.
See Table~\ref{tab:items-compare-country-acc} for the German/rest-of-the-world comparison
for assessing generalizability.

\begin{table}
  \caption{\gentag{tab-acc.tex}Accountability responses from users and non-users. 
    Columns as in Table~\ref{tab:items-users}, but on the agree scale:
    \likertlegendagree.
    Region and the bar therefore read disagree\,/\,neither\,/\,agree.}
  \label{tab:acc}
  \centering\tablefont\setlength{\tabcolsep}{3.5pt}
\begin{tabular}{@{}llclclrrrr@{}}
\toprule
Item & Statement & P & V & Distribution & $M$ [95\,\% CI] & $n$ & \multicolumn{3}{c}{Region (\%)} \\
\cmidrule(lr){8-10}
 & & & & {\scriptsize disagree\,$\leftrightarrow$\,agree} & & & D & N & A \\
\midrule
\addlinespace\multicolumn{10}{@{}l}{\textbf{Users (n = 171)}} \\
R01 & developer as accountable as if they wrote it & Agree & $+$ & \likertbar{1}{1}{8}{29}{61}{0} & 4.47 [4.35, 4.59] & 162 & 2 & 8 & 90 \\
R02 & failure cause can be AI error, human error, or both & Agree & $+$ & \likertbar{2}{3}{9}{35}{51}{0} & 4.28 [4.13, 4.42] & 162 & 5 & 9 & 86 \\
\addlinespace\multicolumn{10}{@{}l}{\textbf{Non-users (n = 69)}} \\
R01 & developer as accountable as if they wrote it & Agree & $+$ & \likertbar{2}{1}{3}{18}{76}{0} & 4.66 [4.47, 4.84] & 67 & 3 & 3 & 94 \\
R02 & failure cause can be AI error, human error, or both & Agree & $+$ & \likertbar{10}{3}{10}{36}{41}{0} & 3.97 [3.66, 4.28] & 63 & 13 & 10 & 77 \\
\bottomrule
\end{tabular}

\end{table}

\begin{table}
	\caption{\gentag{tab-items-compare-country-acc.tex}
		Answers of German and non-German respondents regarding our accountability questions.
		The columns match the ones in Table~\ref{tab:items-compare-country}.}
	\label{tab:items-compare-country-acc}
	\centering\tablefont\setlength{\tabcolsep}{3.5pt}
\begin{tabular}{@{}llccrrcrrl@{}}
\toprule
Item & Statement & P & \multicolumn{3}{c}{Germany} & \multicolumn{3}{c}{elsewhere} & $\Delta M$ [95\,\% CI] \\
\cmidrule(lr){4-6}\cmidrule(lr){7-9}
 & & & O & $M$ & $n$ & O & $M$ & $n$ & \\
\midrule
R01 & developer as accountable as if they wrote it & Agree & \pred{A} & 4.53 & 122 & \pred{A} & 4.47 & 70 & 0.06 [-0.19, 0.31] \\
R02 & failure cause can be AI error, human error, or both & Agree & \pred{A} & 4.18 & 119 & \pred{A} & 4.16 & 69 & 0.02 [-0.31, 0.34] \\
\bottomrule
\end{tabular}

\end{table}

\subsection{Delegation Comfort}\label{app:result2-delegation}

See Figure~\ref{fig:deleg-tiers}.

\begin{figure}
  \caption{\gentag{fig-deleg-tiers.tex}Every cell of the X04\,$\times$\,X05 grid, for the \svy{deleg/n}
    respondents who answered both items.
    Bubble \emph{area} is the number of
    respondents in that cell and the wedges are their exposure tiers:
    \likertswatch{delegtier0}~\svy{exposure/e0/label},
    \likertswatch{delegtier1}~\svy{exposure/e1/label},
    \likertswatch{delegtier2}~\svy{exposure/e2/label},
    \likertswatch{delegtier3}~\svy{exposure/e3/label}. 
    The two blue tiers are the non-users,
    the two orange ones are the users.}
  \label{fig:deleg-tiers}
  \centering
\begin{tikzpicture}
\begin{axis}[delegchart,
  xtick={1,2,3,4,5},
  xticklabels={{none},{mechanical},{modest coding},{substantial coding},{judgment work}},
  ytick={1,2,3,4,5},
  yticklabels={{none},{small, precise},{moderate},{large},{open-ended}}]
\addplot[delegdiagonal, forget plot] coordinates {(1,1) (5,5)};
\delegwedge{1}{1}{15}{0}{11}{0}
\delegwedge{1}{1}{15}{11}{14}{1}
\delegwedge{1}{1}{15}{14}{15}{2}
\delegwedge{1}{2}{7}{0}{2}{0}
\delegwedge{1}{2}{7}{2}{5}{1}
\delegwedge{1}{2}{7}{5}{6}{2}
\delegwedge{1}{2}{7}{6}{7}{3}
\delegwedge{2}{1}{4}{0}{2}{1}
\delegwedge{2}{1}{4}{2}{4}{2}
\delegdisc{2}{2}{6}{1}
\delegdisc{2}{3}{1}{3}
\delegwedge{3}{1}{6}{0}{5}{1}
\delegwedge{3}{1}{6}{5}{6}{2}
\delegwedge{3}{2}{24}{0}{11}{1}
\delegwedge{3}{2}{24}{11}{22}{2}
\delegwedge{3}{2}{24}{22}{24}{3}
\delegwedge{3}{3}{22}{0}{6}{1}
\delegwedge{3}{3}{22}{6}{19}{2}
\delegwedge{3}{3}{22}{19}{22}{3}
\delegdisc{3}{4}{2}{3}
\delegwedge{4}{2}{14}{0}{2}{1}
\delegwedge{4}{2}{14}{2}{8}{2}
\delegwedge{4}{2}{14}{8}{14}{3}
\delegwedge{4}{3}{56}{0}{4}{1}
\delegwedge{4}{3}{56}{4}{25}{2}
\delegwedge{4}{3}{56}{25}{56}{3}
\delegwedge{4}{4}{28}{0}{3}{1}
\delegwedge{4}{4}{28}{3}{15}{2}
\delegwedge{4}{4}{28}{15}{28}{3}
\delegwedge{4}{5}{10}{0}{1}{2}
\delegwedge{4}{5}{10}{1}{10}{3}
\delegdisc{5}{2}{1}{2}
\delegwedge{5}{3}{4}{0}{1}{2}
\delegwedge{5}{3}{4}{1}{4}{3}
\delegwedge{5}{4}{12}{0}{2}{1}
\delegwedge{5}{4}{12}{2}{6}{2}
\delegwedge{5}{4}{12}{6}{12}{3}
\delegdisc{5}{5}{17}{3}
\end{axis}
\end{tikzpicture}
\end{figure}

\section{Step 3: Notions of Collaboration Details}\label{app:step3}

\subsection{\emph{codebook.py} help text}\label{app:method3-codebook-py}

\bgroup\footnotesize
\begin{verbatim}
codebook.py -- database tool for qualitative open coding.
Usage: python3 eval/codebook.py COMMAND [ARGUMENT...]
Specified by eval/prompt.md and .claude/skills/open-coding/SKILL.md.

The environment variable CODEBOOK_BASEPATH must point to the dataset, e.g.
  export CODEBOOK_BASEPATH=data/surveydata_freetexts
which uses data/surveydata_freetexts{.yaml,_codebook.yaml,_memos.yaml,.db,.log}.
Commands shown with "<<'EOF'" read a text body from stdin (a here-document).
Never call this tool concurrently: it locks the DB and fails rather than waits.

Commands:
block-delete-code QUESTIONCODE BLOCKNUMBER CURRENTCODE
    Remove the application of CURRENTCODE from that block.
block-read-by-code [-r] QUESTIONCODE CODE
    Print all blocks using CODE (for constant comparison), as a YAML array.
block-read-by-index [-r] QUESTIONCODE STARTBLOCK NUMBLOCKS
    Print NUMBLOCKS blocks starting at STARTBLOCK, as a YAML array.
block-set-code QUESTIONCODE BLOCKNUMBER CODENAME [SUBSTRINGSPEC]
    Apply CODENAME to that block, to the SUBSTRINGSPEC range or else to all of it.
block-set-memo QUESTIONCODE BLOCKNUMBER  <<'EOF'
    Set that block's memo to the text read from stdin (empty text deletes it).
cb-add-questioncode QUESTIONCODE CODENAME
    Make an existing code of another questioncode available for QUESTIONCODE, too.
cb-create QUESTIONCODE NEWCODE_NAME  <<'EOF'
    Create a new code for QUESTIONCODE with the definition read from stdin.
cb-delete QUESTIONCODE UNUSEDCODE_NAME
    Remove an unused code from QUESTIONCODE's codebook (and, if unused elsewhere, entirely).
cb-list-codes QUESTIONCODE
    Print the names of all codes of QUESTIONCODE's codebook, one per line.
cb-read-code QUESTIONCODE CODENAME...
    Print the full codebook entries of those codes, as a YAML array.
cb-update-definition CODE  <<'EOF'
    Replace that code's definition by the text read from stdin.
cb-update-name OLDNAME NEWNAME
    Rename a code (for all questioncodes it belongs to).
create-db
    Create the DB from the three YAML files. Fails if the DB exists already.
dump-db [-f]
    Write the three YAML files from the DB, then rename the DB to *.db-inactive.
    -f overwrites them even if they were changed manually (losing those changes).
memo-create QUESTIONCODE TITLE
    Create an empty tactical memo with that (quoted) title; prints its number.
memo-delete QUESTIONCODE MEMONUMBER
    Delete that tactical memo.
memo-list QUESTIONCODE
    Print number and title of each tactical memo of QUESTIONCODE, one per line.
memo-read QUESTIONCODE MEMONUMBER...
    Print those tactical memos in full, as a YAML array.
memo-set-body QUESTIONCODE MEMONUMBER  <<'EOF'
    Set that memo's body to the text read from stdin (start it with a MIME type header).
memo-update-title QUESTIONCODE MEMONUMBER NEWTITLE
    Replace that memo's title.
merge-yaml
    Merge manual changes of the three YAML files with the DB contents, then
    leave state DB (so that `create-db` can validate the merged files).
overview QUESTIONCODE
    Print the status report: codes with usage counts, coding progress, memo count.
sql-query [MYQUERY]
    Run a read-only SQL query given on the commandline or on stdin.
write-log TITLE  <<'EOF'
    Append a multiline log entry: the TITLE line plus the indented text from stdin.
\end{verbatim}
\egroup

\subsection{List of memo titles}\label{app:method3-memos-list}

The eventual list of memo titles at the end of the analysis was this:
\begin{enumerate}
  \item Current work status; rules and conventions
  \item Near-term goals
  \item Medium-term goals
  \item Working method: the batched review document
  \item Reusable SQL queries
  \item Conditions on which respondents disagree about the direction
  \item Final outcomes from the analysis
\end{enumerate}

\subsection{Example prompts}\label{app:method3-example-prompts}

Here are two examples of the prompts used in the seven skill calls.
Understanding their content will require reading the results section first.
Prompt for call number 1:
\begin{quotation}
Look at the existing codings of the \Code{col1*}, \Code{col4*}, and \Code{col6*} codes.\\
1. Flag those you think are wrong and explain why.\\
2. Refine the definitions of those codes such that they explain the phenomenon more concretely.\\
If you see several possibilities how to do this for a given code, discuss them with me.
\end{quotation}

Prompt for call number 6:
\begin{quotation}
Let's tackle the DIV4H hypothesis.
Let's first discuss the approach.
My suggestion:

We are looking for a small number of new codes that form a new code family which talks about the holisticness of a block.
Guiding question: If we feel this is a holistic block, talking about an entire process idea,
what are the well-defined and recognizable features that make the block a holistic block?

We initially look at a fraction of the blocks that bear both divide and collab codes, some also having a preference-* code, some not.
We hypothesize such qualities (code ideas).
We then look at another fraction of such blocks for validating these ideas.
If they do not hold up, we iterate.
If they hold up, we look at blocks having collab codes only: They should usually not have these qualities.

If you look at the memos and codes only, not any blocks yet: What do you think of this procedure? Is it likely to work?
What modification do you suggest? Why do you think it is going to be helpful? (I ask this in order to understand how you work, not so much to understand the data.)

Let's discuss this analysis approach:\\
- Does the starting point, "holisticness concepts", make sense to you?\\
- Does such guidance on methodological approach help you or hamper you? Why?\\
- Formulate what YOU expect to be the best approach you'd likely use and why.
  Or can you not do this in advance because you develop your approach only as you go, with a very short planning horizon?

Then do whatever comes out and present me with an analysis document as usual.
\end{quotation}

\subsection{Example Concept Definition}\label{app:method3-codedef-example}

Most definitions of the individual codes have considerable length,
so the complete set is available in the data package only,
but for illustration, we present a single example here.
The definition of \Code{collab-by-humanauthorship}:

\bgroup\itshape
    COLLAB in which the human is the author of the code and the agent answers, suggests, reviews,
    researches, or drafts snippets.\\
    Judge by authorship, not by who technically writes bytes: where the human only gates what the agent
    produced, authorship is the agent's and collab-by-approval applies instead.\\
    Requires clear evidence -- an explicit authorship claim such as ``I'm at the keyboard'', 
    ``I write the driver'', ``me doing most of the actual programming''. 
    Where both sides contribute substantially to the code, 
    this is an ordinary collaboration and the code does not apply.\\
    The agent is heavily involved, so this is not solo-work.
\egroup

\subsection{List of \Code{collab-*} Codes}\label{app:result3-collab-codes}

There will be one subsection per code family below, all using the same notation:
A list of the code names, with number of instances and character length of definition 
in parentheses behind it.

The \Code{collab-by-*} codes each describe a means of collaboration:
\begin{verbatim}
collab-by-approval (13; 415)
collab-by-humanauthority (16; 512)
collab-by-humanauthorship (8; 628)
collab-by-interview (24; 400)
collab-by-iteration (11; 251)
collab-by-partitioning (6; 343)
collab-by-speech (1; 71)
collab-by-tradeoffdialog (6; 334)
\end{verbatim}

The \Code{collab-if-*} codes each describe a condition our respondents mentioned for when they
will use COLLAB mode:
\begin{verbatim}
collab-if-agentunreliable (16; 511)
collab-if-beyondagent (6; 457)
collab-if-errorprone (7; 494)
collab-if-goodcodebase (1; 440)
collab-if-unfamiliar (22; 450)
\end{verbatim}

The \Code{collab-for-*} codes each describe a goal of using COLLAB mode or justification for using it:
\begin{verbatim}
collab-for-codeunderstand (37; 502)
collab-for-exercise (2; 467)
collab-for-learning (16; 319)
collab-for-lowrisk (13; 376)
collab-for-ownership (12; 609)
collab-for-requnderstand (7; 257)
collab-for-split (12; 252)
collab-for-steering (14; 358)
collab-for-togetherness (10; 552)
\end{verbatim}

\subsection{List of \Code{divide-*} Codes}\label{app:result3-divide-codes}

Three families, parallel to the \Code{collab-by-*} codes.
\Code{divide-by-*} codes each describe a means of collaboration:
\begin{verbatim}
divide-by-agentreview (20; 308)
divide-by-guardrails (9; 587)
divide-by-humanauthority (1; 398)
divide-by-orchestration (8; 1590)
divide-by-phase (24; 1381)
divide-by-spec (20; 496)
divide-by-strengthallocation (8; 437)
\end{verbatim}

The \Code{divide-if-*} codes each describe a condition our respondents mentioned for when they
will use DIVIDE mode:
\begin{verbatim}
divide-if-answerable (3; 842)
divide-if-checkable (17; 306)
divide-if-clear (23; 443)
divide-if-lowstakes (10; 319)
divide-if-reviewable (11; 1000)
\end{verbatim}

The residual code \Code{divide-plain} marks bare delegation statements
with no condition, means, or goal attached:
\begin{verbatim}
divide-plain (6; 1138)
\end{verbatim}

The \Code{divide-for-*} codes each describe a goal of using DIVIDE mode or justification for using it:
\begin{verbatim}
divide-for-nondrudgery (2; 967)
divide-for-parallel (9; 462)
divide-for-probe (5; 386)
divide-for-speed (9; 307)
\end{verbatim}

\subsection{List of \Code{constraint-*} Codes}\label{app:result3-constraints}

Codes expressing external influences on the ideal expressed in the response
or on its realization:
\begin{verbatim}
constraint-agentobstacle (9; 788)
constraint-agentsgetbetter (4; 802)
constraint-selfdiscipline (4; 362)
constraint-socialpressure (3; 781)
constraint-unsettledstyle (14; 1051)
\end{verbatim}

\subsection{List of Odd Codes}

Codes for statements that do not fit into the frame of our analysis:
\begin{verbatim}
gg-offtopic (5; 593)
gg-refusal (8; 571)
solo-for-exercise (2; 652)
solo-work (7; 491)
\end{verbatim}

\subsection{List of \Code{preference-*} Codes}\label{app:result3-preferences}

Codes expressing personal preferences, i.e., statements in which something personal remains
once the wish-wording is removed (see Section~\ref{sec:results3-preference-codes}):
\begin{verbatim}
preference-directing (2; 1118)
preference-disengagement (3; 1128)
preference-keepskill (1; 808)
preference-making (2; 872)
preference-worksatisfaction (6; 1024)
\end{verbatim}

\subsection{List of \Code{holistic-*} Codes}\label{app:result3-holistic}

Codes expressing a holistic view of what our question is talking about rather than
a local, process-step-oriented view:
\begin{verbatim}
holistic-fraction (5; 1102)
holistic-process (14; 1518)
holistic-roleshift (6; 1078)
\end{verbatim}

\ifdraftmode
\clearpage
\tableofcontents
\fi

\end{document}